\documentclass[acmsmall]{acmart}

\usepackage{booktabs} 
\usepackage{graphicx}
\usepackage{verbatim}
\usepackage{amsmath}
\usepackage{algpseudocode}
\usepackage{algorithmicx}
\usepackage{tabularx}
\usepackage{amssymb}
\usepackage{multirow}
\usepackage{hhline}
\usepackage{color}
\usepackage{listings}
\usepackage{subcaption}
\usepackage{varwidth}
\usepackage[capitalise]{cleveref}

\usepackage[nolist]{acronym}
\begin{acronym}
	\acro{AAA}{Authentication, Authorization and Accounting}
	\acro{AES}{Advanced Encryption Standard}
	\acro{AP}{Access Point}
	\acro{API}{Application Programming Interface}
	\acro{ASR}{Anonymized Spatial Region}
	\acro{BF}{Bloom Filter}
	\acro{BP}{Baseline Pseudonym}
	\acro{BS}{Base Station} 
    \acro{BSM}{Basic Safety Message}
    \acro{BYOD}{Bring Your Own Device}
	\acro{C2C-CC}{Car2Car Communication Consortium}
	\acro{CA}{Certification Authority}
	\acrodefplural{CA}{Certification Authorities}
	\acro{CBF}{Counting Bloom Filter}
	\acro{CN}{Common Name}
	\acro{CAM}{Cooperative Awareness Message}
	\acro{CIA}{Confidentiality, Integrity and Availability}
	\acro{CRL}{Certificate Revocation List}
	\acro{CDN}{Content Delivery Network}
	\acro{CSR}{Certificate Signing Requests}
	\acro{DAA}{Direct Anonymous Attestation}
	\acro{DDoS}{Distributed Denial of Service}
	\acro{DDH}{Decisional Diffie-Helman}
	\acro{DENM}{Decentralized Environmental Notification Message}
	\acro{DHT}{Distributed Hash Table}
	\acro{DoS}{Denial of Service}
	\acro{DPA}{Data Protection Agency}
	\acro{D2D}{device-to-device}
	\acro{EC}{Elliptic Curve}
	\acro{ECC}{Elliptic Curve Cryptography}
	\acro{ECDSA}{Elliptic Curve Digital Signature Algorithm}
	\acro{ECIES}{Elliptic Curve Integrated Encryption Scheme}
	\acro{ECU}{Electronic Control Unit}
	\acro{ETSI}{European Telecommunications Standards Institute}
	\acro{EVITA}{E-safety Vehicle Intrusion protected Applications}
	\acro{FCFS}{First-Come, First-Served}
	\acro{FOT}{Field Operational Test}
	\acro{FPGA}{Field-Programmable Gate Array}
	\acro{GN}{GeoNetworking}
	\acro{GS}{Group Signatures}
	\acro{GM}{Group Manager}
	\acro{GBA}{Generic Bootstrapping Architecture}
	\acro{GPA}{Global Passive Adversary}
	\acro{GUI}{Graphic User Interface}
	\acro{HBC}{Honest-but-Curious}
	\acro{HP}{Hybrid Pseudonym}
	\acro{HSM}{Hardware Security Module}
	\acro{HTTP}{Hypertext Transfer Protocol}
	\acro{IEEE}{Institute of Electrical and Electronics Engineers}
	\acro{IoT}{Internet of Things}
	\acro{ITS}{Intelligent Transport Systems}
	\acro{IT}{Information Technologies}
	\acro{IMSI}{International Mobile Subscriber Identity}
	\acro{IMEI}{International Mobile Station Equipment Identity}
	\acro{IdP}{Identity Provider}
	\acro{ISP}{Internet Service Provider}
	\acro{LBS}{Location-based Service}
	\acro{LDAP}{Lightweight Directory Access Protocol}
	\acro{LEA}{Law Enforcement Agency}
	\acro{LTC}{Long-Term Certificate}
	\acro{LTCA}{Long-Term \acl{CA}}
	\acrodefplural{LTCA}{Long-Term \aclp{CA}}
	\acro{LTE}{Long Term Evolution}
	\acro{LuST}{Luxembourg SUMO Traffic}
	\acro{MAC}{Message Authentication Code}
	\acro{MCA}{Message \ac{CA}}
	\acro{MEA}{Misbehavior Evaluation Authority}
	\acro{OBU}{On-board Unit}
	\acro{OCSP}{Online Certificate Status Protocol}
	\acro{OSN}{Online Social Network}
	\acro{PC}{Pseudonymous Certificate}
	\acro{PCA}{Pseudonymous \acl{CA}}
	\acrodefplural{PCA}{Pseudonymous \aclp{CA}}
	\acro{PDP}{Policy Decision Point}
	\acro{PEP}{Policy Enforcement Point}
	\acro{PIR}{Private Information Retrieval}
	\acro{PKI}{Public-Key Infrastructure}
	\acro{POI}{Point of Interest}
	\acro{PRECIOSA}{Privacy Enabled Capability in Co-operative Systems and Safety Applications}
	\acro{PRESERVE}{Preparing Secure Vehicle-to-X Communication Systems}
	\acro{PRL}{Pseudonym Revocation List}
	\acro{P2P}{peer-to-peer}
	\acro{RA}{Resolution Authority}
	\acro{REST}{Representational State Transfer}
	\acro{RBAC}{Role Based Access Control}
	\acro{RCA}{Root \acl{CA}}
	\acro{RSU}{Roadside Unit}
	\acro{SAML}{Security Assertion Markup Language}
	\acro{SCORE@F}{Système COopératif Routier Expérimental Français}
	\acro{SeVeCom}{Secure Vehicle Communication}
	\acro{SIS}{Security Infrastructure Server}
	\acro{SP}{Service Provider}
	\acro{SSO}{Single-Sign-On}
	\acro{SoA}{Service-oriented-Approach}
	\acro{SOAP}{Simple Object Access Protocol}
	\acro{SAS}{Sample Aggregation Service}
	\acro{TS}{Task Service}
	\acro{TLS}{Transport Layer Security}
	\acro{TPM}{Trusted Platform Module}
	\acro{TTP}{Trusted Third Party}
	\acro{TVR}{Ticket Validation Repository}
	\acro{URI}{Uniform Resource Identifier}
	\acro{VANET}{Vehicular Ad-hoc Network}
	\acro{V2I}{Vehicle-to-Infrastructure}
	\acro{V2V}{Vehicle-to-Vehicle}
	\acro{V2X}{\ac{V2V} and/or \ac{V2I}}
	\acro{VC}{Vehicular Communication}
	\acro{VM}{Virtual Machine}
	\acro{VSS}{\ac{VC} Security Subsystem}
	\acro{WAVE}{Wireless Access in Vehicular Environments}
	\acro{WSDL}{Web Services Discovery Language}
	\acro{W3C}{World Wide Web Consortium}
	\acro{VANET}{Vehicular Ad-hoc Network}
	\acro{VPKI}{Vehicular Public-Key Infrastructure}
	\acro{WS}{Web Service}
	\acro{WoT}{Web of Trust}
	\acro{WSACA}{\ac{WAVE} Service Advertisement \ac{CA}}
	\acro{XML}{Extensible Markup Language}
	\acro{XACML}{eXtensible Access Control Markup Language}
	\acro{3G}{3rd Generation}
\end{acronym}

\usepackage{algorithm}
\floatname{algorithm}{Protocol}
\PassOptionsToPackage{noend}{algpseudocode}
\usepackage{algpseudocode}

\errorcontextlines\maxdimen

\makeatletter
\newcommand*{\algrule}[1][\algorithmicindent]{\makebox[#1][l]{\hspace*{.5em}\vrule height .75\baselineskip depth .5\baselineskip}}%

\newcount\ALG@printindent@tempcnta
\def\ALG@printindent{%
	\ifnum \theALG@nested>0
	\ifx\ALG@text\ALG@x@notext
	\addvspace{-3pt}
	\else
	\unskip
	\ALG@printindent@tempcnta=1
	\loop
	\algrule[\csname ALG@ind@\the\ALG@printindent@tempcnta\endcsname]%
	\advance \ALG@printindent@tempcnta 1
	\ifnum \ALG@printindent@tempcnta<\numexpr\theALG@nested+1\relax
	\repeat
	\fi
	\fi
}%
\usepackage{etoolbox}
\patchcmd{\ALG@doentity}{\noindent\hskip\ALG@tlm}{\ALG@printindent}{}{\errmessage{failed to patch}}
\makeatother

\algnewcommand\algorithmicdata{\textbf{Data:}}
\algnewcommand\algorithmicresult{\textbf{Result:}}

\algnewcommand\Data{\item[\algorithmicdata]}%
\algnewcommand\Result{\item[\algorithmicresult]}%

\def\BibTeX{{\rm B\kern-.05em{\sc i\kern-.025em b}\kern-.08emT\kern-.1667em\lower.7ex\hbox{E}\kern-.125emX}}
    
\renewcommand\footnotetextcopyrightpermission[1]{} 
\makeatletter
\let\@authorsaddresses\@empty
\makeatother

\fancypagestyle{plain}{
	\fancyfoot[C]{\thepage}
}

\begin{document}

%
\title{Resilient Privacy Protection for Location-based Services through Decentralization}  

%
\author{Hongyu Jin}
\orcid{0000-0003-2022-3976}
\affiliation{%
	\institution{Networked Systems Security Group, KTH Royal Institute of Technology}
	\streetaddress{Kistag{\aa}ngen 16}
	\city{Stockholm}
	\postcode{16440}
	\country{Sweden}}
\email{hongyuj@kth.se}

\author{Panos Papadimitratos}
\orcid{0000-0002-3267-5374}
\affiliation{%
	\institution{Networked Systems Security Group, KTH Royal Institute of Technology, and RISE SICS}
	\streetaddress{Kistag{\aa}ngen 16}
	\city{Stockholm}
	\postcode{16440}
	\country{Sweden}}
\email{papadim@kth.se}

%
\renewcommand{\shortauthors}{H. Jin and P. Papadimitratos}

%
\begin{abstract}
	\acp{LBS} provide valuable services, with convenient features for mobile users. However, the location and other information disclosed through each query to the \ac{LBS} erodes user privacy. This is a concern especially because \ac{LBS} providers can be \emph{honest-but-curious}, collecting queries and tracking users' whereabouts and infer sensitive user data. This motivated both \emph{centralized} and \emph{decentralized} location privacy protection schemes for \acp{LBS}: anonymizing and obfuscating \ac{LBS} queries to not disclose exact information, while still getting useful responses. Decentralized schemes overcome disadvantages of centralized schemes, eliminating anonymizers, and enhancing users' control over sensitive information. However, an insecure decentralized system could create serious risks beyond private information leakage. More so, attacking an improperly designed decentralized \ac{LBS} privacy protection scheme could be an effective and low-cost step to breach user privacy. We address exactly this problem, by proposing security enhancements for mobile data sharing systems. We protect user privacy while preserving accountability of user activities, leveraging pseudonymous authentication with mainstream cryptography. We show our scheme can be deployed with off-the-shelf devices based on an experimental evaluation of an implementation in a static automotive testbed.
\end{abstract}

%

%

%
\maketitle

\section{Introduction}
\label{sec:introduction}

A \acf{LBS} query targets a location/region and expresses one or more specific user interests; the \ac{LBS} server responds with the most up-to-date relevant information, e.g., the latest menu of a restaurant, movies at a cinema, remaining parking slots at a shopping mall, or traffic conditions in the area. During this process, users' current or future whereabouts and interests are disclosed to the \ac{LBS} server through their queries. All submitted information is deemed necessary to best serve users, and the \ac{LBS} server is entrusted with rich data. However, many studies~\cite{barkhuus2003location,fogel2009internet,yun2013understanding} reveal that service providers can be honest-but-curious, aggressively collecting information to profile users, e.g., identifying home or working places or inferring interests for commercial purposes.

\ac{LBS} privacy is studied extensively. Location $k$-anonymity~\cite{gedik2008protecting} ensures that at least $k-1$ other users are involved in an obfuscated region, $\mathcal{R}$, used as the querier's location. Therefore, even in the presence of a local observer in $\mathcal{R}$, the query cannot be linked to a certain user; the \ac{LBS} server only learns the querier is one among $k$ users in $\mathcal{R}$. Protection can be achieved by centralized schemes~\cite{gedik2008protecting,mascetti2007spatial,mokbel2006new} that introduce an anonymizer, a proxy between users and the \ac{LBS} server, which anonymizes user queries before sending them to the \ac{LBS} server. However, the assumed anonymizer trustworthiness merely ``shifts'' the trust from the \ac{LBS} server to the anonymizer, holding rich information the same way that the \ac{LBS} server would. Simply put, an anonymizer could be itself honest-but-curious.

Decentralized approaches~\cite{ghinita2007mobihide,amoeba07} can eliminate the need for an anonymizer and protect user privacy in a collaborative manner: for example, form an obfuscated area with $k$ users within each other's communication range~\cite{ghinita2007mobihide}. However, if such $k$ users are too close, e.g., in a church, a shopping mall, or a cinema, such symbolic ``addresses'' can still be disclosed. Thus, it is hard to define how large $k$ should be to ensure an appropriate level of protection.

An alternative collaborative privacy protection approach is to pass/share \ac{LBS}-obtained information among users, to decrease exposure to the \ac{LBS} server~\cite{shokri2014hiding,jin2015resilient}. The sharing approach requires nodes to cache information received from the \ac{LBS} and pass it to neighbors when requested. Moreover, sharing (and thus reduced exposure to the \ac{LBS} server) is orthogonal to location obfuscation; the two could complement each other.

Nonetheless, opening up the system functionality is a double-edged sword: sharing reduces user exposure to the curious provider (\ac{LBS} or anonymizer) but it also exposes her to possibly faulty, misbehaving or curious peers. In fact, studies on \ac{P2P} systems~\cite{johnson2008evolution,kwok2002peer,zhou2005first} show that insecure decentralized schemes face serious problems: sensitive information could be exposed to peers and malicious nodes could pass on bogus data.

Signed \ac{LBS} server responses can be self-verifiable when passed to peers~\cite{shokri2014hiding} and partially address the second above-mentioned concern regarding malicious nodes. However, queries and cached information from different users could be diverse, making it necessary to share one or multiple complete \ac{LBS}-obtained responses (each with a signature attached), even though only a subset of each \ac{LBS}-obtained response might be needed by the querying peer. For example, the query could be \emph{What is the menu of restaurant A?} and \emph{How many parking spaces are available at B?} while another node could query \emph{What are the opening hours of restaurant A?} and \emph{What stores are in shopping mall B?} It is not straightforward to decide whether the peer responses include information as if the the response were directly obtained from the \ac{LBS} server, thus, guaranteeing quality of service. Even though nodes can assess the completeness of gathered peer responses~\cite{mahin2017crowd} and wait until the required completeness is met, peer responses equivalent to an \ac{LBS} response can be hard to achieve; and, in any case, incur significant delays and overhead, i.e., redundant cryptographic fields and information. Last but not least, peer queries, openly submitted to nodes' neighborhood, could expose users to other nodes and passive eavesdroppers.

These challenges are addressed here by our security architecture for secure and resilient decentralized/collaborative privacy protection for \acp{LBS}. We propose new components orthogonal to the \ac{LBS} functionality. We leverage pseudonymous authentication to provide privacy-enhancing message authentication and integrity for communication with other users/peers and with infrastructure entities. We also leverage proactive caching of \ac{POI} data (e.g., for a region) by a small fraction of users (termed serving nodes) that serve others, sharing the cached \ac{POI} data. This ensures that peer responses can provide the same quality as direct \ac{LBS} responses without extra communication incurred by using an obfuscated area. This also enables cross-checking multiple peer responses to the same query, to detect false responses from malicious nodes. The burden is balanced among users through a periodical randomized role assignment by the infrastructure. While users benefit from the information sharing system, our scheme minimizes their exposure to the \ac{LBS} server and limits exposure to curious serving nodes (thanks to encrypted querying and serving node communication). Our evaluation shows both effective exposure reduction and highly successful valid \ac{POI} provision in a realistic intelligent transportation setting, even with a very high (in practice) fraction (e.g., $20\ \%$) of curious or malicious nodes.

The contributions of this paper are: (1) use of pseudonymous authentication for privacy-preserving user authentication, (2) privacy-enhancing randomized serving node assignment and encrypted \ac{P2P} query-response exchange, (3) cross-checking of \ac{P2P} responses and privacy-preserving proactive \ac{LBS} response validation, (4) qualitative and extensive quantitative simulation-based security and privacy analyses, (5) comparison to the state-of-the-art, MobiCrowd~\cite{shokri2014hiding}, in terms of achieved privacy and security, and efficiency (notably, communication overhead), and (6) experimental evaluation leveraging a field-operational-test grade automotive communication testbed, emulating real-world conditions.

The rest of the paper is organized as follows: We discuss related work in Sec.~\ref{sec:related}. We outline the system and adversarial models, as well as requirements in Sec.~\ref{sec:problem}. Then, we present the proposed scheme in Sec.~\ref{sec:scheme}. We provide a qualitative security and privacy analysis in Sec.~\ref{sec:qualitative}, followed by an extensive quantitative simulation-based evaluation in Sec.~\ref{sec:quantitative}. We show experimental evaluation results with our scheme implemented on an automotive testbed, and discuss design choices for its deployment in real-world scenarios (Sec.~\ref{sec:experiment}) before we conclude (Sec.~\ref{sec:conclusion}).

\section{Related Work}
\label{sec:related}

User queries, including user locations and interests, are essential information for \acp{LBS} to best serve users; this very same information can also be used to breach user privacy. Queries from the same user can be linked and eventually used to reconstruct user traces and infer user activities, e.g., home or work addresses. Anonymization or pseudonymization, i.e., removal of user identities or their replacement with pseudonymous identities, prevents the \ac{LBS} server from correlating real user identities with queries. The \ac{LBS} server only sees a pool of user queries without any information that immediately identifies the querying users. However, studies have shown that simply removing identities from frequent queries could not prevent tracking and even profiling (e.g., inferring real names) of specific users~\cite{gruteser2003anonymous, hoh2005protecting,golle2009anonymity}.

\textbf{Location obfuscation:} Location obfuscation uses cloaked areas or perturbed locations to represent actual user locations. Cloaked areas~\cite{ardagna2007location,chow2006peer} can be used so that the \ac{LBS} server only learns the area the user is in and then responds with the information that pertains to the whole area. However, a determined attacker could learn which users are located in this area. If, in an arguably rare situation, the user is the only one included in this area, the attacker can easily link the query with the only user ``seen'' in this area. Location $k$-anonymity~\cite{kido2005anonymous, gkoulalas2010providing} is used to form an obfuscated area involving at least $k$ users, so that the user would not stand out from the obfuscated area. However, merely using $k$-anonymity could not guarantee an area constructed based on $k$ users large enough to hide symbolic address (e.g.,  church, shopping mall and gas station). Therefore, in densely populated areas, the perturbed/obfuscated region should be large enough so that all nodes within the area are not closely gathered. At the same time, the quality of the \ac{LBS} response would be affected if the location in the query is not accurate enough. Thus, it is hard to define how large should $k$ be to ensure an appropriate level of privacy protection. Geo-indistinguishability~\cite{andres2013geo,eltarjaman2017private,al2017correlation} uses perturbed locations to represent user locations, adding Laplacian noise to user locations based on the required  degree of privacy. This trades off accuracy of \ac{POI} data for higher privacy. Both approaches could fail to protect node privacy when multiple queries can be correlated~\cite{shokri2010unraveling,al2017correlation}.

Moreover, schemes based on $k$-anonymity can be vulnerable to the so-called center of \ac{ASR} attack: the probability of the requesting node being the node closest to the center of \ac{ASR} is larger than $1/k$~\cite{ghinita2007mobihide, kalnis2007preventing}. Such an attack can be prevented by applying a $k$-anonymity algorithm to one of the neighbors. These neighbors are discovered by applying the $k$-anonymity algorithm to itself. This generates a cloaked region that is not necessarily centered around oneself.

\textbf{Centralized and decentralized approaches:} Both centralized and decentralized variants of the above-mentioned privacy enhancing techniques were proposed. Centralized approaches introduce a \ac{TTP} that acts as a so-called anonymizer for the queries sent to the \ac{LBS} server. The anonymizer could find $k-1$ nearest neighbors of the querying user and send the queries from all $k$ users to the \ac{LBS} server~\cite{ardagna2007location} . The anonymizer can also remove user identities from the queries before sending them to the \ac{LBS} server, while keeping information that matches user identities and queries locally, so that it can appropriately respond to each user the needed  information. However, there can be an inherent controversy here: we assume the \ac{LBS} server is honest-but-curious, but the anonymizer could be honest-but-curious too. The (centralized) anonymizer could collect user queries and use the information the same way the \ac{LBS} server would.


Decentralized approaches thwart honest-but-curious infrastructure by relying on users. To achieve $k$-anonymity with decentralized schemes, users can search for $k-1$ neighbors in an ad-hoc network and use the area including these $k$ users as the cloaked area~\cite{hashem2007safeguarding,ghinita2007mobihide,hu2009non,ghinita2008private,chow2011spatial,ghaffari2017p}. However, this incurs high communication overhead among peers: a node needs to initiate the process whenever it wants to query the \ac{LBS} server. To reduce overhead and complexity of peers, anonymizer peers can be used, that act as a temporary proxy for their neighbors~\cite{amoeba07}. Therefore, users do not need to search for neighbors everytime. AMOEBA \cite{amoeba07} protects user privacy by forming groups and delegating \ac{LBS} queries to anonymizer peers, termed group leaders. The group members send the queries to the group leaders that act as the \ac{TTP} in centralized schemes: the group leaders query the \ac{LBS} server with collected queries and distribute the responses to each group member. Such group formation and provision of $k$-anonymity (both centralized and decentralized) are orthogonal to our work here. In fact, they could possibly co-exist with and even be facilitated by our scheme, which explicitly addresses trust assumptions and provides a security architecture. 

\textbf{Caching and sharing:} Caching and sharing \ac{POI} data~\cite{ku2008nearest,shokri2014hiding,niu2015enhancing,liu2016silence,mahin2017crowd} are orthogonal to above-mentioned protection approaches. Cached \ac{POI} data can serve future queries and reduce direct queries to the \ac{LBS} server. Road-side infrastructure can proactively cache \ac{POI} data~\cite{niu2015enhancing,liu2016silence} and serve user queries; however, users would be vulnerable to honest-but-curious road-side infrastructure entities. Nodes can cache information received from the \ac{LBS} server and pass to neighbors when requested~\cite{ku2008nearest,shokri2014hiding,mahin2017crowd}. This is the approach we extend in this paper. Gathered partial \ac{POI} data from neighbors can form  more complete \ac{POI} data~\cite{ku2008nearest,mahin2017crowd}, but it is hard to examine whether the gathered \ac{POI} data covers all that could be obtained directly from the \ac{LBS} server. Moreover, such approaches are vulnerable to data pollution by malicious or semi-honest nodes, providing masqueraded \ac{POI} data or partial data (from available own data) to mislead other nodes. Responses signed by the \ac{LBS} server~\cite{shokri2014hiding} can be self-verifiable allowing detection of manipulation by a node. However, a misbehaving node passing on tampered responses would remain ``invisible'' and continue attacking the system, wasting computational power~\cite{jin2015resilient}. Moreover, decentralizing functionality could provide an easy way for passive adversaries to learn information from neighbors. Thus, it is important for collaborative approaches to consider internal and external passive or active adversaries. This is exactly where this work comes in: providing a secure and privacy-enhancing \ac{LBS} while protecting benign nodes against both honest-but-curious and malicious nodes/peers.

\textbf{Location privacy in other areas:} Location privacy has also been widely studied in relevant areas. SPPEAR~\cite{gisdakis2014sppear} and SHIELD~\cite{gisdakis2015shield} propose a state of the art architecture and protocols for secure and privacy preserving participatory sensing. Users obtain short-term credentials and contribute sensing data pseudonymously to a central aggregator~\cite{gisdakis2014sppear}. The aggregator makes use of the large volume of received sensed data to filter out outliers~\cite{gisdakis2015shield}. The aggregated data is made available through queries submitted to the aggregator; this can be done securely and privately with our proposal here.

In \acp{VANET}, privacy is important, because high rate vehicle-transmitted safety beacons with location information for awareness among neighboring vehicles, expose vehicle activities. More importantly, vehicle messages need to be authenticated, but using traditional certificates would trivially allow linking messages sent from a vehicle. This motivates the use of pseudonymous authentication~\cite{papadimitratos2008secure} for message unlinkability while providing message and entity authentication. The most recent proposals protect users from honest-but-curious credential providers and investigate deployment and scalability aspects~\cite{alexiou2013vespa, gisdakis2013serosa, khodaei2014towards, khodaei2018secmace}. In our scheme, pseudonymous authentication is an important component for providing secure and privacy-preserving \ac{LBS}. We explain how this is customized and integrated in our design in Sec.~\ref{sec:pseudonym}.

\textbf{Query privacy:} Apart from location privacy, query privacy is an important issue for \ac{LBS}: the content of queries could disclose user interests. Location privacy protection schemes based on $k$-anonymity could potentially protect query privacy: with at least $k$ queries sent together to the \ac{LBS} server, the \ac{LBS} server could not figure out which node each query belongs to. However, if a specific \ac{POI} type has higher popularity than other types, then multiple (among the $k$) nodes could be interested in the same \ac{POI} type; this reduces the query diversity of these $k$ queries~\cite{liu2009query,dewri2010query,pingley2011protection}. Query $l$-diversity~\cite{dewri2010query} proposes to form a cloaked region involving user queries that have at least $l$ \ac{POI} types. However, this approach fails to provide location and query privacy protection if a node continuously queries for the same \ac{POI} type during its trip: intersection of multiple (spatially and temporally) consecutive $k$-anonymized and $l$-diversified query sets could result in a pair of distinct node and \ac{POI} type (i.e., the querying node and the queried \ac{POI} type). An anonymizer can take past queries into consideration and form the consecutive cloaking areas involving the same set of \ac{POI} types~\cite{dewri2010query,pingley2011protection}. Private retrieval of \acp{POI}~\cite{yi2016practical,hu2018messages} searches an encrypted \ac{POI} database with anonymized queries that do not disclose any query content to the \ac{LBS} server. We note that such query privacy protection approaches are orthogonal to sharing-based approaches (including our scheme), which decrease the ratio of queries disclosed to the \ac{LBS} server, thus can complement each other.

\section{System Model and Requirements}
\label{sec:problem}

\begin{figure}[t]
	\centering
	\includegraphics[width=0.8\columnwidth]{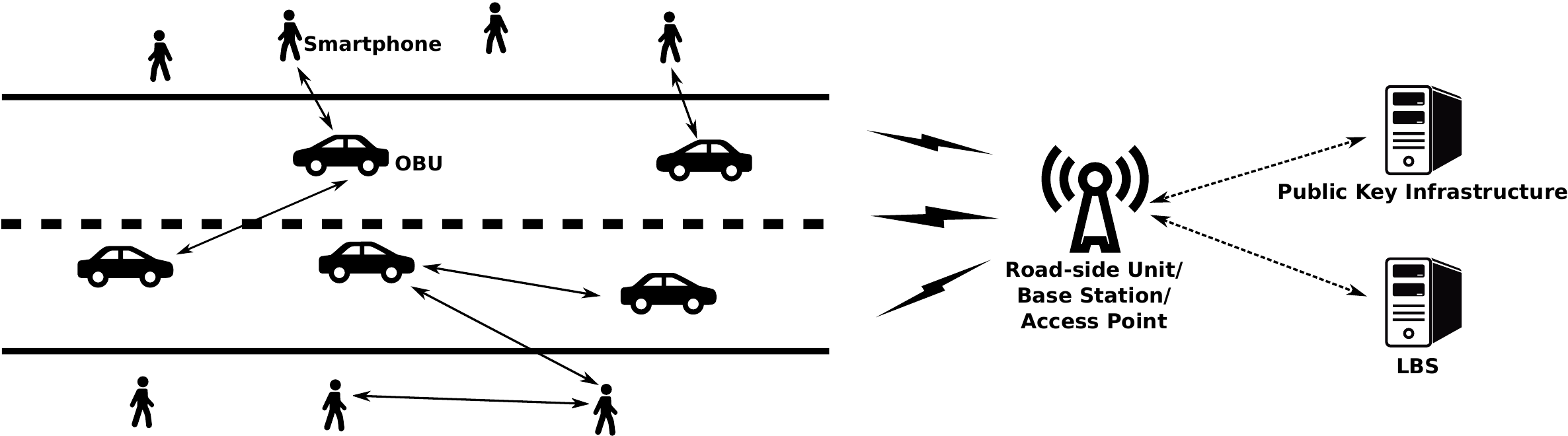}
	\caption{System Architecture \small{(Icons by Freepik, freepik.com)}}
	\label{fig:system}
\end{figure}

\textbf{System Model:} Fig.~\ref{fig:system} illustrates the considered system architecture. Mobile devices (termed \emph{nodes} or \emph{peers} in the rest of the paper), e.g., smartphones and vehicular \acp{OBU}, are equipped with multiple communication interfaces, e.g., Wi-Fi and cellular data. They can access \acp{LBS}, submitting queries regarding their current locations/regions. They also communicate in a \ac{P2P} manner over a wireless ad hoc (e.g., IEEE 802.11p) or cellular (e.g., LTE direct) network. Nodes can share \ac{POI} data and choose to query the \ac{LBS} server only when no satisfactory response is received from their peers. Nodes are registered with an \emph{identity and credential management} system (i.e., a \ac{PKI}) comprising \acp{CA} (see Sec.~\ref{sec:scheme} for more details). The \ac{PKI} issues credentials to the registered nodes and the \acp{SP} (i.e., the \ac{LBS} servers here), so that \acp{SP} and nodes can interact securely. We assume nodes have connectivity to the \ac{LBS} and the \ac{PKI} through Internet (e.g, via cellular data, 3G or 4G, or Wi-Fi networks) throughout their trips.

\textbf{Adversary Model:} We assume \ac{LBS} servers are honest-but-curious: they follow the protocols, responding faithfully to queries, but they may trace the nodes (linking their queries) or even de-anonymize them and infer sensitive data (e.g., home and work sites). This is because queries sent to the \ac{LBS} servers expose user locations and interests, and can be used to infer additional user information. We maintain and extend the honest-but-curious assumption to cover any \ac{TTP}, including the ones we introduce in our scheme, notably the \ac{PKI} (Sec.~\ref{sec:scheme}).

Nodes can be also honest-but-curious or deviate from the protocols. Honest-but-curious nodes could stay in the network and collect peer queries and responses. A single honest-but-curious node could have negligible effect on the privacy of legitimate nodes. However, if such honest-but-curious nodes collaborate and merge their transcripts of all queries they overheard from peers in the system, it is very likely they are powerful enough to threaten user privacy. An adversary could recruit otherwise benign participants (e.g., with some incentives) and then fetch and merge their query and response transcripts; or user equipment (\acp{OBU}, smartphones) could be compromised by malware that propagates across the the network and enable the adversary access such transcripts. Furthermore, nodes can deviate from the collaborative protocol functionality and policies and attack the system, notably their peers. They can forge or tamper with responses and masquerade other nodes. This could, in turn, affect the quality of service, i.e., the timeliness and authenticity of \ac{POI} data obtained from peers, and thus force benign nodes, which may get no useful \ac{POI} data, to query directly (i.e., expose themselves to) the \ac{LBS} server(s).

\textbf{Requirements:} We require that peer-provided information be verifiable and nodes be accountable for their actions (messages). Nodes should be able to efficiently obtain \ac{POI} data from their peers with the same quality as that obtained directly from the \ac{LBS} server. While nodes benefit from \ac{P2P} \ac{POI} data sharing, node exposure to neighboring, both assisting and non-assisting, peers and the \ac{LBS} server should be minimized. Towards these objectives, the following security and privacy requirements need to be met:

\emph{Authentication and integrity} - Node messages should allow their receivers to corroborate the legitimacy of their senders (Note: not their identity; please see anonymity/pseudonymity below.) and verify they were not modified or replayed.

\emph{Accountability} - Message senders should not be able to deny having sent a message (non-repudiation). Any node can be tied to its actions, and, if need arises, be held accountable and possibly have its long-term identity revealed and have itself evicted from the system.

\emph{Anonymity/Pseudonymity and unlinkability} - Node actual (long-term) identities should not be linked to their \ac{P2P} and node-to-LBS messages. Anonymity should be conditional, allowing the system to identify a misbehaving node and evict it. Ideally, it should be impossible for any observer to link any two or more messages (e.g., queries) by the same node. However, for efficiency reasons, messages can be linkable at most over a protocol selectable period, $\tau$. 

\emph{Confidentiality and reduced exposure} - \ac{POI} data and sensitive user information (e.g., node queries) should be accessible only by authorized entities; the amount of information revealed to peers and the \ac{LBS} server should be minimal.

\emph{Resilience} - Nodes should be resilient to (compromised) malicious nodes that actively disrupt their operations and deviate from protocol definitions. In particular, they should be able to validate correctness of information to reject bogus \ac{POI} data generated by malicious nodes.			

\emph{Sybil-resistance} - A registered node should be able to participate only with a single identity (pseudonym) at any point in time. It should not be able to be present in the system with multiple legitimate identities (thus, as multiple participants) and inappropriately affect the outcome of protocols.

\section{Our Scheme}
\label{sec:scheme}

In this section, we introduce our decentralized privacy protection scheme in detail. Our design is driven by privacy, resilience and efficiency considerations. Our approach significantly extends \ac{P2P} data sharing \ac{LBS} privacy schemes~\cite{shokri2014hiding,jin2015resilient}, addressing a broad(er) set of requirements (Sec.~\ref{sec:problem}) and contributing the following main ideas: (1) Each node is equipped with short-term anonymous credentials, to authenticate all node-to-node and node-to-\ac{LBS} interactions. (2) Peer-provided \ac{POI} data can be drawn from a large volume of \ac{POI} data, proactively distributed by the \ac{LBS} server to a small fraction of randomly chosen nodes, termed \emph{serving nodes}. (3) Nodes submit queries to serving nodes, which periodically announce their presence and available \ac{POI} data (i.e., \ac{POI} data for their regions). (4) Encrypted peer query and response messages, to minimize revealed information to neighboring nodes. (5) Cross-checking of peer responses and proactive checking with the \ac{LBS} server, to validate peer responses and detect/evict malicious nodes. Table \ref{table:notation} summarizes the used notation.

\begin{table}[htp]
	\caption{Notation}
	\centering
	\renewcommand{\arraystretch}{1}
	\footnotesize
	\begin{tabular}{l | *{1}{c} r}
		\hline \hline
		$U$ & \emph{A node(/user)} \\\hline
		$LTCA$ & \emph{Long-Term Certification Authority} \\\hline
		$PCA$ & \emph{Pseudonymous Certification Authority} \\\hline
		$RA$ & \emph{Resolution Authority} \\\hline
		$Lk/LK$ & \emph{Long-term Private/Public Key} \\\hline
		$LTC$  & \emph{\acl{LTC}} \\\hline
		$Sk/SK$ & \emph{Short-term Private/Public Key} \\\hline
		$PC$ & \emph{Short-term/Pseudonymous Certificate} \\\hline
		$K_{s}$ & \emph{A symmetric session key} \\\hline
		$SN$ & \emph{Serial Number} \\\hline
		$(n/q/r)id$ & \emph{(Node/Query/Region) identifier} \\\hline
		$L$ & \emph{Square region side length} \\\hline
		$G$ & \emph{Number of POI type groups} \\\hline
		$N$ & \emph{Maximum peer requests per node query} \\\hline
		$\Gamma$/$\tau$ & \emph{Pseudonym request interval/Pseudonym lifetime} \\\hline
		$Pr_{serve}$ & \emph{Probability of serving node assignment} \\\hline
		$T_{serve}$ & \emph{Serving period of each serving node assignment} \\\hline
		$Pr_{check}$ & \emph{Probability of checking with the LBS server} \\\hline
		$T_{beacon}$ & \emph{Beacon interval} \\\hline
		$T_{POI}$ & \emph{POI update interval} \\\hline
		$T_{wait}$ & \emph{Beacon waiting time before requesting the LBS server} \\\hline
		$Q_{{rid}}$ & \emph{A query for \ac{POI} data in a region, $rid$} \\\hline 
		$POI_{{rid}}$ & \emph{\ac{POI} data covering an entire region, $rid$} \\\hline
		$t/t_{now}$ & \emph{Timestamp/Fresh timestamp indicating current time} \\\hline
		$\{msg\}_{\sigma_{_{LTC/PC}}}$ & \emph{Signed message with signature and LTC/PC attached} \\\hline
		$E_K(msg)$ & \emph{Encryption of message with K} \\\hline
		$Sign(msg, LTC/PC)$ & \emph{Signature on a message with corresponding Lk/Sk under LTC/PC} \\\hline
		$H()$ & \emph{Hash function} \\\hline
		$Send(msg, E)$ & \emph{Send a message to an entity E} \\\hline
		$Receive(msg, E)$ & \emph{Receive a message from an entity E} \\\hline
		\hline
	\end{tabular}
	\renewcommand{\arraystretch}{1}
	\label{table:notation}
\end{table}

\subsection{Overview}

Nodes are interested in diverse \ac{POI} data throughout their trips. The \ac{LBS} server maintains a database of real-time \ac{POI} data, subject to change over time. Without loss of generality, we assume: \ac{POI} data is refreshed/updated every $T_{POI}$; the area (e.g., of a city) is divided into a number of equally sized regions; nodes are interested in \ac{POI} data for regions they are currently in. 

The \ac{LBS} server and the nodes in the system are registered with an \emph{identity and credential management facility}, i.e., a \ac{PKI}.  A \emph{\ac{LTCA}} issues a \emph{\ac{LTC}} for each registered node, used as a long-term identity for the node. With the \ac{LTC}, a node can obtain an \ac{LTCA}-issued ticket, presented to a \emph{\ac{PCA}} for obtaining \emph{\acp{PC}/pseudonyms} (if the ticket is validated). The ticket is authenticated by the \ac{LTCA} but it is \emph{anonymized}: it does not reveal the long-term (real) node identity (refereed as \emph{node identity} in the rest of the paper) to the \ac{PCA}~\cite{khodaei2014towards, khodaei2018secmace}. Therefore, a single \ac{LTCA} or a single \ac{PCA} cannot link the node identity to the issued pseudonyms. Thus, the messages signed under \acp{PC} are also unlinkable to the node identity. To ensure unlinkability after a change of pseudonym, the node can randomly reset its IP and MAC address. This duty separation is based on work done in the context of \ac{VC} systems~\cite{gisdakis2013serosa,khodaei2014towards, khodaei2018secmace}. The ticket and pseudonym acquisition protocols are presented in Sec.~\ref{sec:pseudonym}.

Pseudonyms are used by nodes to authenticate themselves to the system entities, including the \ac{LBS} server and other nodes (peers), and to establish secure communication channel between system entities. Messages are signed under (i.e., with the private keys corresponding to) pseudonyms in order to hide node identities while providing message authentication and integrity.

\begin{figure}[t]
	\centering
	\begin{subfigure}[b]{.24\columnwidth}
		\includegraphics[width=\columnwidth]{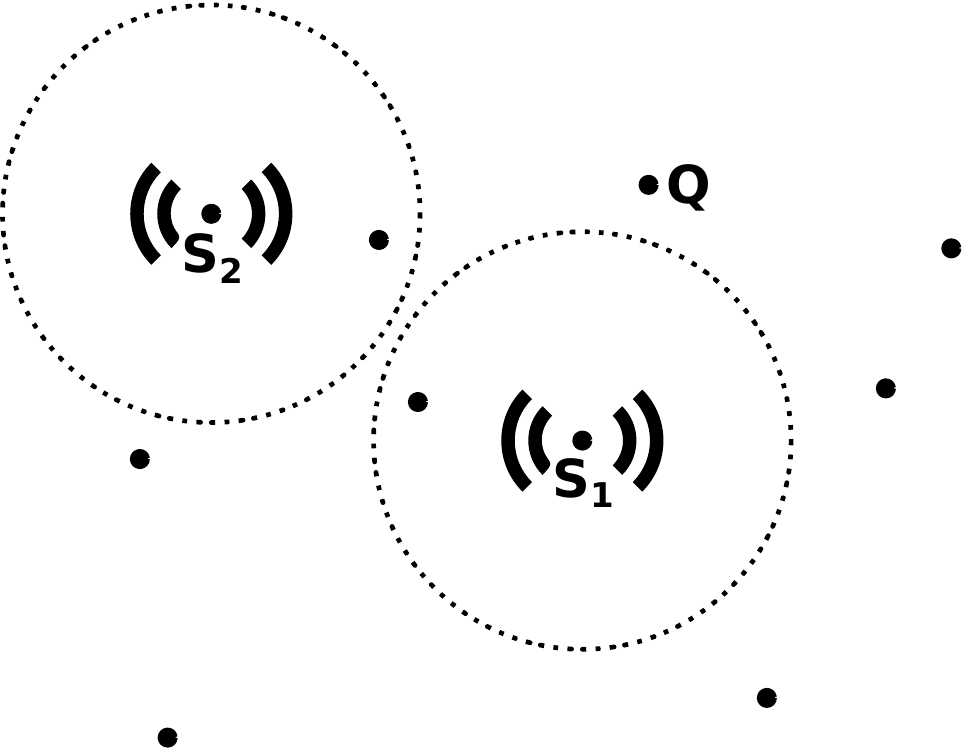}
		\caption{At $t_1$, $Q$ listens for beacons from serving nodes}
		\label{subfig_operation1}
	\end{subfigure}\hspace{2em}
	\begin{subfigure}[b]{.21\columnwidth}
		\includegraphics[width=\columnwidth]{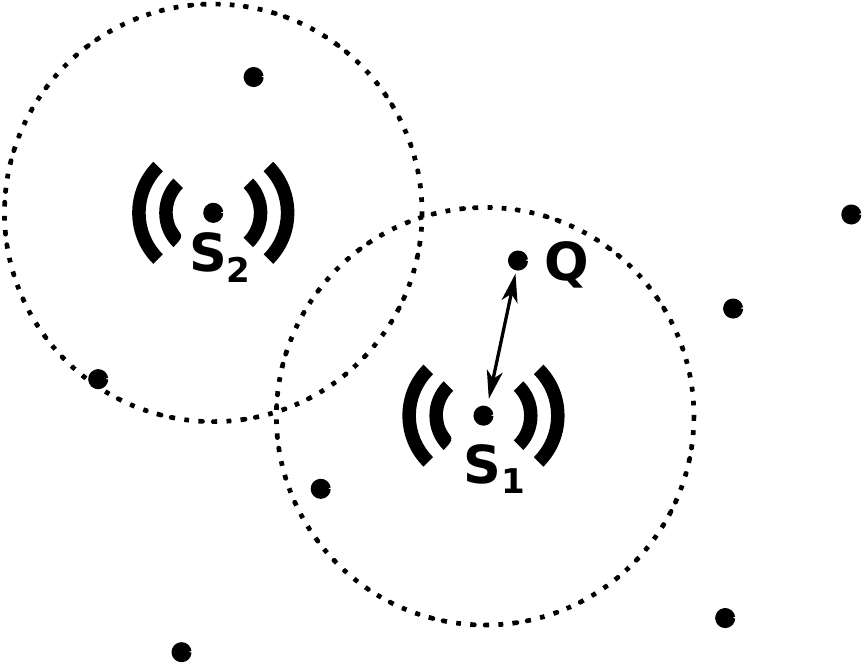}%
		\caption{At $t_2$, $Q$ receives beacons broadcasted by $S_1$ and queries $S_1$}
		\label{subfig_operation2}
	\end{subfigure}\hspace{2em}
	\begin{subfigure}[b]{.19\columnwidth}
		\includegraphics[width=\columnwidth]{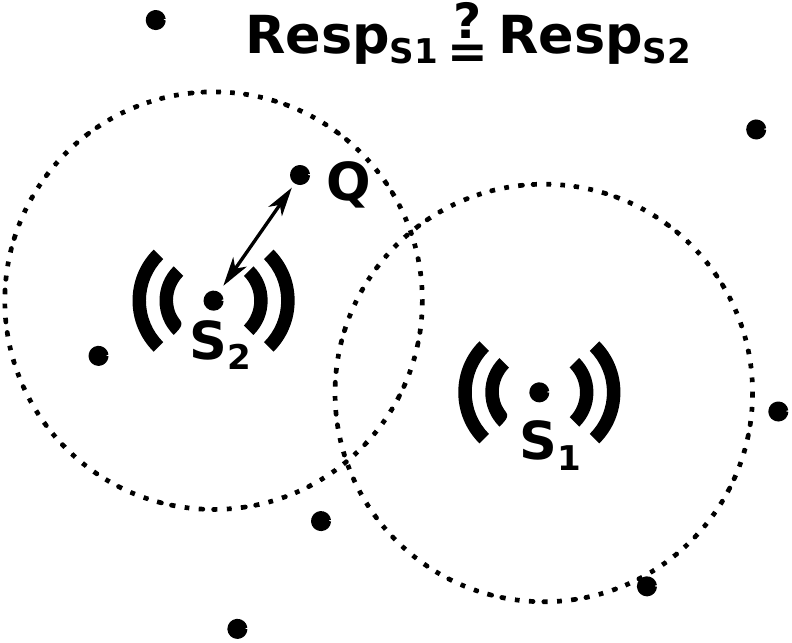}%
		\caption{At $t_3$, $Q$ receives beacons broadcasted by $S_2$ and queries $S_2$}
		\label{subfig_operation3}
	\end{subfigure}
	\caption{Illustration of peer query and response collection.}
	\label{fig:operation}
\end{figure}

We assign to the \ac{PCA} the responsibility for probabilistically choosing the serving nodes. A \emph{serving node} is responsible for requesting extensive \ac{POI} data, covering an entire region (termed \emph{regional \ac{POI} data}), consisting of \emph{\ac{POI} entries} for that region, from the \ac{LBS} server; storing the \ac{POI} data in its local storage; and serving other nodes whose requests it receives. The \ac{PCA} explicitly assigns/binds such a role to \acp{PC} (i.e., by setting a specific field in each \ac{PC}) it issues; thus nodes cannot masquerade serving nodes. Each serving node broadcasts beacons periodically, messages indicating the region it is in. The beacon is signed and the pseudonym is attached, so that any receiver can establish a secure communication channel with the serving node. This prevents exposure of all the communication to nearby nodes, thus disclosure of nodes' interests and presence in the region. A node in need of \ac{POI} data queries multiple serving nodes; with multiple redundant responses, it can cross-check and validate the obtained data and, at the same time, detect peer node misbehavior (e.g., provision of false, forged, or outdated POI data). \cref{fig:operation} shows an example of the operation concerning peer node queries. At $t_1$, $Q$ becomes interested in some \ac{POI} data and it starts listening for serving node beacons (\cref{subfig_operation1}). At $t_2$, $Q$ moves within the communication range of $S_1$; it queries $S_1$ upon a receipt of one of its broadcasted beacons (\cref{subfig_operation2}); at $t_3$, $Q$ encounters another serving node, $S_2$, it queries $S_2$, and then cross-checks the two responses (\cref{subfig_operation3}).

The \ac{LBS} server is also issued an \ac{LTC}, used to authenticate itself to the nodes and infrastructure entities. The \ac{LBS} server and the nodes are involved in (i.e., registered with) the same \ac{PKI} architecture, so that the \ac{LBS} server is also able to authenticate the nodes with their \acp{PC}. The \ac{LBS} server authenticates itself with the \ac{LTC} and provides message integrity and authentication by signing the messages. 

Furthermore, to prevent abuse of the node anonymity, our scheme provides \emph{conditional anonymity} and allows \emph{revocation of anonymity and eviction of nodes}. The node interactions with the facility entities are explained throughout this section below (Sec.~\ref{sec:pseudonym}). While we assume the certificates of system entities (i.e., authorities and the \ac{LBS} server) are pre-installed in each node, all the messages signed under \acp{PC} should be attached with the corresponding \acp{PC}. We mandate that all signatures must be verified before the messages can be processed. However, for simplicity, we skip the steps for signature verifications in the algorithms and protocols: progress in protocol execution indicates signatures are already validated.

\subsection{Pseudonymous Authentication}
\label{sec:pseudonym}

{\small

}



\begin{figure}[h!]
	\centering
	\includegraphics[width=0.7\columnwidth]{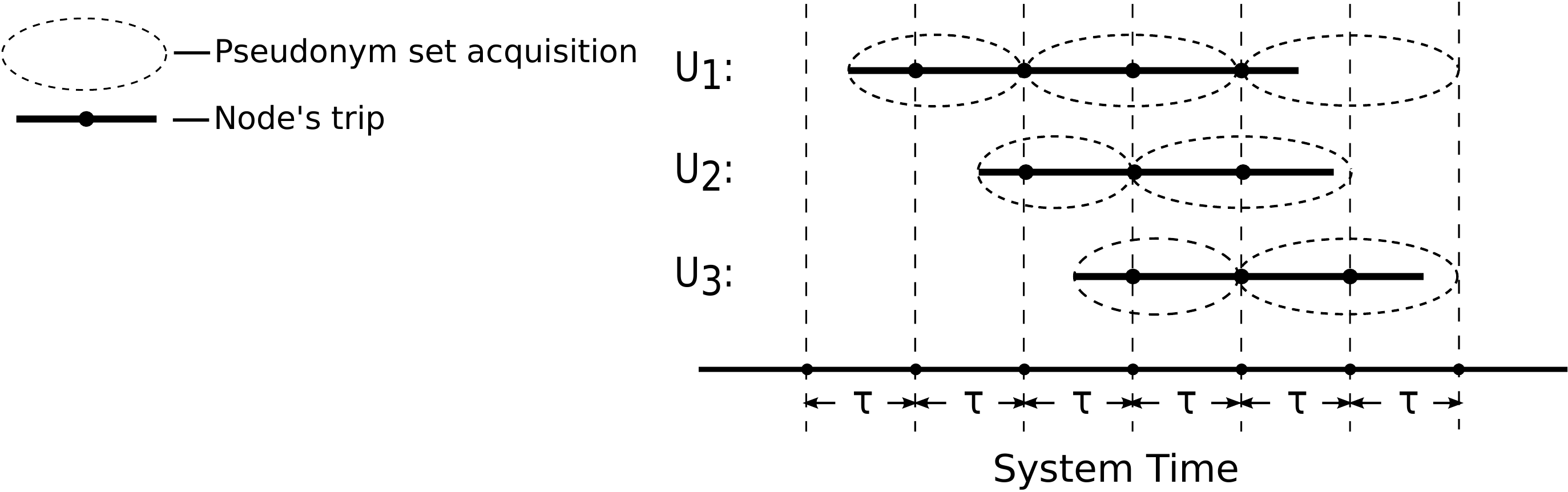}
	\caption{Example of ticket and pseudonym set acquisitions with $\Gamma = 2\tau$ for various node trips}
	\label{fig:acquisition}
\end{figure}

\textbf{Ticket and Pseudonym Acquisition}: We refer the reader to~\cite{khodaei2018secmace}, for a detailed protocol description of \ac{LTC} issuance. Here, we focus on our customized pseudonym issuance protocol. We follow the privacy-enhancing pseudonym issuance policy proposed in~\cite{khodaei2014towards,khodaei2018secmace}: independently of the time of contacting the \ac{LTCA}/\ac{PCA}, all pseudonym lifetimes are aligned. This implies that nodes change their pseudonyms at the same time, and as a result, the successive pseudonyms of a specific node would not be linked based on pseudonym lifetimes. In our scheme, we do not restrict the ticket lifetimes to be started at same time, since the anonymity set (all nodes registered with an \ac{LTCA}) for ticket acquisition is large enough considering an expected large number of nodes in the system. \cref{fig:acquisition} shows an example of pseudonym issuance for nodes with various trips, for $\Gamma = 2\tau$. For each pseudonym request, a node always requests $\Gamma/\tau$ pseudonyms. In general, a node is more likely to start its trip in the middle of a $\tau$ period. As shown in \cref{fig:acquisition}, the first pseudonym of a node is not valid from the beginning of a $\tau$ period, but from the beginning of the trip until the end of that $\tau$ period. For the rest of the trip, the node always requests $\Gamma/\tau$ pseudonyms from the \ac{PCA}. Although a node (e.g., $U_1$) could finish its trip shortly after its last pseudonym acquisition, the node always requests pseudonyms with lifetimes covering the entire $\Gamma$ period: there is a practical uncertainty for the trip duration and this hides the trip ending time from both the \ac{LTCA} and the \ac{PCA}.

{\small

\begin{align}
U \rightarrow LTCA &: ticket\_req\{t_{s}\}_{\sigma_{_{LTC_U}}} \label{eq:pc-1} \\
LTCA &: check(nid_{_U}, t_{s}) \label{eq:pc-2} \\
LTCA \rightarrow C &: ticket = \{SN_{ticket},t_{s}\}_{\sigma_{_{LTC{_{LTCA}}}}} \label{eq:pc-3} \\
U &: n = \lceil \frac{\Gamma}{\tau} \rceil \label{eq:pc-4} \\
U &: \{Sk_i, SK_i\},\ i \in [0, n) \label{eq:pc-5} \\
U \rightarrow PCA &: pseudonym\_req\{ticket, \{SK_0\}_{\sigma_{_{Sk_1}}} ... \{SK_{n-1}\}_{\sigma_{_{Sk_{n-1}}}}\} \label{eq:pc-6} \\
PCA &: t_{s_0} = t_{s}, \ t_{e_0} = t_{s} - (t_{s}\ mod\ \tau) + \tau \label{eq:pc-7} \\
PCA &: t_{s_i} = t_{e_{i-1}}, \ t_{e_i} = t_{s_i} + \tau,\ i \in [1, n) \label{eq:pc-8} \\
PCA &: s = 1 \text{ with probability } Pr_{serve},\ s = 0 \text{ otherwise.} \label{eq:pc-9} \\
PCA \rightarrow U &: PC_i = \{SN_{PC_i}, SK_i, s, t_{s_i}, t_{e_i}\}_{\sigma_{_{LTC_{PCA}}}},\ i \in [0, n) \label{eq:pc-10}
\end{align}

}

When a node needs to request pseudonyms, it sends a request to the \ac{LTCA} with a pseudonym validity starting time, $t_{s}$ (\ref{eq:pc-1}). For a node starting its new trip, $t_{s}$ is the trip beginning (e.g., the beginning of the first dashed circle for $U_1$ in Fig.~\ref{fig:acquisition}); while for a node in the middle of its trip, $t_{s}$ is the starting point of next $\Gamma$ period (e.g., the beginning of the second dashed circle). The pseudonyms are issued for the remaining time of each $\Gamma$, as we described earlier. The \ac{LTCA} checks if a ticket was issued with an overlapping lifetime, based on the locally stored ticket issuance history for each node (\ref{eq:pc-2}); if not, it issues a ticket valid from $t_{s}$ (\ref{eq:pc-3}). With the ticket in hand, the node obtains a set of pseudonyms from the \ac{PCA}: the node generates the needed public/private key pairs and sends self-signed public keys with the ticket to the \ac{PCA} (\ref{eq:pc-4}-\ref{eq:pc-6}). The anonymized ticket does not reveal the node identity to the \ac{PCA}. The \ac{PCA} calculates the lifetimes of pseudonyms according to the pseudonym lifetime policy (\ref{eq:pc-7}-\ref{eq:pc-8}). The assignment of serving nodes is done during the pseudonym issuance process. The \ac{PCA} assigns the requesting node as a serving node with probability $Pr_{serve}$. This is achieved by setting a field, $s$, in each \ac{PC} (\ref{eq:pc-9}-\ref{eq:pc-10}). It indicates that the node possessing the pseudonym is responsible for requesting \ac{POI} data from the \ac{LBS} server and serving other nodes. We align the serving period, $T_{serve}$, of a serving node with $\Gamma$, thus all pseudonyms for the $\Gamma$ period will be assigned the same $s$. Both ticket and pseudonym acquisitions are done across secure channels. We explain our resilient and privacy-enhancing \ac{LBS} functionality in more detail in Sec.~\ref{sec:privacy_lbs}. 

{\small

}

\textbf{Reporting Misbehavior}: When misbehavior (of any type) is detected by a node, the node sends to the \ac{RA} the messages related to the misbehavior, with the corresponding pseudonyms attached. In case the messages are proved to be related to misbehavior, the \ac{LTC} of the misbehaving node is exposed (and the node possibly evicted from the system) by the \ac{RA} with the help of the \ac{LTCA} and \ac{PCA}. The revoked \acp{LTC}/\acp{PC} are published through a \ac{CRL}. We describe the response validation process in Sec.~\ref{sec:privacy_lbs} in more detail.


\subsection{Privacy-enhancing LBS}
\label{sec:privacy_lbs}

\textbf{LBS Query Model:} We divide a large area (e.g., a city) into equally-sized regions, each with a unique identifier, $rid$. A node is interested in \ac{POI} data, of various types, around its current location. This can be any subset of the regional \ac{POI} data, which comprise various \ac{POI} data types. For example, a node query could be \emph{What are the menus of nearby restaurants?} and \emph{How many parking spaces are available nearby?} Without loss of generality, we assume that \ac{POI} data in the \ac{LBS} server is updated every $T_{POI}$, with the expiry time explicitly indicated in returned \ac{POI} data; which have varying lifetimes, depending on their types. Basic information of some \ac{POI} entries (e.g., name and address of a restaurant) could remain unchanged for several years, while others, e.g., weather and traffic conditions or parking availability could be updated on a frequent basis (e.g., every 10 minutes). We capture a generalized \ac{POI} update frequency with a constant $T_{POI}$ and we are mainly concerned with frequently updated \ac{POI} data. \ac{POI} data with longer lifetimes could actually be preloaded to user devices. For example, Google Maps\footnote{https://www.google.com/maps} provide offline caching of \ac{POI} data (e.g., names and addresses), so that they can be queried locally, without an Internet connection. However, frequently updated \ac{POI} data (e.g., traffic or road conditions, or available parking slots) cannot be preloaded and should be obtained on-demand. Each node can cache \ac{POI} data it received and use the information as long as it is not expired: a node does not initiate a query for the same piece of data it cached before the data expires; new queries seeks \ac{POI} data for a different location or \ac{POI} type than those of non-expired \ac{POI} data in its cache.

\textbf{Regional \ac{POI} Data}: Once cached by a serving node, they can be used to serve other nodes. For region-based queries~\cite{shokri2014hiding,niu2015enhancing}, nodes are assumed to be interested in \ac{POI} data for their current regions. Thus, regional \ac{POI} data includes \ac{POI} data within the corresponding region. A limitation of such a query model is that, for some \ac{POI} types, no response is available in the querier's current region, although the nearest out-of-region \ac{POI} data can be still valuable. Typically, in this situation, a direct query to the \ac{LBS} server could be serviced with the nearest \ac{POI} entries in other regions. Our scheme can be readily adapted to address this issue, by generating and distributing to the serving nodes regional \ac{POI} data consisting of additional nearest \ac{POI} entries (when some \ac{POI} types are unavailable in a region or some out-of-region \ac{POI} entries are closer to the queriers at the border of the region), instead of strictly including only \ac{POI} entries for the current region of the serving node(s). A $k$-nearest-neighbor query model~\cite{yi2016practical} can be natural here; returning the $k$ nearest \ac{POI} data, including  possibly out-of-region \ac{POI} data.

\begin{algorithm}[h]
	\caption{Beacon, repeat every $T_{beacon}$}
	\label{alg:beacon}
	\small
	\begin{algorithmic}[1]
		
		\State Node is in region $rid$, possessing a valid serving pseudonym, $PC$ with $s=1$
		\State $Broadcast(\{rid, t_{exp}\}_{\sigma_{_{PC}}})$
		\State \textbf{return}
	\end{algorithmic}
\end{algorithm}

\begin{algorithm}[h]
	\caption{\ac{POI} data update}
	\label{alg:region_update}
	\small
	\begin{algorithmic}[1]
		\State Serving node, $U$, entering a new region $rid$,\\
		or reaching a \ac{POI} update point (i.e., $t_{exp}$ for $POI_{rid}$) for its current region, $rid$.
		\If {$Cache \neq \{\}$}
		\For {each $\{POI_{rid_i}, t_{exp_i}\} \in Cache$}
		\If {$rid_i == rid$ and $t_{exp_i} > t_{now}$}
		\State \textbf{return}
		\EndIf
		\EndFor
		\EndIf
		\State $Send(\{qid, rid\}_{\sigma_{_{PC}}}, LBS)$  *
		\State $Receive(\{qid, \{POI_{_{rid}}, t_{exp}\}_{\sigma_{_{LTC_{LBS}}}}\}, LBS)$  *
		\State Update $Cache$ with $\{POI_{_{rid}}, t_{exp}\}$
		\State \textbf{return}
	\end{algorithmic}
\end{algorithm}

\textbf{Serving Node}: As mentioned earlier, the \ac{PCA} assigns the role of serving node with probability, $Pr_{serve}$, when a node requests pseudonyms from the \ac{PCA}. Each pseudonym, $PC$, of a serving node has the field $s$ set to $1$ (Sec.~\ref{sec:pseudonym}). After a node becomes a serving one, it broadcasts a beacon every $T_{beacon}$, as per Protocol~\ref{alg:beacon}. A beacon, signed under \ac{PC}, includes an $rid$, indicating the region the serving node is in, thus the \ac{POI} data it can provide, and the expiry time of the \ac{POI} data it has (i.e., obtained from \ac{LBS} and stored locally) for that region.

Serving nodes retrieve regional \ac{POI} data from the \ac{LBS} server whenever they enter new regions (including the case they become serving nodes in their current regions). As shown in Protocol~\ref{alg:region_update}, when a serving node enters a new region, it checks whether the latest \ac{POI} data, covering that entire region, has been cached (during a previous visit to the same region). If not (or the cached \ac{POI} data is expired), it requests from the \ac{LBS} server the latest \ac{POI} data of that region. The serving node is also responsible for updating \ac{POI} data from the \ac{LBS} server to have the latest \ac{POI} data cached locally and serve other nodes. The communication between nodes and the \ac{LBS} server is carried out over a secure channel, e.g., a TLS channel; we mark steps over a secure channel with asterisk ($*$).\footnote{In general, the format of \acp{PC} does not need to comply with the format of a standard X.509 certificate, because \acp{PC} can only include minimal necessary information to keep communication overhead low and to reduce node identifiability~\cite{calandriello2011performance, khodaei2014towards, khodaei2018secmace}, while \acp{LTC} can be standard X.509 certificates. As a result, \acp{PC} are not compatible with SSL/TLS protocols, but nodes are required to provide message integrity and authentication by signing the messages with their private keys corresponding to the \acp{PC}. When a node is required to present its \ac{LTC} (e.g., for a ticket acquisition), an SSL/TLS channel can be established with the \ac{LTC}.}

\begin{algorithm}[ht]
	\caption{A node query}
	\label{alg:lbs_query}
	\small
	\begin{algorithmic}[1]
		\State Node, $U$, interested in \ac{POI} data in region $rid$; interest expressed by query $Q_{_{rid}}$
		\State $Result = Search(Cache, Q_{_{rid}})$
		\If {$Result \neq \phi$ }
		\State \Return
		\EndIf
		\State $j=0$, $t = t_{now} + T_{wait}$
		\While {$j < N$}
		\State Listen for beacons; wait until $t$; if $t$ passed, \textbf{break}
		\State Received a beacon $\{rid^\prime, t_{exp}^\prime\}_{\sigma_{PC_{U_j}}}$ from $U_j$.
		\If {$s == 1$ in $PC_{U_j}$, $rid^\prime == rid$ and $t_{exp}^\prime > t_{now}$}
		\State $Send(\{E_{_{K_{s_j}}}(qid_{j}, rid, Q_{_{rid}}, PC_{U}), E_{_{SK_{U_j}}}(K_{s_j}), t_{now}\}_{\sigma_{ PC_{U}}}, U_j)$
		\State $Receive(\{E_{K_{s_j}}(qid_{j}, Resp_j), Auth_j, t_j\}_{\sigma_{ PC_{U_j}}}, U_j)$
		\State $j = j + 1$
		\EndIf
		\EndWhile
		\If {$j == 0$}
		\State $Send(\{qid, Q_{_{rid}}\}_{\sigma_{PC_U}}, LBS)$  *
		\State $Receive(\{qid, Resp\}_{\sigma_{LTC_{LBS}}}, LBS)$  *
		\ElsIf {$j == 1$}
		\State $b = 1$ with probability $Pr_{check}$ or $b = 0$ with probability $1 - Pr_{check}$
		\If {$b == 1$}
		\State $Send(\{qid_0, Q_{rid}, H(Resp_0),  Auth_0, t_0, PC_{U_0}\}, LBS)$  *
		\State $Receive(\{qid_0, Auth_0, result\}_{\sigma_{LTC_{LBS}}}, LBS)$  *  
		\If {$result$ is negative}
		\State $Report = \{K_{s_0}, \{E_{K_{s_0}}(qid_{0}, Resp_0), Auth_0, t_0)\}_{\sigma_{ PC_{U_1}}}\}$
		\State $Send(\{\{Q_{rid}, Report\}, t_{now}\}_{\sigma_{PC_U}}, RA)$  *
		\EndIf
		\EndIf
		\ElsIf {Conflict exists among $\{Resp_0 ... Resp_{j-1}\}$}
		\State $Report = \{\{K_{s_0}, \{E_{K_{s_0}}(qid_{0}, Resp_0), Auth_0, t_0)\}_{\sigma_{ PC_{U_0}}}\} ...$
		\State $\quad\quad\quad\quad \{K_{s_{j-1}}, \{E_{K_{s_{j-1}}}(qid_{{j-1}}, Resp_{j-1}), Auth_{j-1}, t_{j-1})\}_{\sigma_{ PC_{U_{j-1}}}}\}\}$
		\State $Send(\{\{Q_{rid}, Report\}, t_{now}\}_{\sigma_{PC_U}}, RA)$  *
		\EndIf
		\State \textbf{return}
	\end{algorithmic}
\end{algorithm}

\begin{algorithm}[h]
	\caption{Serving of a node query}
	\label{alg:lbs_resp}
	\small
	\begin{algorithmic}[1]
		\State A serving node, $U$, waits for incoming peer query in region $rid$
		\State $Receive(\{E_{_{K_s}}(qid, rid^\prime, Q_{_{rid^\prime}}, PC_{U^\prime}), E_{_{SK_{U}}}(K_s), t_{U^\prime}\}_{\sigma_{PC_{U^\prime}}}, U^\prime)$
		\If {$rid == rid^\prime$}
		\State $Resp = Search(POI_{rid}, Q_{_{rid^\prime}})$
		\If {$Resp \neq \phi$ }
		\State $Auth = Sign(\{qid, Q_{_{rid}}, H(Resp), t_{now}\}, PC_{U})$
		\State $Send(\{E_{K_s}(qid, Resp), Auth, t_{now}\}_{\sigma_{_{PC_{U}}}},  U^\prime)$
		\EndIf
		\EndIf
		\State \textbf{return}
	\end{algorithmic}
\end{algorithm}

\textbf{LBS Query}: Protocol~\ref{alg:lbs_query} and Protocol~\ref{alg:lbs_resp} illustrate the querying and the serving process respectively. When a node is interested in \ac{POI} data (for its current region $rid$), it first searches in the local cache whether the data is cached (while it is/was a serving node) and is still valid (i.e., not expired). If yes, then the node interest (i.e., the node query) is fulfilled and the process is finished. Otherwise, the node starts listening to beacons in the network for the period of $T_{wait}$ at most. If a beacon including $rid$ is received from a serving node (i.e., $s=1$ in \ac{PC}), the regular node sends the query to and receives the response from the beacon sender. If a beacon from a non-serving node is received, this is considered as misbehavior and reported to the \ac{RA}.

The query-response process is encrypted with a session key, $K_s$, generated by the querying node for each peer query, encrypted with the public key in the serving node's \ac{PC}. Thus, the communication is kept confidential between the two nodes. The node listens and waits for at most $N$ beacons with the region identifier ($rid$) of interest, and requests information from the beacon senders. The sought number of responses, $N$, is a protocol selectable parameter, used to provide redundancy for response validation. We explain it in more detail (in \emph{response validation}) below.

Protocol~\ref{alg:lbs_resp} shows the action of a serving node when a peer query is received. It first decrypts the session key, $K_{s}$, with its private key, and then decrypts the query and the pseudonym of the querying node with $K_{s}$. Once the query is verified, it searches in the local cache according to the query. If the search yields a result, then the serving node encrypts the response with $K_{s}$, and generates an \emph{authenticator}, \emph{Auth}, which can be used to check the correctness of the response with the \ac{LBS} server. 

\textbf{Response Validation}: Steps 16-33 in Protocol~\ref{alg:lbs_query} show the response validation approach. To protect the querying node from false information, forged by malicious serving nodes, and detect/reveal such misbehavior, the querying node queries $N>1$ (discovered) serving nodes, with the same query within a $T_{wait}$ period. If the waiting timer expires before $N$ responses are received, the node concludes this process. In an apparently benign system setting (e.g., no malicious node detected recently), the node could use the information received from the first request and use the rest for cross-checking. Each (legitimate) serving node by default cached the same (\ac{LBS}-obtained) \ac{POI} data for that region, thus their responses to the same query should be the same (i.e., a given search on the same data should return the same result, as long as the responders are legitimate). In order to make sure that any two honest responses from the serving nodes are identical, we assume a query won't span across a \ac{POI} data update point: if a node receives two responses from two serving nodes before and after the POI data originating from the LBS server were updated, these two honest responses could result in conflicting data. Without loss of generality, we assume a node waits (i.e., does not initiate a query in order to obtain fresher \ac{POI} data) until the \ac{POI} is updated (based on the $t_{exp}$ value learned from beacons) if the remaining time is less than $T_{wait}/2$; in contrast, it initiates the query after $t_{exp}$ is passed. If the remaining time is within $[T_{wait}/2, T_{wait}]$, the node initiates the query immediately but it only waits until the end of the current $T_{POI}$ period (i.e., $t_{exp}$), because any peer response after $t_{exp}$ could potentially conflict with previous peer responses.

At the very least, a node query succeeds if at least one serving node is discovered and queried successfully (i.e., it produces an authenticated response). Nonetheless, this does not provide any protection in the event this responder is malicious. \ac{POI} data from additional serving nodes can be used for cross-checking: any conflicting responses will be reported (with the originally attached signatures and pseudonyms) to the \ac{RA}. The higher $N$ is, the faster the detection of malicious nodes can be expected, especially in an environment with a high malicious node ratio. However, there is no guarantee that a querying node will discover $N$ serving nodes (before $T_{wait}$ expires). In a less densely populated area, the number of serving nodes would be also low. Thus, a querying node might be able to discover only one (in some cases, possibly) serving node. In this case, it will check the correctness of the single peer response with the \ac{LBS} server with probability $Pr_{check}$.

For each peer query, the serving node generates an authenticator and send to the querying node with the peer response. An authenticator is a signature on the concatenation of query id ($qid$), the \ac{LBS} query ($Q_{rid}$), the hash value of the response ($H(Resp)$) and the timestamp on peer response ($t$). Once the \ac{LBS} server receives the request, it first verifies the signature (i.e., the $Auth$ itself) with the attached serving node's pseudonym. Once the signature is verified, it fetches the response based on $Q_{rid}$ from its own database and check the correctness of $H(Resp)$. We use $H(Resp)$ instead of $Resp$ in order to decrease the communication overhead (typically the size of $Resp$ could be much larger than a hash value). Again, such validation is possible because the serving node has requested the \ac{POI} data covering the entire region; thus, the peer response and the \ac{LBS} response should be the same. If they are different, the querying node receives a negative result and reports the peer response to the \ac{RA}.

\begin{algorithm}[t]
	\caption{Processing of a misbehavior report}
	\label{alg:mis_process}
	\small
	\begin{algorithmic}[1]
		\State $Receive(\{\{Q_{rid}, Report\}, t\}_{\sigma_{PC_U}}, U)$  *
		\State $Send(\{qid, Q_{_{rid}}\}_{\sigma_{LTC_{RA}}}, LBS)$  *
		\State $Receive(\{qid, Resp\}_{\sigma_{LTC_{LBS}}}, LBS)$  *
		\For {Each $\{K_{s_i}, \{E_{K_{s_i}}(qid_{i}, Resp_i), Auth_i, t_i)\}_{\sigma_{ PC_{U_i}}}\}$ in $Report$}
		\If {$Resp_i \neq Resp$}
		\State Initiates a pseudonym resolution for $PC_{U_i}$
		\EndIf
		\EndFor
		\State \textbf{return}
	\end{algorithmic}
\end{algorithm}

Once the \ac{RA} receives any report on misbehavior, the \ac{RA} continues with processing the report as shown in Protocol~\ref{alg:mis_process}. The \ac{RA} checks with the \ac{LBS} server, submitting the same query as that by node reporting the misbehavior. If any dishonest peer response is discovered, the \ac{RA} reveals misbehaving node(s) through pseudonym resolution (Sec.~\ref{sec:pseudonym}).\footnote{A malicious serving node could increase the beacon rate attempting to increase the probability to be chosen by a querying node. Therefore, during the collection of responses from $N$ serving nodes, if the querying node detects any abnormally high beacon rate from a specific serving node (under the same \ac{PC}), such misbehavior (i.e., the authenticated beacons at a high rate) should be reported to the \ac{RA}.}

\textbf{POI Type Division}: The overhead for obtaining and caching complete regional \ac{POI} data could vary, depending on \ac{POI} density and number of available \ac{POI} types. A parameter, $G$, can be used to adjust the regional \ac{POI} data overhead for each serving node. $G$ indicates the number of groups available POI types are divided into. By assigning randomly one of the groups to each serving node at the time of \ac{PC} issuance (i.e., role assignment), each serving node is only responsible for that subset of \ac{POI} data. To achieve this, \acp{PC} of serving nodes should be augmented with the assigned group(s) (i.e., with an index of each group). The division of \ac{POI} types can be pre-configured and be known to all participating nodes (that can map a specific \ac{POI} type to a specific group). Therefore, when a querying node receives a beacon signed under a \ac{PC}, it knows, by looking at the group index in the \ac{PC}, whether the beacon sender can respond to its query.

\section{Security and Privacy Evaluation}
\label{sec:evaluation}

In this section, we analyze the achieved security and privacy properties with our scheme. We first provide a qualitative analysis as per the requirements in Sec.~\ref{sec:problem}. Then, we provide an extensive quantitative analysis on protecting node privacy against the \ac{LBS} server and honest-but-curious nodes and on resilience to malicious nodes.

\subsection{Qualitative Analysis}
\label{sec:qualitative}

\textbf{Authentication and integrity:} Entity and message authentication and message integrity are achieved thanks to message digital signature verifications. Message (e.g., beacon) timestamps prevent replays of old messages. A peer query is encrypted and bound to a specific serving node, thus it cannot be meaningfully replayed and any other serving node would fast reject it. More important, the query and response identifiers can trivially allow the (given) same serving (querying) node to reject and not serve (accept) them, at the expense of modest local memory.

\textbf{Accountability:} In spite of the pseudo-/ano-nymity, upon detection of misbehavior reported to the \ac{RA}, the \ac{RA} can reveal the actual, long-term identity of the reported misbehaving node, through pseudonym resolution and possibly evict the node from the system. The \ac{RA} can verify the truthfulness of the reported misbehavior with the help of the \ac{LBS} server.

\textbf{Confidentiality:} Communication among nodes and infrastructure entities (\ac{LTCA}, \ac{PCA} and \ac{LBS}) is kept confidential by using public key cryptography and (symmetric) session keys.

\textbf{Sybil-resilience:} The \ac{LTCA} and the \ac{PCA} issue $ticket$s and pseudonyms with non-overlapping lifetimes, ensuring a node is equipped with only one valid pseudonym at any point in time.

\textbf{Response validation:} Our scheme prevents malicious nodes from providing false information, using probabilistic serving node assignment, cross-checking and proactive \ac{LBS} checking. First, each node is chosen as serving node probabilistically by the \ac{PCA} and such role is explicitly bound to the provided pseudonyms. Therefore, a malicious node has no control on whether it becomes a serving node, which is the only possibility to provide false information to its peers. For a non-serving node, the waiting time to be assigned as serving node follows a geometric distribution with parameter $Pr_{serve}$. Thus, the expected waiting time is $\frac{1-Pr_{serve}}{Pr_{serve}} \cdot \Gamma$. For example, when $Pr_{serve}=0.05$ and $T_{serve} = \Gamma$, an adversary has to wait for $19*\Gamma$ on average before being selected as a serving node. Second, a malicious serving node has to be chosen by the querying node; again, this selection is not under the control of any malicious node. There may exist multiple serving nodes around a querying node that takes the initiative to choose one or multiple ($N$) serving node(s). The use of multiple (redundant) serving nodes can reveal a malicious node by cross-checking their responses, given all (benign) serving nodes have the same \ac{POI} data for the region. Moreover, with the help of the proactive \ac{LBS} checking mechanism, benign nodes can protect themselves against malicious nodes when only one peer response is received. Such proactive check does not expose long-term or short-term identities of the querying node, because the authenticator in a peer response is a signature generated by the serving node. We provide quantitative evaluation of resilience of our scheme in Sec.~\ref{sec:quantitative}.

\textbf{Exposure reduction:} Our scheme reduces the exposure to the \ac{LBS} server by sharing information among the peers, as its most closely related predecessors~\cite{shokri2014hiding,jin2015resilient} did. In addition, through the controlled selection of serving nodes and encrypted \ac{P2P} communication, only the selected serving nodes learn queries. Queries by one node, $U_{q}$, to a certain serving node, $U_{s}$, can be synthetically linked only while $U_{q}$ uses the same pseudonym. Moreover, $U_{q}$ could choose among several $U_{s_{i}}$ nodes even within a given region for successive queries. Mobility of all nodes (serving or not), short-lived pseudonyms, encryption of query-response process, and rotating assignment of serving nodes minimize exposure to any curious serving node acting alone.

\begin{table}[htp!]
	\footnotesize
	\caption{Linked queries for different collusion cases}
	\centering
		\begin{tabular}{| l | l | l | l |}
			\hline 
			\textbf{\emph{Case}} &	\textbf{\emph{Linked queries}} & \multicolumn{2}{l|}{\textbf{\emph{Colluding entities}}} \\\hline
			$C_1$ & Same $Id_{PC}$ & \multicolumn{2}{p{4.5cm}|}{\raggedright  No collusion with CA} \\\hline
			$C_2$ & Same $Id_{ticket}$ & \multicolumn{2}{p{4.5cm}|}{\raggedright Collusion with \ac{PCA}} \\\hline
			$C_3$ & Same $Id_{LTC}$ & \multicolumn{2}{p{4.5cm}|}{\raggedright Collusion with \ac{PCA} and \ac{LTCA}} \\\hline
	\end{tabular}
	\label{table:exposure}
\end{table}

However, colluding serving nodes could merge the queries they received, attempting to link them the same way the curious \ac{LBS} server would do. Moreover, collusion with the \acp{CA} could allow further linking of queries by linking pseudonyms of the same node. Given the honest-but-curious assumption for the security infrastructure, the \ac{RA} is not able to initiate a pseudonym resolution without any reported and confirmed misbehavior (through a check with the \ac{LBS} server). Table~\ref{table:exposure} shows the queries that can be linked for different collusion cases. We provide quantitative evaluation on node exposure for different collusion cases in Sec.~\ref{sec:quantitative}. We refer to~\cite{khodaei2014towards,khodaei2018secmace} for the information disclosed to honest-but-curious \ac{PKI} entities.

The \ac{PCA}, when colluding with honest-but-curious nodes, could assign the role of serving nodes to the colluding nodes or increase $Pr_{serve}$ within a reasonable range (e.g., from $0.06$ to $0.08$), attempting to collect more node queries through the colluding nodes. However, this deviates from the honest-but-curious assumption for the security infrastructure, because the honest \ac{PCA} should randomly assign serving nodes with a given $Pr_{serve}$ (defined by system policy).

\textbf{Jamming:} An attacker could jam beacons from benign serving nodes to hinder the provision of correct responses. This is possible because, by default, beacons from a specific serving node are predictable, given a specific $T_{beacon}$ (e.g., observed based on previously overheard beacons or specified by system policy). Jamming, i.e., ``erasing'' benign serving node beacons makes it harder for malicious serving nodes to be detected through response cross-checks. This can be mitigated with a random (e.g., uniformly distributed) beaconing interval, so that upcoming beacons from a specific serving node are not predictable. We show that a random beacon interval has negligible effect on the performance of our scheme in Sec.~\ref{sec:quantitative}.

\subsection{Quantitative Analysis}
\label{sec:quantitative}

We further evaluate our scheme through simulations, to show its effectiveness in a real-world scenario. \emph{Exposure} to the \ac{LBS} server and honest-but-curious nodes is quantitatively evaluated through two metrics: peer hit ratio and exposure degree. The peer hit ratio shows the ratio of node queries that are hidden from the \ac{LBS} server, while the exposure degree indicates the accuracy of reconstructed user trajectories based on exposed user locations. The resiliency of our scheme, in the presence of malicious nodes, is also evaluated, based on the ratio of affected node queries by (forged) false responses. We also compare, thorough the two quantitative privacy measurements, with MobiCrowd~\cite{shokri2014hiding}.

We find that with $6\%$ of the peer nodes acting as serving nodes (i.e., $Pr_{serve} = 0.06$), around $50\%$ of the queries can be hidden from the \ac{LBS} server. At the same time, the exposure degree to the \ac{LBS} server is significantly decreased, from around $0.6$ to $0.16$. This is achieved thanks to the aforementioned low ratio of query exposure to the \ac{LBS} server and the use of pseudonymous authentication with relatively short-lived pseudonyms. Even in the presence of a high ratio (e.g., $20\%$) of colluding honest-but-curious nodes, the exposure degree is kept very low thanks to the encrypted peer query-response process. From a different viewpoint, a high ratio (e.g., $20\%$) of colluding malicious nodes (i.e., malicious nodes acting as serving nodes providing the same false response to a node query, in an attempt to defeat the response validation) can affect only less than $1.5\%$ of the node queries (submitted to nearby serving nodes); which can be further mitigated or even eliminated by proactive \ac{LBS} checking mechanism with different $Pr_{check}$. Moreover, when achieving a similar peer hit ratio to that of MobiCrowd, we show our scheme achieve a much lower exposure degree to honest-but-curious nodes and impose a much lower communication overhead.

\begin{table}[htp!]
	\caption{Simulation Parameters (\textbf{\emph{Bold}} for Default Settings)}
	\centering
	\footnotesize
	\begin{minipage}[t]{.4\textwidth}
	
	\begin{tabular}{ l | c }
		\hline
		$L$ & 1, \textbf{\emph{2}}, 3  $km$ \\\hline
		$T_{serve}$ & \textbf{\emph{10}} $min$ \\\hline
		$\Gamma$ & \textbf{\emph{10}} $min$ \\\hline
		$\tau$ & 1, 2.5, \textbf{\emph{5}}, 10 $min$ \\\hline
		$T_{poi}$ & \textbf{\emph{20}}, 30, 40, 50, 60 $min$ \\\hline
		$T_{wait}$ & 5, 10, 20, 30, \textbf{\emph{60}}, 90, 120 $s$ \\\hline
		$T_{beacon}$ & 5, \textbf{\emph{10}}, 15, 20, uniform(5, 15) $s$ \\\hline
	\end{tabular}
\end{minipage}%
\begin{minipage}[t]{.4\textwidth}
	\begin{tabular}{ l | c }
		\hline
		$T_{query}$ & \textbf{\emph{3}} $min$ \\\hline
		$N$ & 2, \textbf{\emph{3}} \\\hline
		$Pr_{serve}$ & 0.02, 0.04, \textbf{\emph{0.06}}, 0.08, 0.1, 0.12, 0.18 \\\hline
		$Ratio_{adv}$ & 0.05, 0.1, \textbf{\emph{0.2}}, 0.3, 0.4, 0.5 \\\hline
		$Pr_{check}$ & \textbf{\emph{0}}, 0.1, 0.2, 0.3, 0.5, 0.7, 1 \\\hline
		$G$ & \textbf{\emph{1}}, 2, 3 \\\hline
		$Ratio_{coop}$ & \textbf{\emph{0.5}}, 1 (MobiCrowd Only) \\\hline
	\end{tabular}
\end{minipage}%
	\renewcommand{\arraystretch}{1}
	\label{table:parameter}

\end{table}

\textbf{Simulation setup:} We simulate our scheme with OmNET++\footnote{omnetpp.org} and the TraCI mobility interface in Veins~\cite{sommer2011bidirectionally}, connected to the SUMO~\cite{SUMO2012} traffic simulator. We use the \ac{LuST} internal mobility scenario (i.e., with source, destination, or both points internal to the city)~\cite{codeca2015luxembourg} and the TAPAS Cologne mobility scenario~\cite{uppoor2014generation}. We assume a penetration ratio of 40\%, i.e., 40\% of the mobile nodes use \acp{LBS} and participate in our collaborative scheme. For each mobility scenario, we use the traces for the 12:30 pm -- 2:00 pm period and use the 1:00 pm -- 2:00 pm part for the evaluation. We assume ideal wireless connection for ad-hoc node-to-node communication, with a range of 200 $m$. The results are averaged over 5 seeded simulation runs. In the simulation, we divide the whole area into equally sized regions, with a region size $L \times L$. Table~\ref{table:parameter} shows the parameters of our simulation, bold values indicating default simulation settings, where $T_{query}$ is the query interval of a node. For example, if $Pr_{serve}$ is the parameter we evaluate, then we set the rest of the parameters as: $L = 2\ km$, $T_{serve} = \Gamma = 10$ $min$, $\tau = 5$ $min$, $T_{poi} = 20$ $min$, $T_{wait} = 60$ $s$, $T_{beacon} = 10$ $s$, $T_{query} = 3$ $min$, $N = 3$, and $G=1$. For the evaluation of exposure to honest-but-curious nodes and resilience to malicious nodes, we set $Ratio_{adv} = 20\%$ and $Pr_{check} = 0$ by default.


For MobiCrowd, we assume a querying node broadcasts its query every $10\ s$ and consider collaboration/cooperation ratio ($Ratio_{coop}$) values of $0.5$ and $1$. We set default $Ratio_{coop}$ to 0.5 for MobiCrowd: a node would respond with a probability of $0.5$ to peer queries. In our scheme, we do not consider $Ratio_{coop}$, because $Pr_{serve}$ can be considered as a combination of collaboration ratio and actual serving node assignment probability. For example, an actual serving node assignment probability of $0.12$ and a collaboration ratio of $0.5$ could result in the equivalent $Pr_{serve} = 0.06$.

\begin{figure*}[tp]
	\begin{subfigure}[b]{.24\columnwidth}
		\includegraphics[width=\columnwidth]{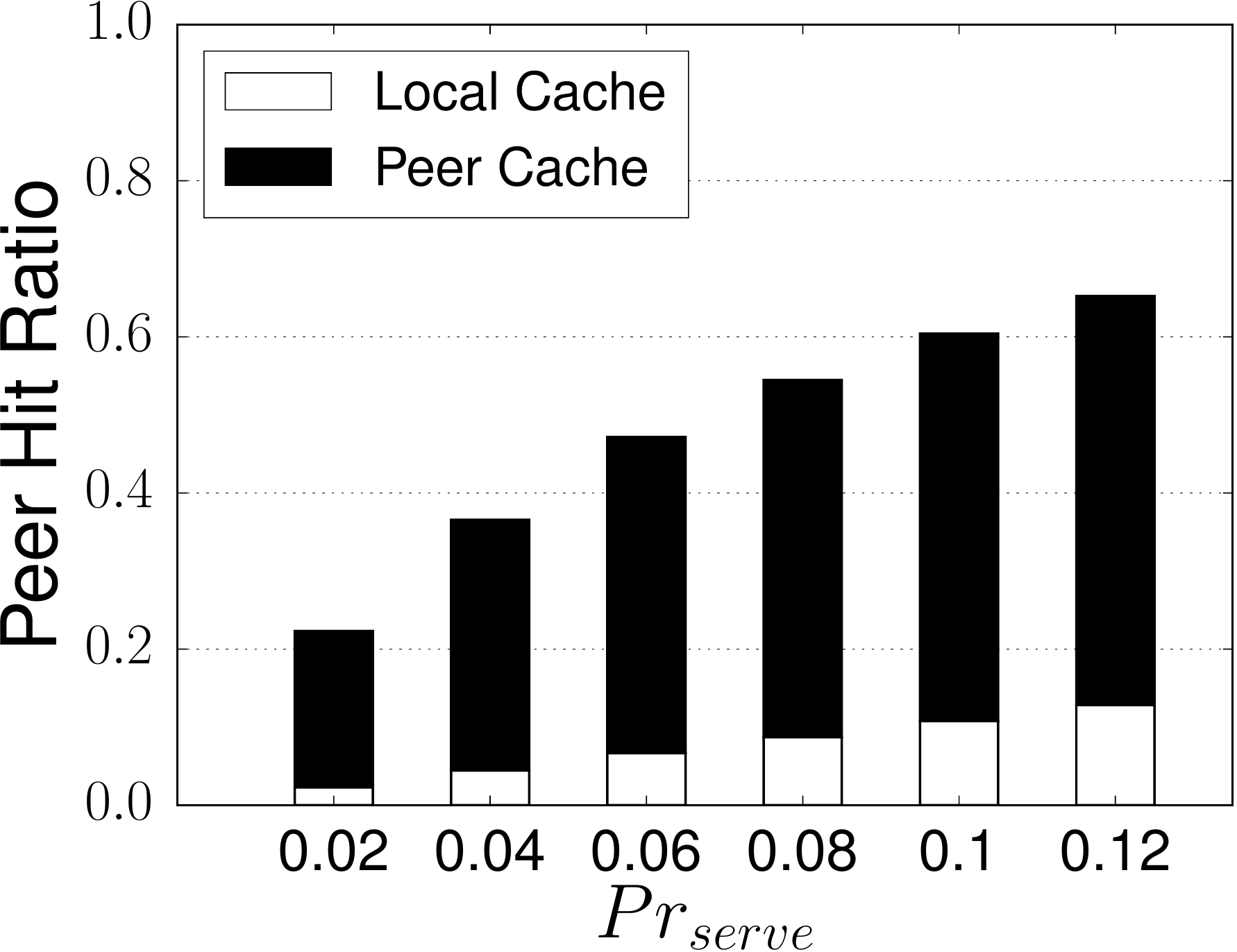}%
		\caption{}
		\label{subfig_chr_prserve_lust_1km}
	\end{subfigure}
	\begin{subfigure}[b]{.24\columnwidth}
		\includegraphics[width=\columnwidth]{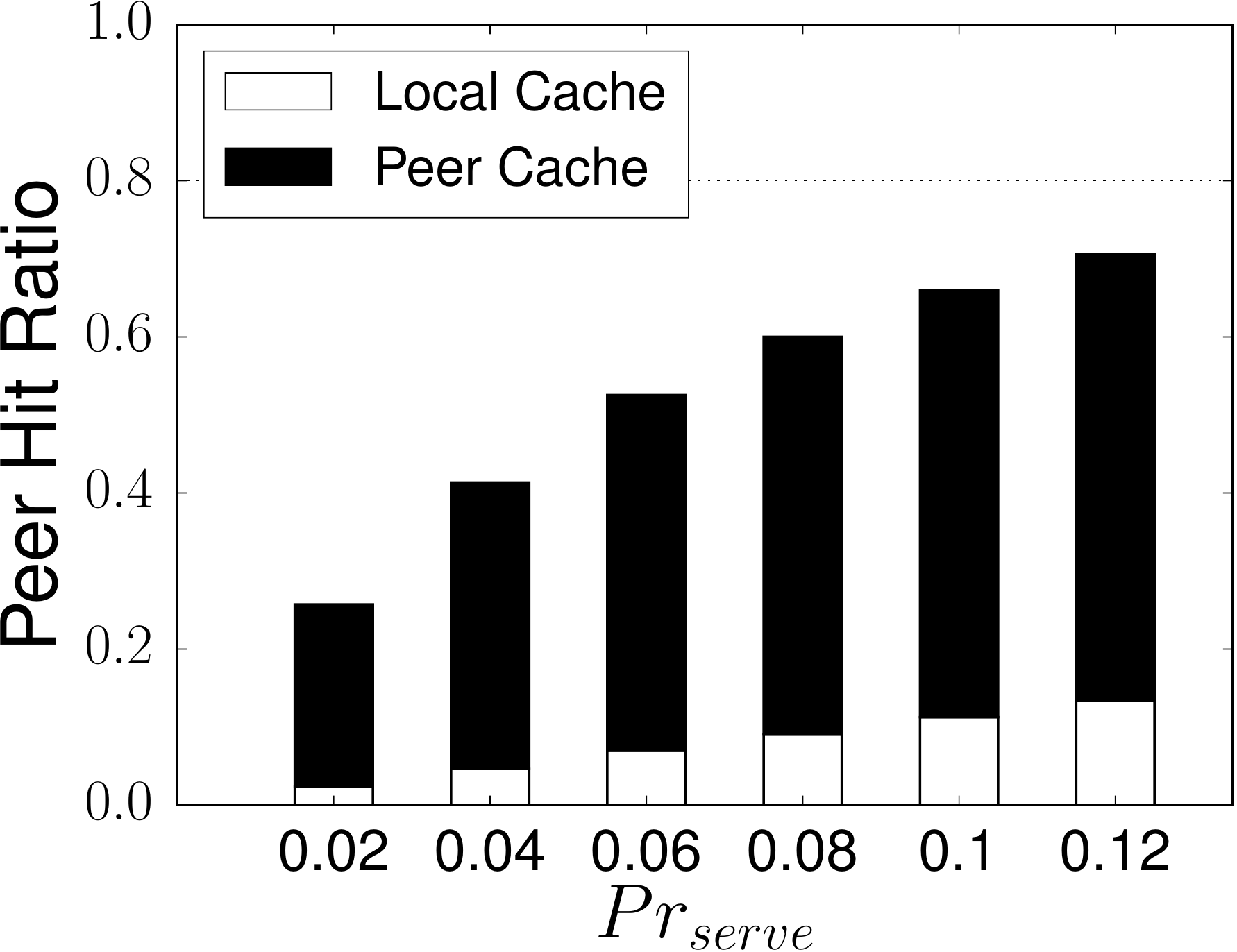}%
		\caption{}
		\label{subfig_chr_prserve_lust_2km}
	\end{subfigure}
	\begin{subfigure}[b]{.24\columnwidth}
		\includegraphics[width=\columnwidth]{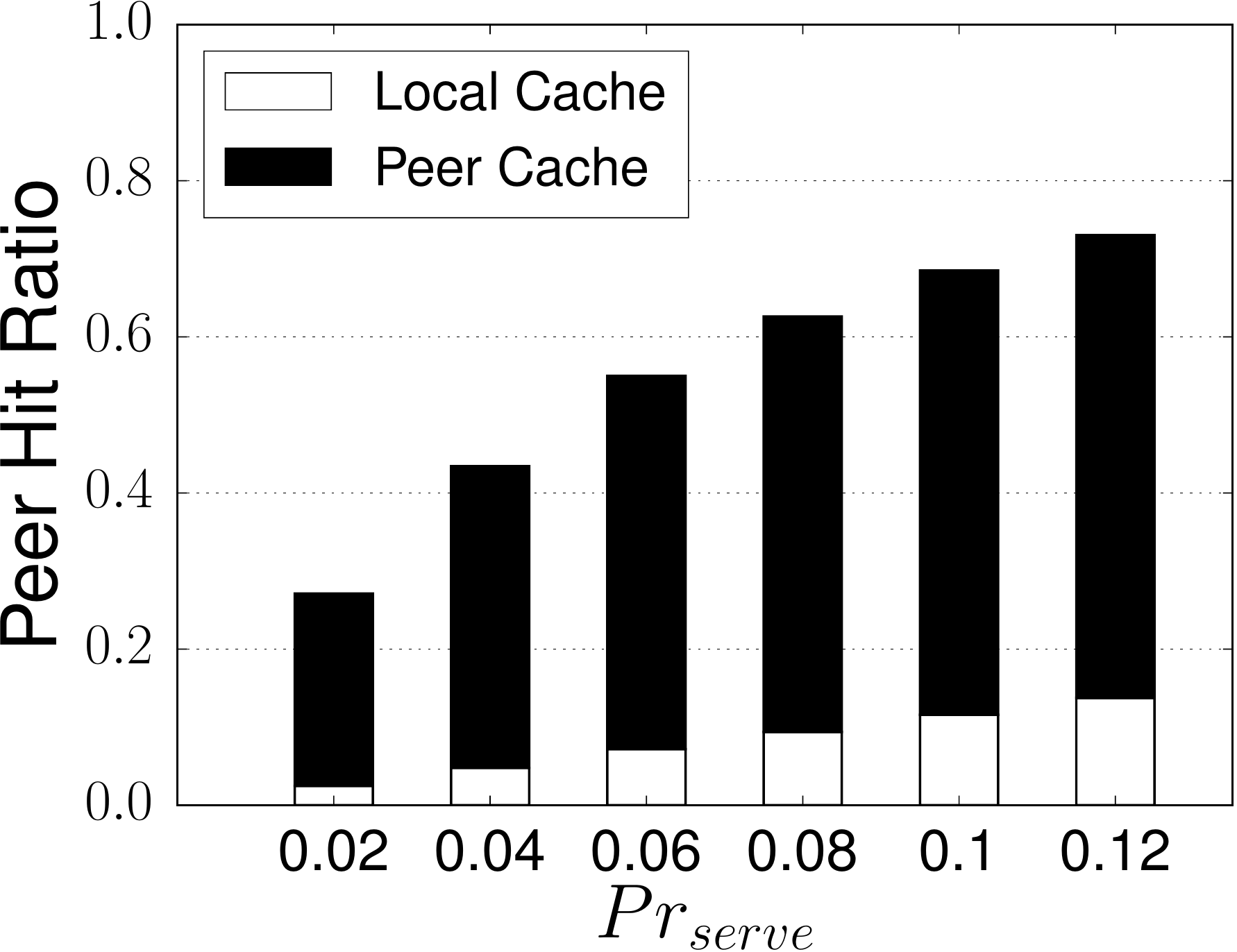}%
		\caption{}
		\label{subfig_chr_prserve_lust_3km}
	\end{subfigure}
	\begin{subfigure}[b]{.24\columnwidth}
		\includegraphics[width=\columnwidth]{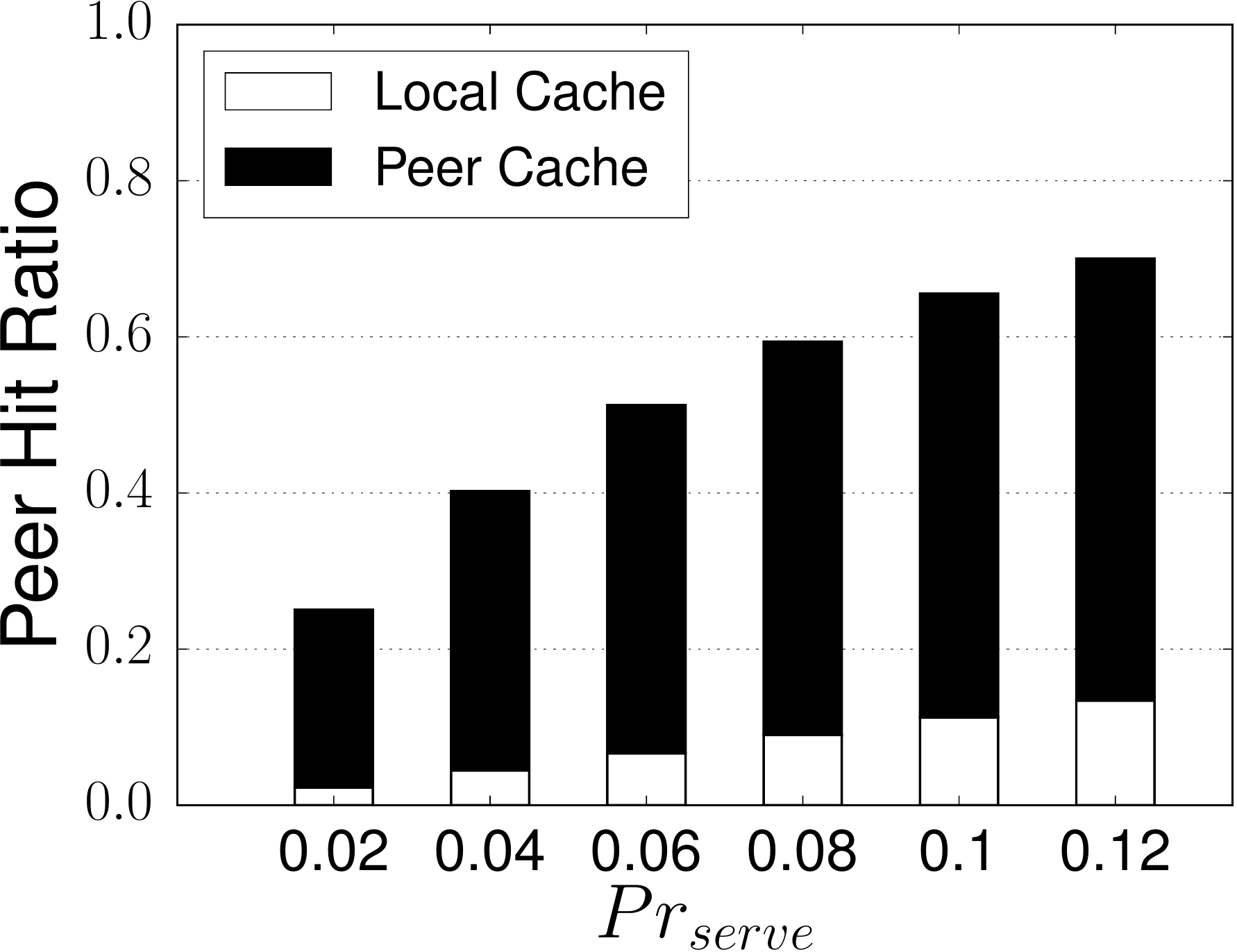}%
		\caption{}
		\label{subfig_chr_prserve_lust_rnd}
	\end{subfigure}
	\caption{LuST: Peer hit ratio as a function of $Pr_{serve}$ with  $L =$ (\subref{subfig_chr_prserve_lust_1km}) $1$ $km$, (\subref{subfig_chr_prserve_lust_2km}) $2$ $km$, (\subref{subfig_chr_prserve_lust_3km}) $3$ $km$, and (\subref{subfig_chr_prserve_lust_rnd}) $T_{beacon} \sim uniform(5, 15) s$.}
	\label{fig_chr_prserve_lust}
\end{figure*}

\textbf{Peer Hit Ratio:} The peer hit ratio reflects the ratio of hidden node queries from the honest-but-curious \ac{LBS} server. It is defined as the ratio of the node queries for which a response is obtained using local or peer caches (of serving nodes), while the remaining is responded by the \ac{LBS} server itself. \cref{fig_chr_prserve_lust} shows the peer hit ratio as a function of $Pr_{serve}$ for the LuST scenario. \cref{subfig_chr_prserve_lust_1km,subfig_chr_prserve_lust_2km,subfig_chr_prserve_lust_3km} show peer hit ratio with $L = 1, 2, 3$ $km$ respectively and the rest of the parameters having the default values in Table~\ref{table:parameter}. Peer hit ratios improve with increasing $Pr_{serve}$. For example, we see a significant increase in peer hit ratio when $Pr_{serve}$ increases from $0.02$ to $0.04$, while such improvement becomes moderate for high $Pr_{serve}$ values (e.g., from $0.1$ to $0.12$). Thus, modest $Pr_{serve}$ (e.g., $0.06$ or $0.08$) is enough to hide a significant amount (e.g., more than $50\%$) of queries from the \ac{LBS} server. We see a slight increase in peer hit ratio with larger region sizes, because more serving nodes exist in each region. Moreover, with a small region size, more queries are initiated at the borders of regions, thus querying nodes are more likely to cross to other regions while waiting for a response. Fig.~\ref{subfig_chr_prserve_lust_rnd} shows that a uniformly distributed $T_{beacon}$ does not affect the peer hit ratio; while it protects benign serving nodes from being jammed (Sec.~\ref{sec:qualitative}).

\begin{figure*}[t]
	\begin{subfigure}[b]{.24\columnwidth}
		\includegraphics[width=\columnwidth]{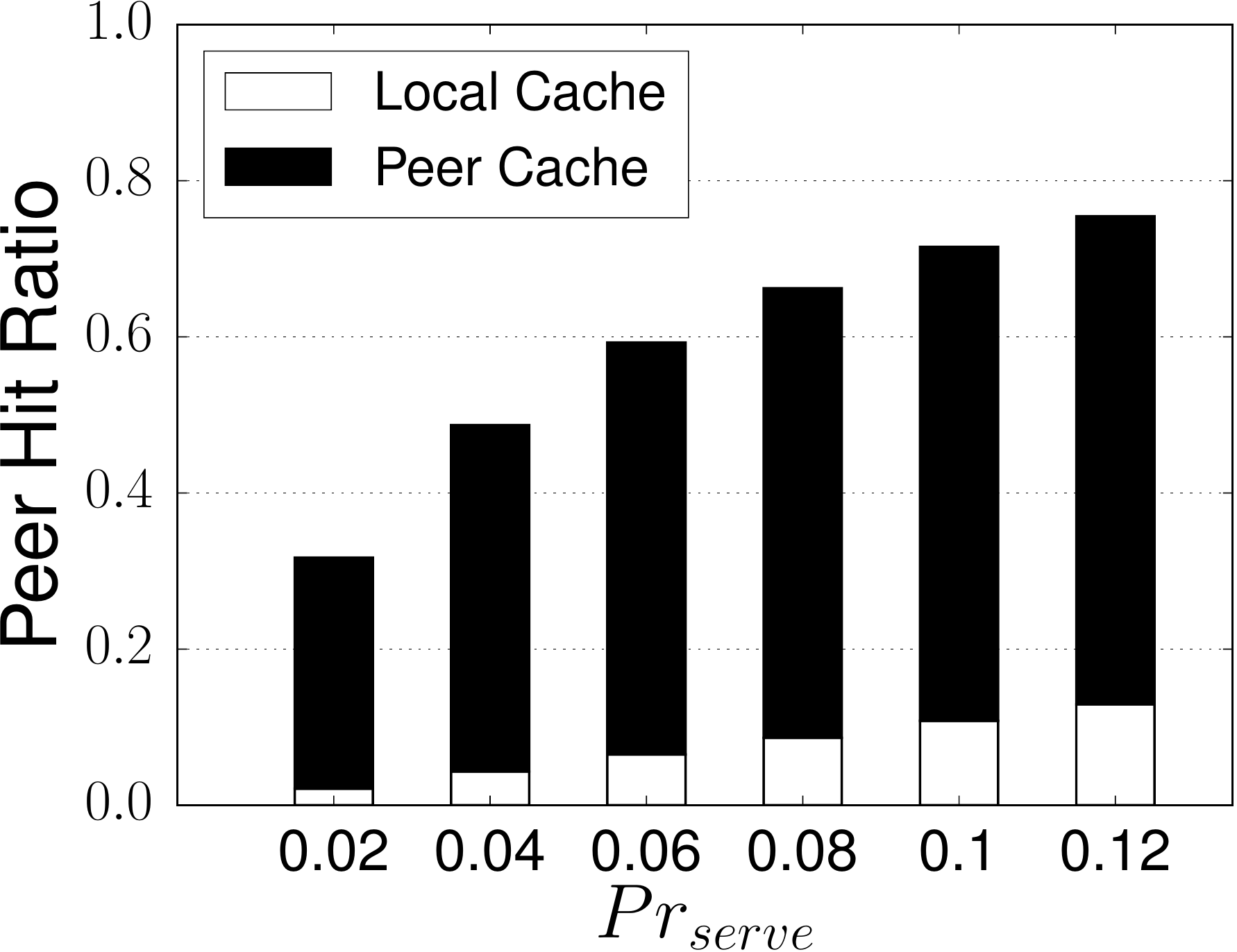}%
		\caption{}
		\label{subfig_chr_prserve_koln_1km}
	\end{subfigure}
	\begin{subfigure}[b]{.24\columnwidth}
		\includegraphics[width=\columnwidth]{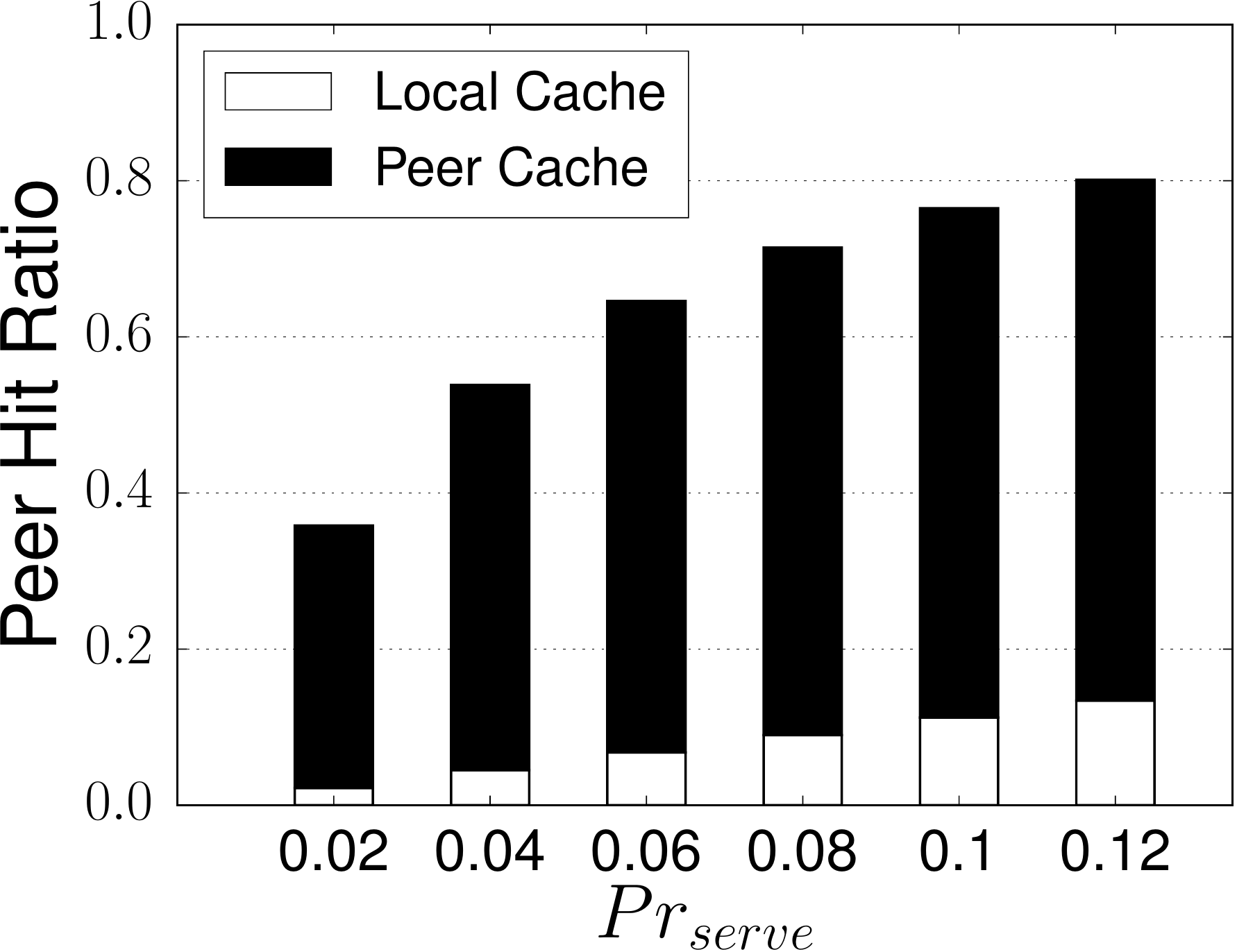}%
		\caption{}
		\label{subfig_chr_prserve_koln_2km}
	\end{subfigure}
	\begin{subfigure}[b]{.24\columnwidth}
		\includegraphics[width=\columnwidth]{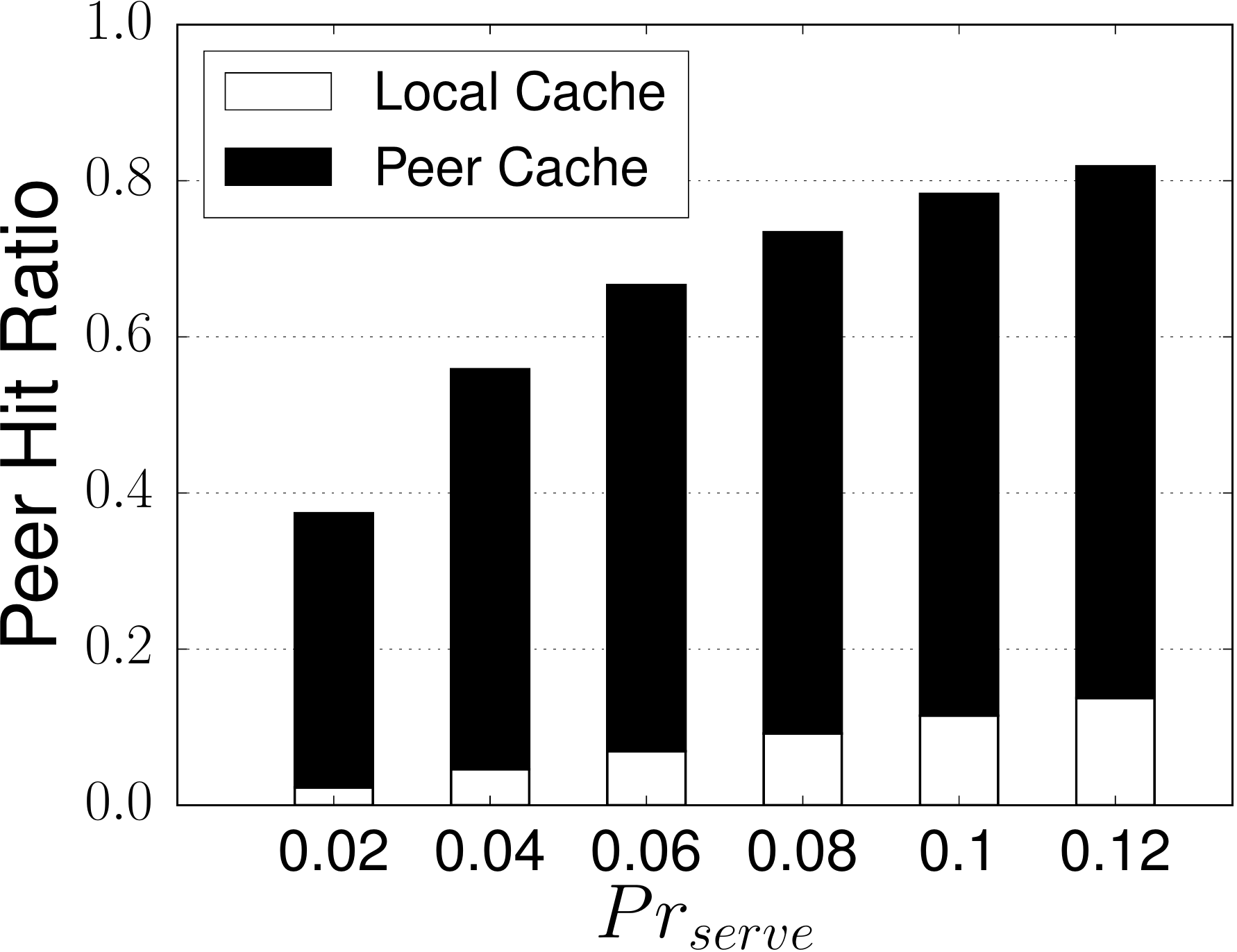}%
		\caption{}
		\label{subfig_chr_prserve_koln_3km}
	\end{subfigure}
	\caption{TAPASCologne: Peer hit ratio as a function of $Pr_{serve}$ with $L =$ (\subref{subfig_chr_prserve_koln_1km}) 1 $km$, (\subref{subfig_chr_prserve_lust_2km}) $2$ $km$, and (\subref{subfig_chr_prserve_lust_3km}) $3$ $km$.}
	\label{fig_chr_prserve_koln}
\end{figure*}

\begin{figure}[htp!]
	\begin{subfigure}[b]{0.24\columnwidth}
		\includegraphics[width=\columnwidth]{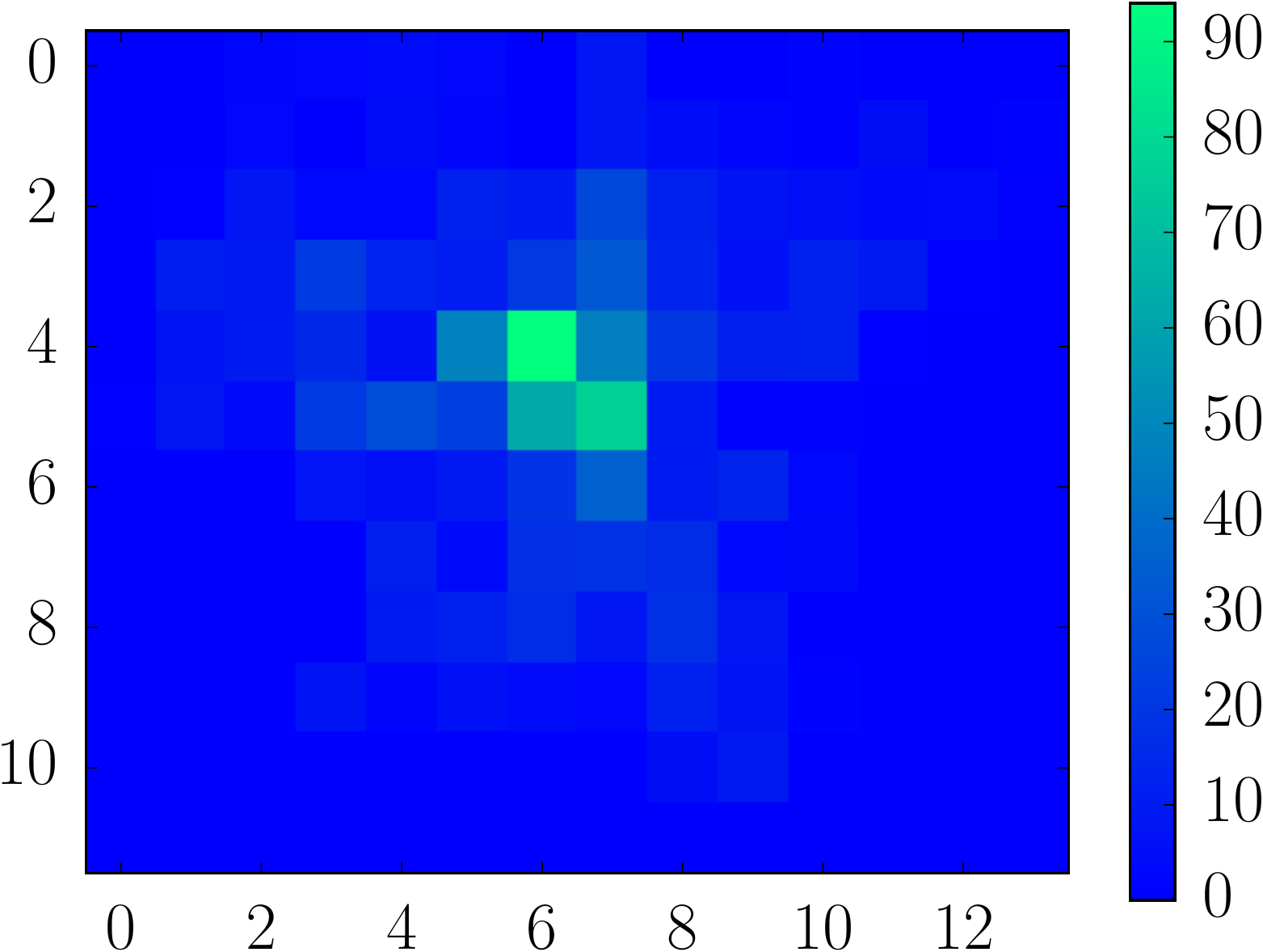}
		\caption{}
		\label{subfig_density_lust}
	\end{subfigure}
	\begin{subfigure}[b]{0.24\columnwidth}
		\includegraphics[width=\columnwidth]{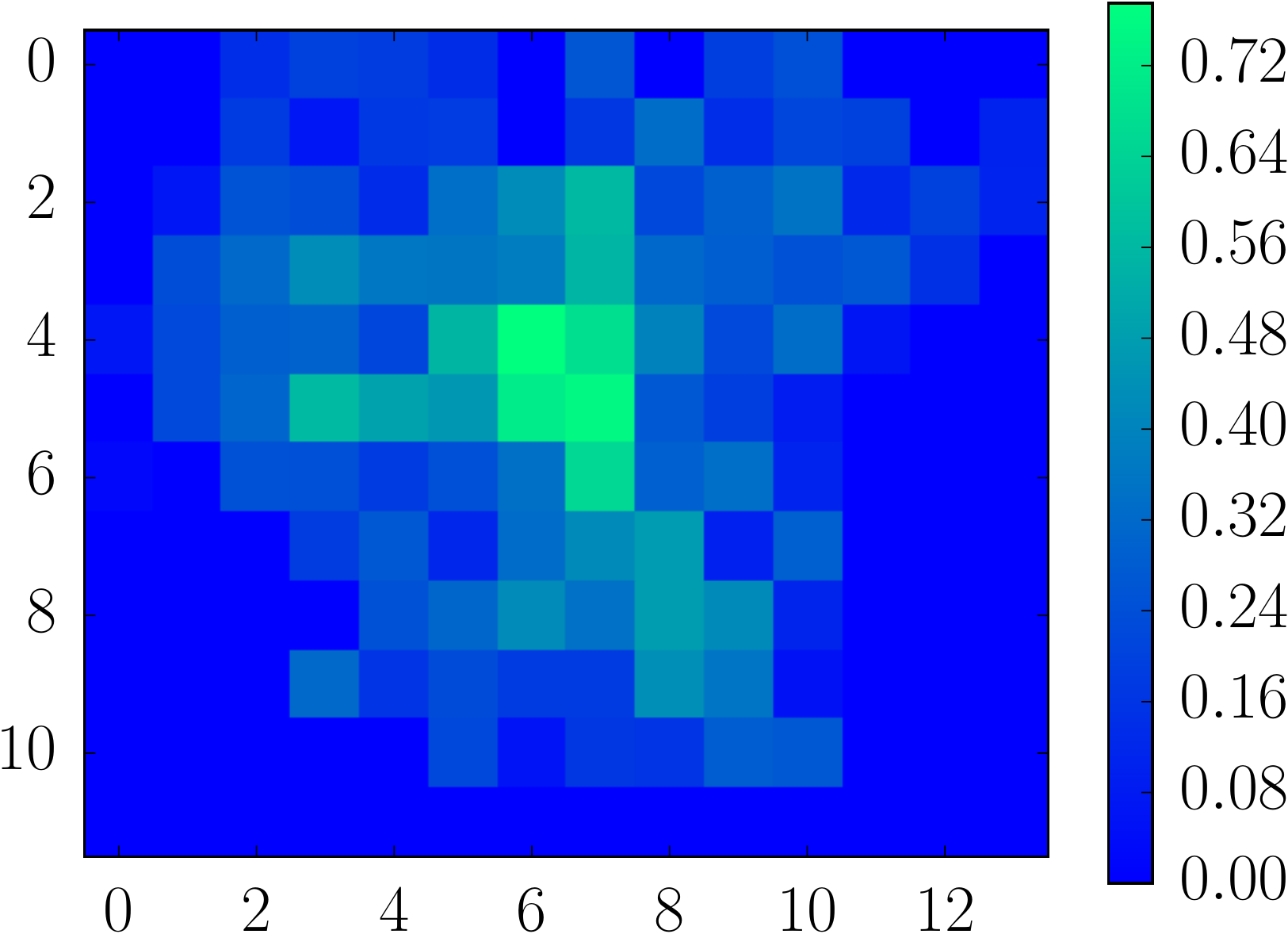}%
		\caption{}%
		\label{subfig_bygrid_1km_lust}
	\end{subfigure}
	\begin{subfigure}[b]{0.24\columnwidth}
		\includegraphics[width=\columnwidth]{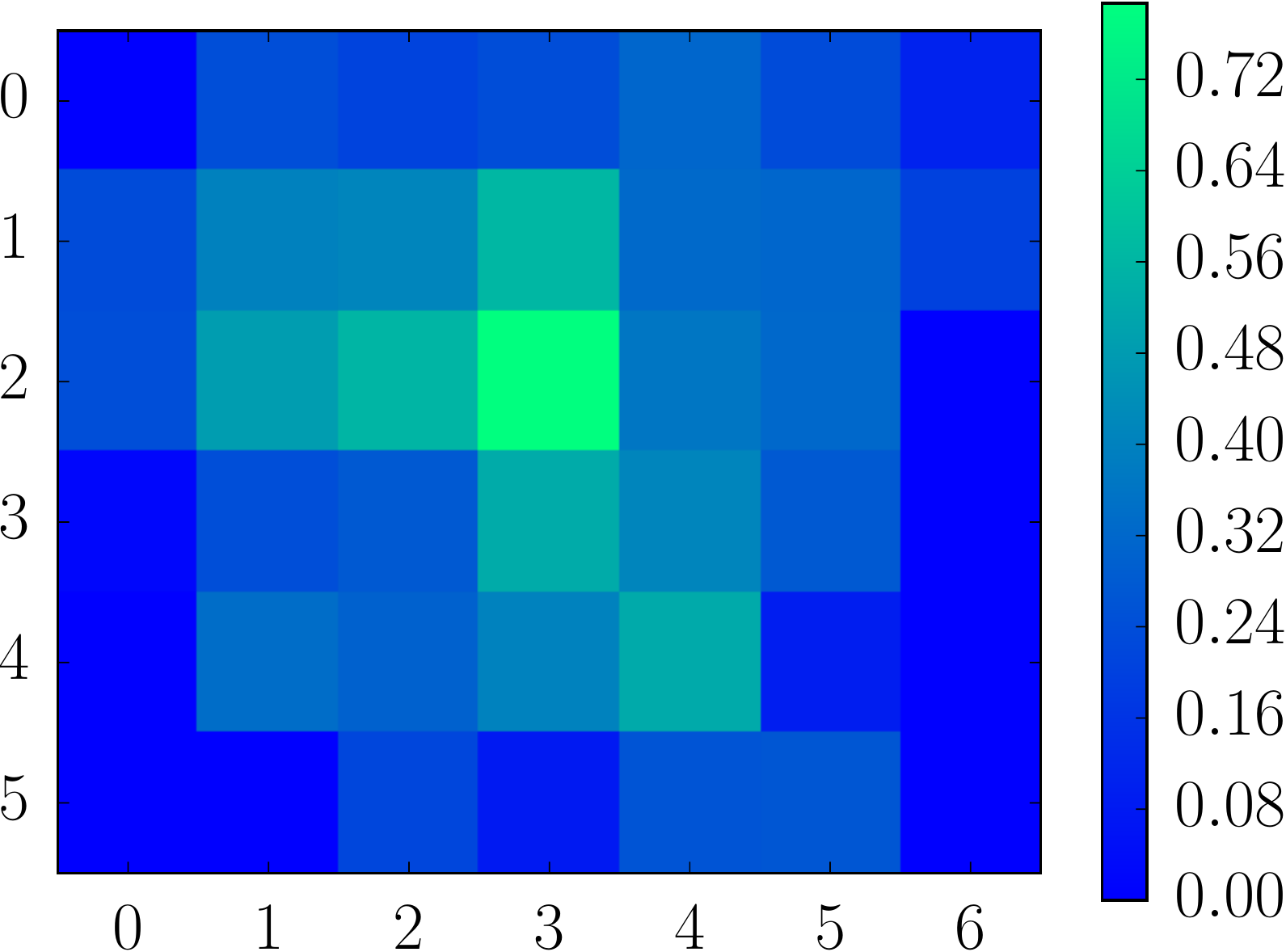}
		\caption{}
		\label{subfig_bygrid_2km_lust}
	\end{subfigure}
	\begin{subfigure}[b]{0.24\columnwidth}
		\includegraphics[width=\columnwidth]{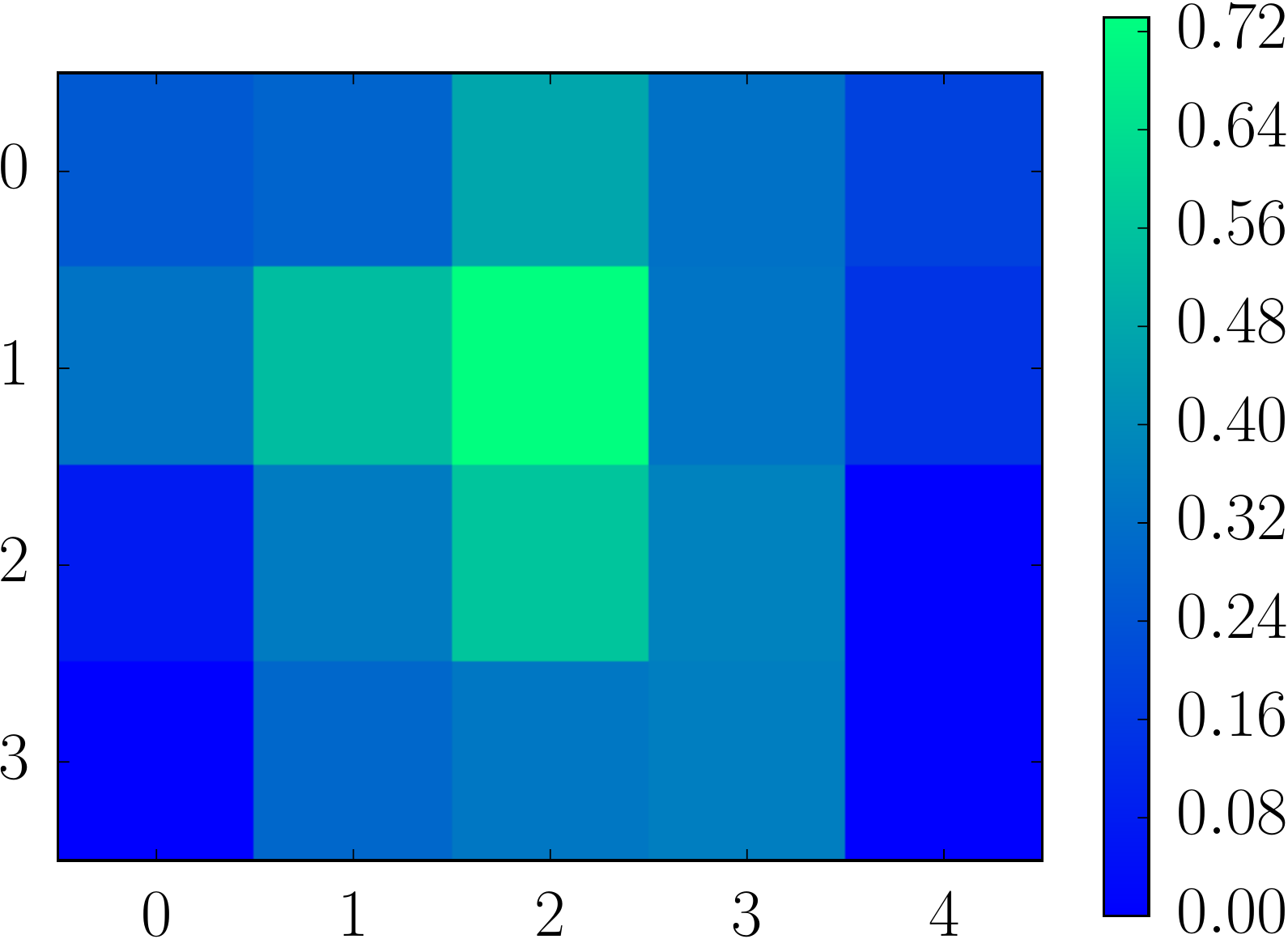}%
		\caption{}%
		\label{subfig_bygrid_3km_lust}
	\end{subfigure}
	\caption{LuST: (\subref{subfig_density_lust}) Node density for each region at 1 pm with $L = 1$ $km$. Peer hit ratio for each region with (\subref{subfig_bygrid_1km_lust}) $L = 1$ $km$, (\subref{subfig_bygrid_2km_lust}) $L = 2$ $km$, and (\subref{subfig_bygrid_3km_lust}) $L = 3$ $km$.}
	\label{fig_density_lust}
\end{figure}

\begin{figure}[h]
	\begin{subfigure}[b]{0.24\columnwidth}
		\includegraphics[width=\columnwidth]{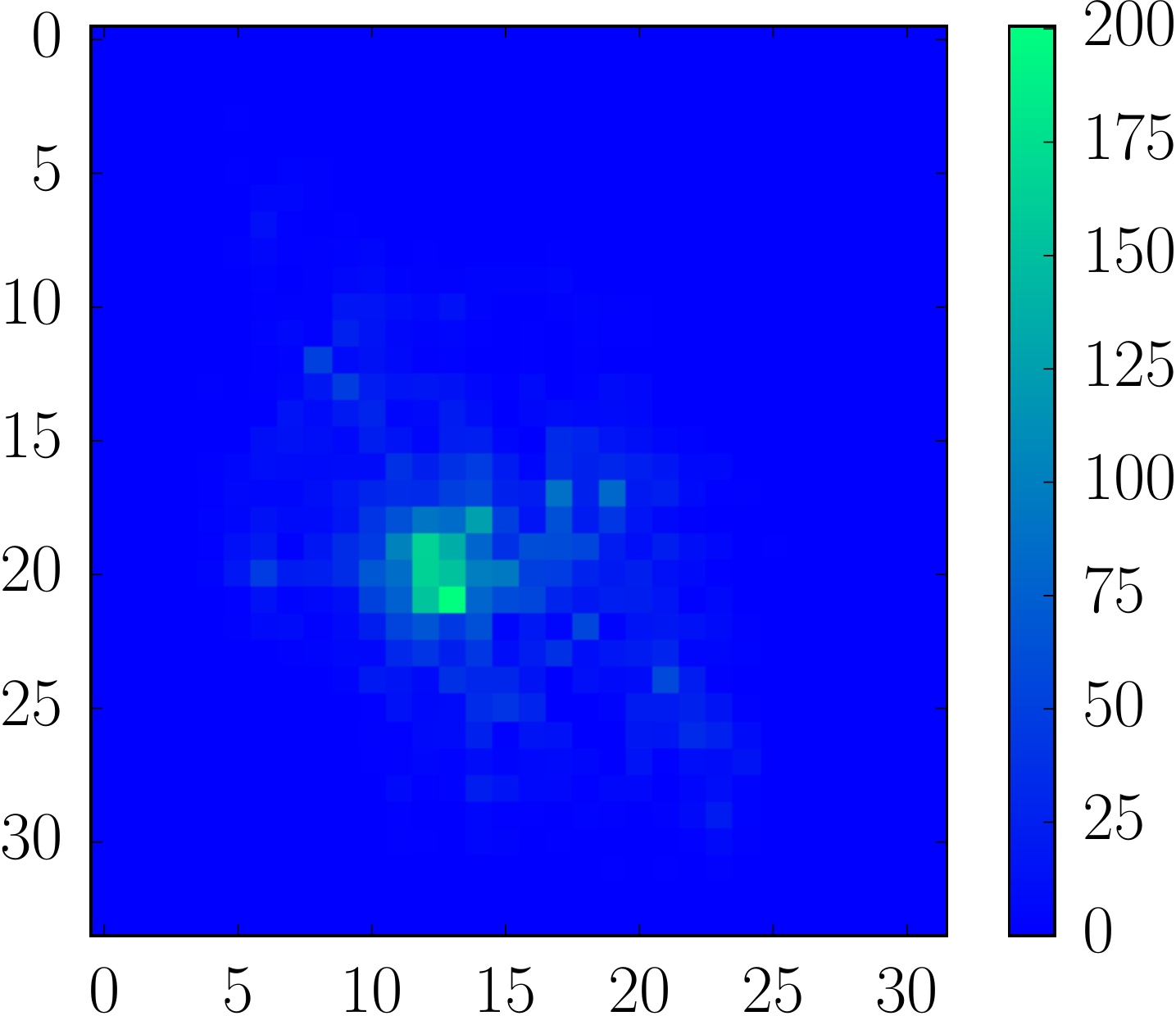}
		\caption{}
		\label{subfig_density_koln}
	\end{subfigure}
	\begin{subfigure}[b]{0.24\columnwidth}
		\includegraphics[width=\columnwidth]{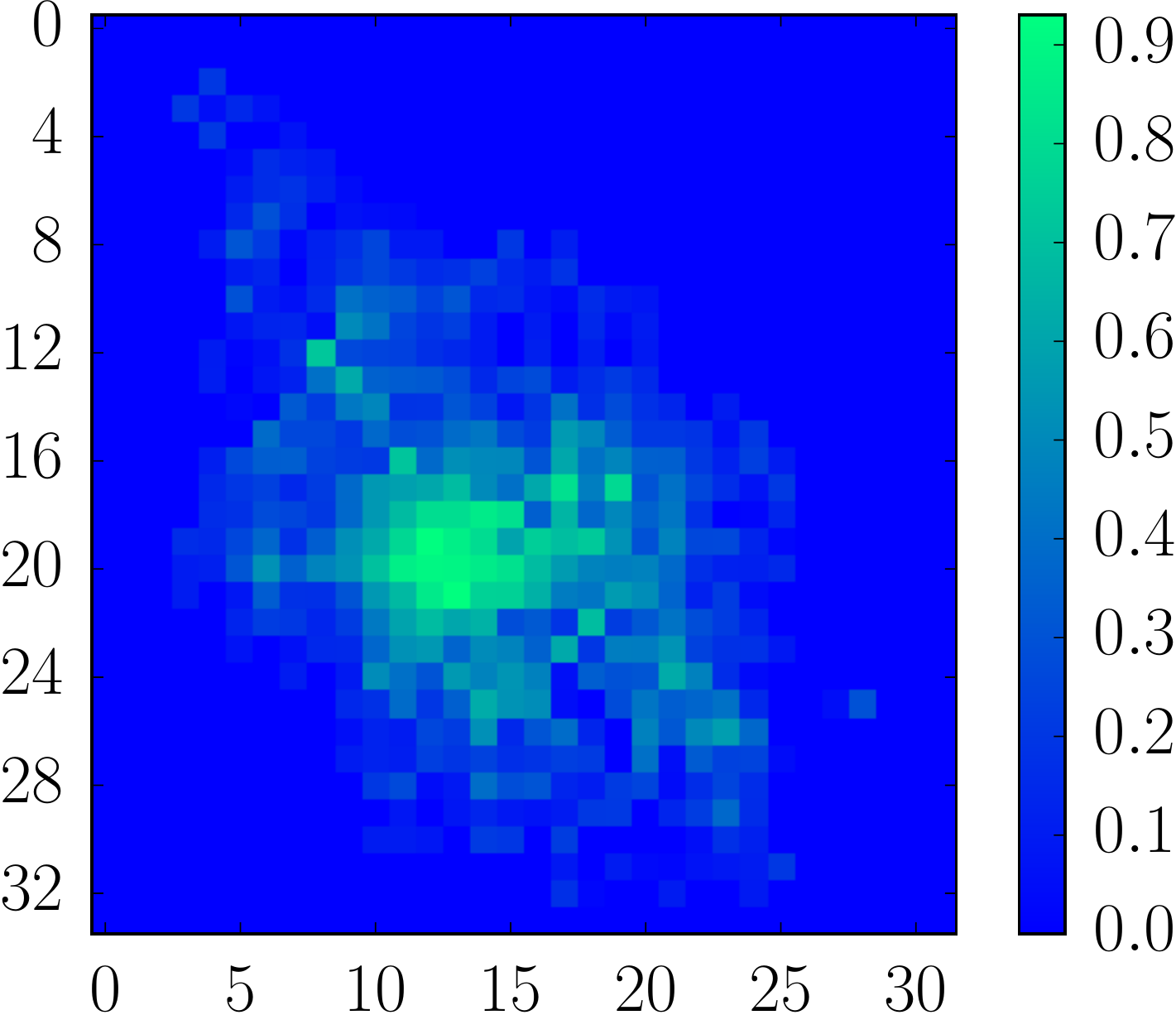}%
		\caption{}%
		\label{subfig_bygrid_1km_koln}
	\end{subfigure}
	\begin{subfigure}[b]{0.24\columnwidth}
		\includegraphics[width=\columnwidth]{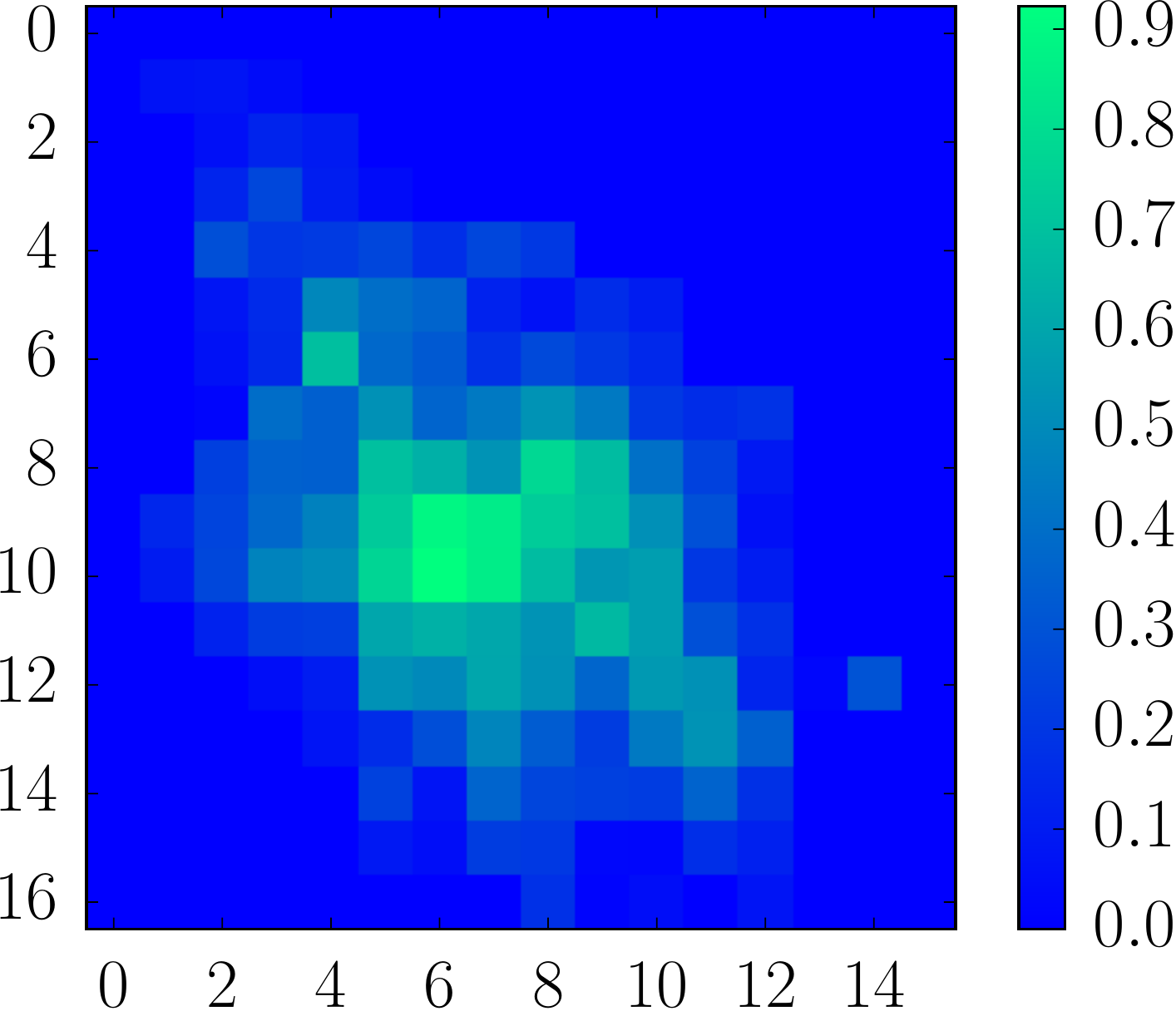}
		\caption{}
		\label{subfig_bygrid_2km_koln}
	\end{subfigure}
	\begin{subfigure}[b]{0.24\columnwidth}
		\includegraphics[width=\columnwidth]{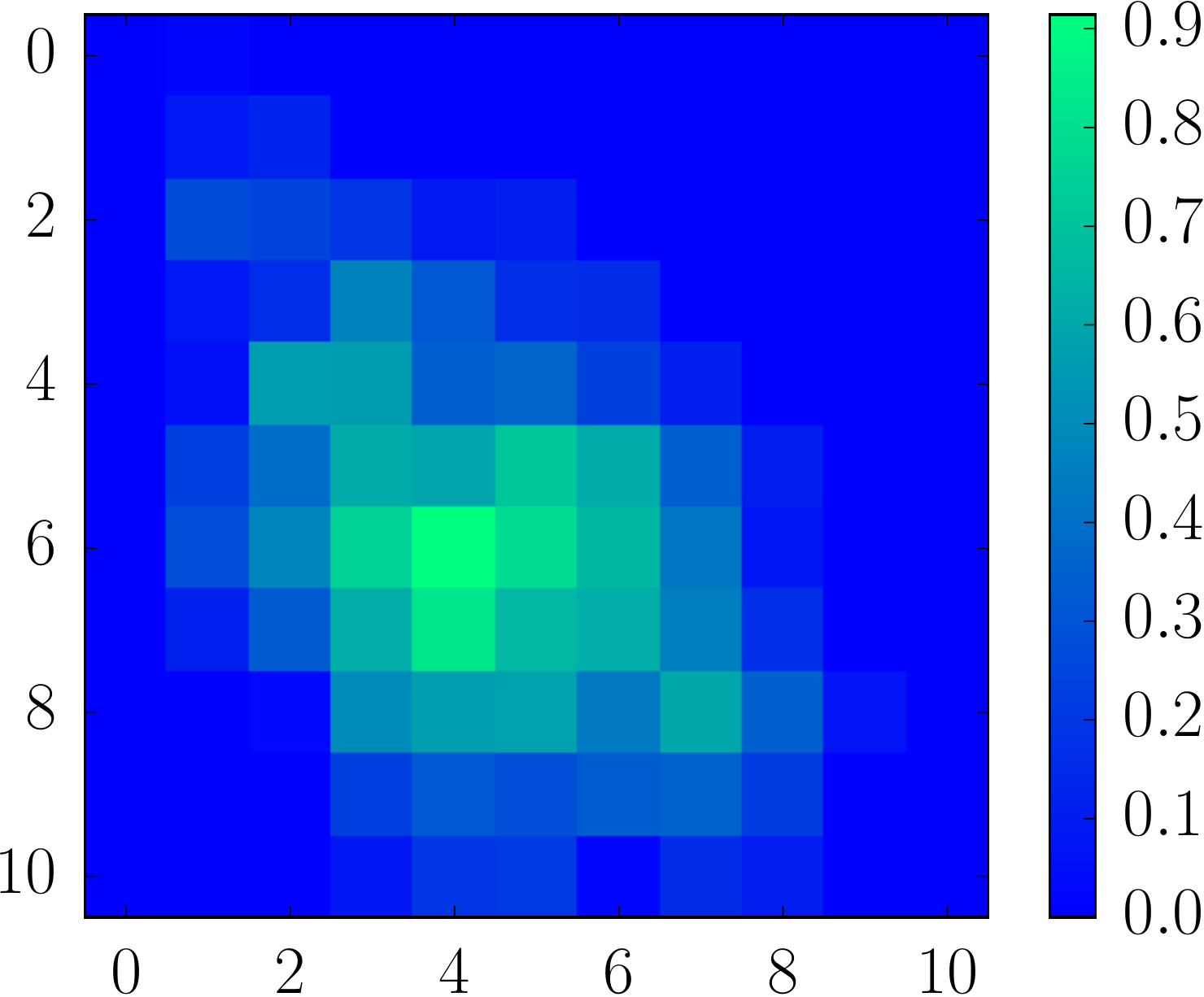}%
		\caption{}%
		\label{subfig_bygrid_3km_koln}
	\end{subfigure}
	\caption{TAPASCologne: (\subref{subfig_density_lust}) Node density for each region at 1 pm with $L = 1$ $km$. Peer hit ratio for each region with (\subref{subfig_bygrid_1km_koln}) $L = 1$ $km$, (\subref{subfig_bygrid_2km_koln}) $L = 2$ $km$, and (\subref{subfig_bygrid_3km_koln}) $L = 3$ $km$.}
	\label{fig_density_koln}
\end{figure}

\cref{fig_chr_prserve_koln} shows the peer hit ratio for the TAPASCologne scenario: it follows the same trend as in \cref{fig_chr_prserve_lust}, but values are slightly higher for the TAPASCologne scenario than those for the LuST scenario. This is because the node density in the central part of TAPASCologne is higher than those in the central part of LuST: higher density results in higher numbers of serving nodes given a $Pr_{serve}$. This is confirmed through \cref{fig_density_lust} and \cref{fig_density_koln}. \cref{subfig_density_lust} and \cref{subfig_density_koln} show node density maps (nodes per $1\ km \times 1\ km$ region) at 1 pm for the LuST and TAPASCologne scenarios, and \cref{subfig_bygrid_1km_lust,subfig_bygrid_2km_lust,subfig_bygrid_3km_lust} and \cref{subfig_bygrid_1km_koln,subfig_bygrid_2km_koln,subfig_bygrid_3km_koln} show maps of the peer hit ratio with different $L$. We see a clear correlation between node density and peer hit ratio in each region; as expected, the peer hit ratio is roughly proportional to the node density. For the LuST scenario and the default settings, in the central area, the peer hit ratio exceeds $0.5$; it approaches $0.8$ in the densest region: the higher the node density, the higher the exposure reduction to the \ac{LBS} server. For TAPASCologne, the peer hit ratio even approaches $95\%$ in the densest areas. For a low-density region, a local relative increase of the system parameter, $Pr_{serve}$, could improve the peer hit ratio with a modest increase in overhead. 


\begin{figure*}[htp!]
	\begin{subfigure}[b]{.24\columnwidth}
		\includegraphics[width=\columnwidth]{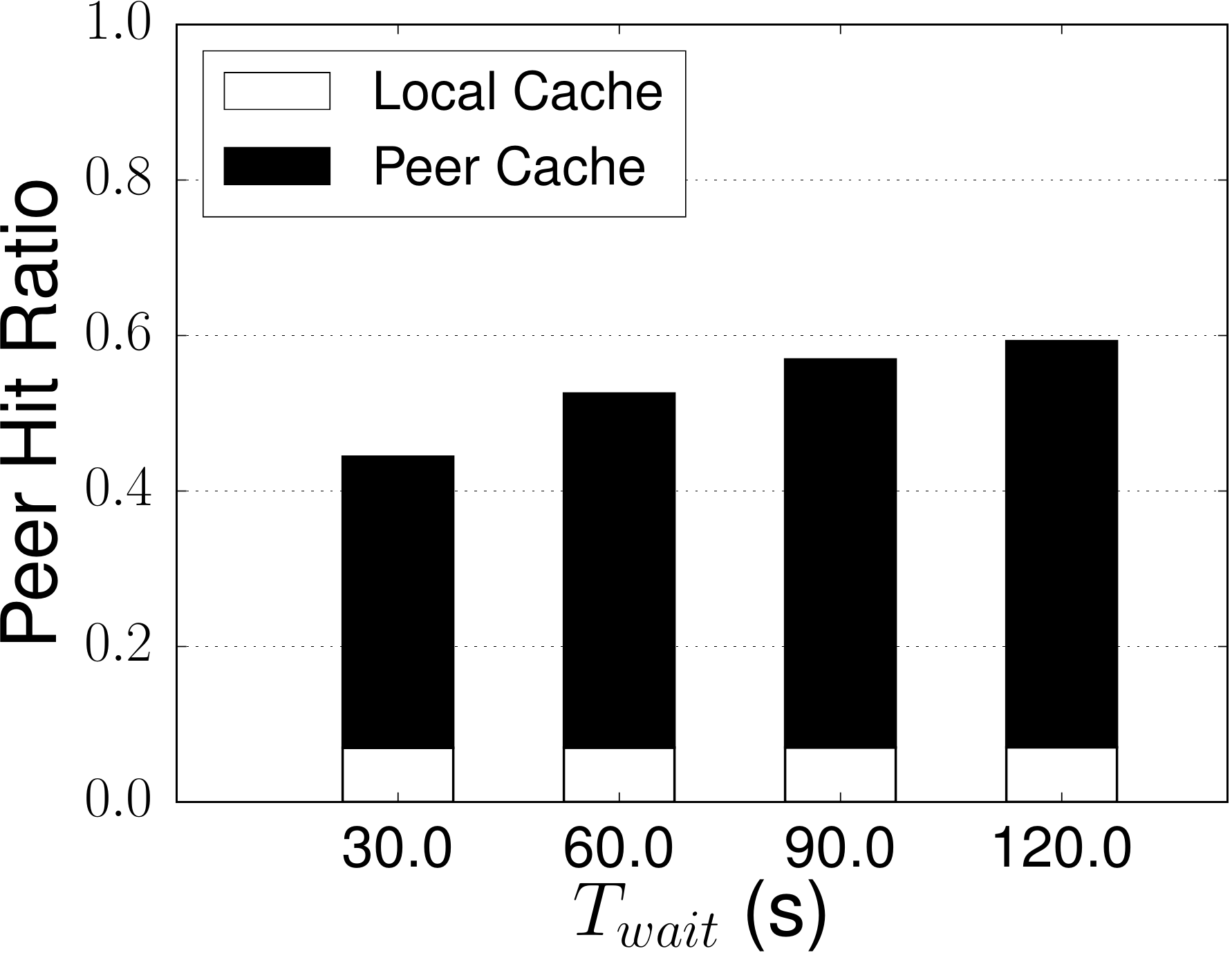}%
		\caption{}
		\label{subfig_chr_wait}
	\end{subfigure}
	\begin{subfigure}[b]{.24\columnwidth}
		\includegraphics[width=\columnwidth]{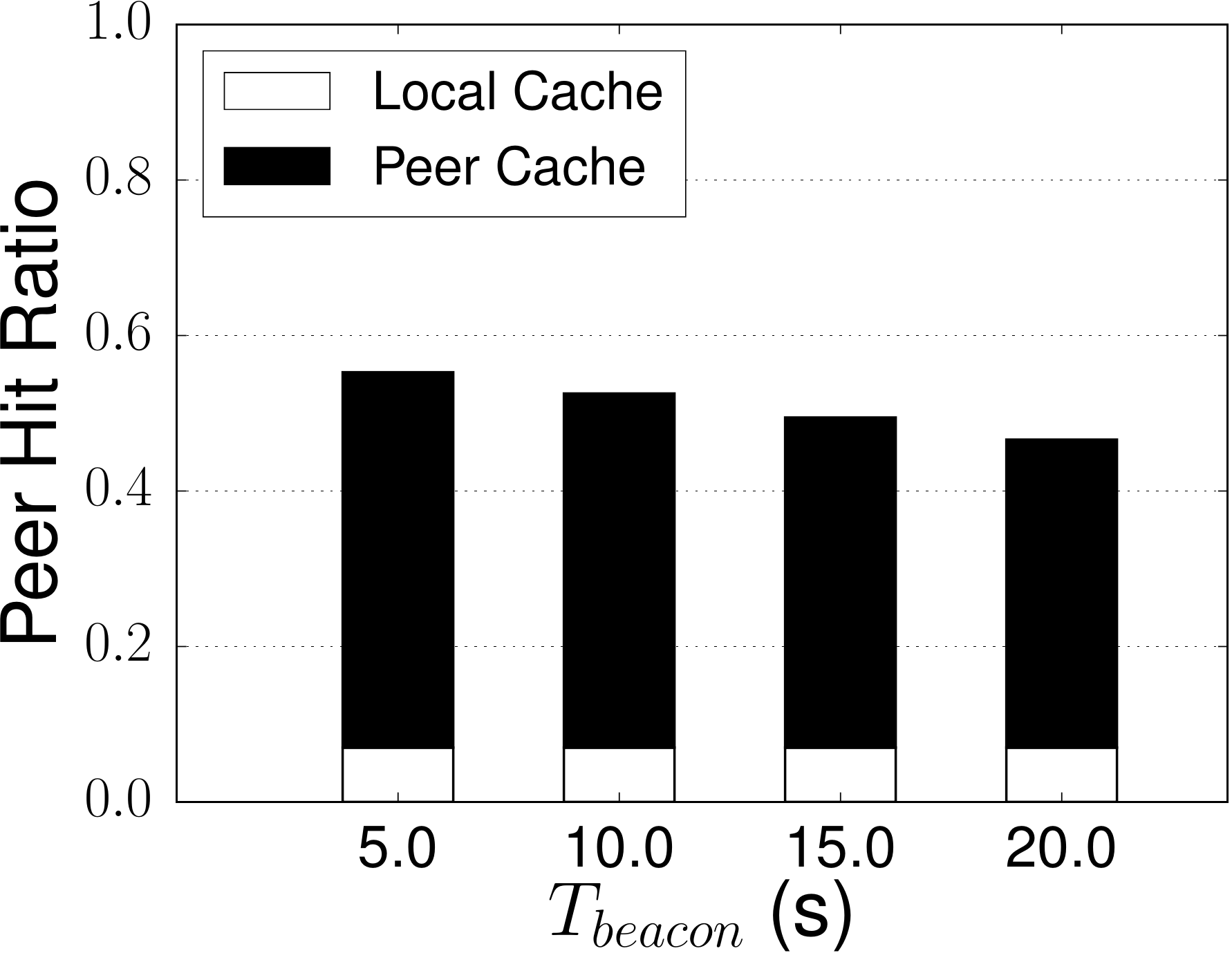}%
		\caption{}
		\label{subfig_chr_beacon}
	\end{subfigure}
	\begin{subfigure}[b]{.24\columnwidth}
		\includegraphics[width=\columnwidth]{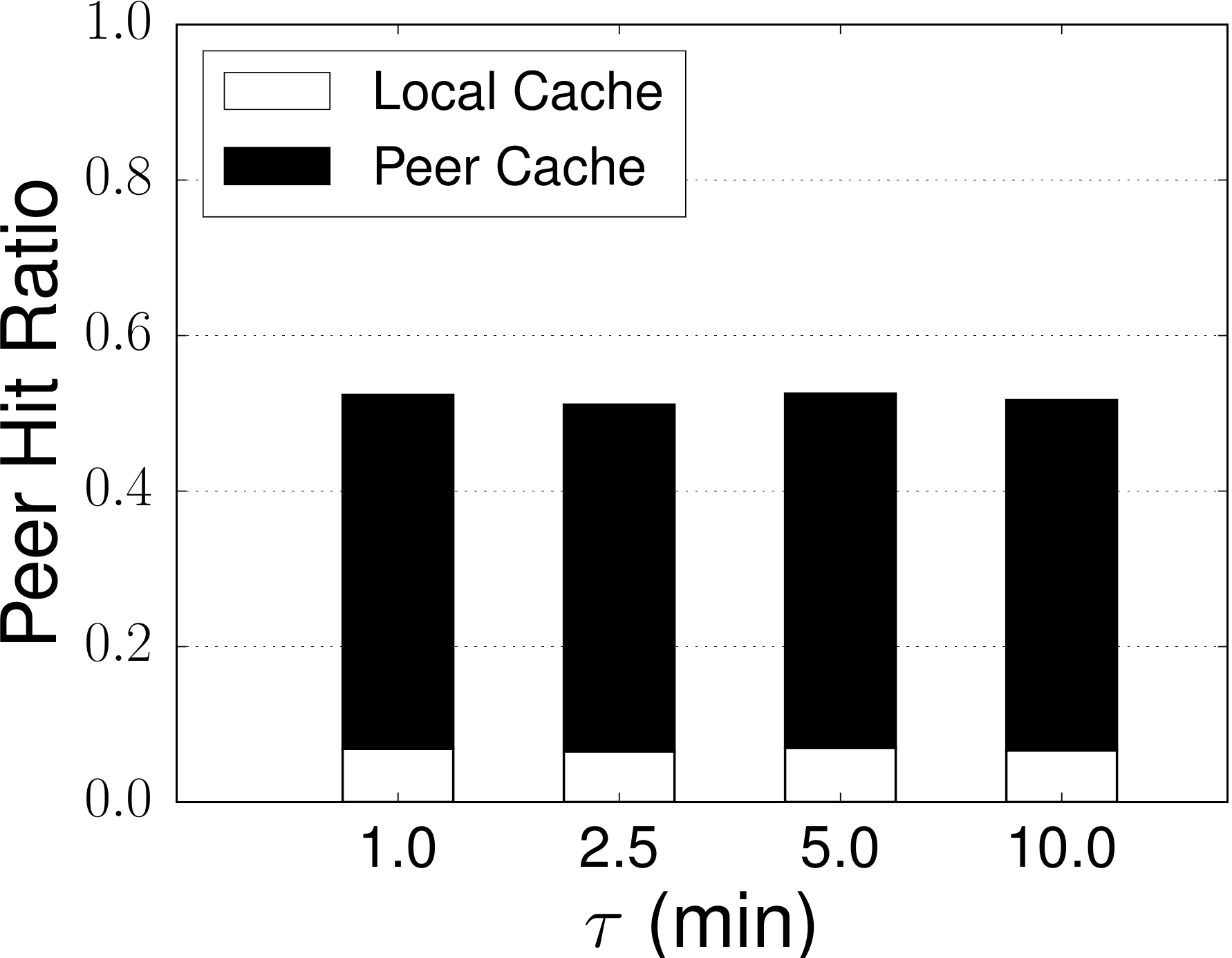}%
		\caption{}
		\label{subfig_chr_tau}
	\end{subfigure}
	\begin{subfigure}[b]{.24\columnwidth}
		\includegraphics[width=\columnwidth]{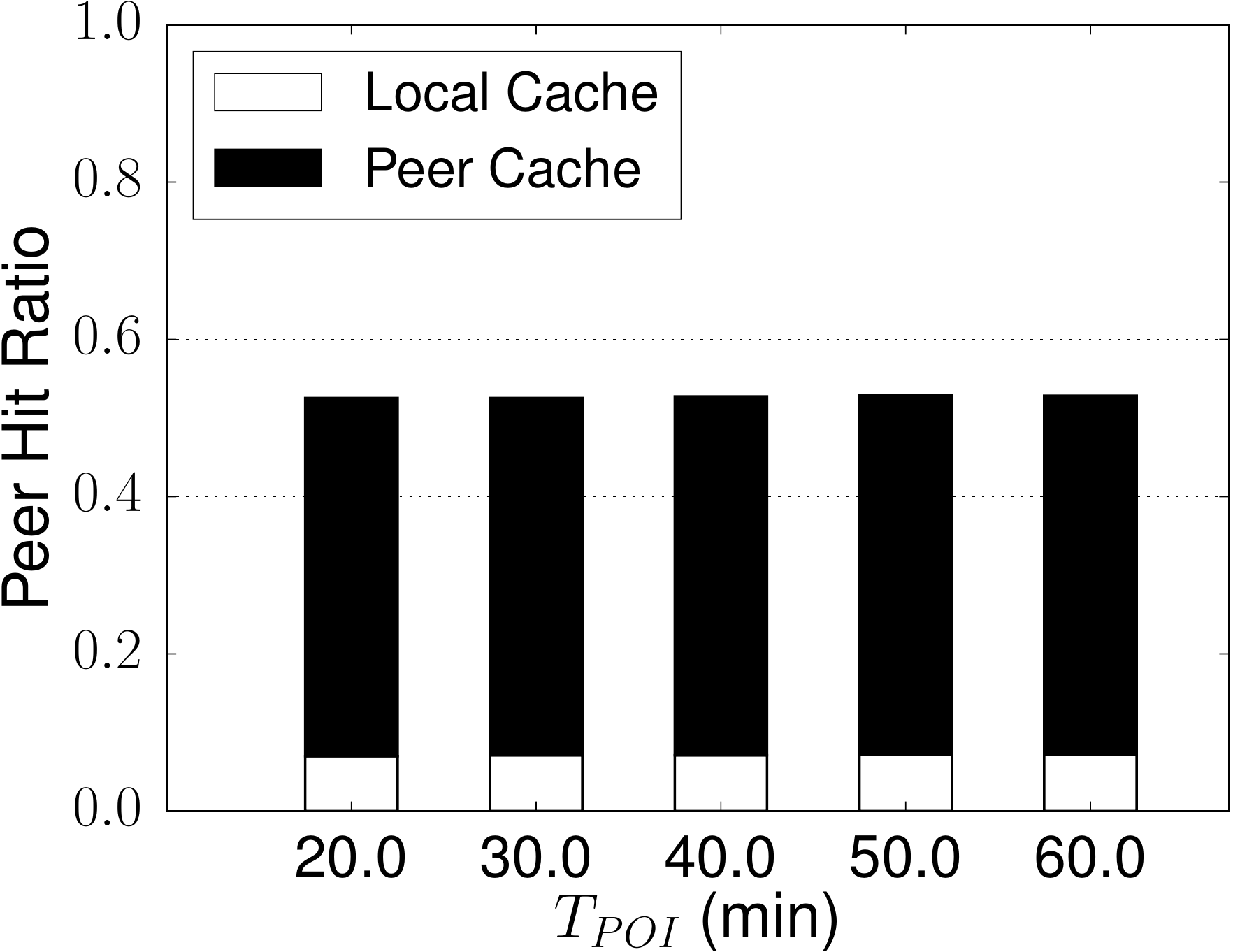}%
		\caption{}
		\label{subfig_chr_poi}
	\end{subfigure}
	\caption{LuST: Peer hit ratio as a function of (\subref{subfig_chr_wait}) $T_{wait}$, (\subref{subfig_chr_beacon}) $T_{beacon}$, (\subref{subfig_chr_tau}) $\tau$, and (\subref{subfig_chr_poi}) $T_{POI}$. (Default: $T_{wait}=60\ s$, $T_{beacon}=10\ s$, $\tau=5\ min$ and $T_{POI}=20\ min$.)}
	\label{fig_chr_param_lust}
\end{figure*}

\cref{subfig_chr_wait,subfig_chr_beacon,subfig_chr_tau,subfig_chr_poi} show peer hit ratios as a function of $T_{wait}$, $T_{beacon}$, $\tau$ and $T_{POI}$, respectively. Higher $T_{wait}$ allows a querying node more time to discover serving nodes, and lower $T_{beacon}$ for a serving node results in a higher probability to be discovered by nearby querying nodes; thus both result in a higher peer hit ratio. However, a change in $\tau$ or $T_{POI}$ does not affect the number of serving nodes in the system or the probability of discovering serving nodes, thus the peer hit ratio remains roughly the same as $\tau$ or $T_{POI}$ changes.

\begin{figure}[htp!]
	\begin{subfigure}[b]{0.24\columnwidth}
		\includegraphics[width=\columnwidth]{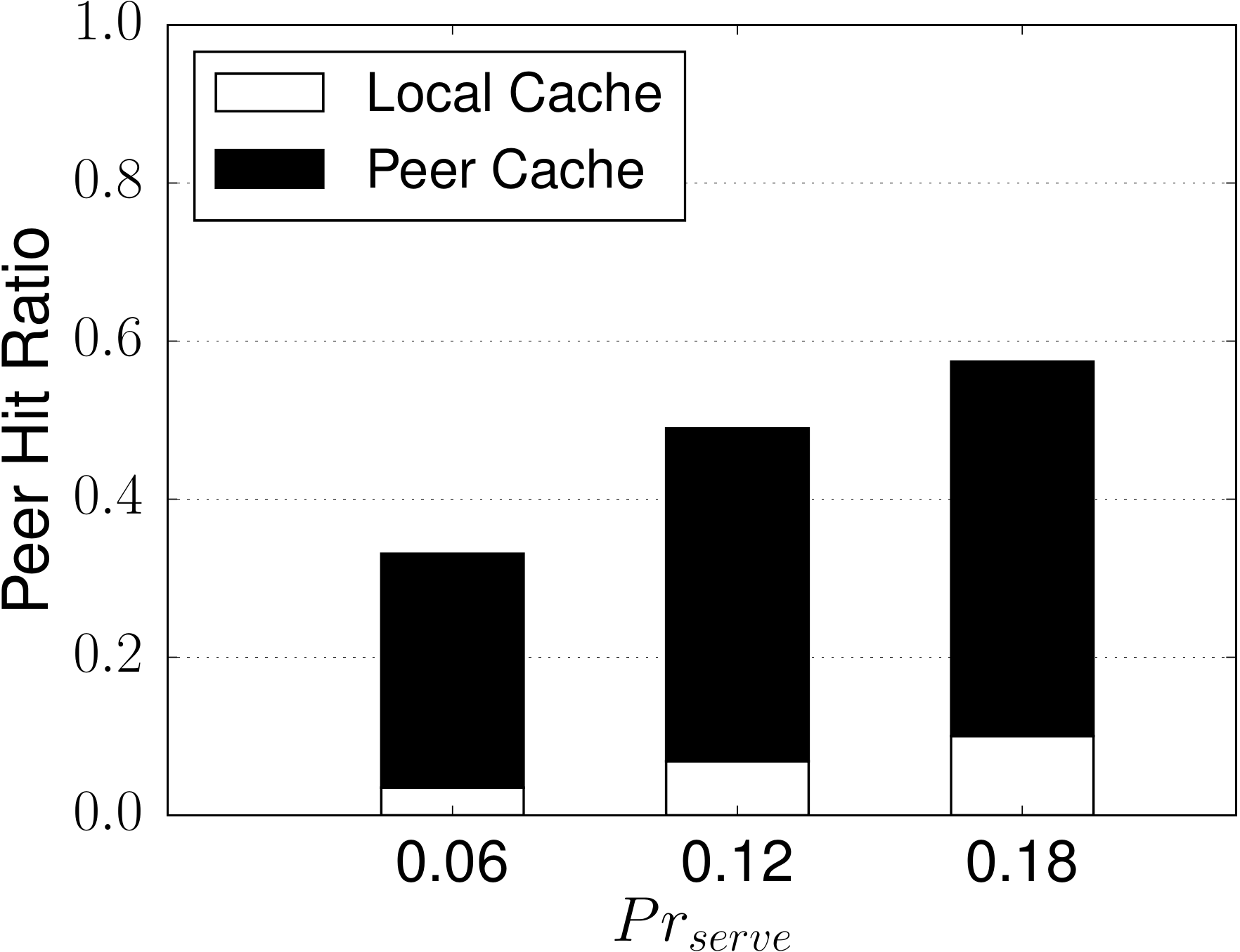}
		\caption{}
		\label{subfig_chr_type_2_lust}
	\end{subfigure}
	\begin{subfigure}[b]{0.24\columnwidth}
		\includegraphics[width=\columnwidth]{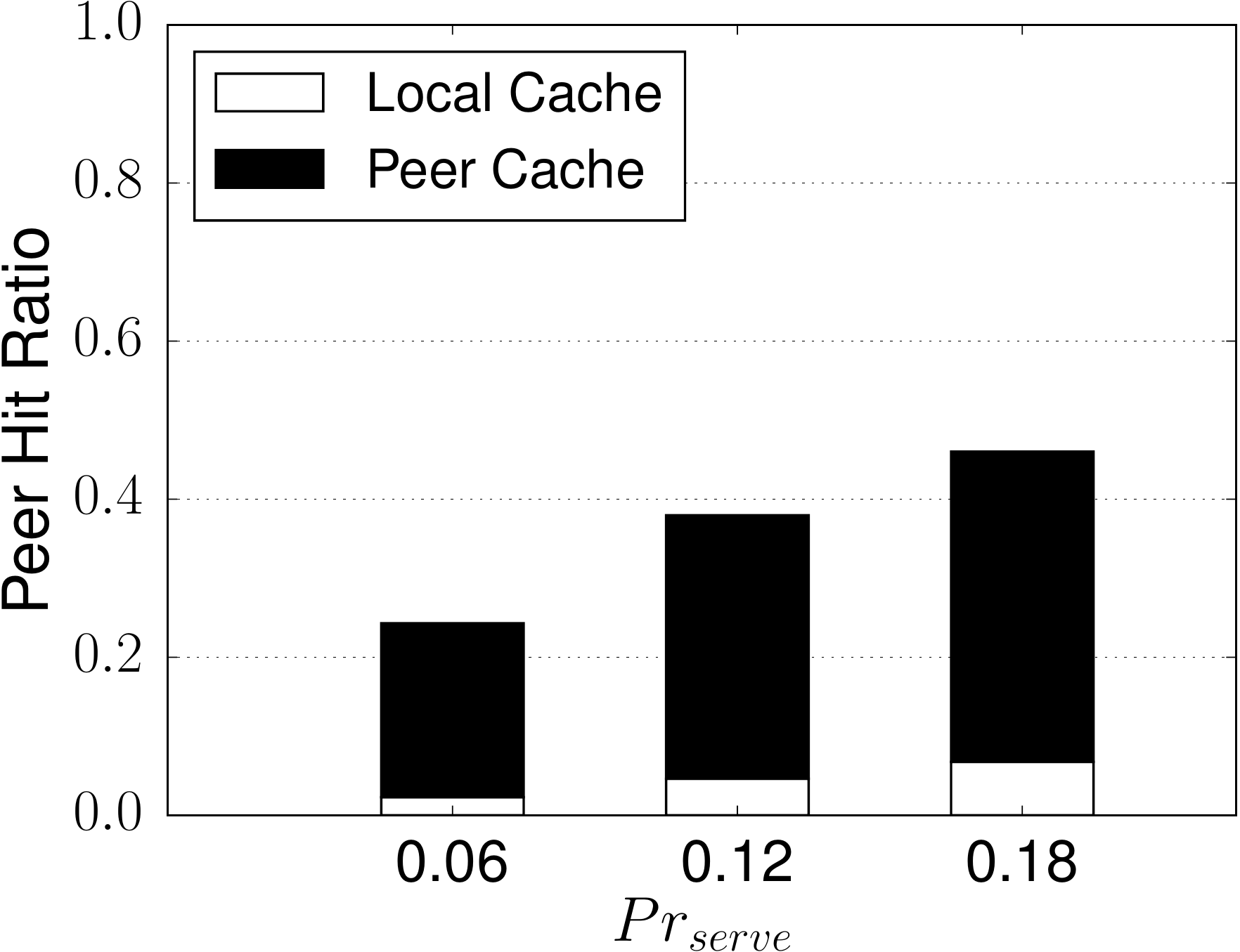}%
		\caption{}%
		\label{subfig_chr_type_3_lust}
	\end{subfigure}
	\begin{subfigure}[b]{0.24\columnwidth}
		\includegraphics[width=\columnwidth]{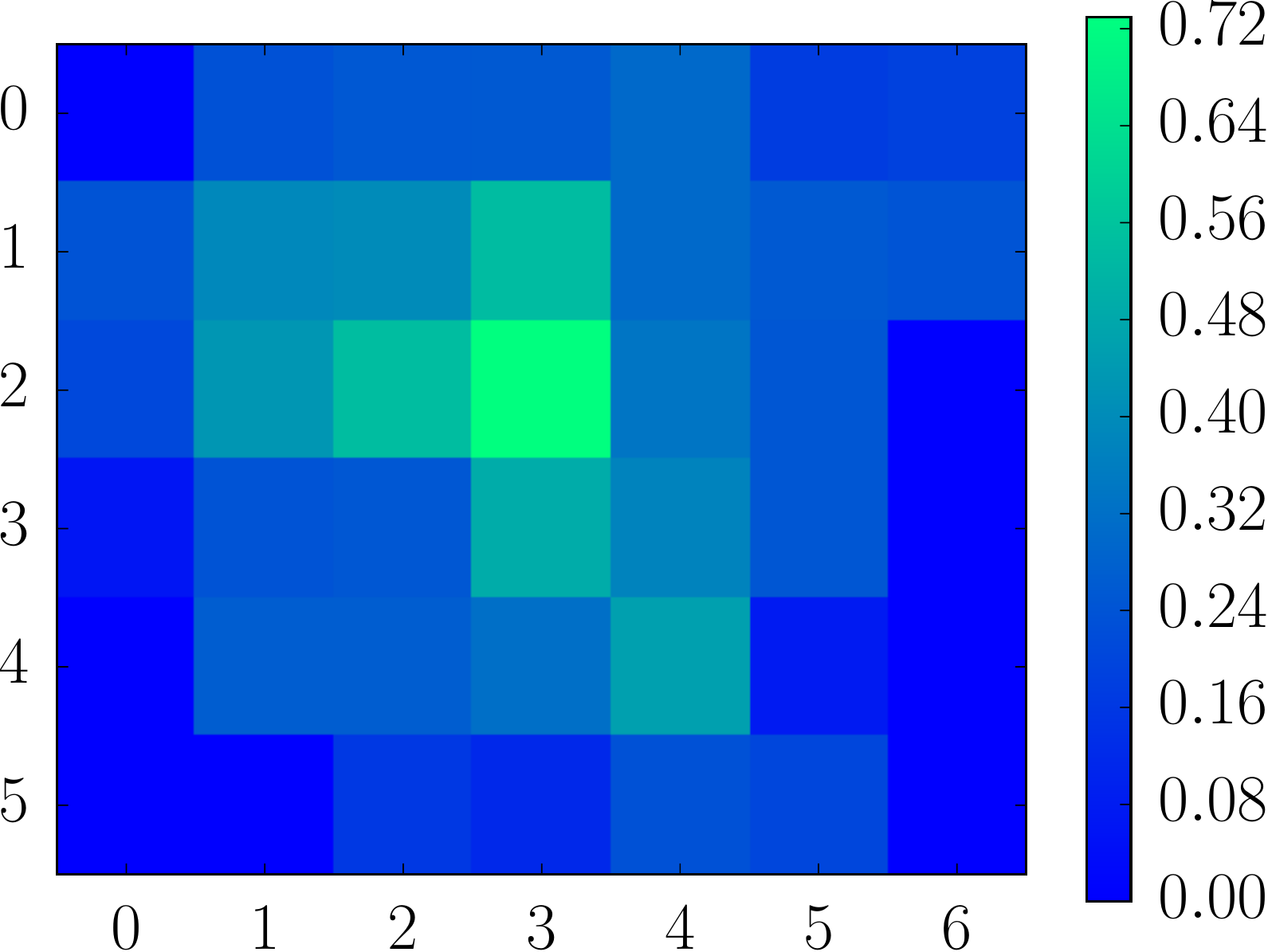}
		\caption{}
		\label{subfig_bygrid012_2_lust}
	\end{subfigure}
	\begin{subfigure}[b]{0.24\columnwidth}
		\includegraphics[width=\columnwidth]{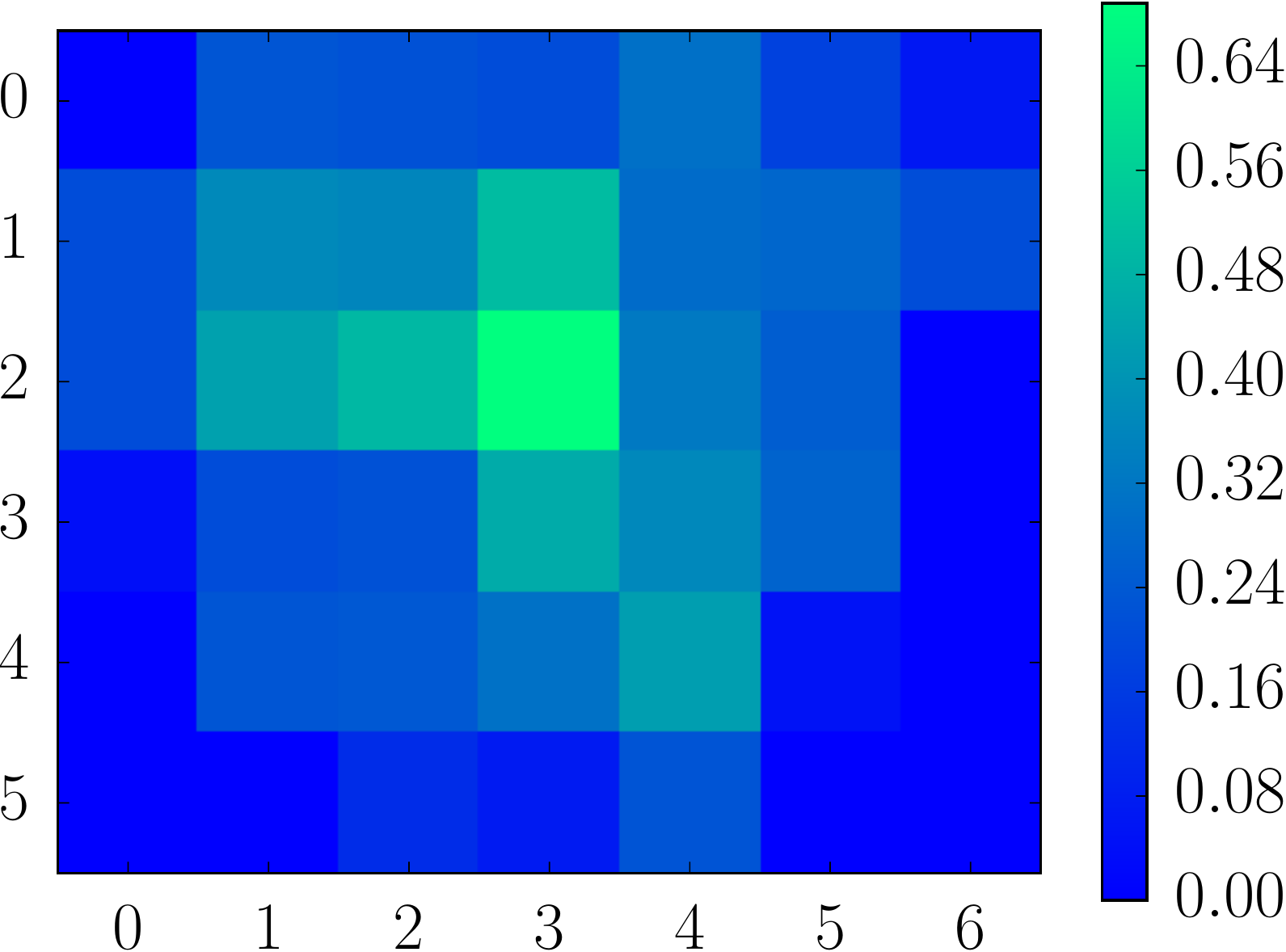}%
		\caption{}%
		\label{subfig_bygrid018_3_lust}
	\end{subfigure}
	\caption{LuST:  Peer hit ratio as a function of $Pr_{serve}$ with (\subref{subfig_chr_type_2_lust}) $G=2$, and (\subref{subfig_chr_type_3_lust}) $G=3$. Peer hit ratio for each region with (\subref{subfig_bygrid012_2_lust}) $G=2$ and $Pr_{serve}=0.12$, and (\subref{subfig_bygrid018_3_lust}) $G=3$ and $Pr_{serve}=0.18$.}
	\label{fig_chr_type_lust}
\end{figure}

\begin{figure*}[htp!]
	\begin{subfigure}[b]{.24\columnwidth}
		\includegraphics[width=\columnwidth]{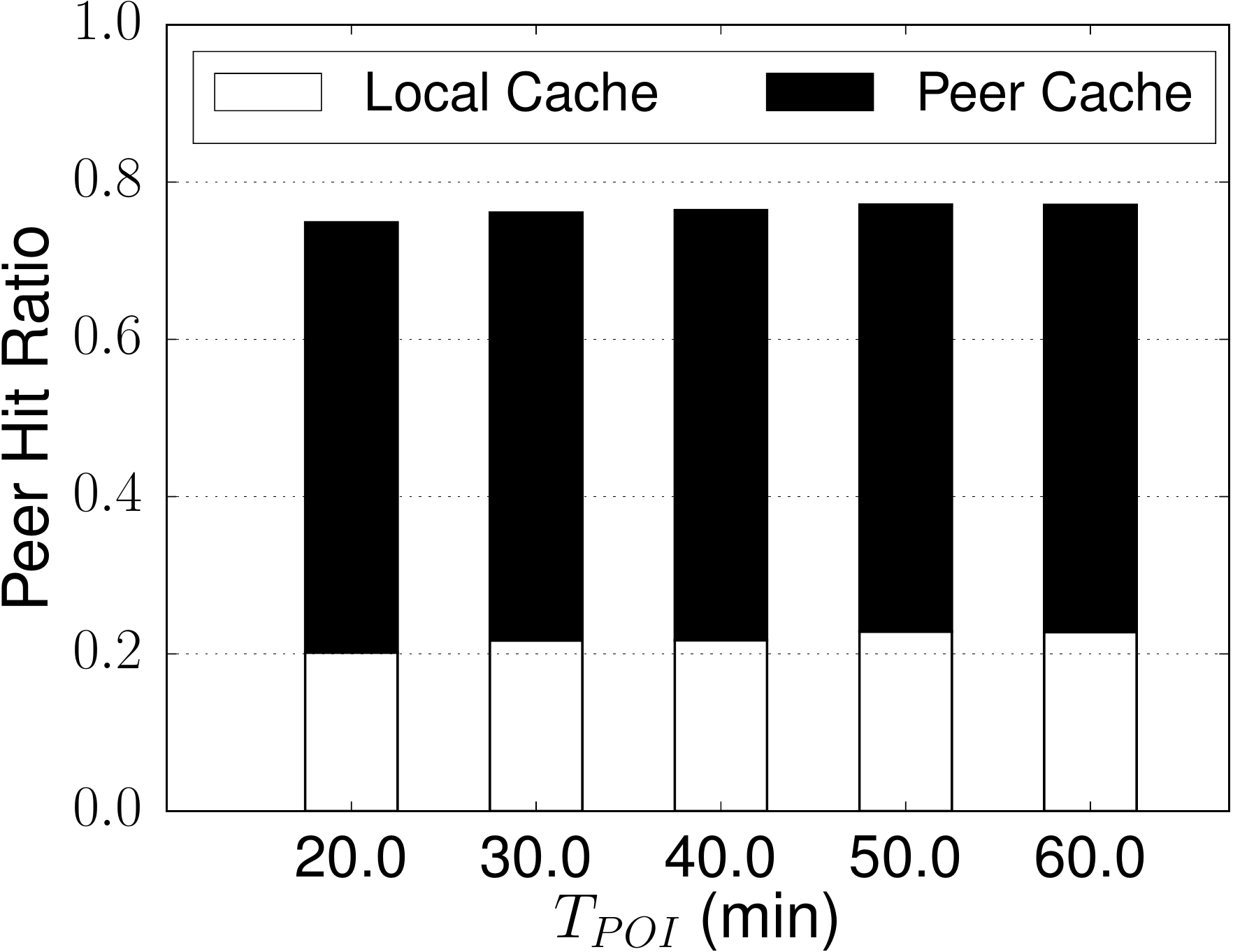}%
		\caption{}
		\label{subfig_poi_mc}
	\end{subfigure}
	\begin{subfigure}[b]{.24\columnwidth}
		\includegraphics[width=\columnwidth]{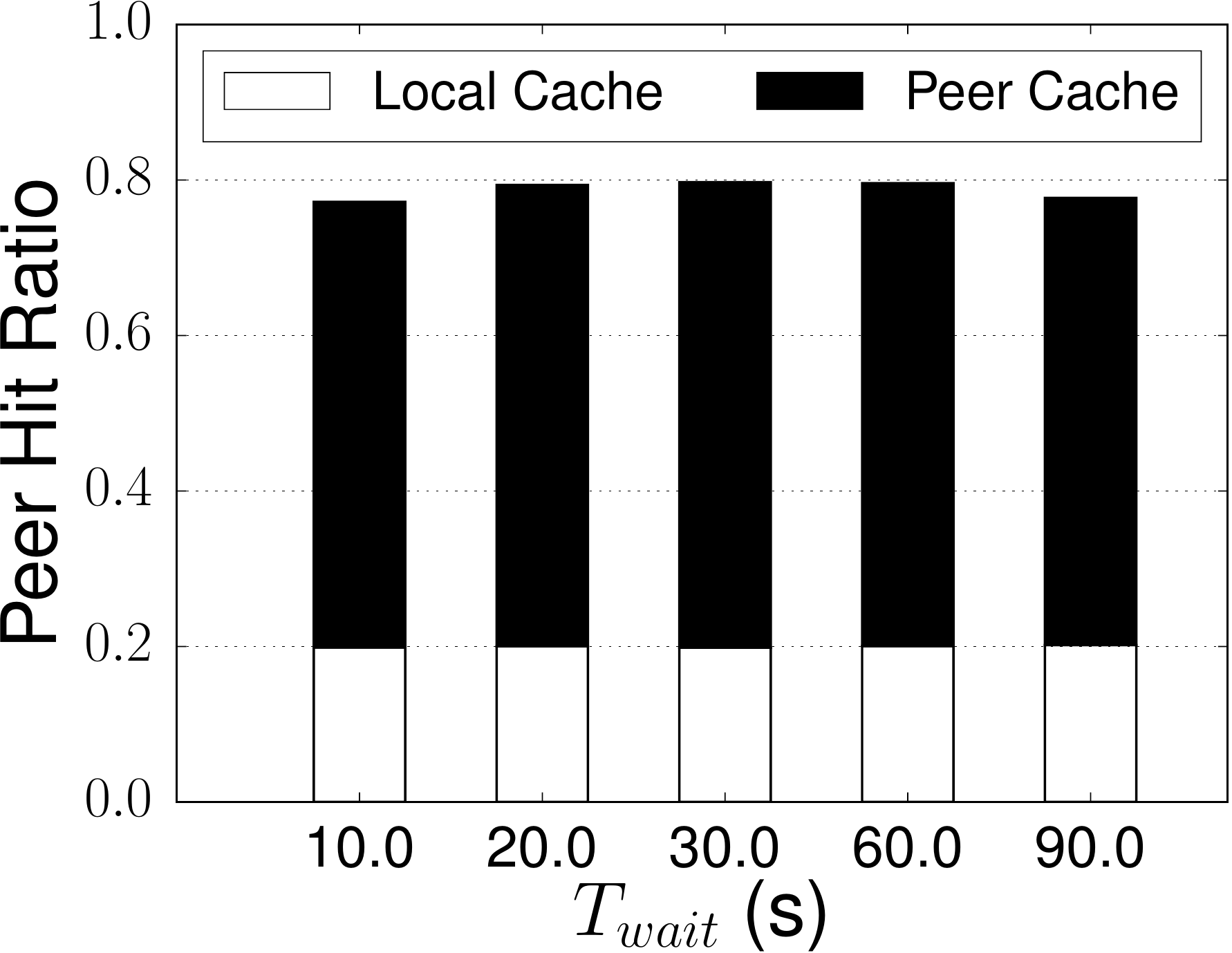}%
		\caption{}
		\label{subfig_chr_mc}
	\end{subfigure}
	\begin{subfigure}[b]{.24\columnwidth}
		\includegraphics[width=\columnwidth]{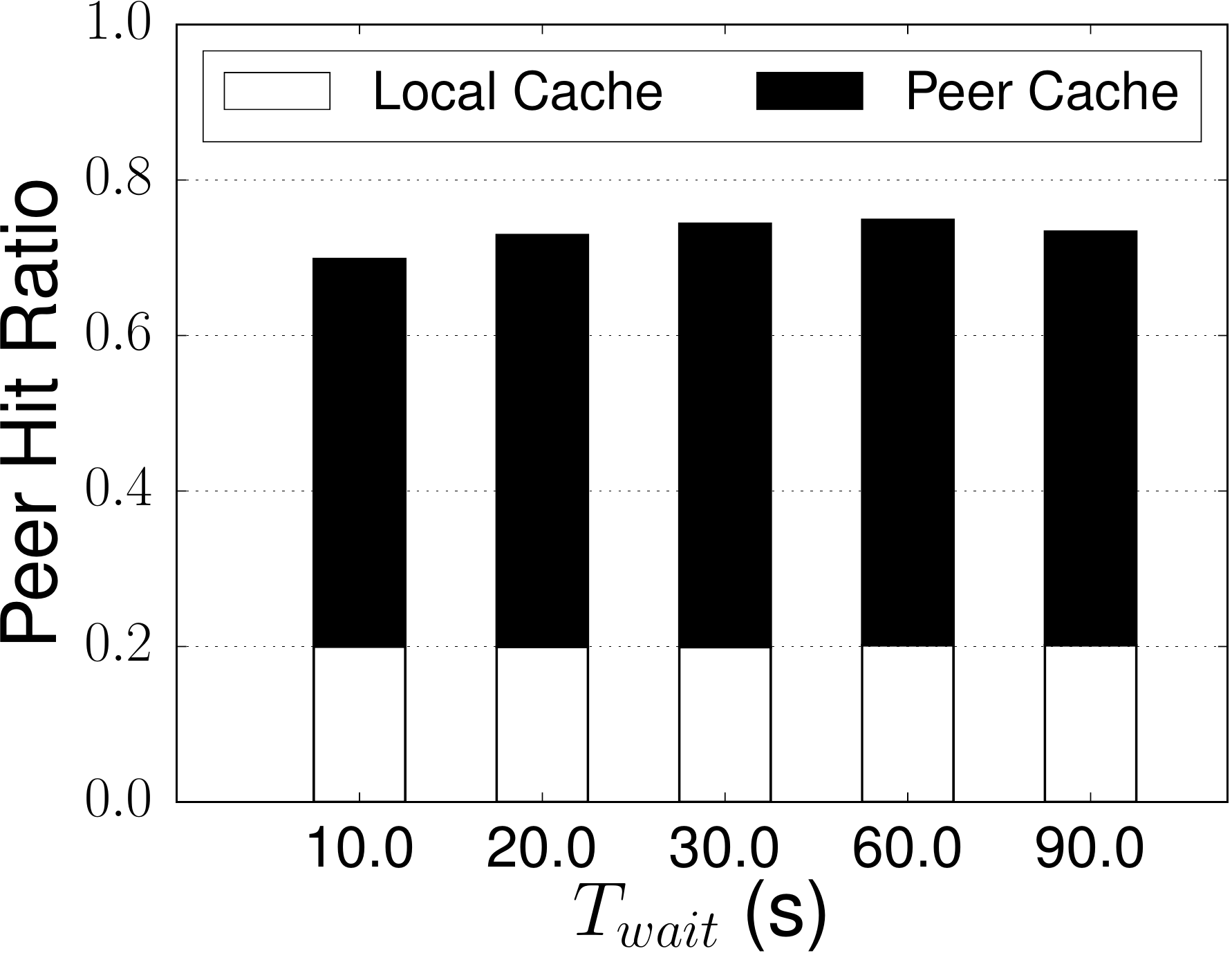}%
		\caption{}
		\label{subfig_chr_mc_05}
	\end{subfigure}
	\caption{LuST with MobiCrowd: Peer hit ratio as a function of (\subref{subfig_poi_mc}) $T_{POI}$, and $T_{wait}$ with (\subref{subfig_chr_mc}) $Ratio_{coop} = 1$ and (\subref{subfig_chr_mc_05}) $Ratio_{coop} = 0.5$. (Default: $T_{POI}=20\ min$, $T_{wait}=60\ s$ and $Ratio_{coop} = 0.5$.)}
	\label{fig_chr}
\end{figure*}

We evaluate further our scheme considering multiple groups of \ac{POI} types (i.e., $G>1$). \cref{fig_chr_type_lust} shows the peer hit ratio as a function of $Pr_{serve}$, with $G=2,3$. Given a $Pr_{serve}$, the peer hit ratio decreases as $G$ grows (\cref{subfig_chr_type_2_lust,subfig_chr_type_3_lust}), because less serving nodes exist for a queried \ac{POI} type. In order to achieve the same level of cache hit ratio, $Pr_{serve}$ should increase accordingly with $G$. When a serving node is interested in a \ac{POI} type that it is not responsible for (thus, not cached), it queries the \ac{LBS} server directly, because it had already exposed itself to the \ac{LBS} server while obtaining regional \ac{POI} data. Therefore, the cache hit ratio with $G=3$ and $Pr_{serve}=0.18$ is lower than that with $G=2$ and $Pr_{serve}=0.12$; the reason is that more serving node queries are sent directly to the \ac{LBS} server with $G=3$ and $Pr_{serve}=0.18$.

\cref{fig_chr} shows the peer hit ratio achieved by  MobiCrowd for the LuST scenario. From \cref{subfig_poi_mc}, we see that the peer hit ratio slightly increases as $T_{POI}$ increases, because cached \ac{POI} data is valid for a longer period, thus serving more peer (and own) queries. \cref{subfig_chr_mc,subfig_chr_mc_05} show the peer hit ratio as a function of $T_{wait}$ when the collaboration ratios are $1$ and $0.5$ respectively. A higher collaboration ratio results in a higher peer hit ratio. However, a longer $T_{wait}$ period does not necessarily improve the peer hit ratio. For example, it slightly increases as $T_{wait}$ increases from $10\ s$ to $60\ s$, but it decreases for $T_{wait}=90\ s$. Long $T_{wait}$ increases the probability to encounter nodes that cached the required \ac{POI} data, but it decreases the chance to serve other nodes with its own cached \ac{POI} data (that could have been obtained earlier from the \ac{LBS} server with a shorter $T_{wait}$). The peer hit ratio is around $0.7$ with MobiCrowd when $Ratio_{coop}=0.5$ and $T_{wait} = 10\ s$. Our scheme provides roughly the same peer hit ratio when $Pr_{serve} = 0.12$ (\cref{subfig_chr_prserve_lust_2km}) with the default simulation settings. However, our scheme significantly improves over MobiCrowd in terms of exposure and overhead. MobiCrowd trades off higher node exposure to neighboring honest-but-curious nodes and higher communication overhead for increased peer hit ratio.

\textbf{Node Exposure:} After gauging node privacy through the peer hit ratio, we refine the measurement of node exposure to curious \ac{LBS} servers and curious peer nodes with the help of the \emph{exposure degree ($ExpoDeg$)}, defined for any node as:

{
\small
\begin{align*}
ExpoDeg(Id_{LTC}, C) = \sum\limits_{Id_i \in ID(Id_{LTC},C)}\frac{T(Id_i)}{T(Id_{LTC})}*\frac{R_{H}(Id_i)}{R(Id_{LTC})}. \label{eq:expo_deg}
\end{align*}
}

$Id_{LTC}$ is the node long-term identity, corresponding to a whole series of node actions in the system. $ID(Id_{LTC}, C)$ is a set of identities, corresponding to $Id_{LTC}$, exposed to the honest-but-curious (possibly colluding) entities for collusion case $C$ (Table~\ref{table:exposure}). $ID(Id_{LTC},C)$ differs for different collusion cases. $T(Id_i)$ is the corresponding trip duration of a node under identity $Id_i \in ID(Id_{LTC},C)$. $R(Id_i)$ is the number of regions the node visits during its trip under identity $Id_i$ and $R_{H}(Id_i)$ is the number of visited regions exposed to honest-but-curious entities under the same identity $Id_i$. $\frac{R_{H}(Id_i)}{R(Id_{LTC})}$ indicates the exposure degree under a single identity $Id_i$. To derive the exposure degree of a node, the exposure degrees under each $Id_i$ are weighted by a time parameter $\frac{T(Id_i)}{T(Id_{LTC})}$: the ratio of the (partial) trip duration under identity $Id_i$ over the total trip time. The exposure degree indicates the accuracy of reconstructed node trajectories based on recorded node queries, taking into consideration the effect of pseudonymous authentication on location privacy protection. We measure exposure to colluding curious nodes through the aggregation of the recorded queries.

\begin{figure*}[h]
	\begin{subfigure}[b]{.24\columnwidth}
		\includegraphics[width=\columnwidth]{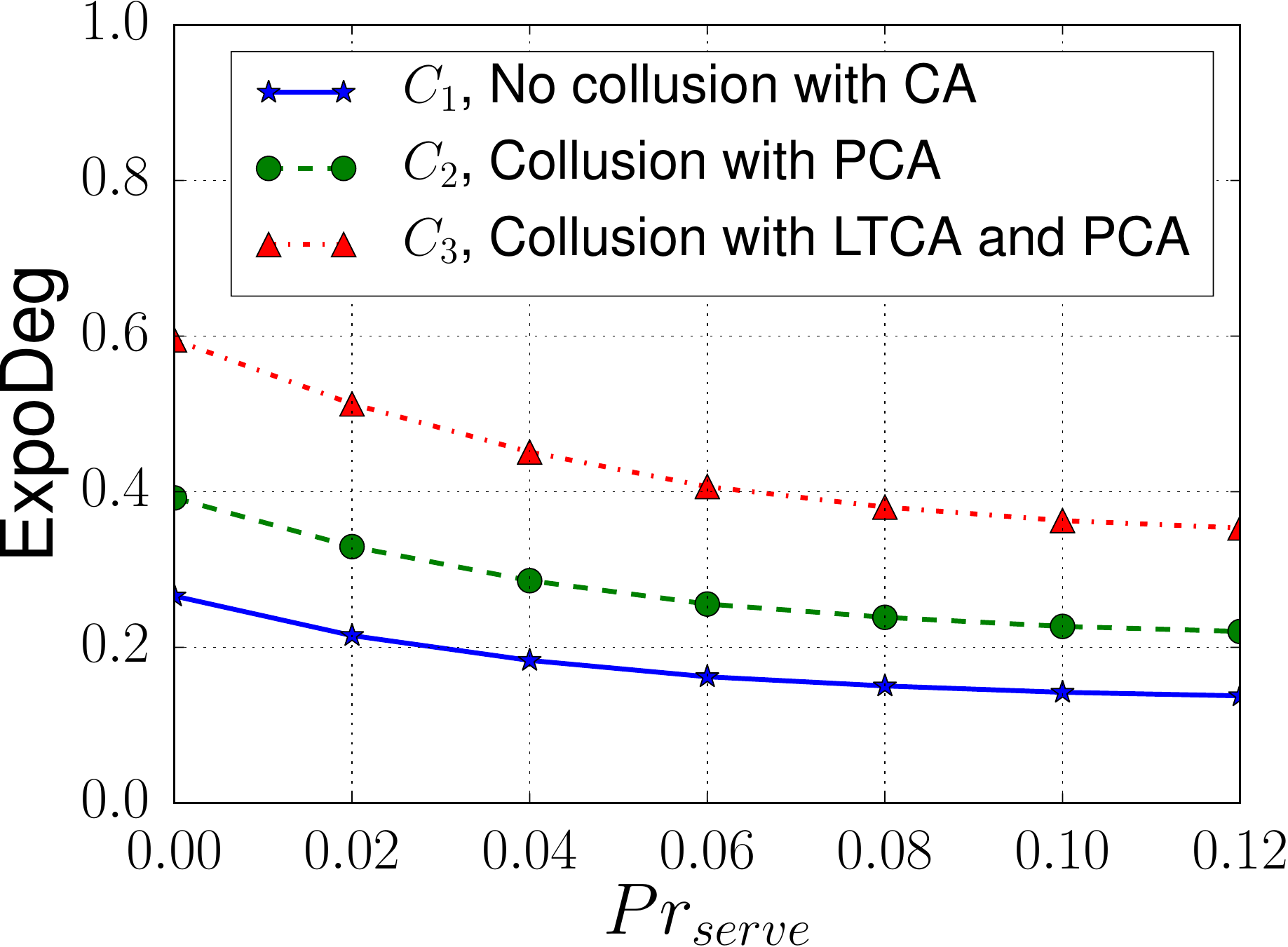}
		\caption{}
		\label{subfig_lbs_lust}
	\end{subfigure}
	\begin{subfigure}[b]{.24\columnwidth}
		\includegraphics[width=\columnwidth]{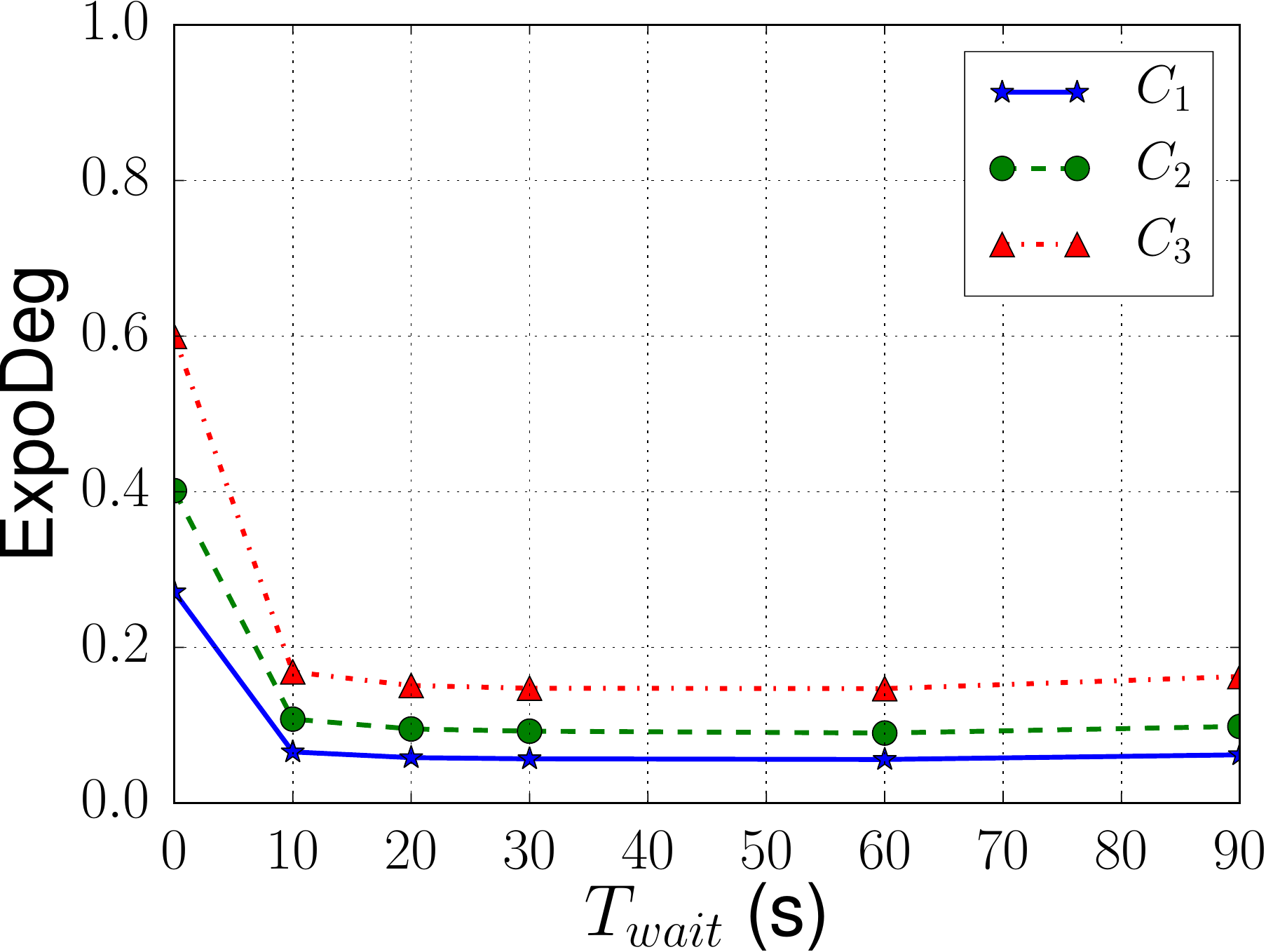}%
		\caption{}%
		\label{subfig_mc_lbs_lust}%
	\end{subfigure}
	\begin{subfigure}[b]{.24\columnwidth}
		\includegraphics[width=\columnwidth]{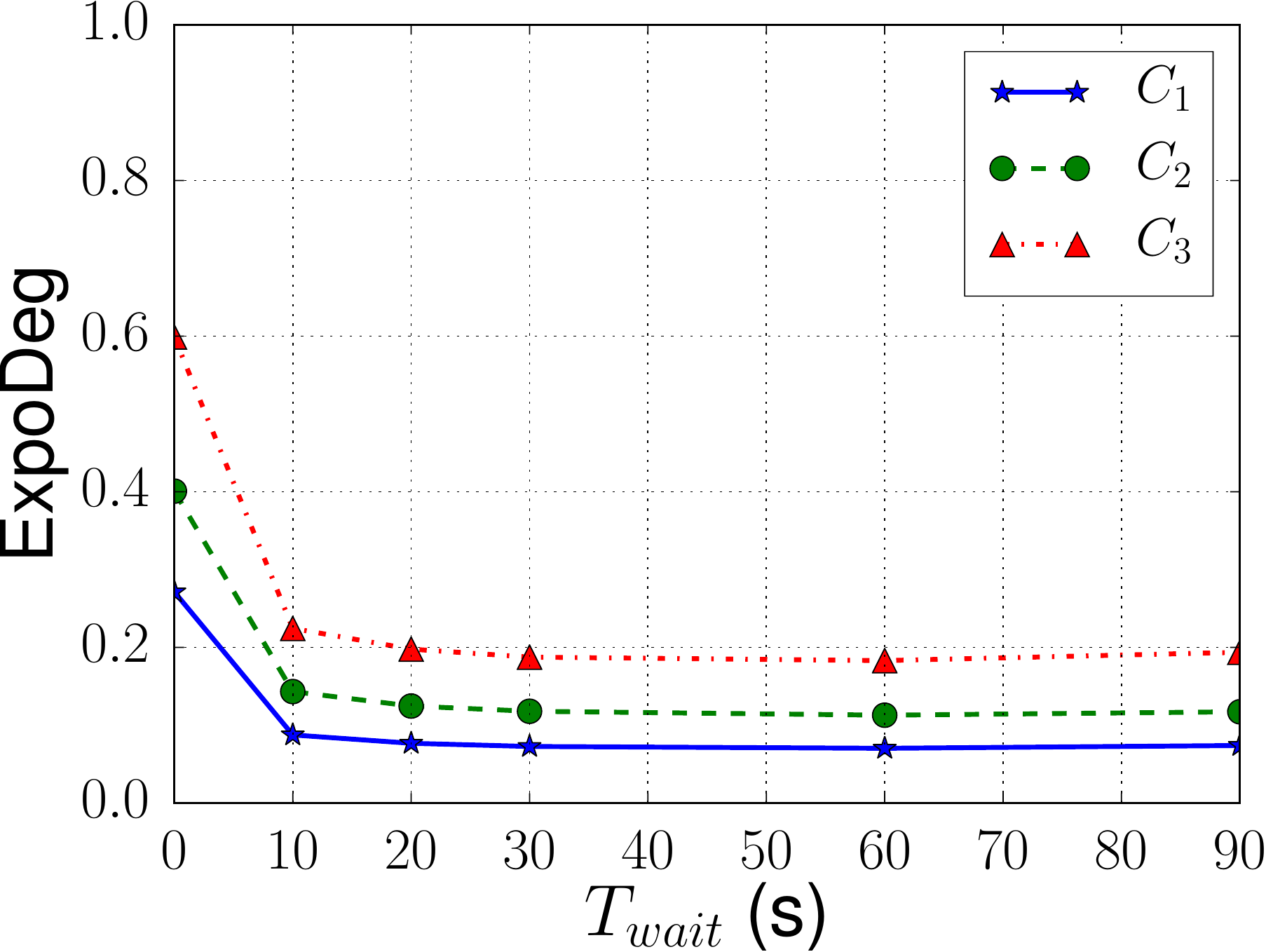}%
		\caption{}%
		\label{subfig_mc_05_lbs_lust}%
	\end{subfigure}
	
	\begin{subfigure}[b]{.24\columnwidth}
		\includegraphics[width=\columnwidth]{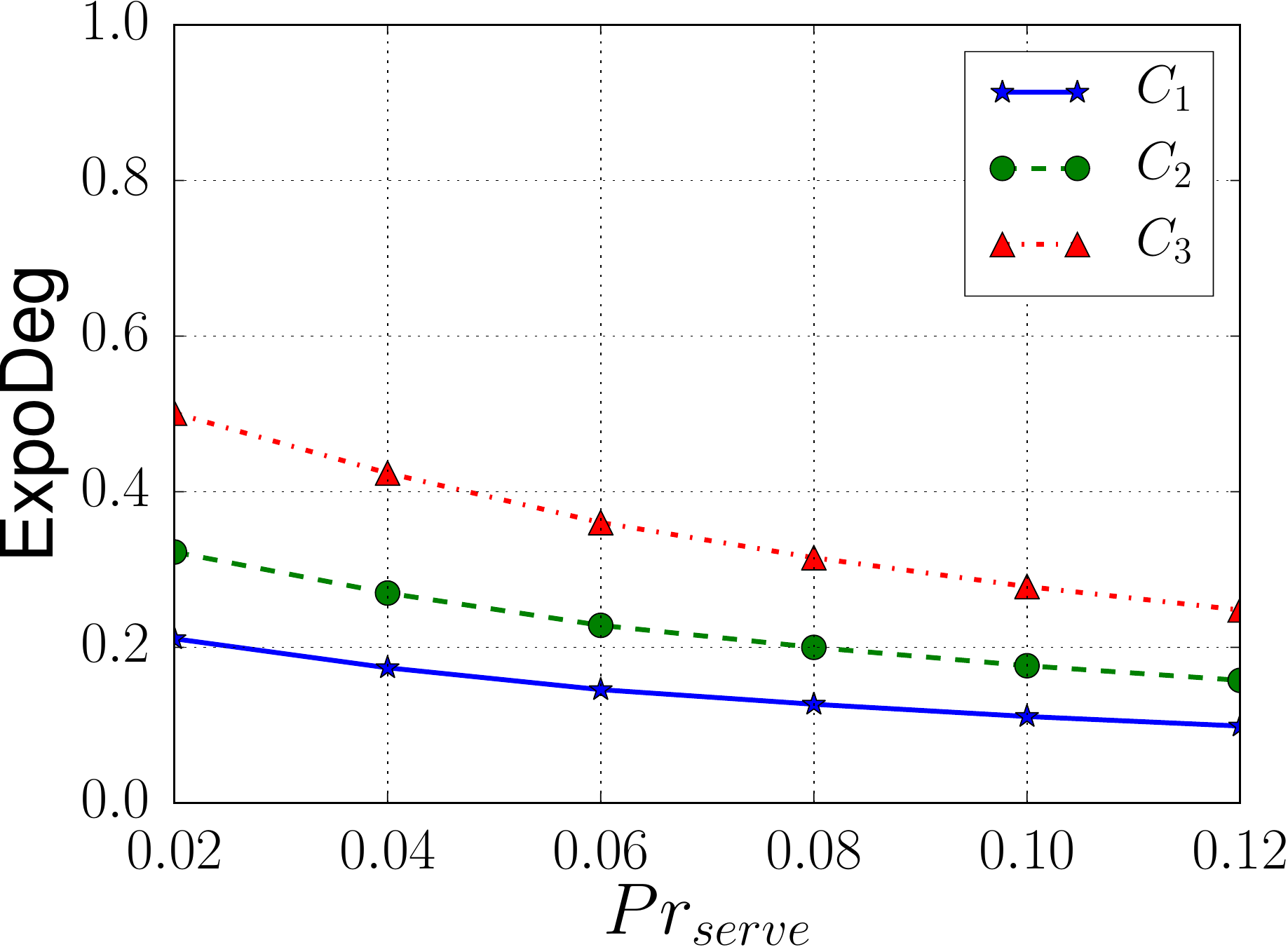}%
		\caption{}%
		\label{fig_expo_lbs_norsu}%
	\end{subfigure}
	\begin{subfigure}[b]{.24\columnwidth}
		\includegraphics[width=\columnwidth]{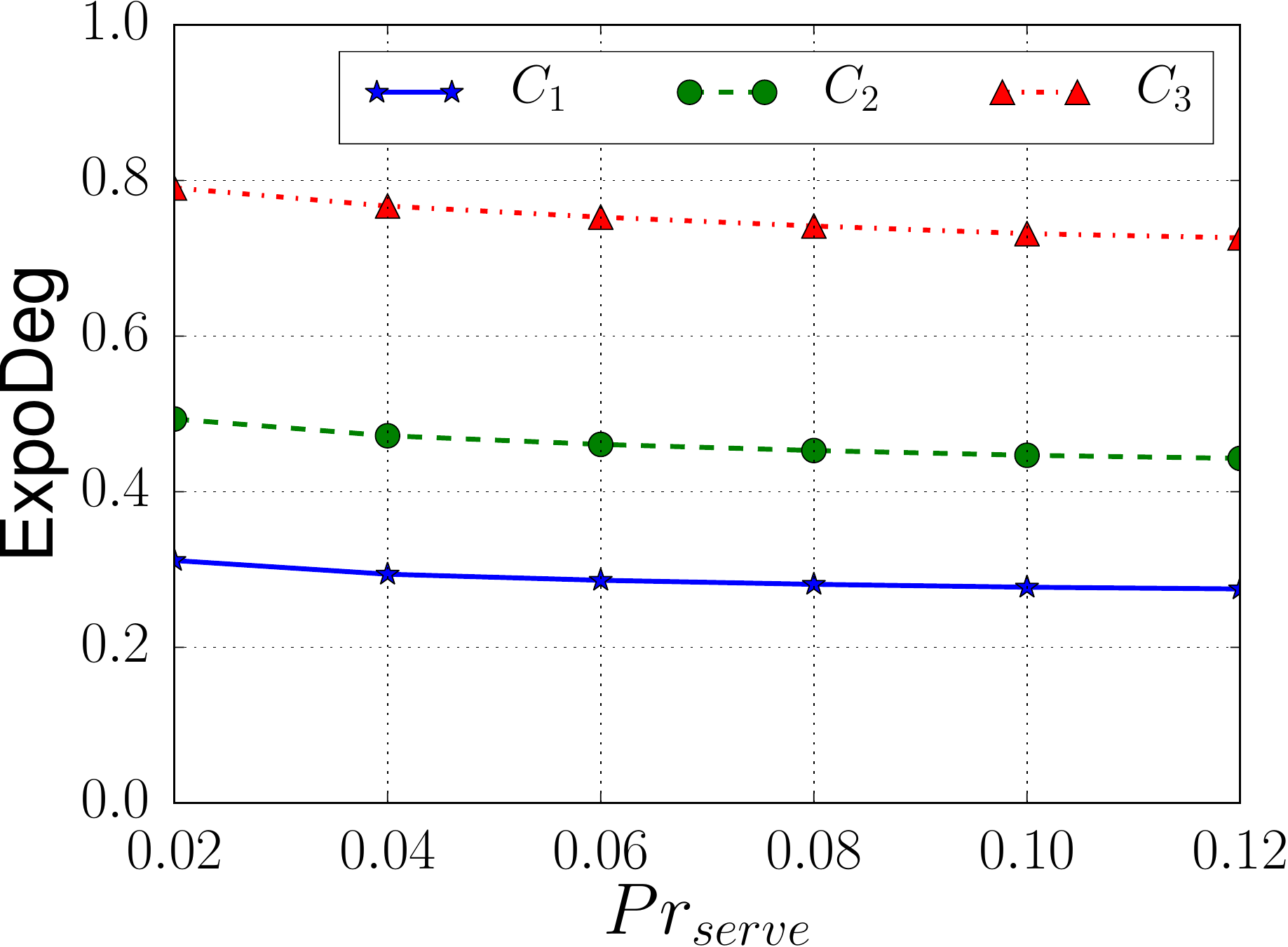}%
		\caption{}%
		\label{fig_expo_lbs_rsu}%
	\end{subfigure}
	\begin{subfigure}[b]{.24\columnwidth}
		\includegraphics[width=\columnwidth]{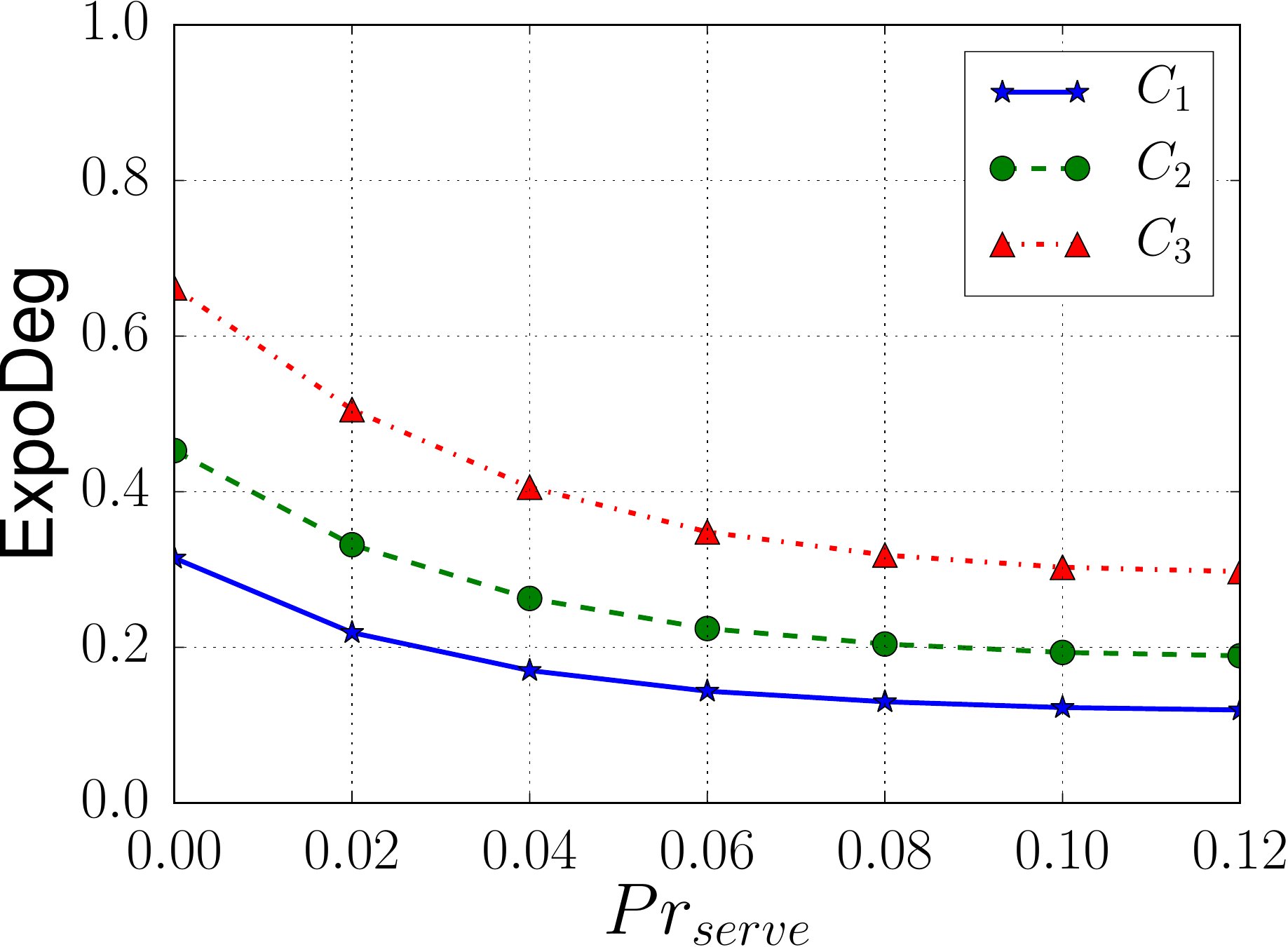}%
		\caption{}
		\label{subfig_lbs_koln}
	\end{subfigure}
	\caption{LuST: $ExpoDeg$ to the \ac{LBS} server as a function of (\subref{subfig_lbs_lust}) $Pr_{serve}$. $ExpoDeg$ to the \ac{LBS} server as a function of $T_{wait}$ with MobiCrowd when (\subref{subfig_mc_lbs_lust}) $Ratio_{coop} = 1$ and (\subref{subfig_mc_05_lbs_lust}) $Ratio_{coop} = 0.5$. $ExpoDeg$ of (\subref{fig_expo_lbs_norsu}) non-serving and (\subref{fig_expo_lbs_rsu}) serving nodes to the \ac{LBS} server as a function of $Pr_{serve}$. TAPASCologne: (\subref{subfig_lbs_koln}) $ExpoDeg$ to the \ac{LBS} server as a function of $Pr_{serve}$. (Default: $T_{wait}=60\ s$ and $Pr_{serve}=0.06$.)}
	\label{fig_lbs_expo}
\end{figure*}

\cref{subfig_lbs_lust,subfig_lbs_koln} show the average $ExpoDeg$ (i.e., the average over all nodes) for exposure to the \ac{LBS} server as a function of $Pr_{serve}$, for different collusion cases, for the LuST and the TAPASCologne scenarios respectively. The collusion case $C_3$ is equivalent to the case that messages are authenticated with node \acp{LTC} (i.e., no pseudonymous authentication). $Pr_{serve}=0$ is equivalent to the case that all queries are sent to the \ac{LBS} server. For the LuST scenario, $ExpoDeg$ is around $0.6$ without any protection in place: even if all the queries are sent to the \ac{LBS} server, $ExpoDeg$ is not $1$ because a node enters and exits one or more regions between two successive queries. With pseudonymous authentication only, $ExpoDeg$ drops below $0.3$ and bounces back to around $0.4$ for the collusion case $C_2$. With the decentralized information sharing scheme in use, the exposure decreases further. For example, when $Pr_{serve}=0.06$, $ExpoDeg$ is around $0.16$ for $C_1$, but rises to around $0.4$ for $C_3$.

Fig.~\ref{subfig_mc_lbs_lust} and fig.~\ref{subfig_mc_05_lbs_lust} show $ExpoDeg$ to the \ac{LBS} server with MobiCrowd when $Ratio_{coop} = 1$ and $Ratio_{coop} = 0.5$ respectively. We see a lower $Ratio_{coop}$ could result in higher $ExpoDeg$. However, $ExpoDeg$ with $T_{wait} = 10\ s$ and $Ratio_{coop} = 0.5$ is lower than $ExpoDeg$ with $Pr_{serve} = 0.12$ in our scheme, although these two settings result in a similar peer hit ratio (\cref{subfig_chr_prserve_lust_2km} and \cref{subfig_chr_mc_05}). This is because serving nodes in our scheme need to continuously expose their locations to obtain regional \ac{POI} data. \cref{fig_expo_lbs_norsu,fig_expo_lbs_rsu} show the $ExpoDeg$ for non-serving nodes and serving nodes respectively. The $ExpoDeg$ for non-serving nodes in our scheme is roughly the same as the $ExpoDeg$ for MobiCrowd in the above two settings. Although serving nodes have much higher $ExpoDeg$, a probabilistic and periodical assignment of serving nodes balances high exposure among nodes throughout their trips. However, the $ExpoDeg$ to honest-but-curious nodes for MobiCrowd is much higher than the $ExpoDeg$ to honest-but-curious nodes for our scheme.

\begin{figure}[h!]
	\begin{subfigure}[b]{0.24\columnwidth}
		\includegraphics[width=\columnwidth]{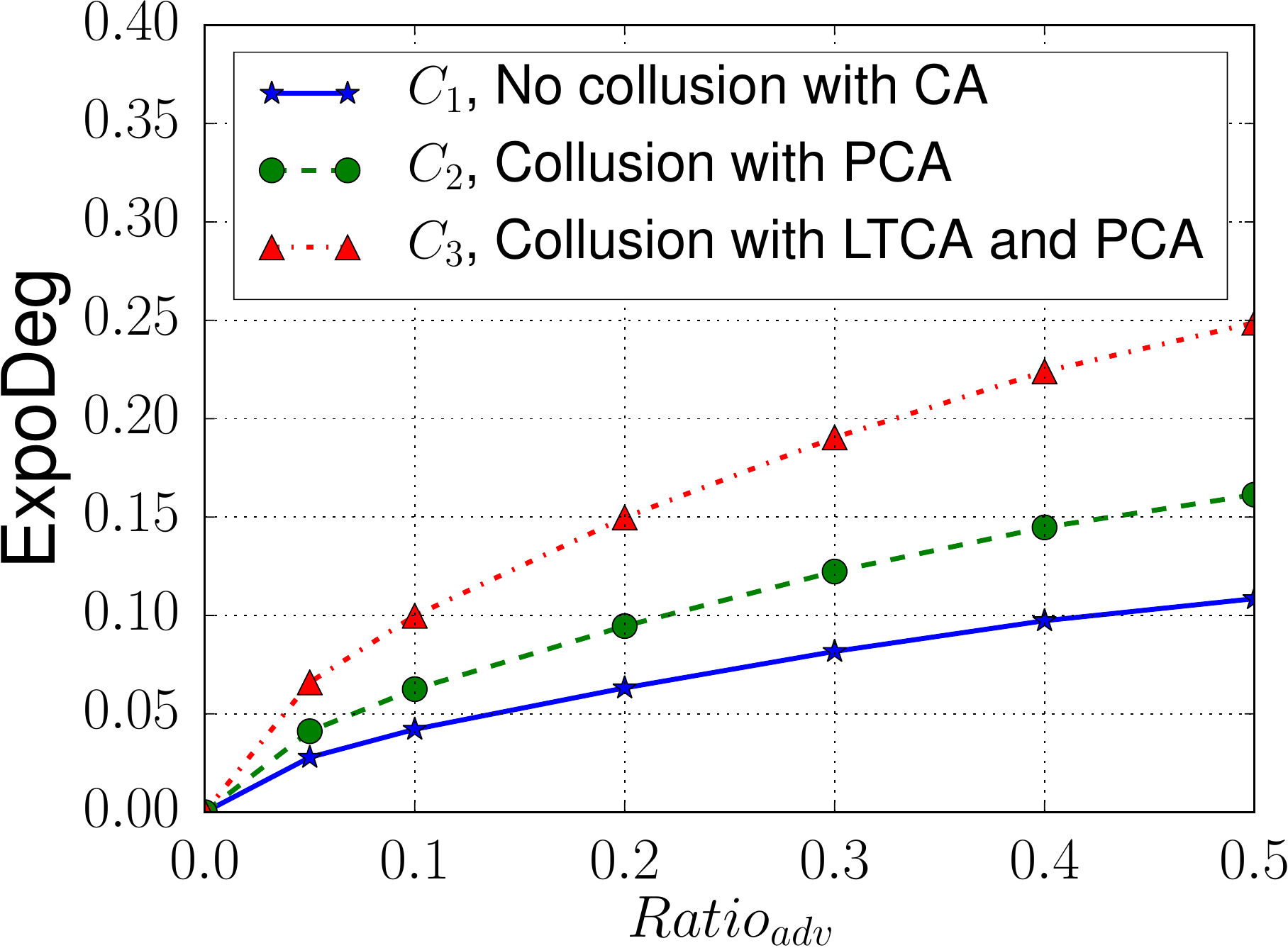}%
		\caption{}%
		\label{fig_expo_ratio_2_enc}%
	\end{subfigure}
	\begin{subfigure}[b]{0.24\columnwidth}
		\includegraphics[width=\columnwidth]{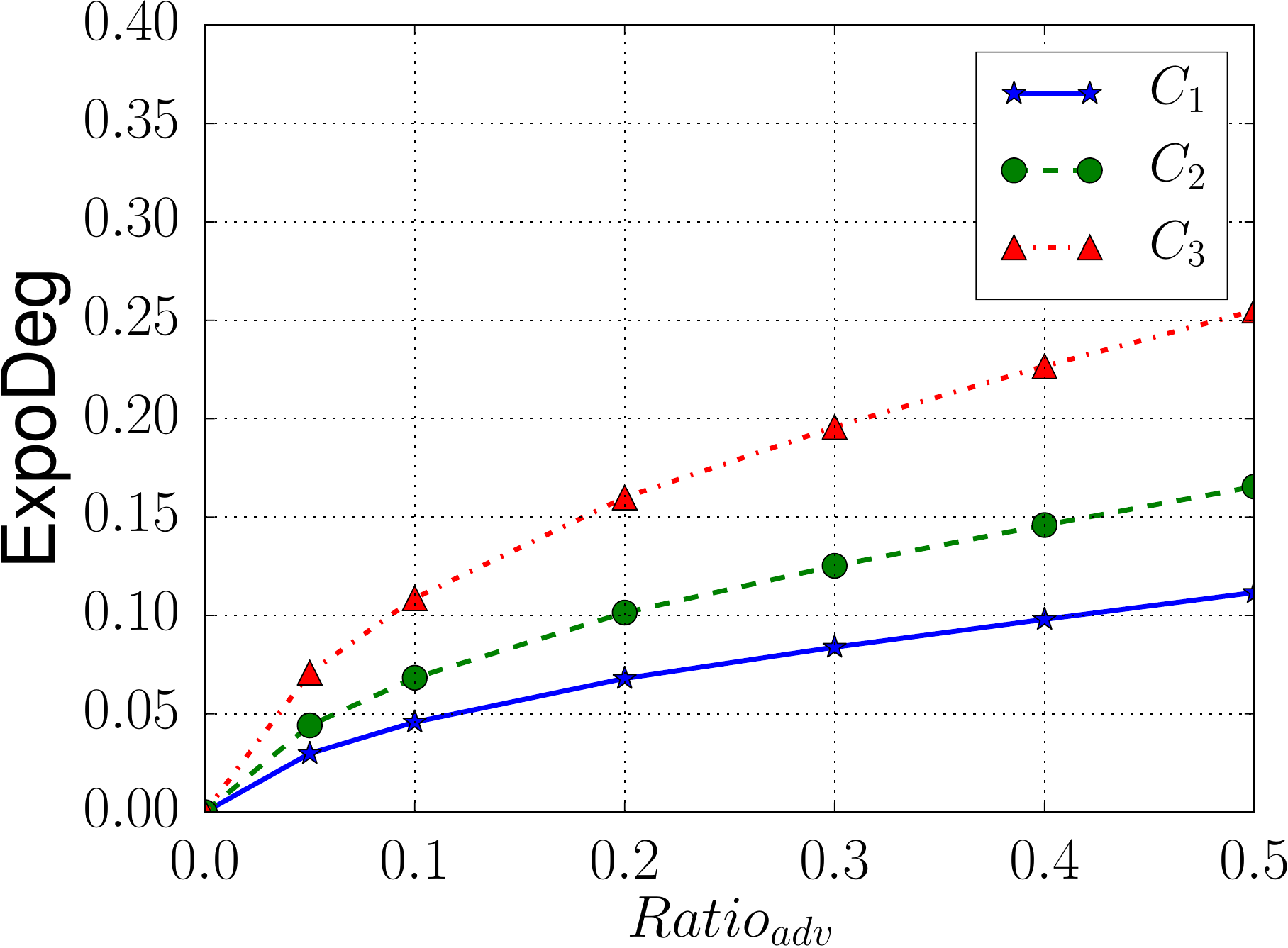}%
		\caption{}%
		\label{fig_expo_ratio_3_enc}%
	\end{subfigure}
	\begin{subfigure}[b]{0.24\columnwidth}
		\includegraphics[width=\columnwidth]{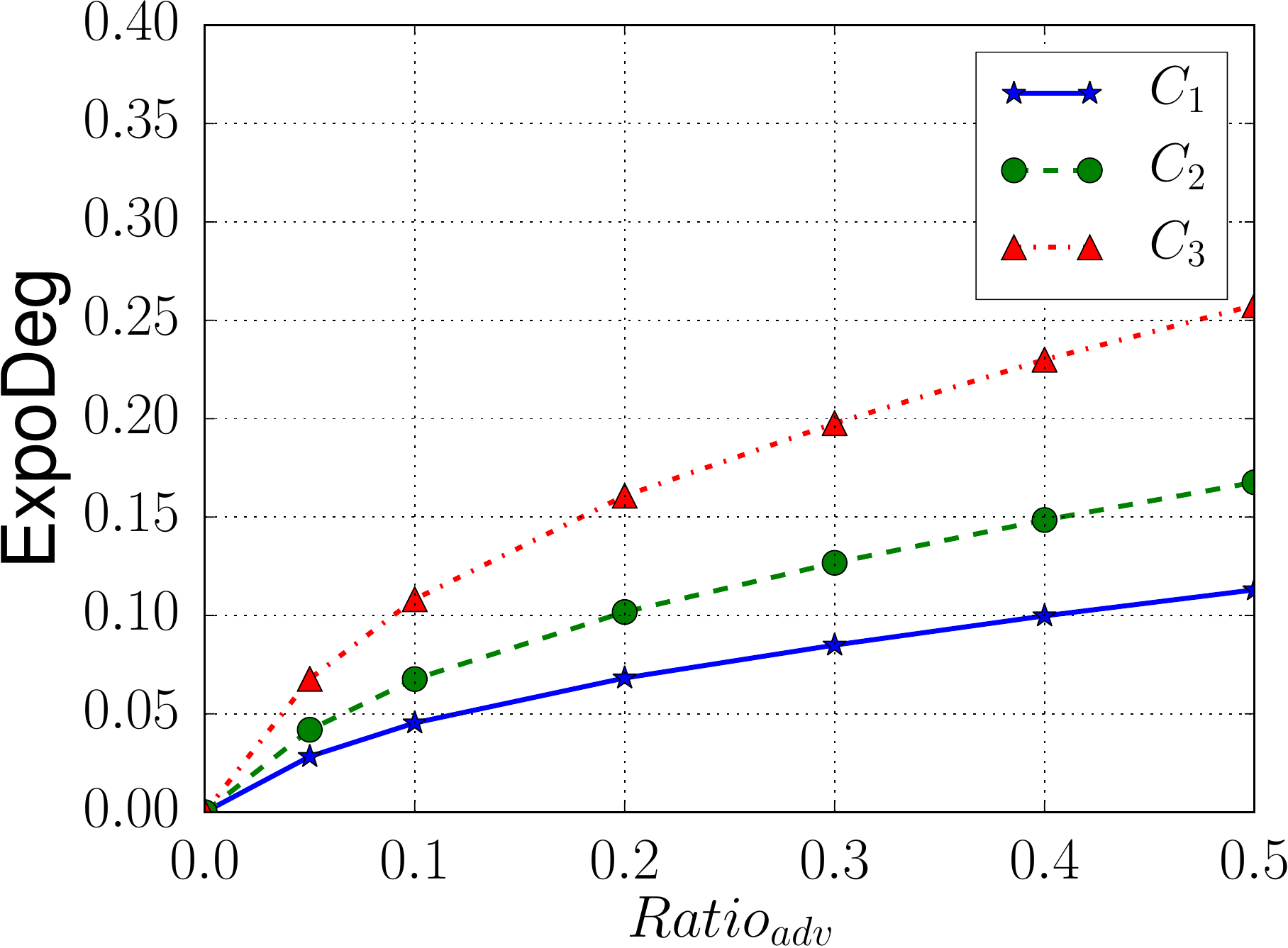}%
		\caption{}%
		\label{fig_expo_ratio_3_enc_r}%
	\end{subfigure}
	
	\begin{subfigure}[b]{0.24\columnwidth}
		\includegraphics[width=\columnwidth]{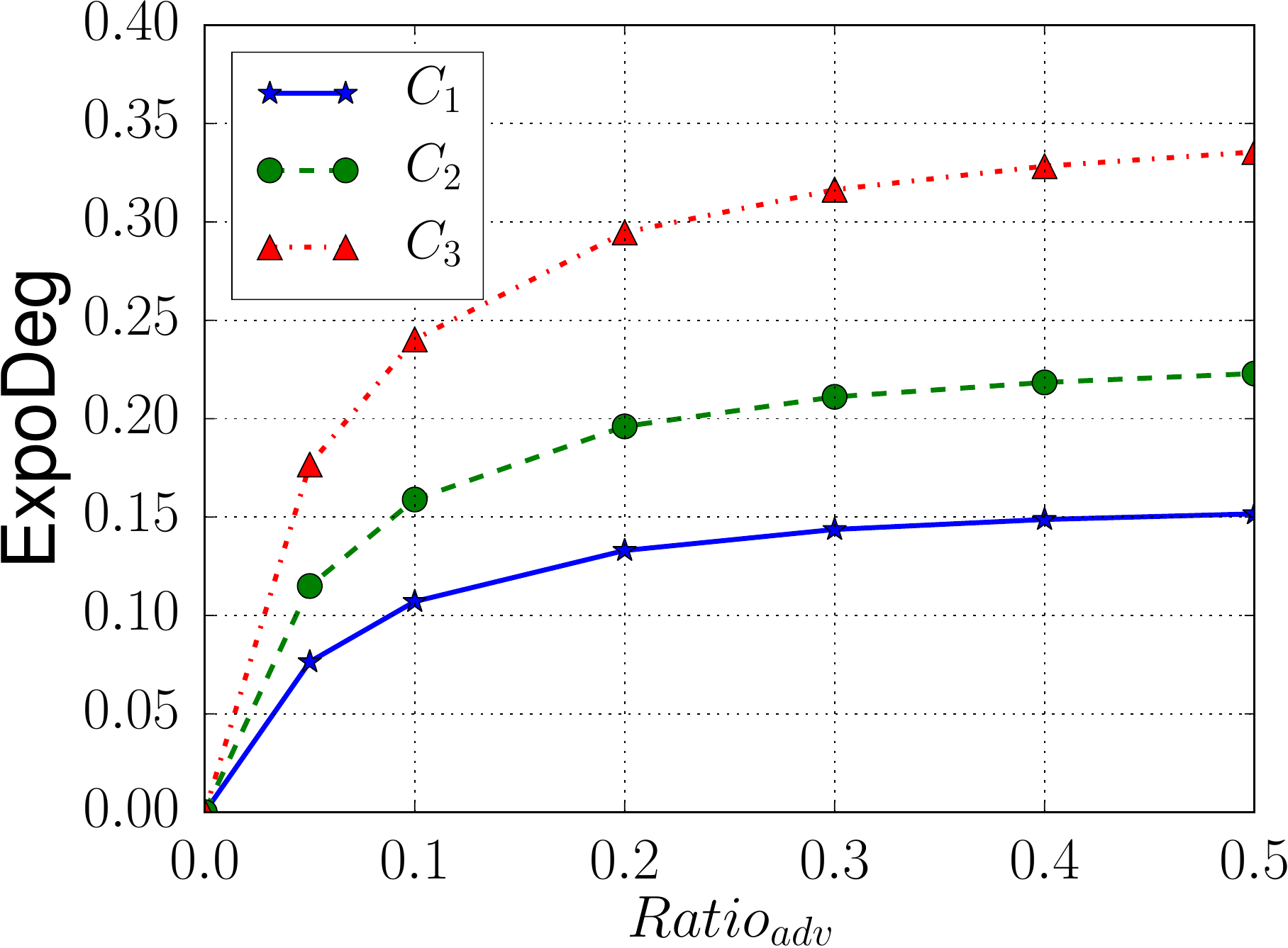}%
		\caption{}%
		\label{fig_expo_ratio_2_noenc}%
	\end{subfigure} 
	\begin{subfigure}[b]{0.24\columnwidth}
		\includegraphics[width=\columnwidth]{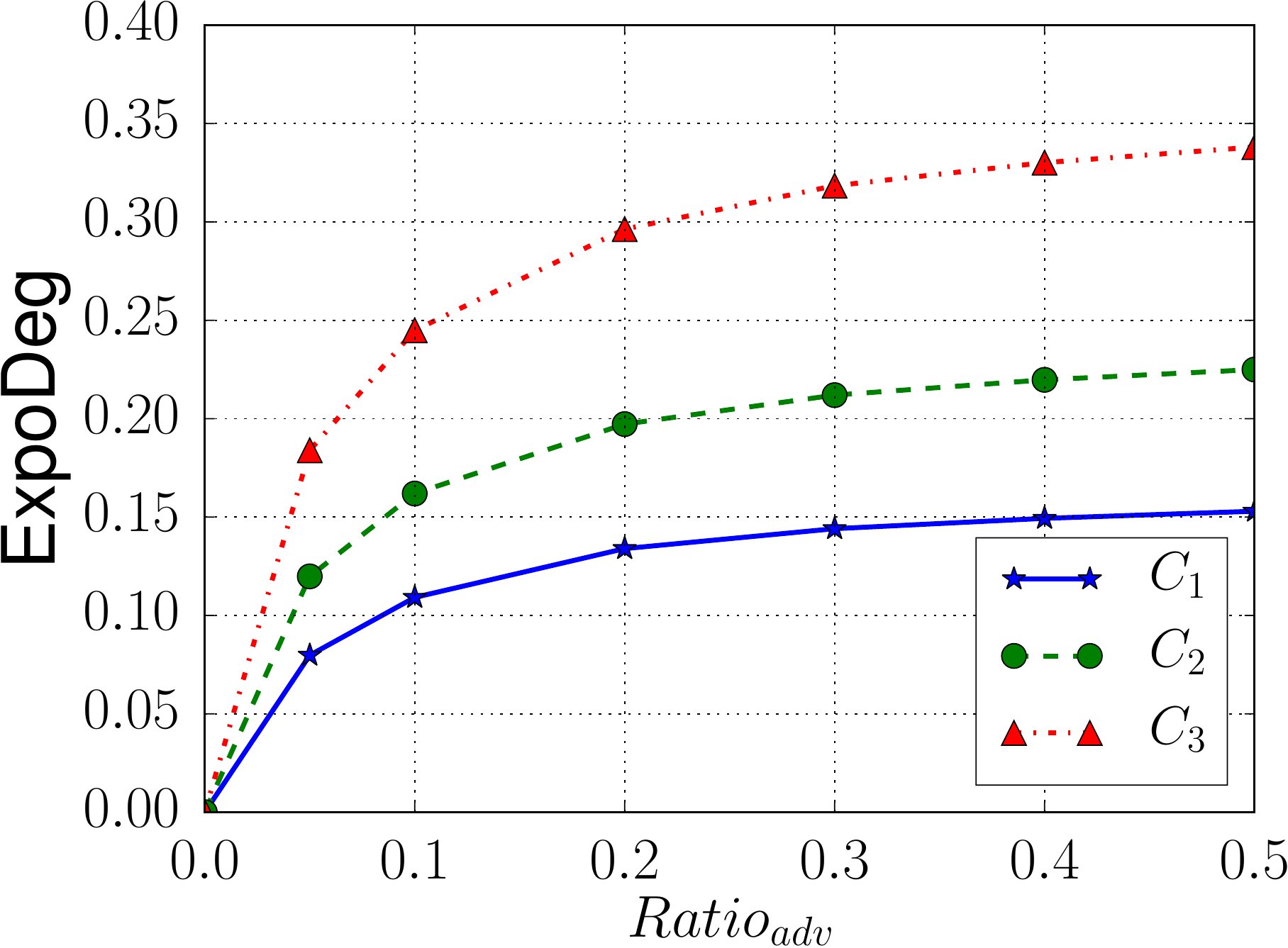}%
		\caption{}%
		\label{fig_expo_ratio_3_noenc}%
	\end{subfigure}
	\begin{subfigure}[b]{0.24\columnwidth}
		\includegraphics[width=\columnwidth]{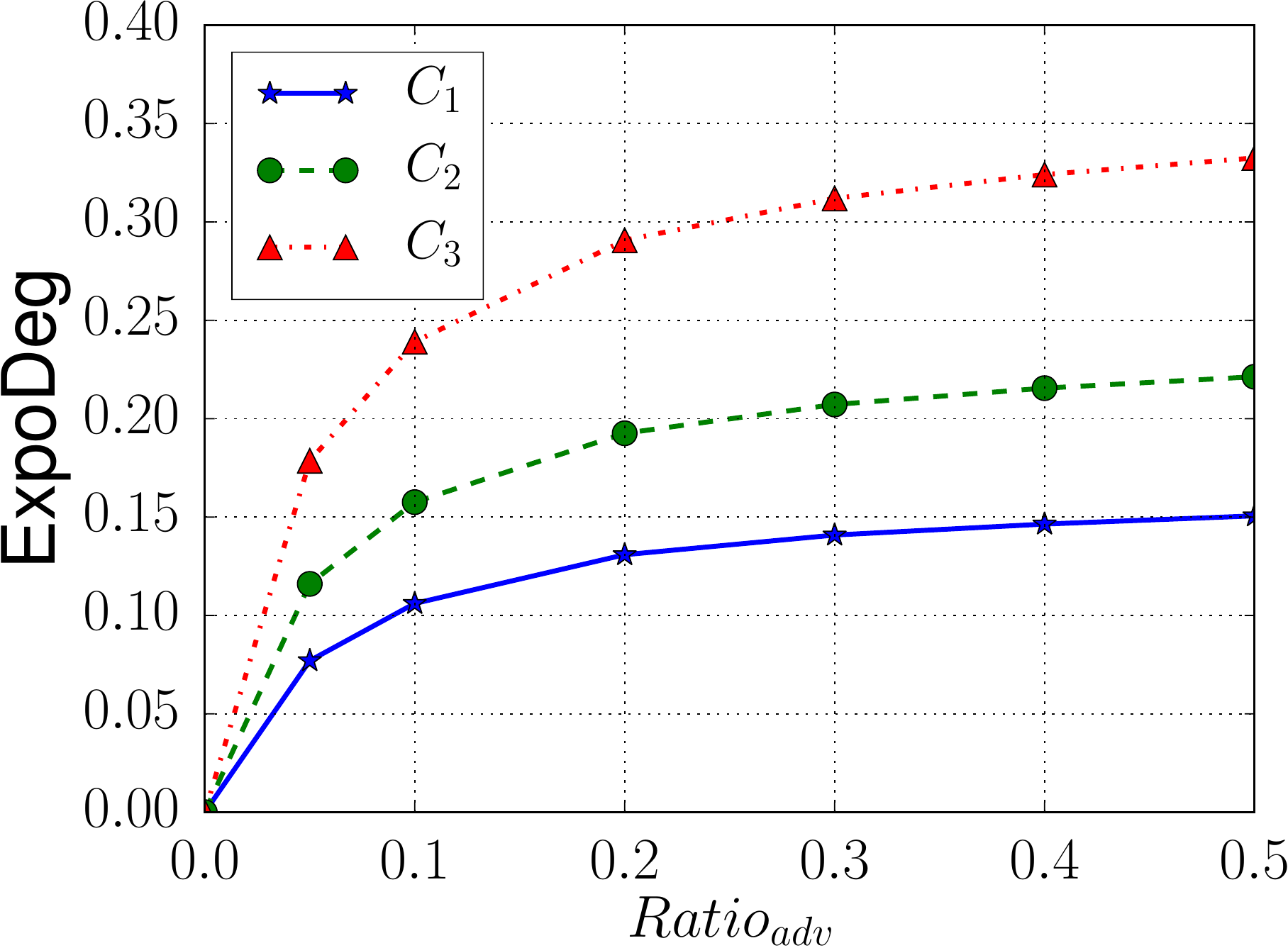}%
		\caption{}%
		\label{fig_expo_ratio_3_noenc_r}%
	\end{subfigure}
	\caption{LuST: $ExpoDeg$ to colluding honest-but-curious nodes as a function of $Ratio_{adv}$ with (first row) and without (second row) \ac{P2P} encryption when $N=2$ (first column), $N=3$ (second column) and $N=3$ with $T_{beacon} \sim uniform (5, 15)s$ (third column).}
	\label{fig_expo_ratio}
\end{figure}

\begin{figure}[h!]
	\begin{subfigure}[b]{.24\columnwidth}
		\includegraphics[width=\columnwidth]{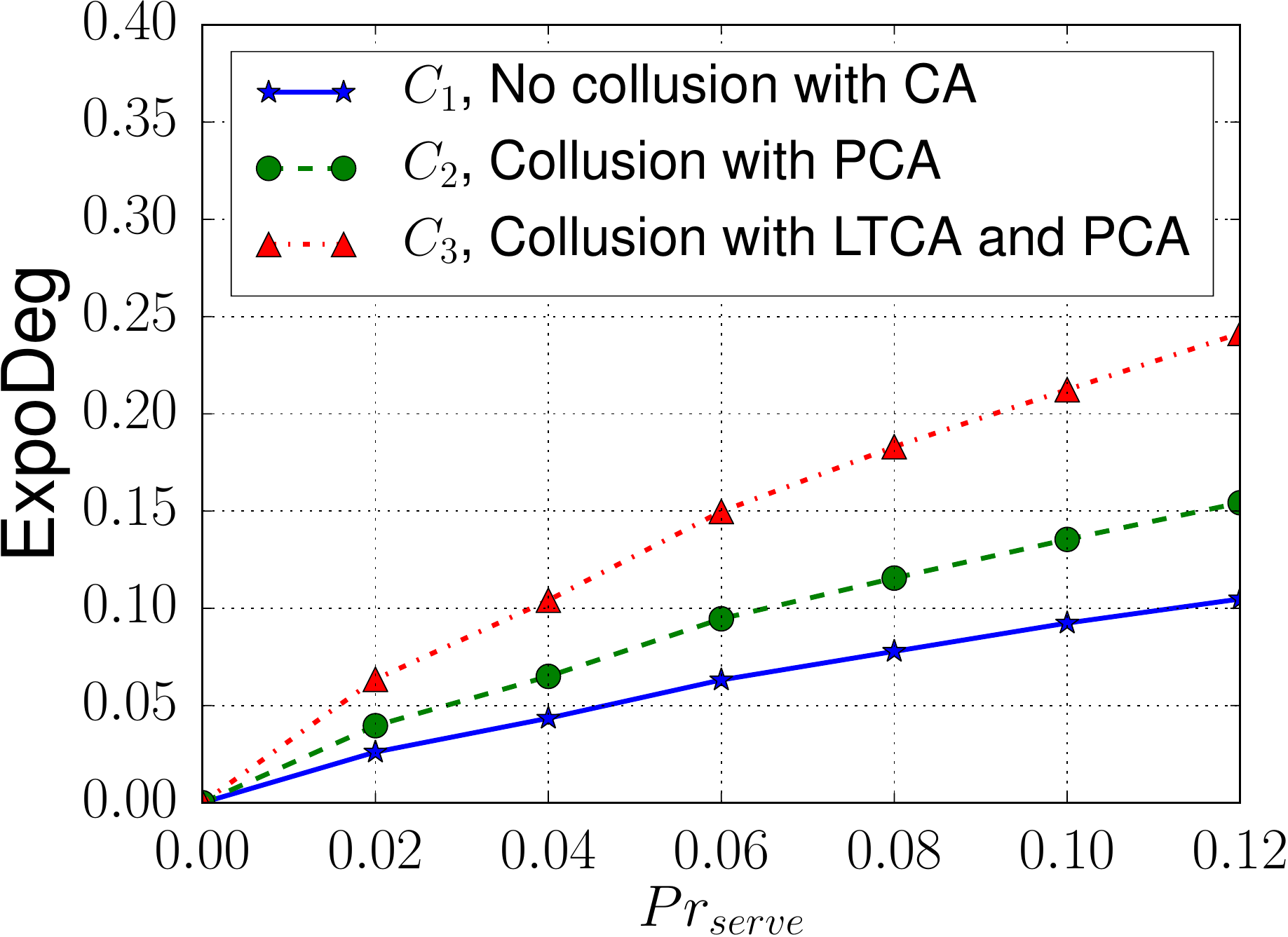}%
		\caption{}%
		\label{fig_expo_prserve_2_enc}%
	\end{subfigure}
	\begin{subfigure}[b]{.24\columnwidth}
		\includegraphics[width=\columnwidth]{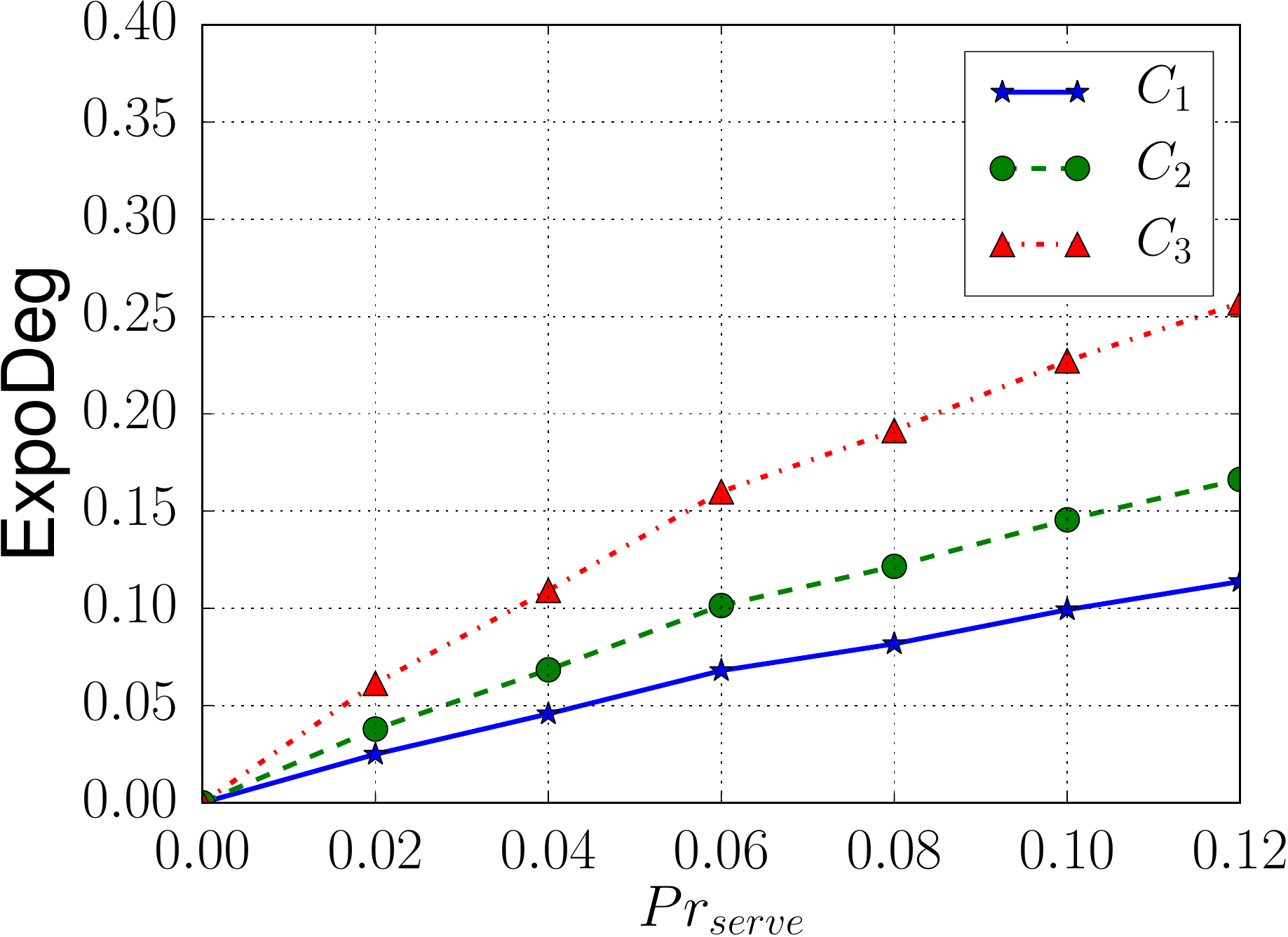}%
		\caption{}%
		\label{fig_expo_prserve_3_enc}%
	\end{subfigure}
	\begin{subfigure}[b]{.24\columnwidth}
		\includegraphics[width=\columnwidth]{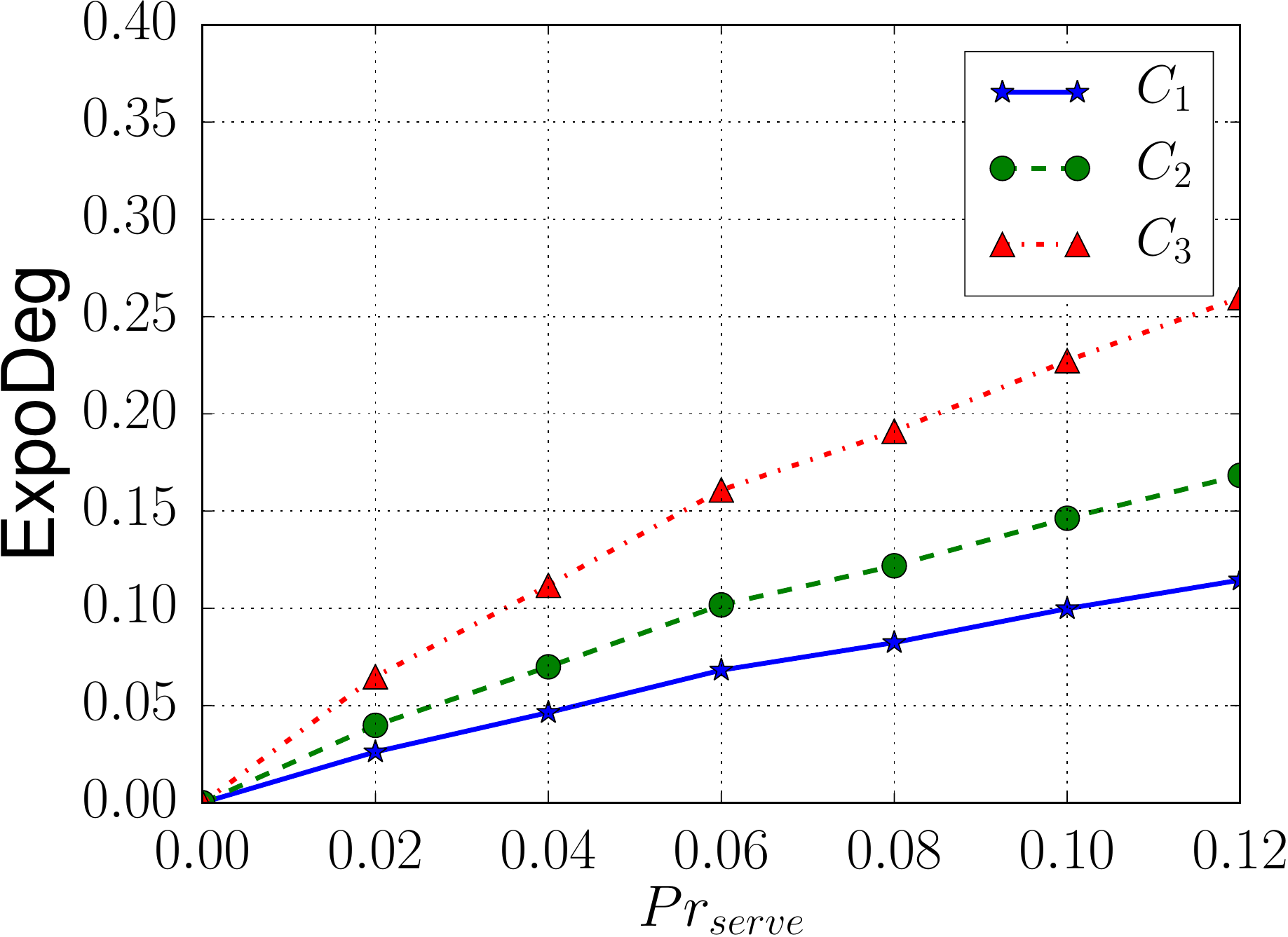}%
		\caption{}%
		\label{fig_expo_prserve_3_enc_r}%
	\end{subfigure}
	
	\begin{subfigure}[b]{0.24\columnwidth}
		\includegraphics[width=\columnwidth]{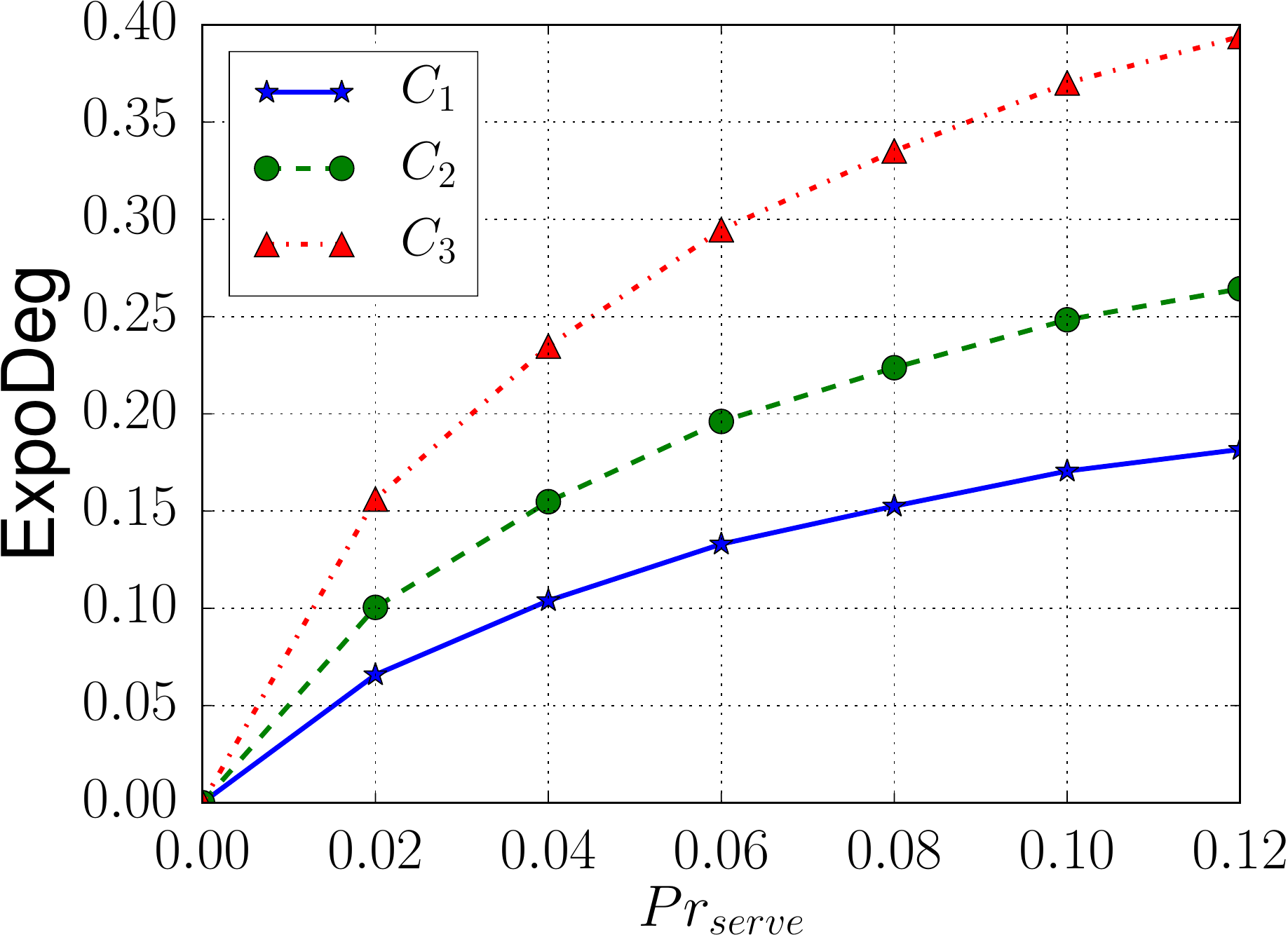}%
		\caption{}%
		\label{fig_expo_prserve_2_noenc}%
	\end{subfigure}
	\begin{subfigure}[b]{0.24\columnwidth}
		\includegraphics[width=\columnwidth]{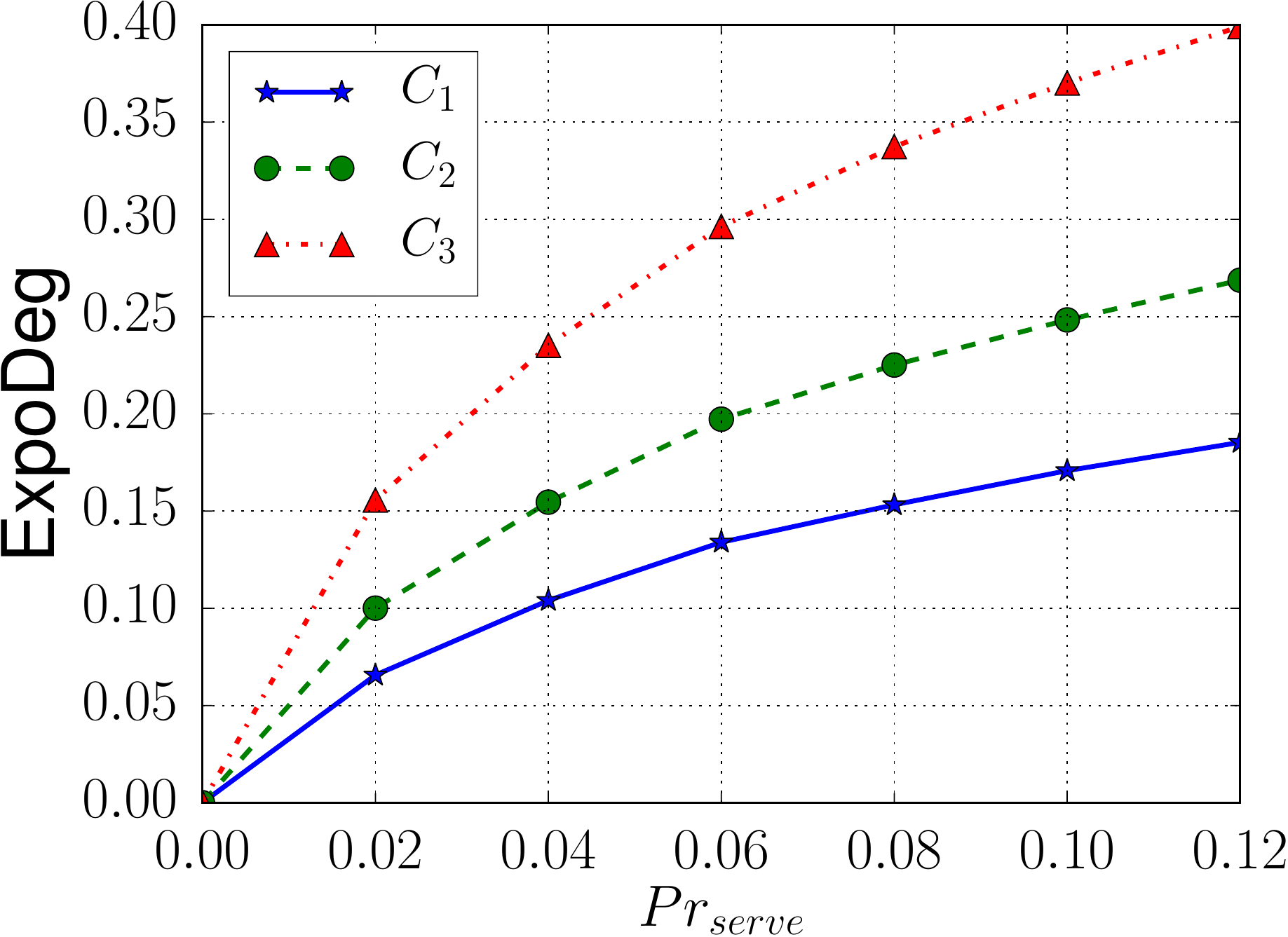}%
		\caption{}%
		\label{fig_expo_prserve_3_noenc}
	\end{subfigure}
	\begin{subfigure}[b]{0.24\columnwidth}
		\includegraphics[width=\columnwidth]{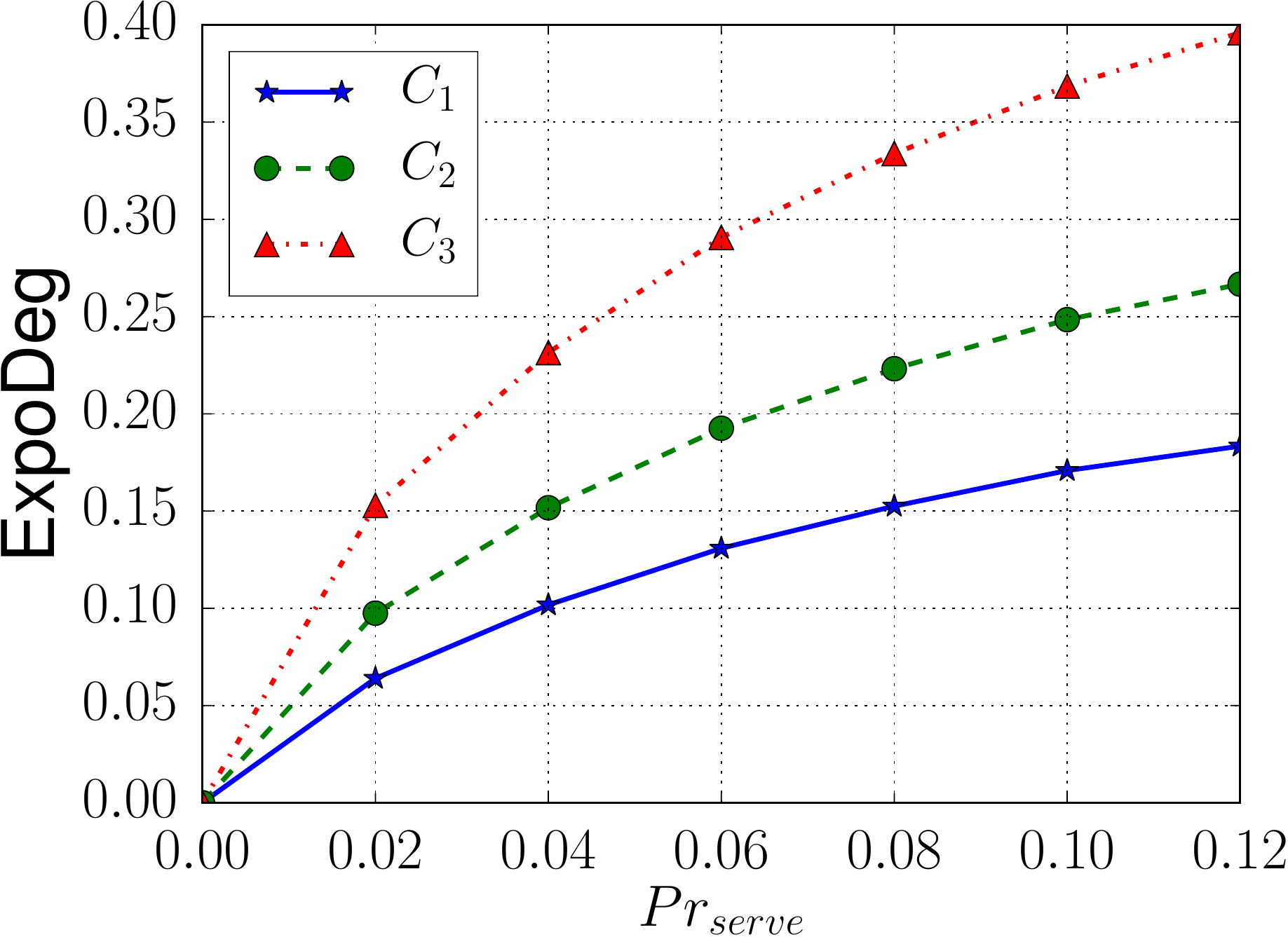}%
		\caption{}%
		\label{fig_expo_prserve_3_noenc_r}
	\end{subfigure}
	\caption{LuST: $ExpoDeg$ to colluding honest-but-curious nodes as a function of $Pr_{serve}$ with (first row) and without (second row) \ac{P2P} encryption when $N=2$ (first column), $N=3$ (second column), and $N=3$ with $T_{beacon} \sim uniform (5, 15)s$ (third column).}
	\label{fig_expo_prserve}
\end{figure}

Fig.~\ref{fig_expo_ratio} and Fig.~\ref{fig_expo_prserve} show $ExpoDeg$ to honest-but-curious nodes. We assume that the honest-but-curious nodes could collude, merge recorded queries and link the queries (as the \ac{LBS} could do). Fig.~\ref{fig_expo_ratio} shows $ExpoDeg$ as a function of $Ratio_{adv}$. Fig.~\ref{fig_expo_ratio_2_enc} and Fig.~\ref{fig_expo_ratio_2_noenc} show $ExpoDeg$ when $N=2$, with and without \ac{P2P} encryption respectively. Fig.~\ref{fig_expo_ratio_2_enc} shows that a modest realistic $Ratio_{adv}$ (e.g., $0.05$ and $0.1$) results in relatively low $ExpoDeg$. For example, when $Ratio_{adv}=0.05$, $ExpoDeg$ is lower than $0.05$ for $C_1$ and $C_2$, and it is slightly higher for $C_3$. However, without \ac{P2P} encryption (Fig.~\ref{fig_expo_ratio_2_noenc}), $ExpoDeg$ significantly increases, because all queries within the communication range of an honest-but-curious (serving or non-serving) node can be recorded. For example, $ExpoDeg$ for $Ratio_{adv}=0.05$ and $Ratio_{adv}=0.1$ without encryption are almost the same as those for $Ratio_{adv}=0.3$ and $Ratio_{adv}=0.5$ with encryption, respectively. This shows the importance of query encryption in terms of reducing node exposure. We see the same effect (mentioned above for $N=2$) in Fig.~\ref{fig_expo_ratio_3_enc} and Fig.~\ref{fig_expo_ratio_3_noenc}, when $N=3$. Fig.~\ref{fig_expo_ratio_2_noenc} and Fig.~\ref{fig_expo_ratio_3_noenc} show that $ExpoDeg$ steadily increases as $Ratio_{adv}$ increases, and rather ``flattens'' out as $Ratio_{adv}$ grows from $0.2$ to $0.5$. This shows that non-encrypted \ac{P2P} communication significantly increases $ExpoDeg$, as it strengthens the capability of honest-but-curious nodes even in settings with relatively low $Ratio_{adv}$.  $ExpoDeg$ only slightly increases from $N=2$ (i.e., Fig.~\ref{fig_expo_ratio_2_enc} and Fig.~\ref{fig_expo_ratio_2_noenc}) to $N=3$ (i.e., Fig.~\ref{fig_expo_ratio_3_enc} and Fig.~\ref{fig_expo_ratio_3_noenc}), because a querying node cannot always discover exactly $N$ serving nodes within $T_{wait}$. An increase in $N$ could help evicting malicious nodes from the system more efficiently with a slightly higher overhead (see the evaluation for $resilience$ below).


{\color{blue}

	\begin{figure}[h!]
		\begin{subfigure}[b]{.24\columnwidth}
			\includegraphics[width=\columnwidth]{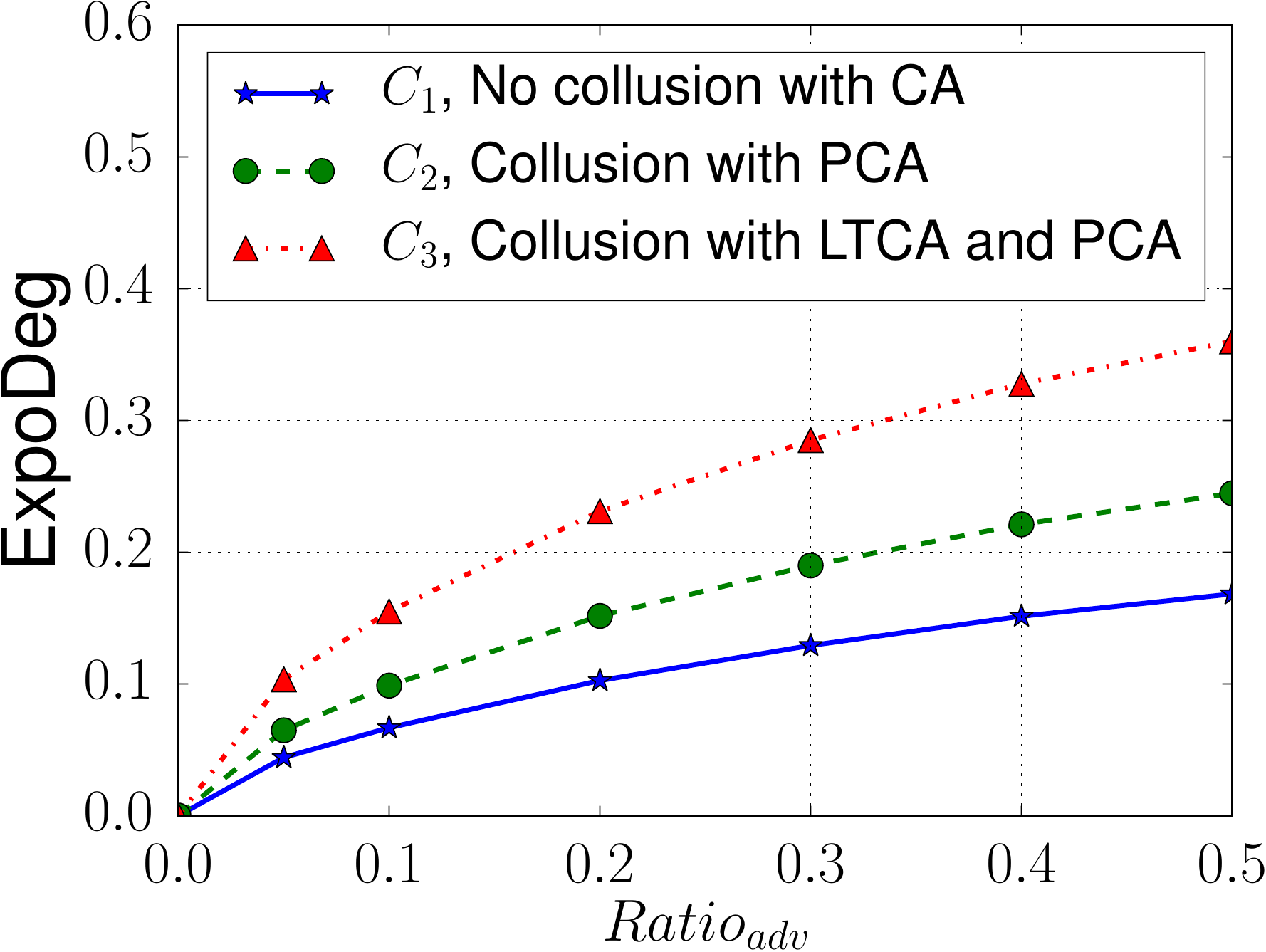}%
			\caption{}%
			\label{fig_expo_adv_3_enc_2km_koln}%
		\end{subfigure}
		\begin{subfigure}[b]{.24\columnwidth}
			\includegraphics[width=\columnwidth]{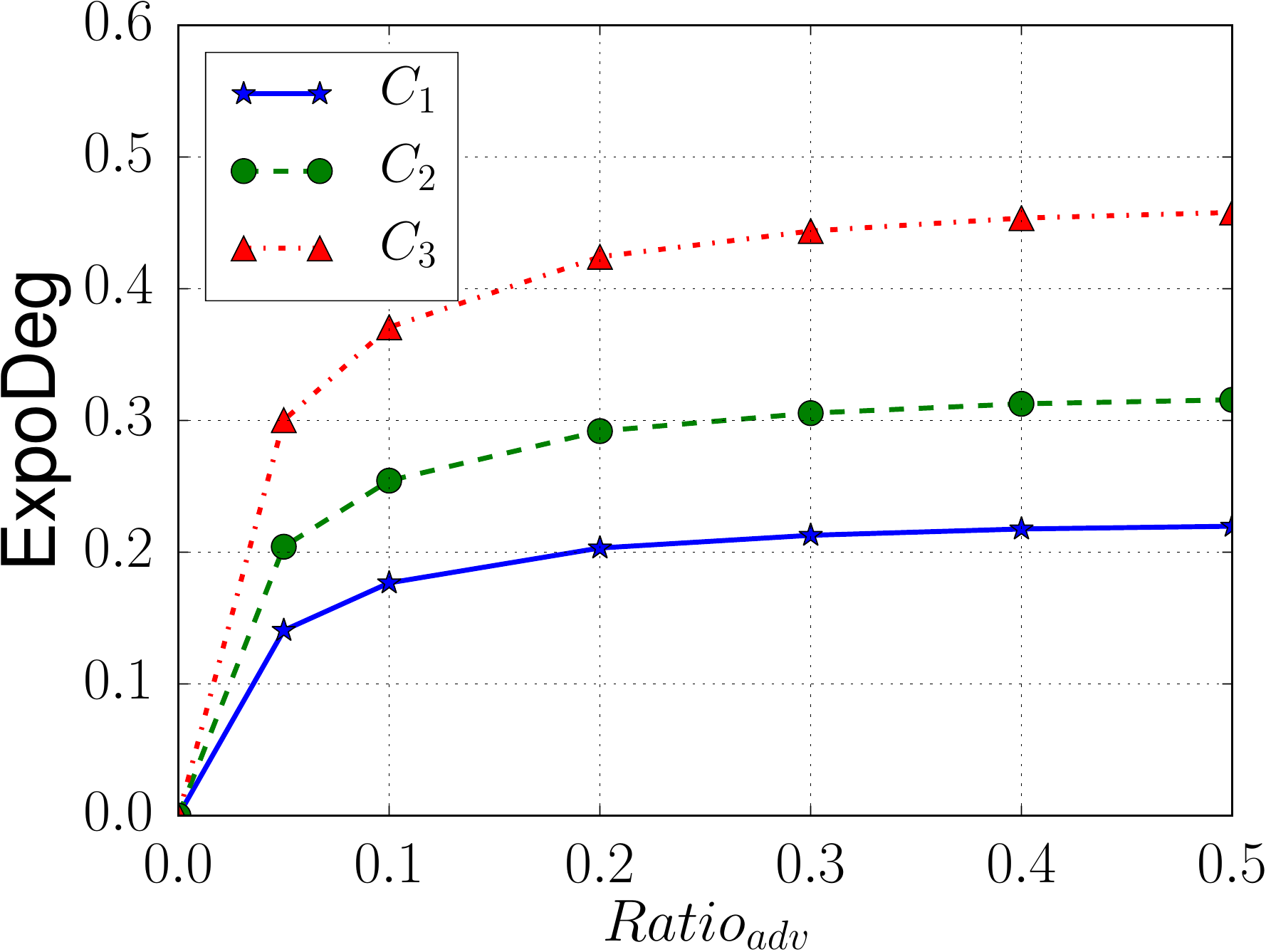}%
			\caption{}%
			\label{fig_expo_adv_3_noenc_2km_koln}%
		\end{subfigure}
		\begin{subfigure}[b]{.24\columnwidth}
			\includegraphics[width=\columnwidth]{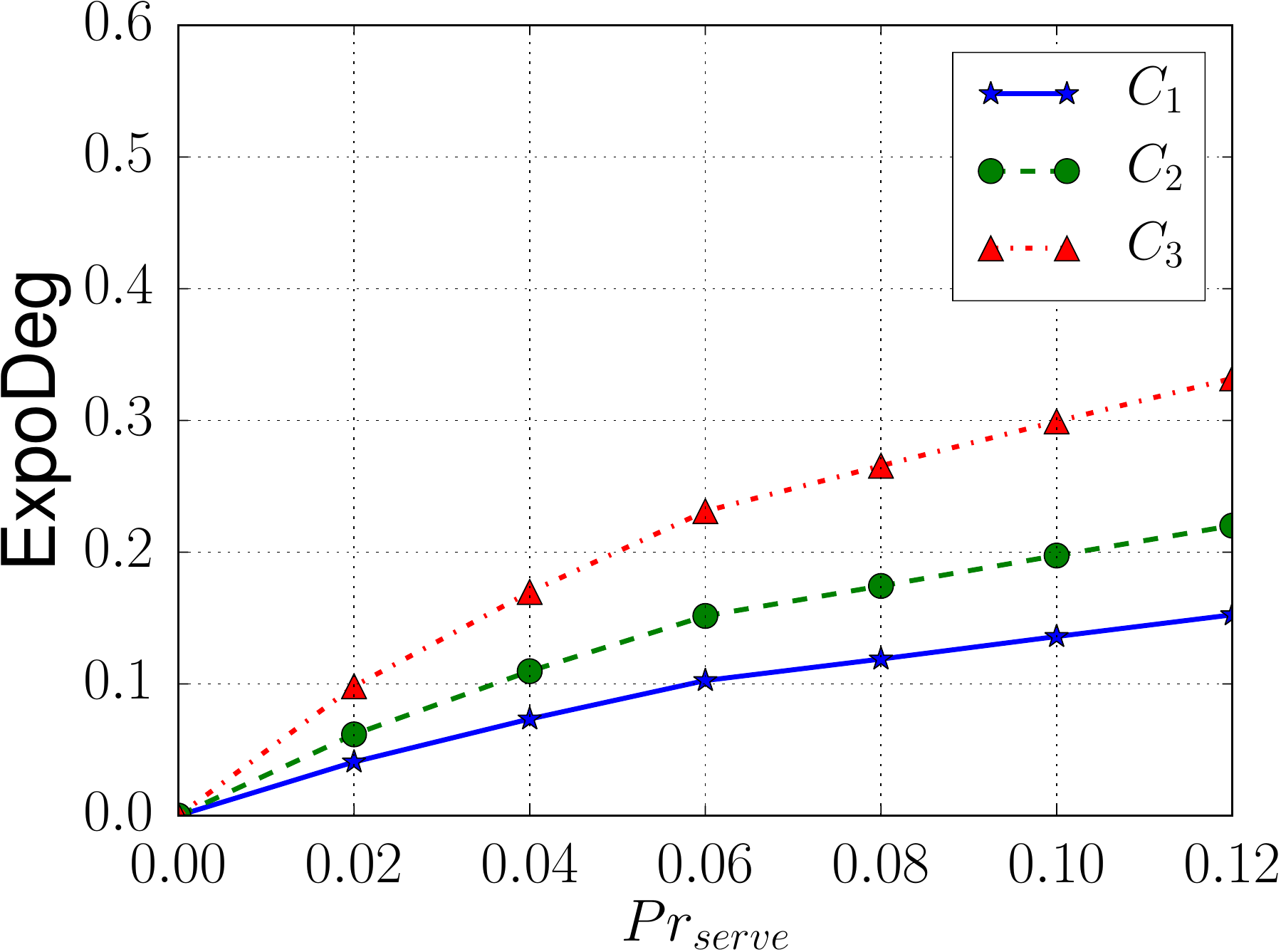}%
			\caption{}%
			\label{fig_expo_prserve_3_enc_2km_koln}%
		\end{subfigure}
		\begin{subfigure}[b]{.24\columnwidth}
			\includegraphics[width=\columnwidth]{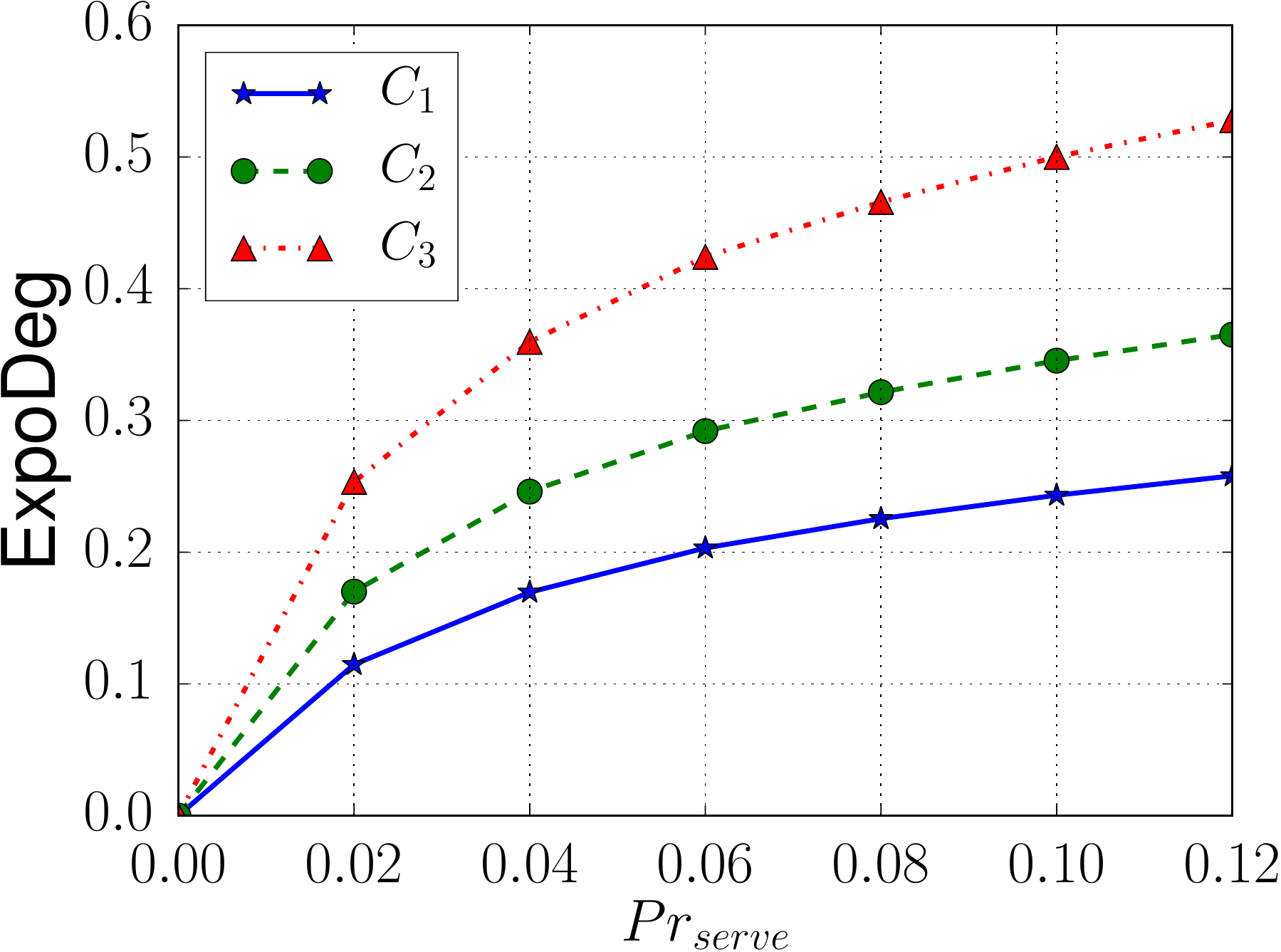}%
			\caption{}%
			\label{fig_expo_prserve_3_noenc_2km_koln}%
		\end{subfigure}
		\caption{TAPASCologne: $ExpoDeg$ to colluding honest-but-curious nodes with and without \ac{P2P} encryption as a function of (\subref{fig_expo_adv_3_enc_2km_koln}, \subref{fig_expo_adv_3_noenc_2km_koln}) $Ratio_{adv}$ and (\subref{fig_expo_prserve_3_enc_2km_koln}, \subref{fig_expo_prserve_3_noenc_2km_koln}) $Pr_{serve}$. (Default: $Ratio_{adv}=0.2$ and $Pr_{serve}=0.06$.)}
		\label{fig_expo_adv_region_koln}
	\end{figure}
	
	\begin{figure}[h!]
		\begin{subfigure}[b]{.24\columnwidth}
			\includegraphics[width=\columnwidth]{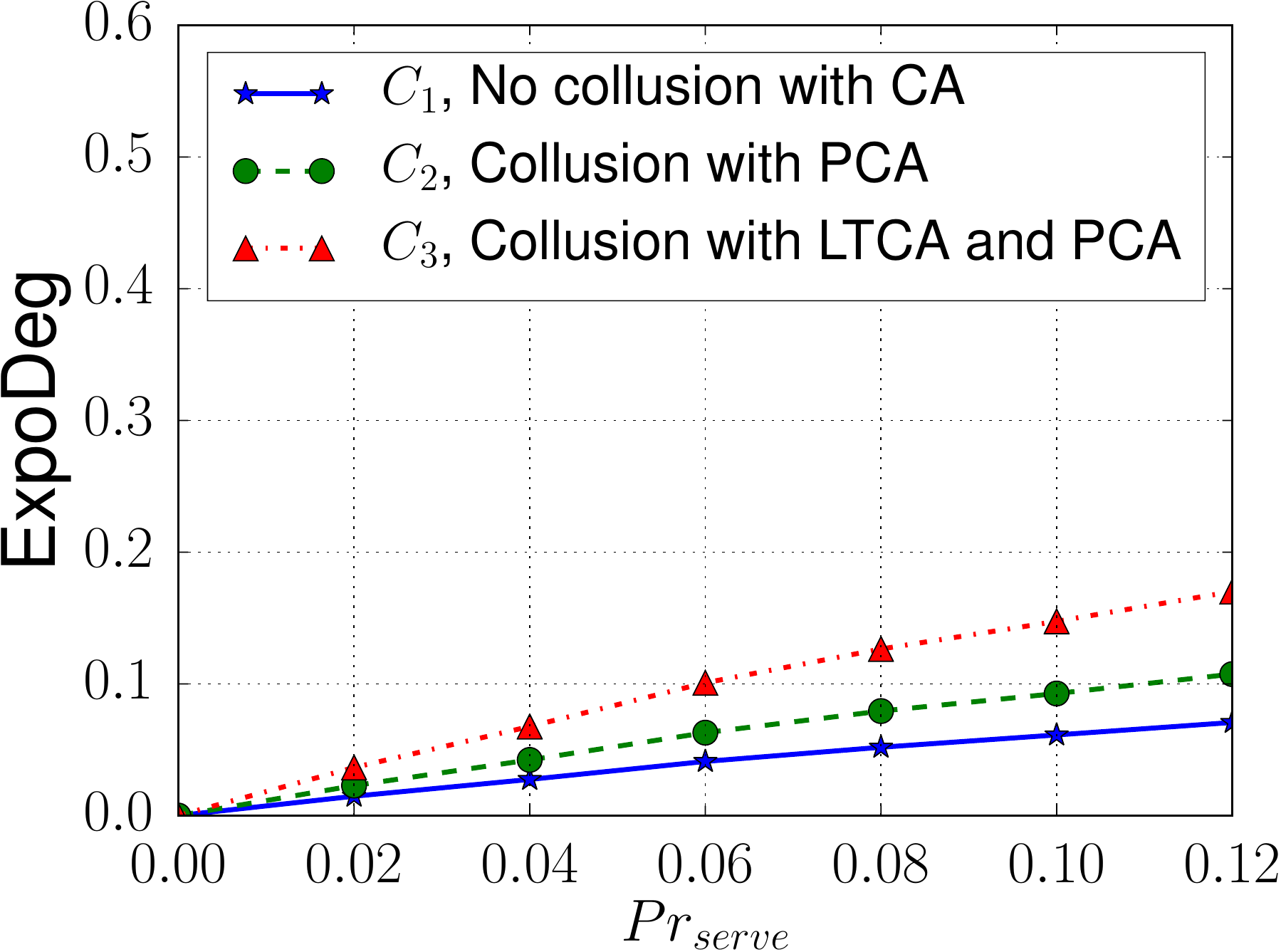}%
			\caption{}%
			\label{fig_expo_prserve_3_enc_1km_lust}%
		\end{subfigure}
		\begin{subfigure}[b]{.24\columnwidth}
			\includegraphics[width=\columnwidth]{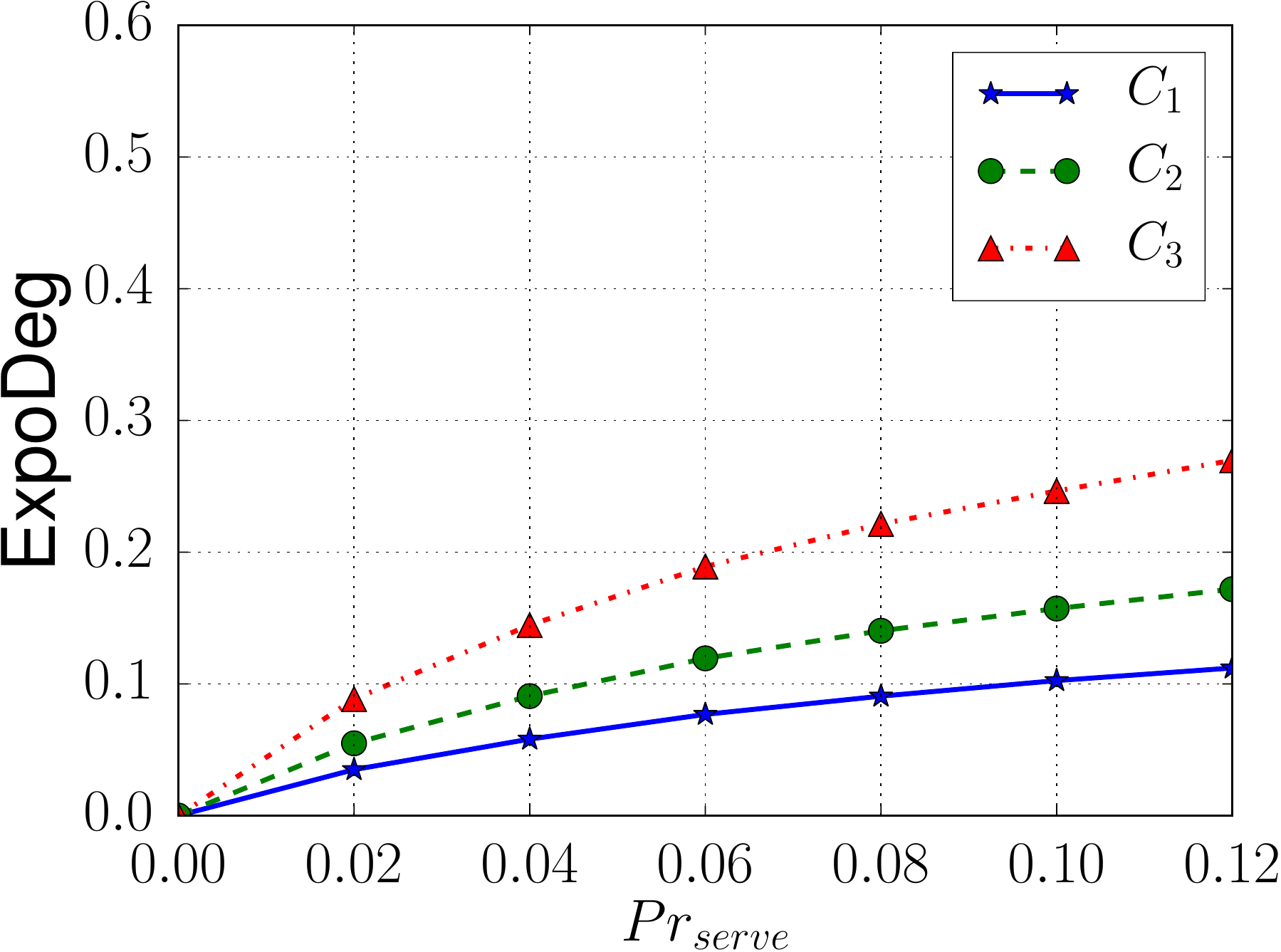}%
			\caption{}%
			\label{fig_expo_prserve_3_noenc_1km_lust}%
		\end{subfigure}
		\begin{subfigure}[b]{.24\columnwidth}
			\includegraphics[width=\columnwidth]{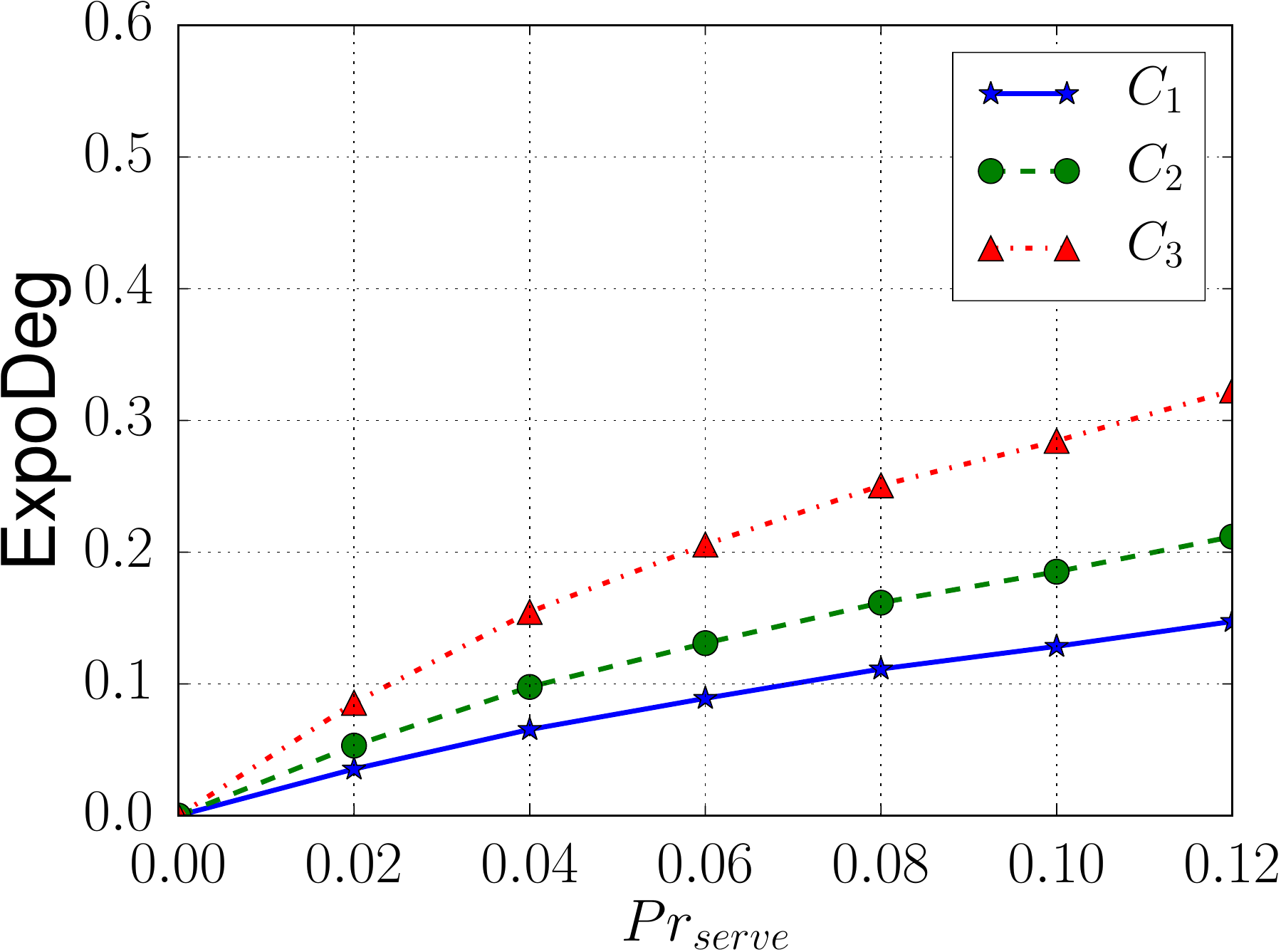}%
			\caption{}%
			\label{fig_expo_prserve_3_enc_3km_lust}%
		\end{subfigure}
		\begin{subfigure}[b]{.24\columnwidth}
			\includegraphics[width=\columnwidth]{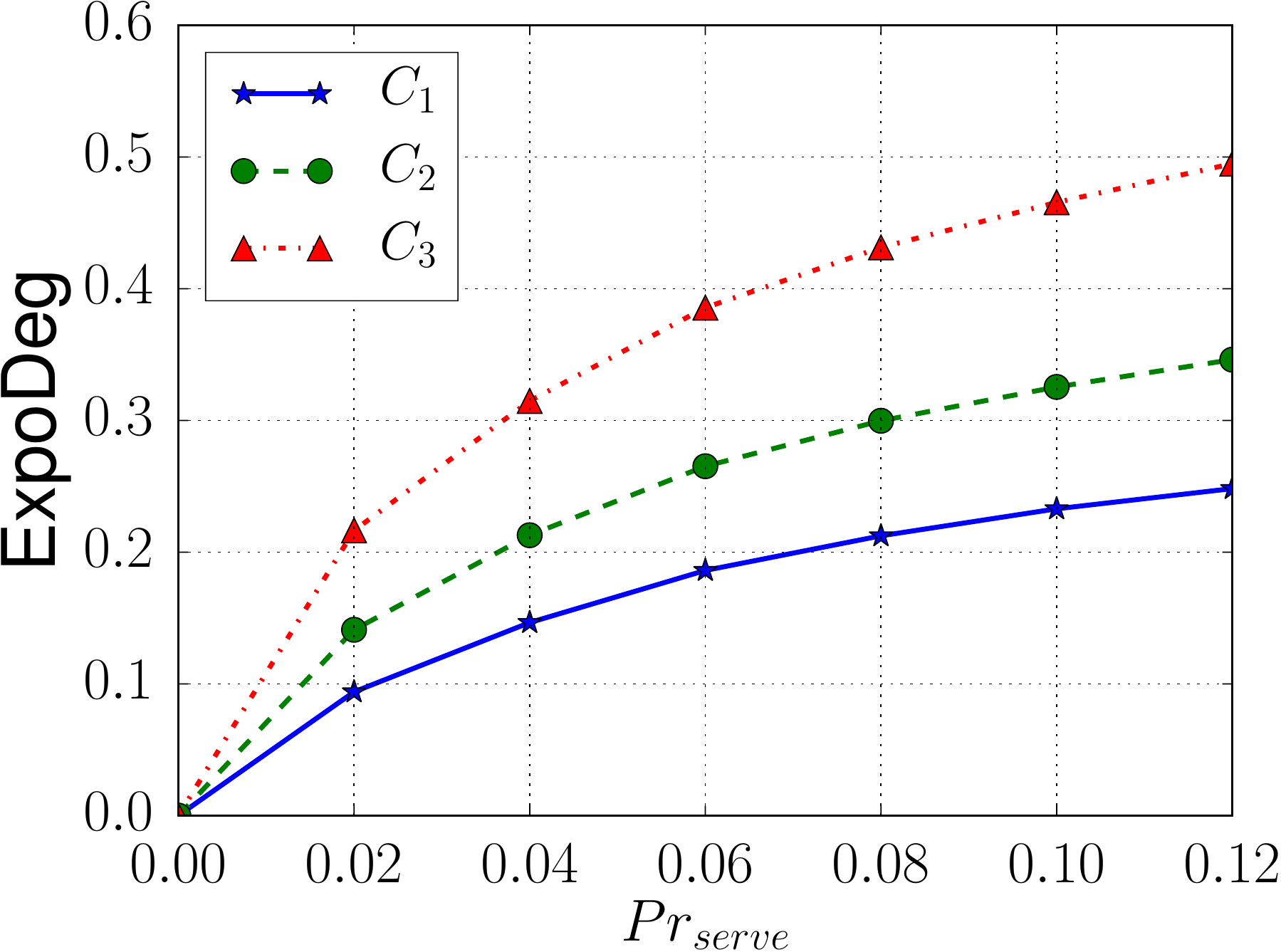}%
			\caption{}%
			\label{fig_expo_prserve_3_noenc_3km_lust}%
		\end{subfigure}
		\caption{LuST: $ExpoDeg$ to colluding honest-but-curious nodes as a function of $Pr_{serve}$ with and without \ac{P2P} encryption for (\subref{fig_expo_prserve_3_enc_1km_lust}, \subref{fig_expo_prserve_3_noenc_1km_lust}) $L = 1$ $km$ and (\subref{fig_expo_prserve_3_enc_3km_lust}, \subref{fig_expo_prserve_3_noenc_3km_lust}) $L = 3$ $km$.}
		\label{fig_expo_prserve_region_lust}
	\end{figure}
	\begin{figure}[htp!]
		\begin{subfigure}[b]{0.24\columnwidth}
			\includegraphics[width=\columnwidth]{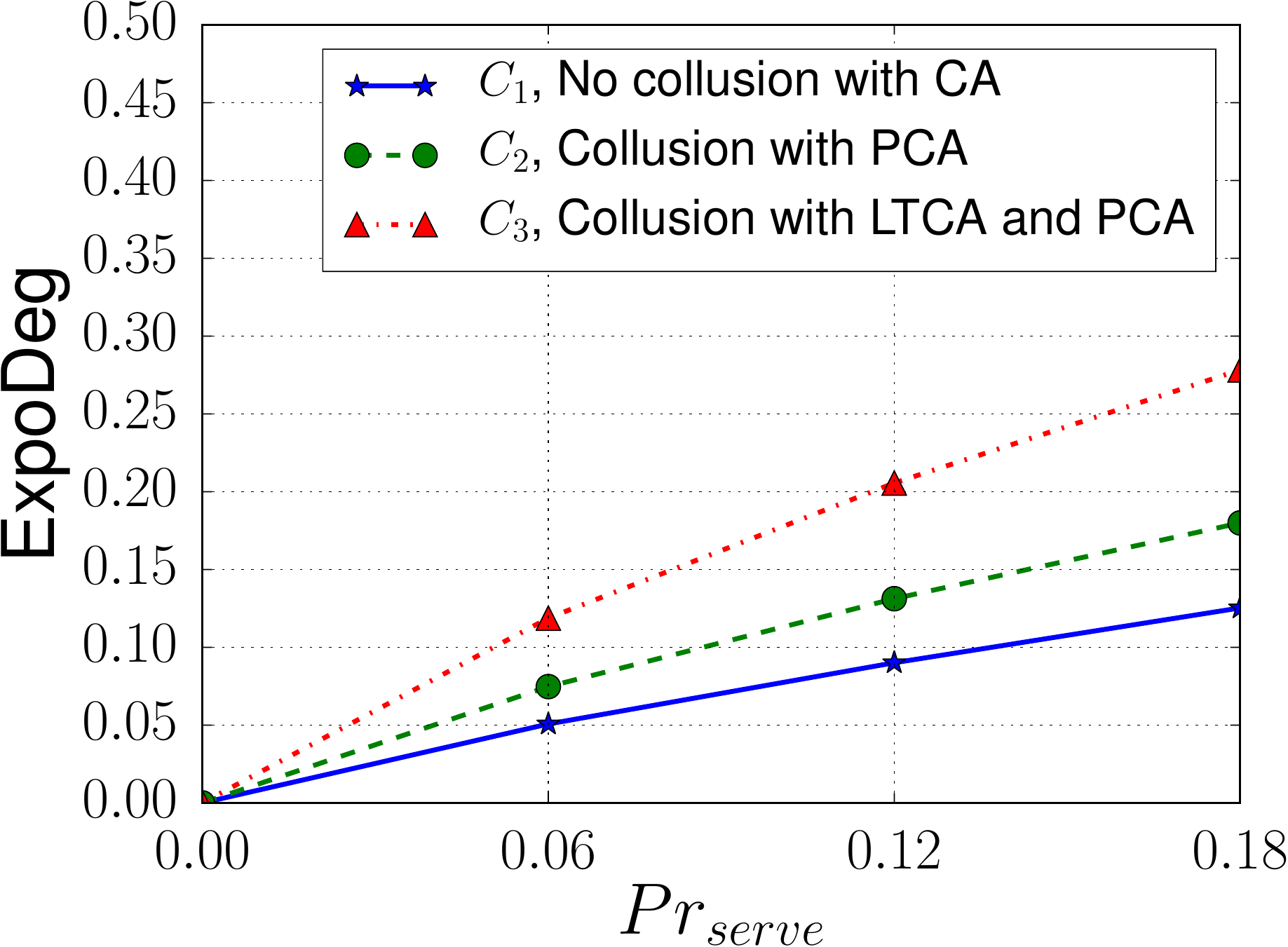}
			\caption{}
			\label{subfig_expodeg_type_2_enc_lust}
		\end{subfigure}
		\begin{subfigure}[b]{0.24\columnwidth}
			\includegraphics[width=\columnwidth]{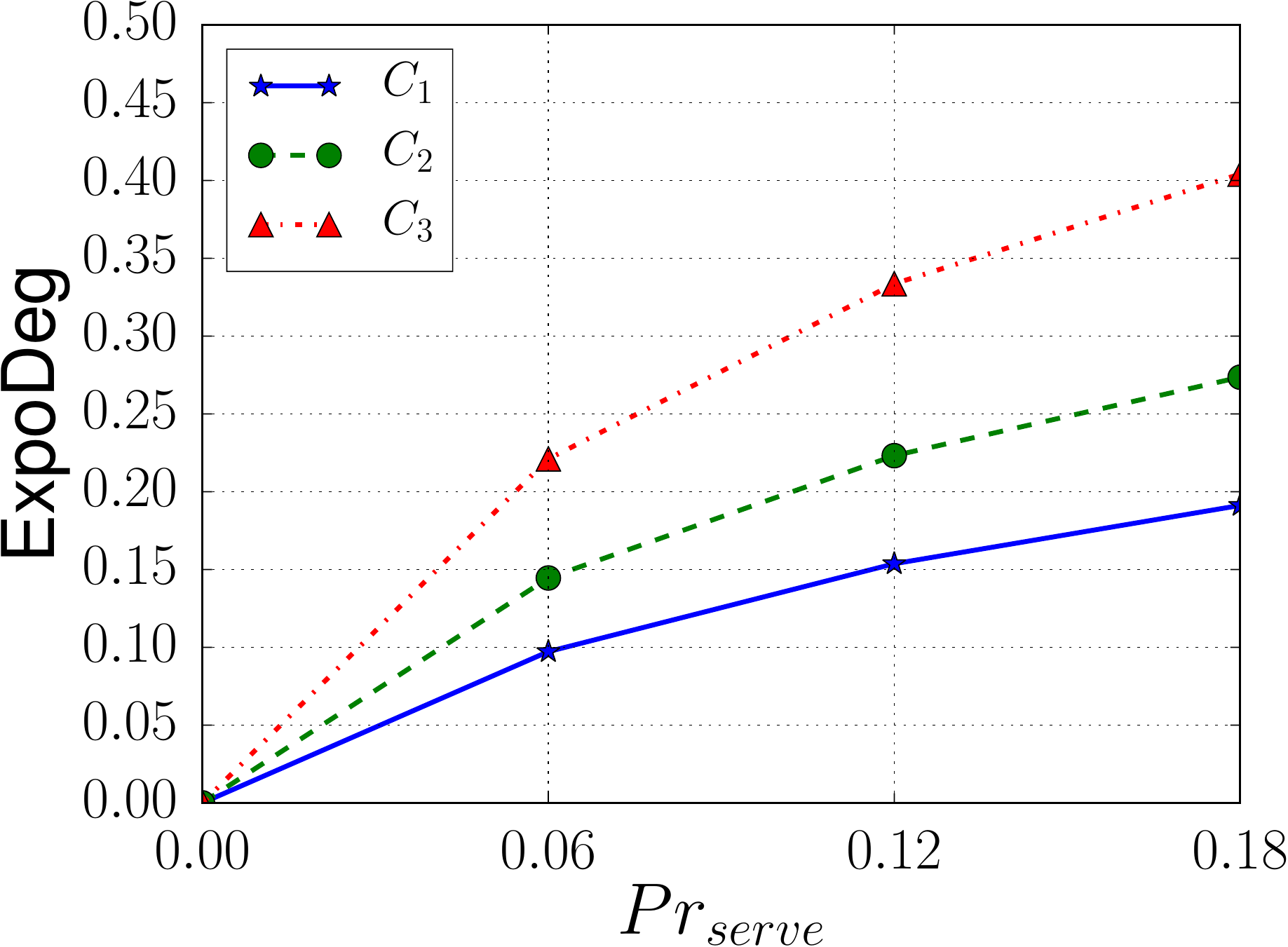}%
			\caption{}%
			\label{subfig_expodeg_type_2_noenc_lust}
		\end{subfigure}
		\begin{subfigure}[b]{0.24\columnwidth}
			\includegraphics[width=\columnwidth]{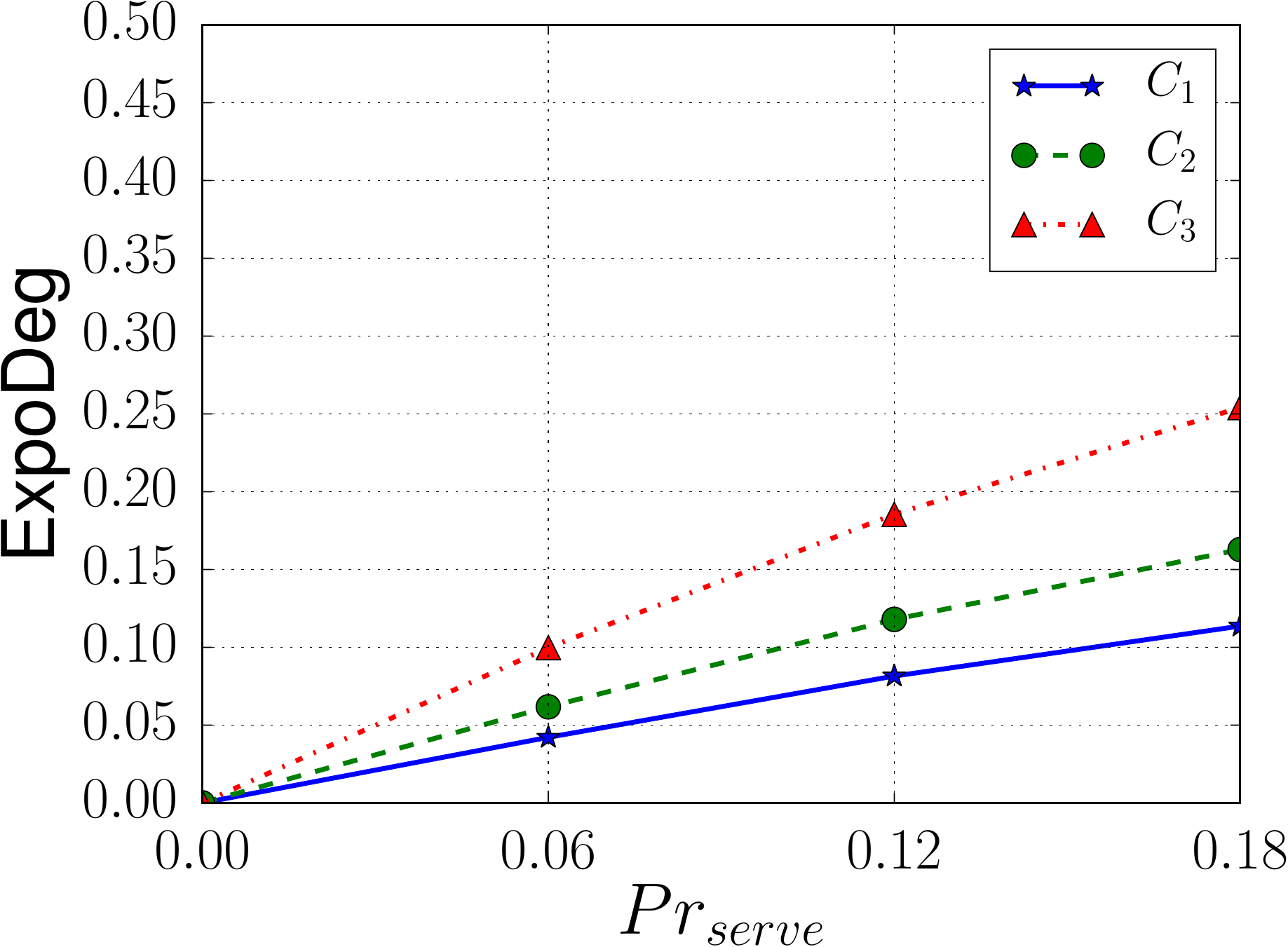}
			\caption{}
			\label{subfig_expodeg_type_3_enc_lust}
		\end{subfigure}
		\begin{subfigure}[b]{0.24\columnwidth}
			\includegraphics[width=\columnwidth]{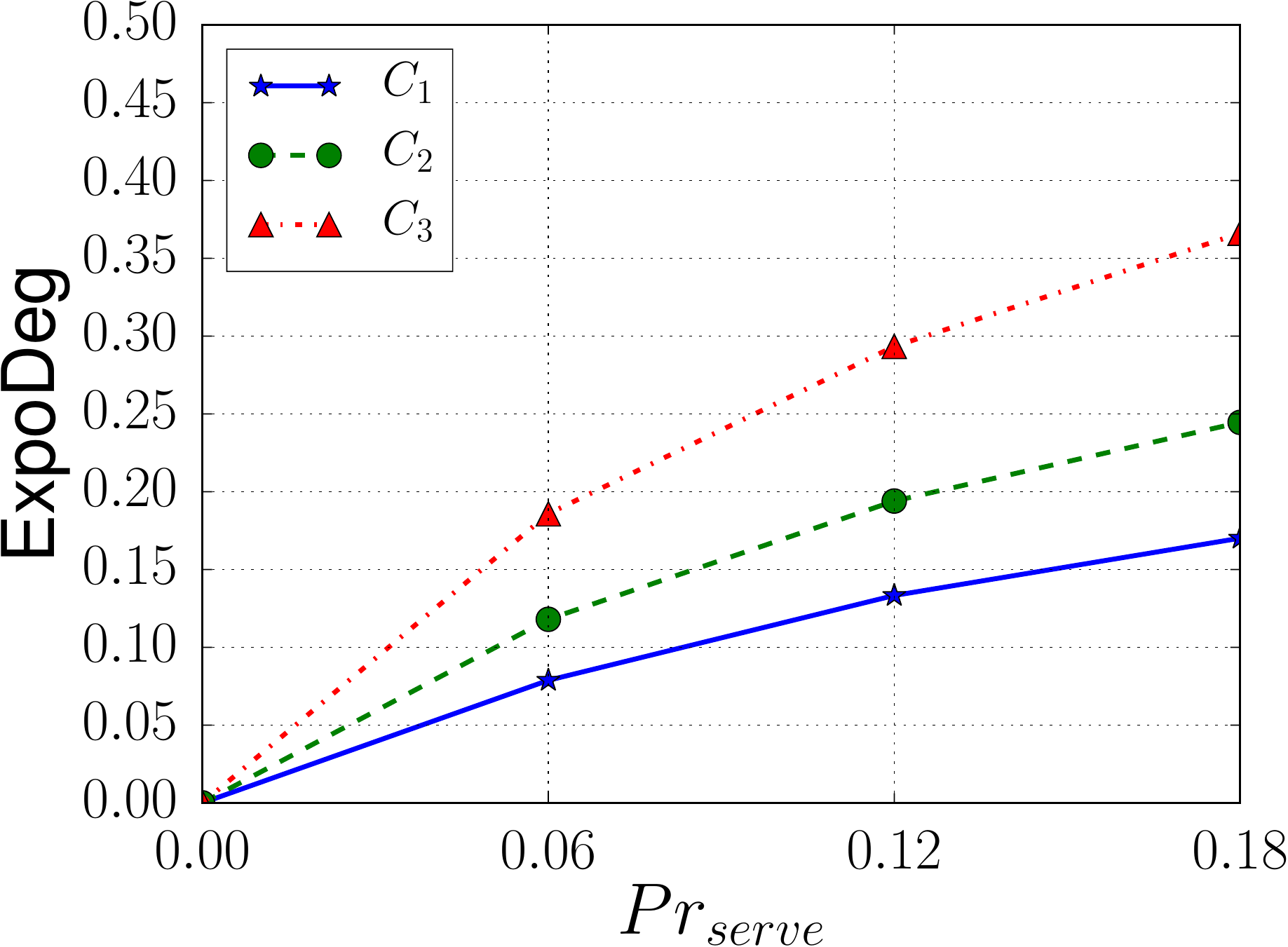}%
			\caption{}%
			\label{subfig_expodeg_type_3_noenc_lust}
		\end{subfigure}
		\caption{LuST: $ExpoDeg$ to colluding honest-but-curious nodes as a function of $Pr_{serve}$ with and without \ac{P2P} encryption for (\subref{subfig_expodeg_type_2_enc_lust}, \subref{subfig_expodeg_type_2_noenc_lust}) $G = 2$ and (\subref{subfig_expodeg_type_3_enc_lust}, \subref{subfig_expodeg_type_3_noenc_lust}) $G = 3$.}
		\label{fig_expodeg_type_lust}
	\end{figure}
	
}

\cref{fig_expo_prserve} shows $ExpoDeg$ to colluding honest-but-curious nodes as a function of $Pr_{serve}$. Again, the encrypted \ac{P2P} communication has a significant effect on reducing $ExpoDeg$. $ExpoDeg$ for both encrypted and non-encrypted \ac{P2P} communication increases with increased $Pr_{serve}$, because higher $Pr_{serve}$ results in a higher number of honest-but-curious serving nodes in the system given a $Ratio_{adv}$. Therefore, more peer queries would be disclosed to honest-but-curious serving nodes. The third columns of \cref{fig_expo_ratio} and \cref{fig_expo_prserve} show $ExpoDeg$ with uniformly distributed $T_{beacon}$ for the default settings; again, this has negligible effect on $ExpoDeg$. \cref{fig_expo_adv_region_koln} shows the same trend for $ExpoDeg$ for the TAPASCologne scenario with $N=3$, with a similar $ExpoDeg$ increase from the LuST scenario to the TAPASCologne scenario, which can be observed from~\cref{subfig_lbs_lust} and \cref{subfig_lbs_koln}. \cref{fig_expo_prserve_region_lust} show $ExpoDeg$ for different $L$. A smaller region decreases $ExpoDeg$, because the simulated area is divided into more regions for smaller $L$ (thus more regions a node visits during its trip), while the number of exposed regions could remain roughly the same.

We continue the evaluation with $G>1$. \cref{fig_expodeg_type_lust} shows that $ExpoDeg$ decreases as $G$ increases for the same $Pr_{serve}$, because less queries are responded by the serving nodes. $ExpoDeg$ increases with higher $G$ under the comparable settings (i.e., $G=1$ and $Pr_{serve}=0.06$, $G=2$ and $Pr_{serve}=0.12$, and $G=3$ and $Pr_{serve}=0.18$), because more serving nodes expose themselves continuously with their beacons. For example, for these settings, $ExpoDeg$ for $C_1$ is around 0.07, 0.09 and 0.11 respectively in \cref{fig_expo_prserve_3_enc}, \cref{subfig_expodeg_type_2_enc_lust}, \cref{subfig_expodeg_type_3_enc_lust}.

\begin{figure}[h!]
	\begin{subfigure}[b]{.24\columnwidth}
		\includegraphics[width=\columnwidth]{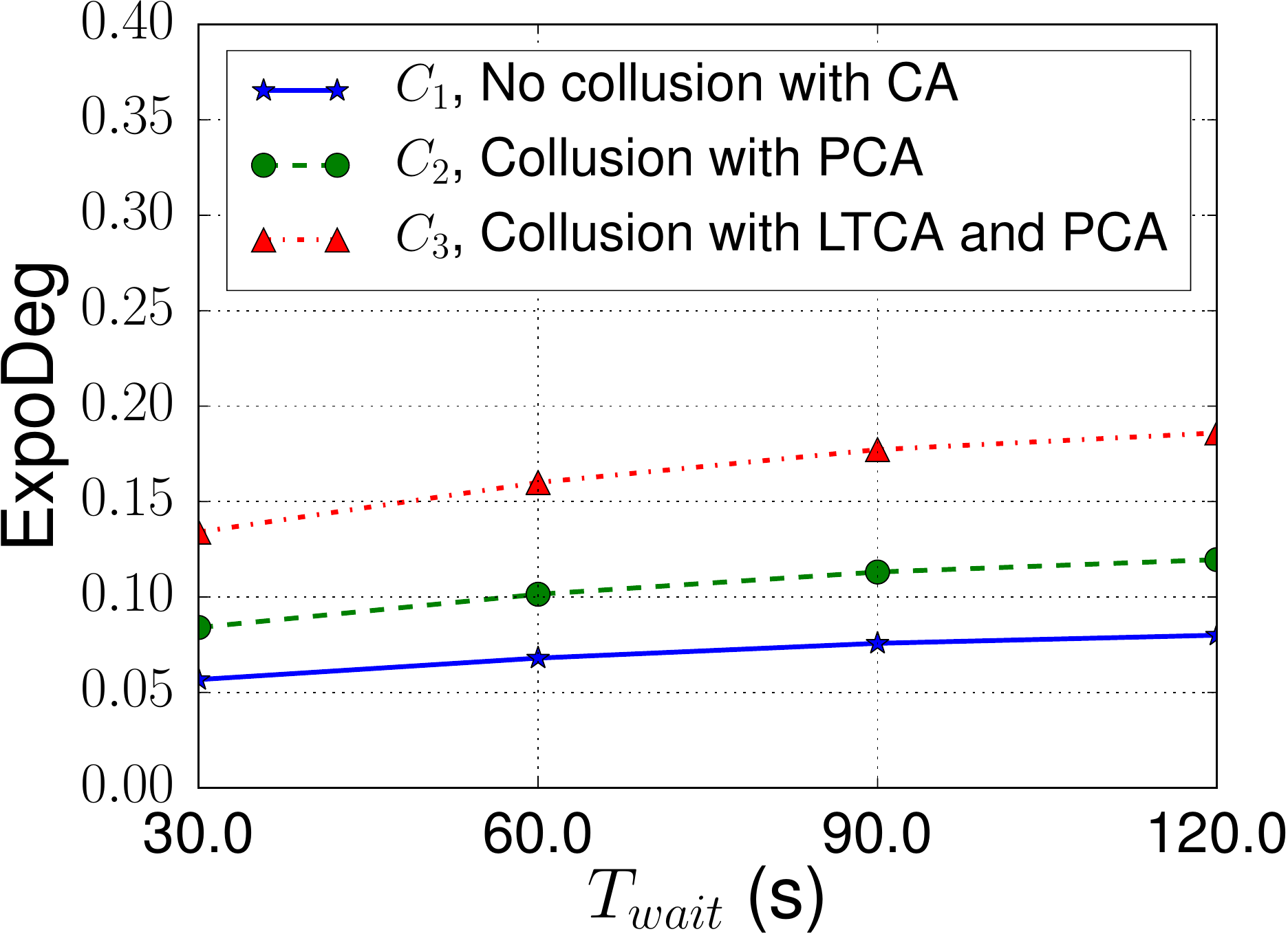}%
		\caption{}%
		\label{fig_expo_wait_enc}%
	\end{subfigure}
	\begin{subfigure}[b]{.24\columnwidth}
		\includegraphics[width=\columnwidth]{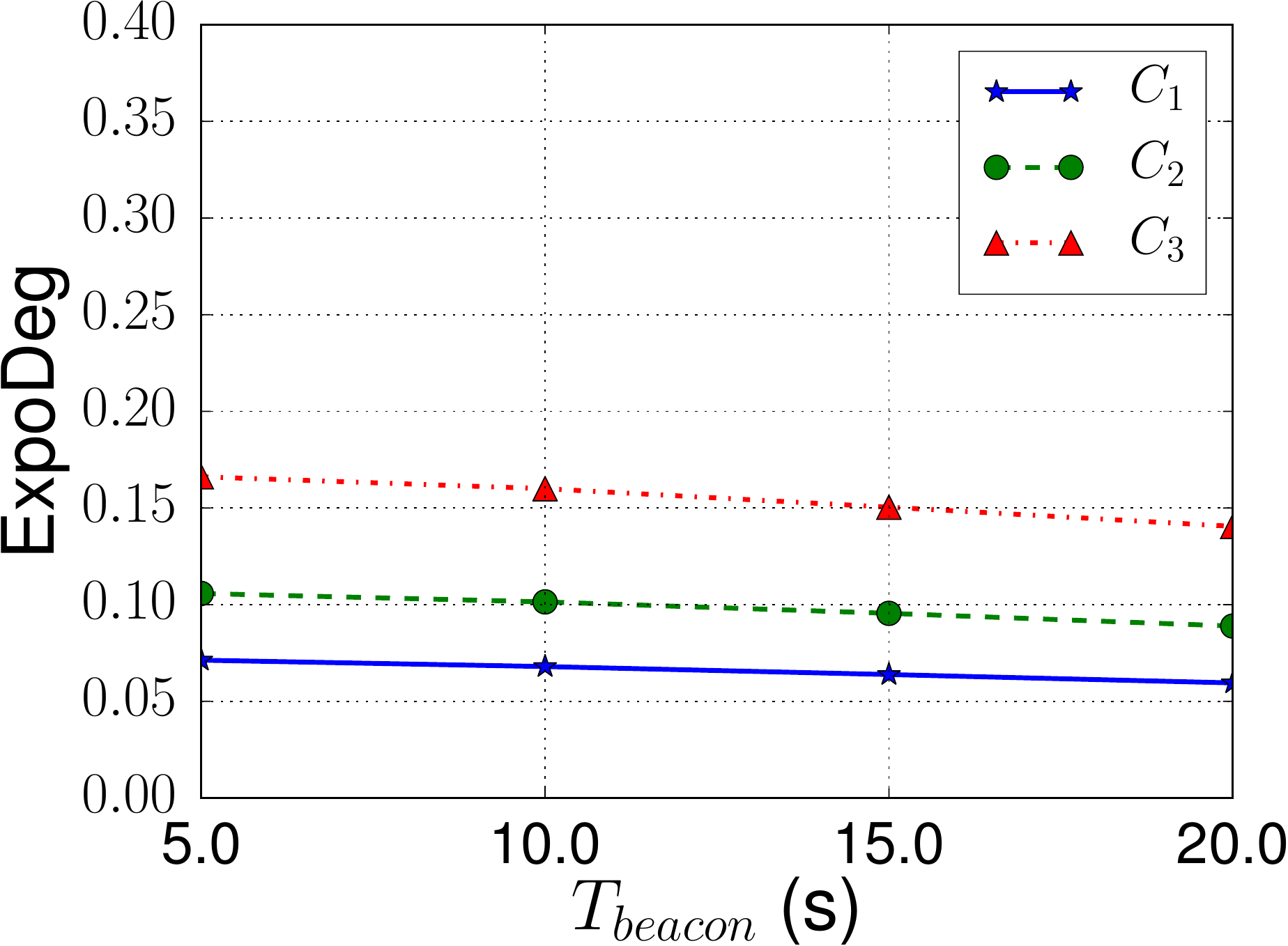}%
		\caption{}%
		\label{fig_expo_beacon_enc}%
	\end{subfigure}
	\begin{subfigure}[b]{.24\columnwidth}
		\includegraphics[width=\columnwidth]{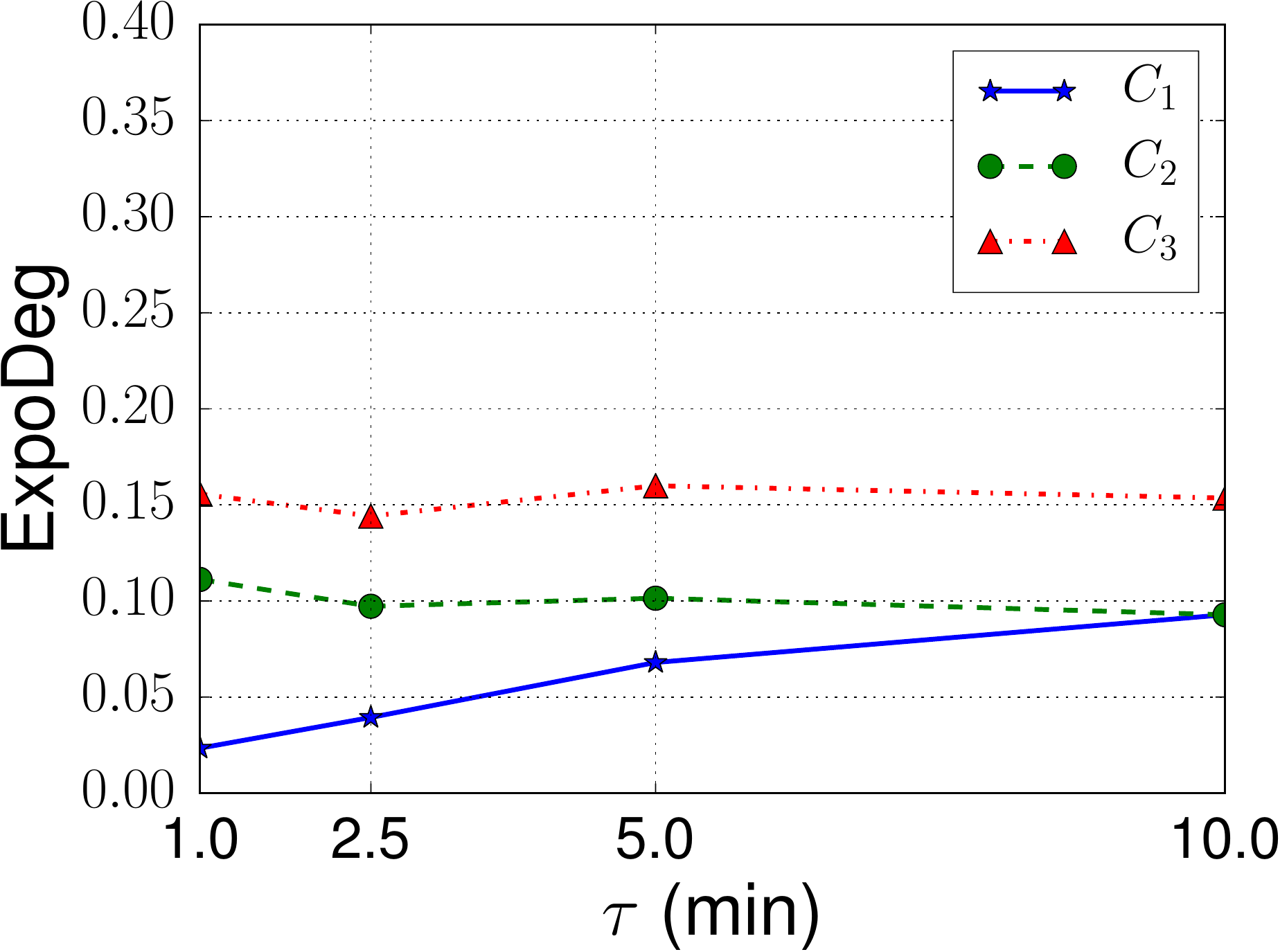}%
		\caption{}%
		\label{fig_expo_tau_enc}%
	\end{subfigure}
	\begin{subfigure}[b]{0.24\columnwidth}
		\includegraphics[width=\columnwidth]{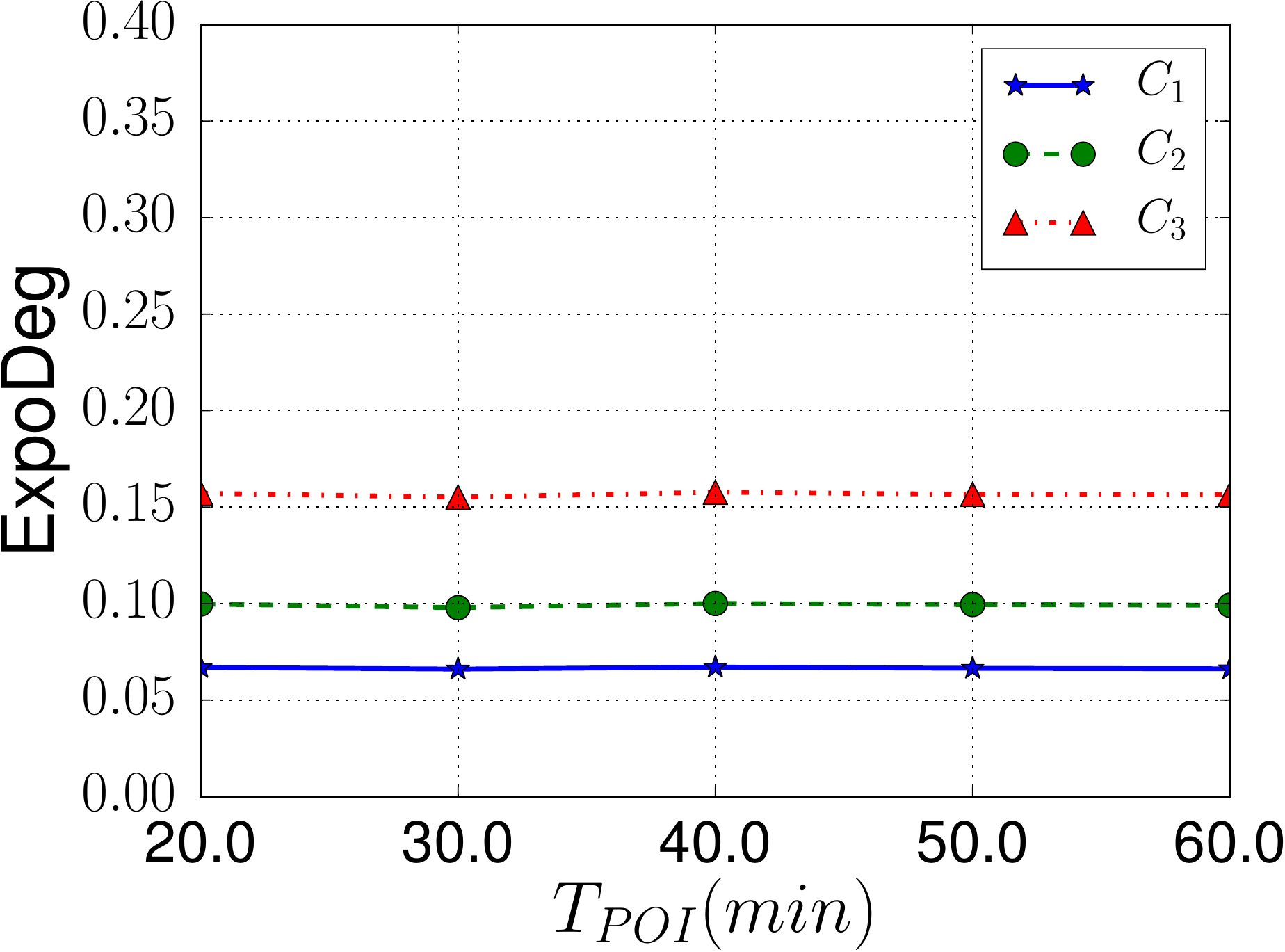}%
		\caption{}%
		\label{fig_expo_poi_enc}%
	\end{subfigure}
	\caption{LuST: $ExpoDeg$ to colluding honest-but-curious nodes with \ac{P2P} encryption as a function of (\subref{fig_expo_wait_enc}) $T_{wait}$, (\subref{fig_expo_beacon_enc}) $T_{beacon}$, (\subref{fig_expo_tau_enc}) $\tau$  and (\subref{fig_expo_poi_enc}) $T_{POI}$. (Default: $T_{wait}=60\ s$, $T_{beacon}=10\ s$, $\tau=5\ min$ and $T_{POI}=20\ min$.)}
	\label{fig_expo_param}
\end{figure}
\begin{figure}[h!]
	\begin{subfigure}[b]{.24\columnwidth}
		\includegraphics[width=\columnwidth]{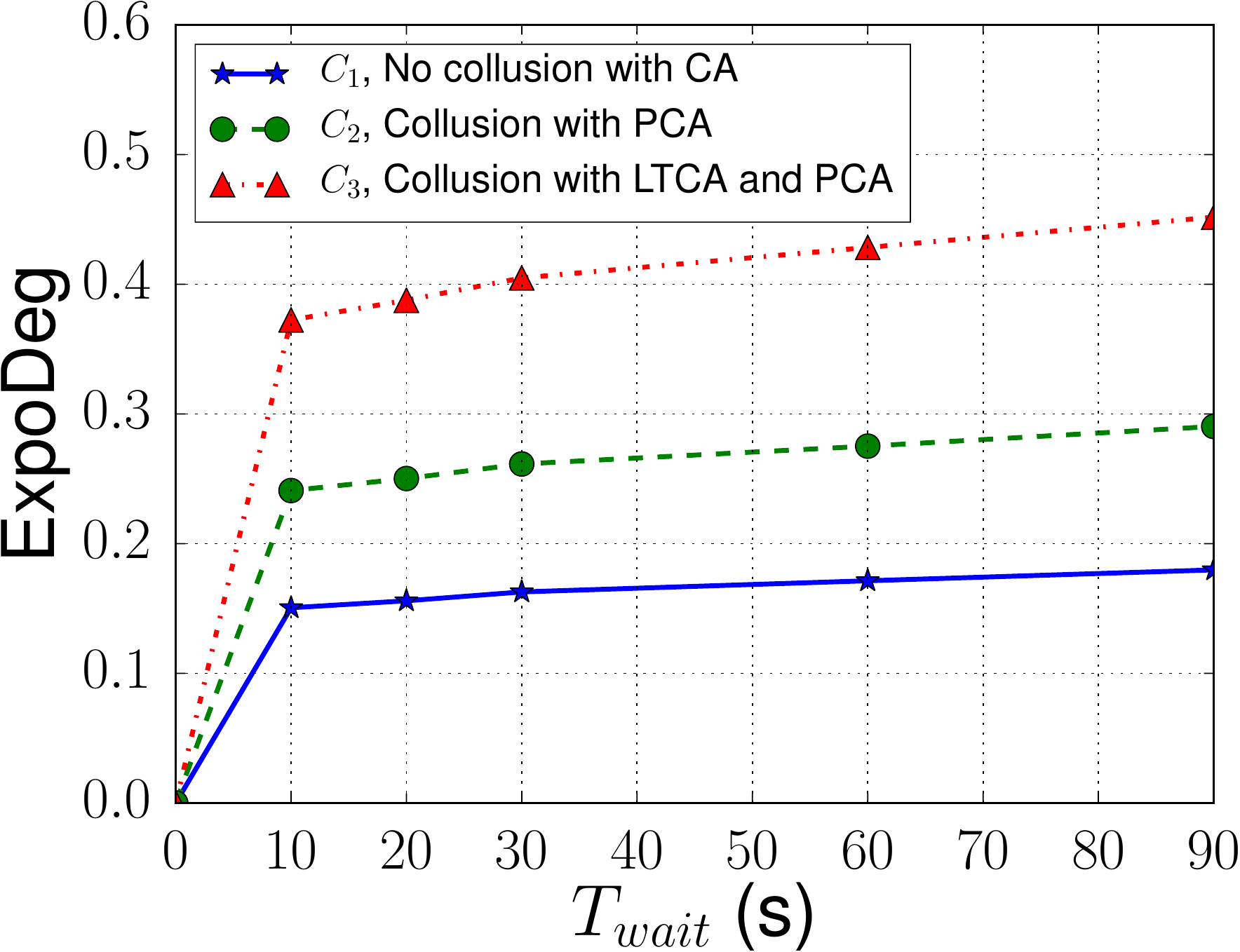}%
		\caption{}%
		\label{fig_expo_wait_lust}%
	\end{subfigure}
	\begin{subfigure}[b]{.24\columnwidth}
		\includegraphics[width=\columnwidth]{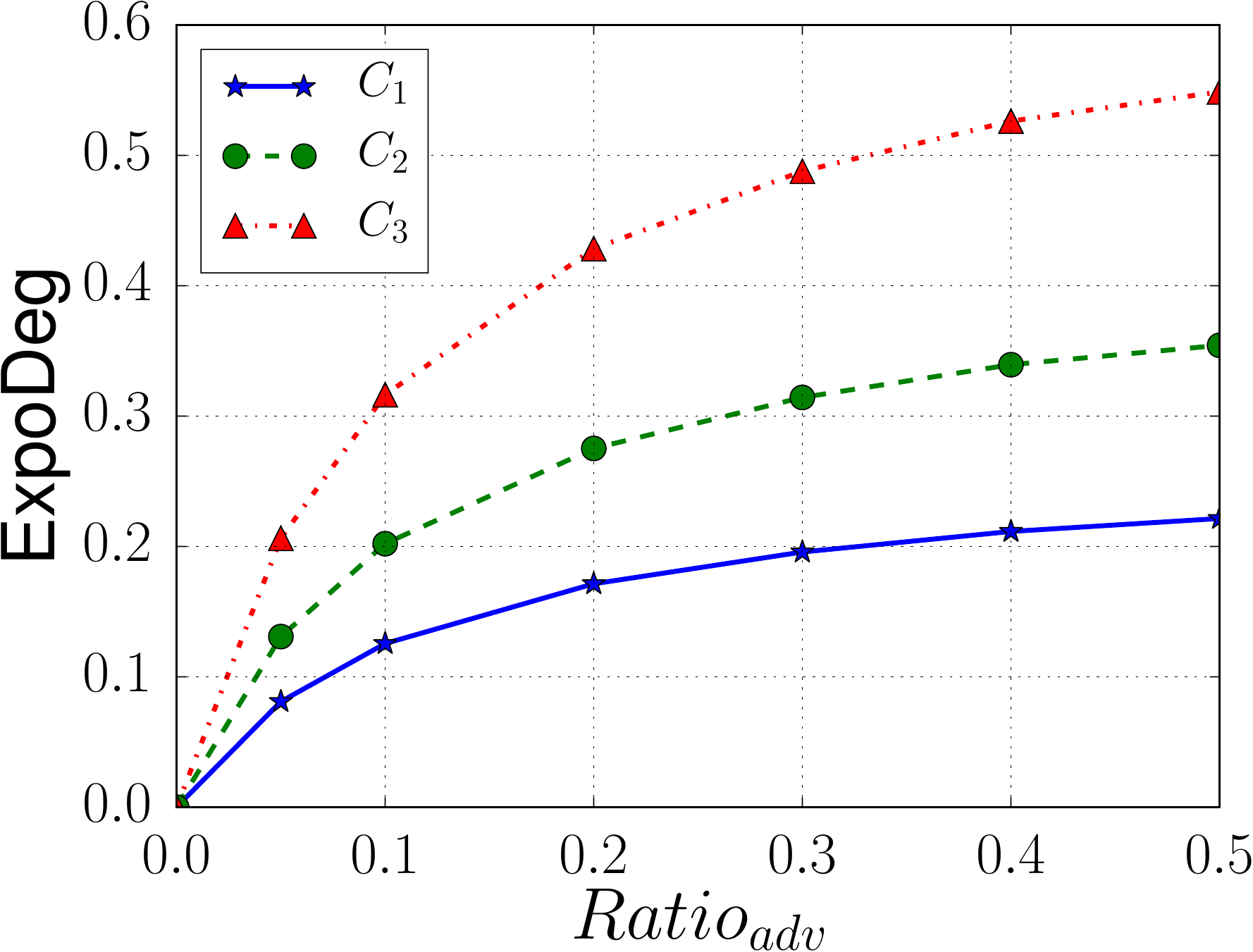}%
		\caption{}%
		\label{fig_expo_adv_lust}%
	\end{subfigure}
	\begin{subfigure}[b]{.24\columnwidth}
		\includegraphics[width=\columnwidth]{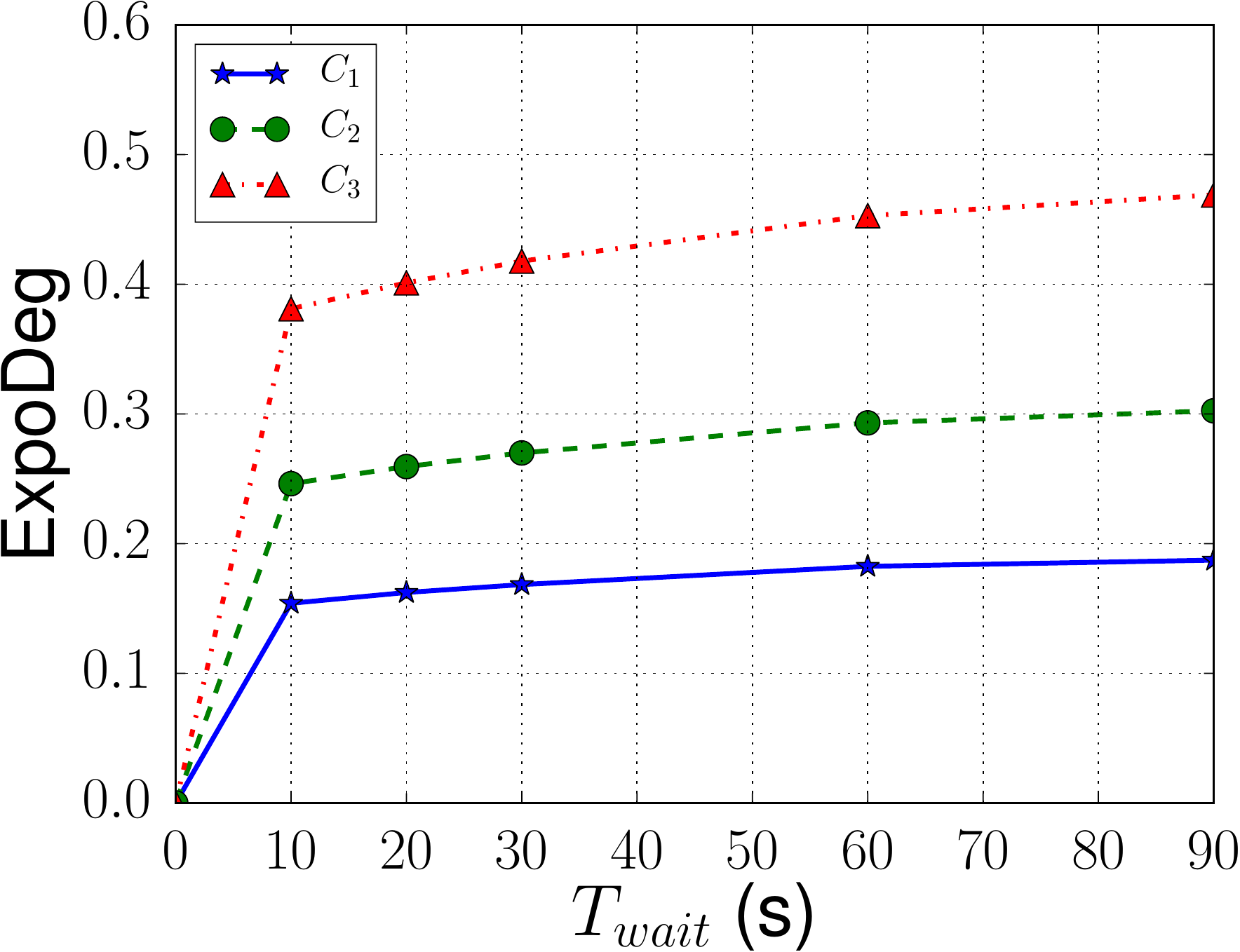}%
		\caption{}%
		\label{fig_expo_wait_05_lust}%
	\end{subfigure}
	\begin{subfigure}[b]{.24\columnwidth}
		\includegraphics[width=\columnwidth]{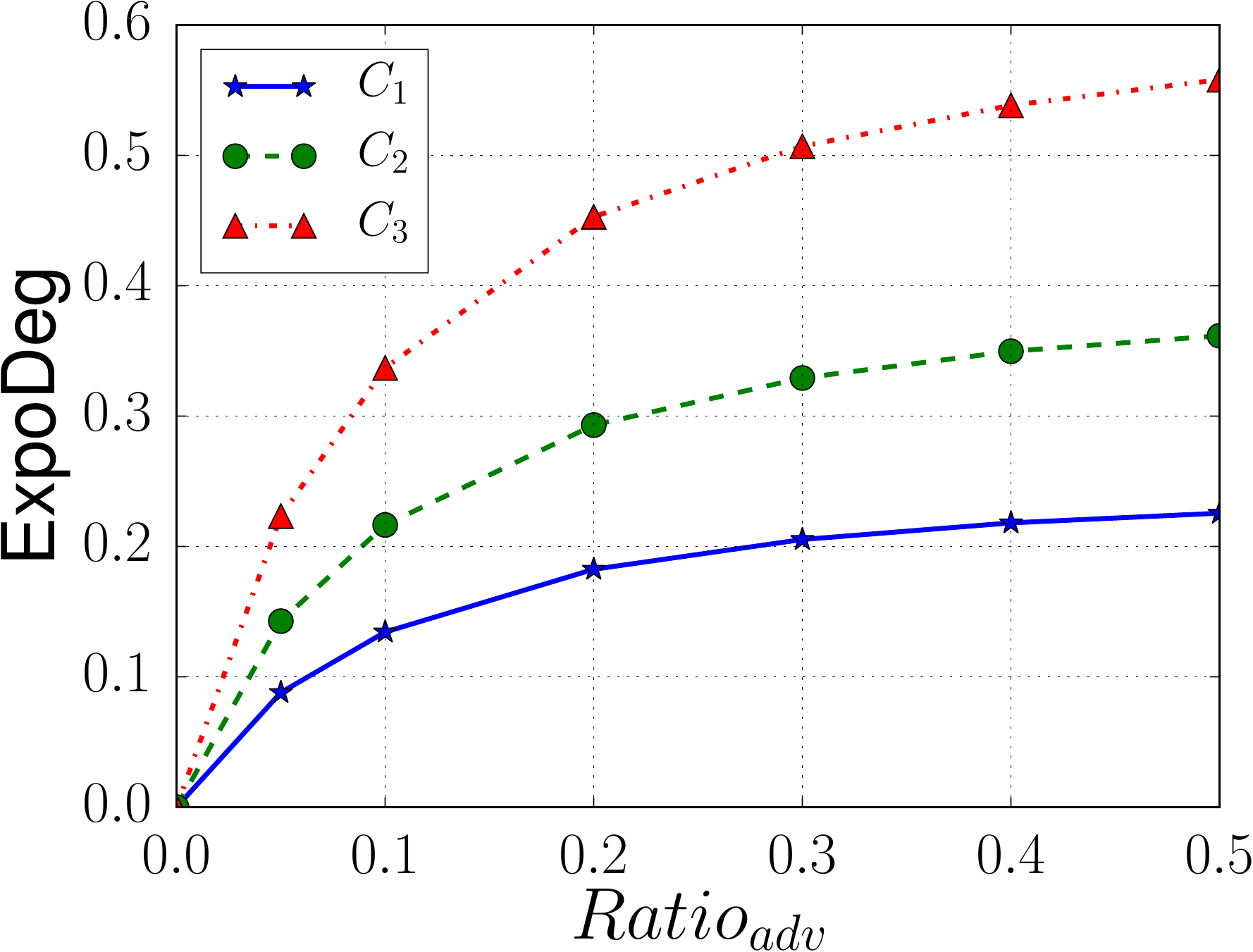}%
		\caption{}%
		\label{fig_expo_adv_05_lust}%
	\end{subfigure}
	\caption{LuST with MobiCrowd: $ExpoDeg$ to colluding honest-but-curious nodes with (\subref{fig_expo_wait_lust}, \subref{fig_expo_adv_lust}) $Ratio_{coop} = 1$ and (\subref{fig_expo_wait_05_lust}, \subref{fig_expo_adv_05_lust}) $Ratio_{coop} = 0.5$ as a function of $T_{wait}$ (\subref{fig_expo_wait_lust}, \subref{fig_expo_wait_05_lust}) and $Ratio_{adv}$ (\subref{fig_expo_adv_lust}, \subref{fig_expo_adv_05_lust}). (Default: $T_{wait}=60\ s$ and $Ratio_{adv}=0.2$.)}
	\label{fig_expo_mc_lust}
\end{figure}

\cref{fig_expo_wait_enc,fig_expo_beacon_enc} show $ExpoDeg$ to colluding honest-but-curious nodes with P2P encryption as a function of $T_{wait}$ and $T_{beacon}$ respectively. $ExpoDeg$ slightly increases with longer $T_{wait}$ and shorter $T_{beacon}$; in accordance with the peer hit ratio shown in \cref{subfig_chr_wait,subfig_chr_beacon}. \cref{fig_expo_tau_enc} show $ExpoDeg$ as a function of $\tau$ with a fixed $\Gamma = 10\ min$. $ExpoDeg$ significantly increases with increased $\tau$ for $C_1$, because more queries can be linked together under the same pseudonym when its lifetime is longer. For $C_2$,  the trend is that $ExpoDeg$ slightly decreases with increased $\tau$, given the same $\Gamma$, because lower $\tau$ results in less pseudonym requests per trip (see Sec.~\ref{sec:pseudonym}). For example, consider a node trip duration of 15 $min$. With a fixed $\Gamma = 10$ $min$, the nodes needs two pseudonym requests when $\tau = 1$ $min$. For example, the first pseudonym request covers a duration of $9.5$ $min$ and the second pseudonym request covers the rest of $5.5$ $min$. However, when $\tau = 10$ $min$, if there is, e.g., $2$ $min$, remaining in the current $\tau$, three pseudonym requests would be needed for the same example. The node would use the pseudonyms obtained through three requests for $2$ $min$, $10$ $min$ and $3$ $min$ respectively. Thus, the latter case results in lower $ExpoDeg$, because the node exposure is ``split'' under three $Id_{ticket}$ rather than two for the former case. $ExpoDeg$ remains the same under different $T_{POI}$ (\cref{fig_expo_poi_enc}), because this only affects the communication overhead of serving nodes for updating \ac{POI} data and does not change the probability of encountering serving nodes by querying nodes.

\cref{fig_expo_mc_lust} shows $ExpoDeg$ for MobiCrowd. $ExpoDeg$ increases with longer $T_{wait}$, because more queries are broadcasted before the \ac{LBS} server is queried. Higher $Ratio_{adv}$ also increases $ExpoDeg$ as expected. \cref{subfig_chr_prserve_lust_2km} and \cref{subfig_chr_mc_05} showed that MobiCrowd with $Ratio_{coop}=0.5$ and $T_{wait} = 10\ s$ provides a similar peer hit ratio as our scheme with $Pr_{serve} = 0.12$ and $T_{wait} = 60\ s$ for the LuST scenario. However, from \cref{fig_expo_wait_05_lust,fig_expo_adv_05_lust}, we see that for MobiCrowd, $ExpoDeg$ is higher than for our scheme. For example, with the above settings and $Ratio_{adv} = 0.2$, $ExpoDeg$ for MobiCrowd is around $0.16$, $0.25$ and $0.38$ for $C_1$, $C_2$ and $C_3$ respectively: that is, higher than the $0.12$, $0.17$ and $0.26$ values achieved by our scheme.

\textbf{Resilience:} We evaluate the resilience of our scheme against malicious nodes. We consider the same $Ratio_{adv}$ values as those for the evaluation of the exposure. Although a ratio e.g., $20\ \%$, of adversarial nodes (thus, node owners) may be unrealistic, we consider such a rather harsh setting to capture situations with extensive node infection by malware~\cite{ramachandran2006modeling,cheng2011modeling} while node owners can be benign and unsuspecting. We assume that, node owners can be notified and malicious nodes be reinstated as benign nodes through, e.g., diagnostics and updates, once misbehavior is reported and malicious nodes are identified. A recovered benign node would be issued a new \ac{LTC} and obtain new \acp{PC}. Thus, old \acp{PC} from detected malicious nodes will be revoked and published through the latest \ac{CRL}. In the simulation, we assume the malicious serving nodes collude to provide identical false responses (based on falsified data stored locally) to a specific node query, so that the false responses appear consistent and thus be accepted by the querying node. Otherwise, even the responses from independently acting malicious serving nodes could be conflicting and be easily detected.

\begin{figure}[tbp!]
	\centering
	\begin{subfigure}[b]{0.24\columnwidth}
		\includegraphics[width=\columnwidth]{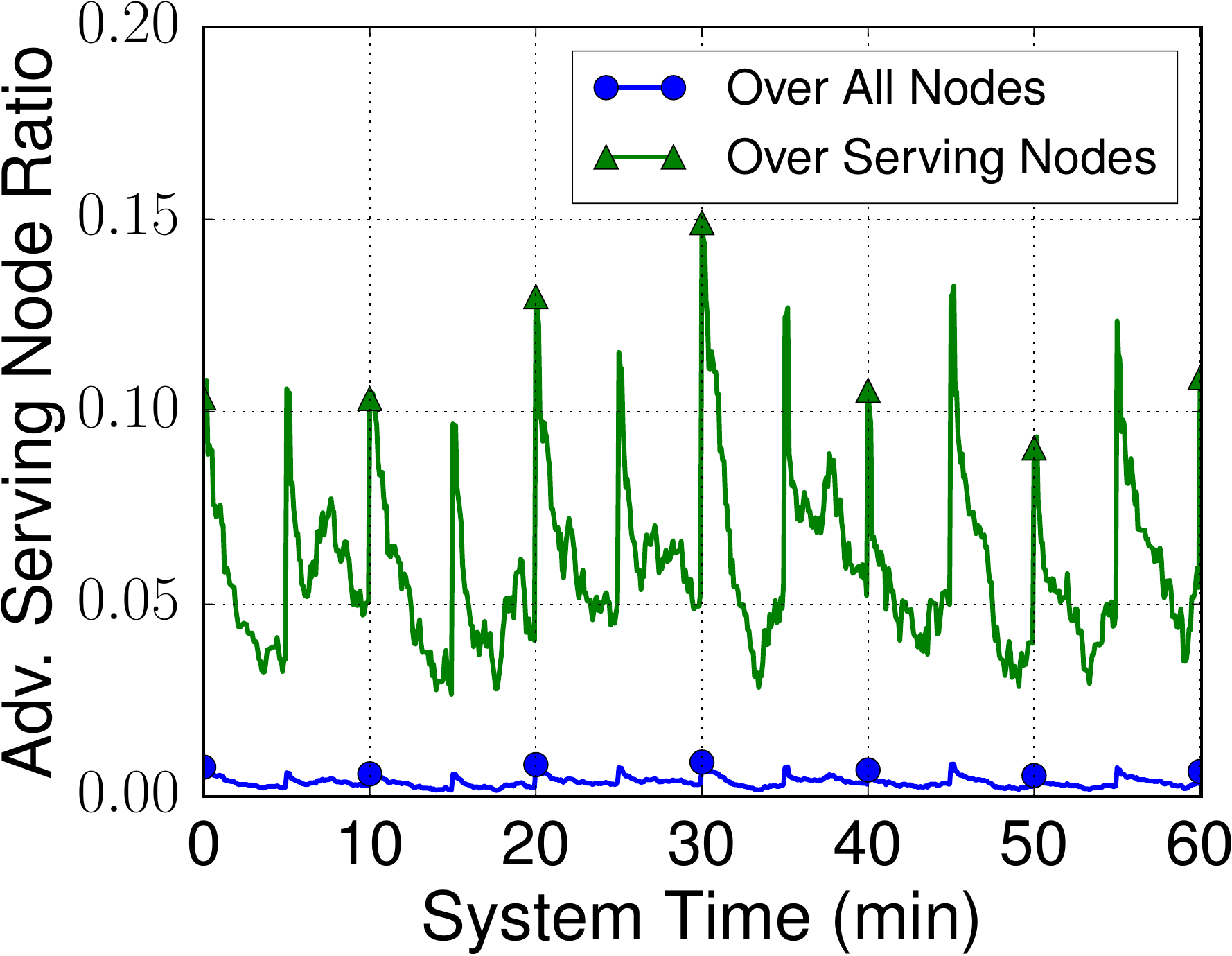}%
		\caption{}%
		\label{subfig_ratio_node_2}
	\end{subfigure}
	\begin{subfigure}[b]{0.24\columnwidth}
		\includegraphics[width=\columnwidth]{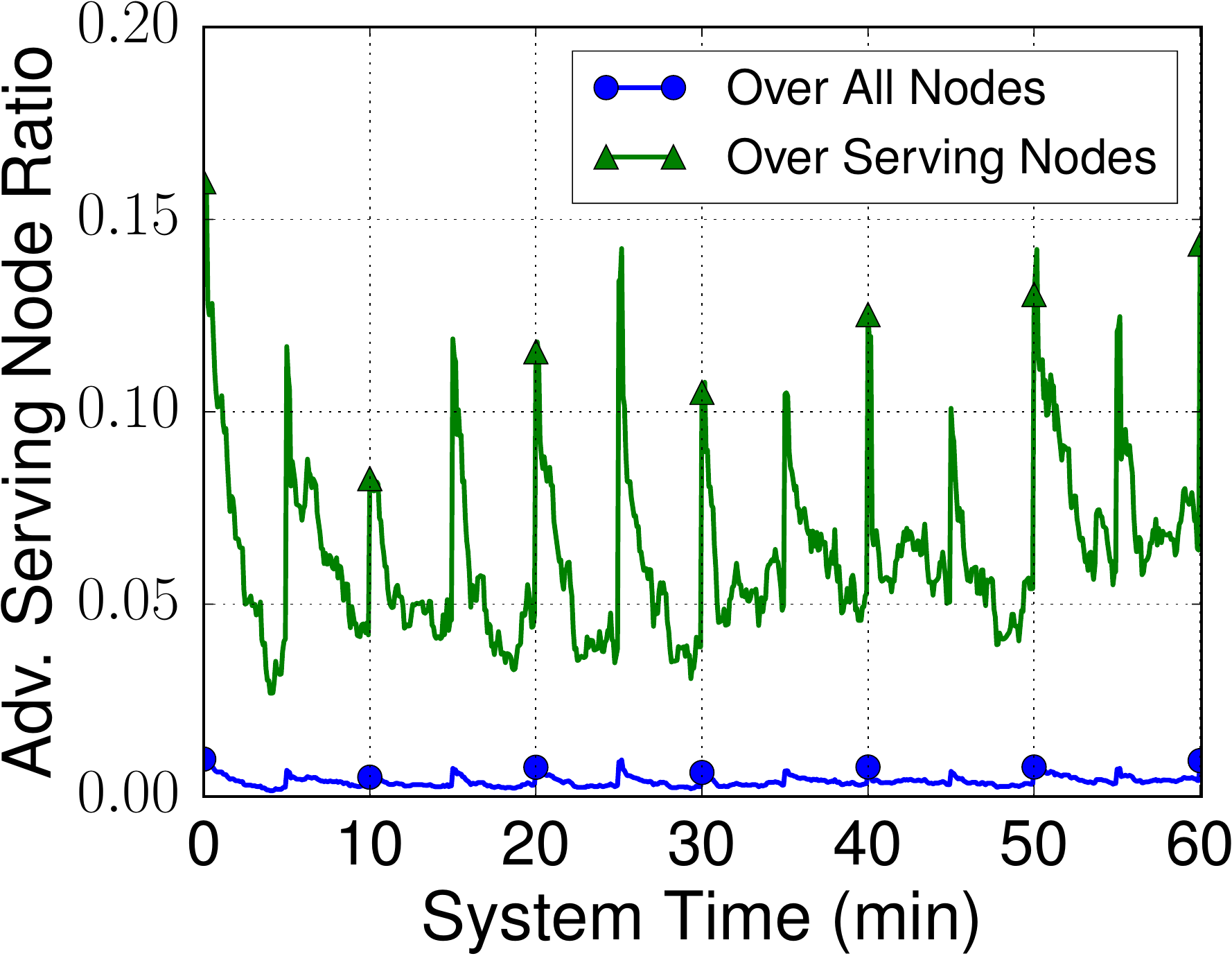}%
		\caption{}%
		\label{subfig_ratio_node_3_1km}
	\end{subfigure}
	\begin{subfigure}[b]{0.24\columnwidth}
		\includegraphics[width=\columnwidth]{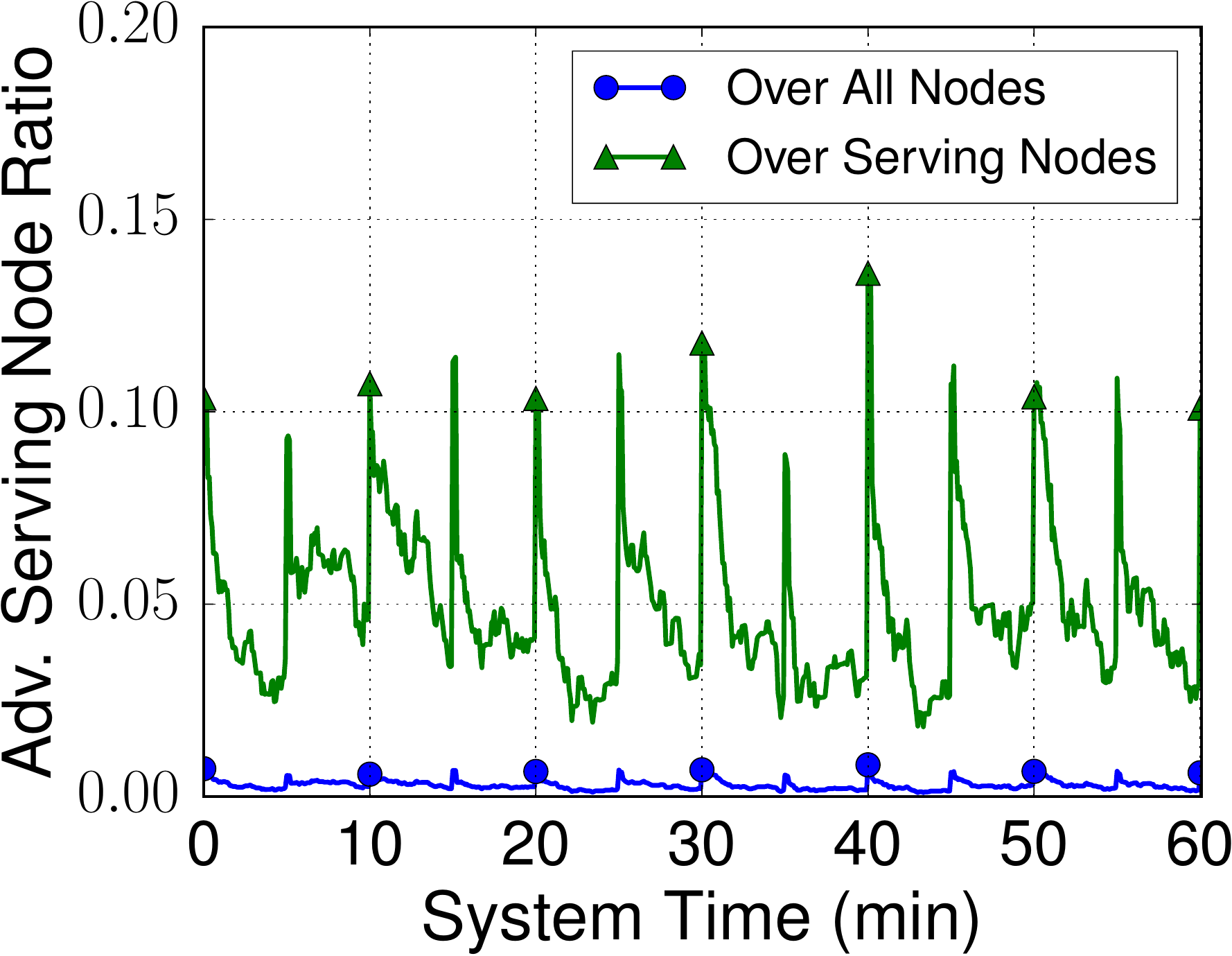}%
		\caption{}%
		\label{subfig_ratio_node_3_2km}%
	\end{subfigure}
	\begin{subfigure}[b]{0.24\columnwidth}
		\includegraphics[width=\columnwidth]{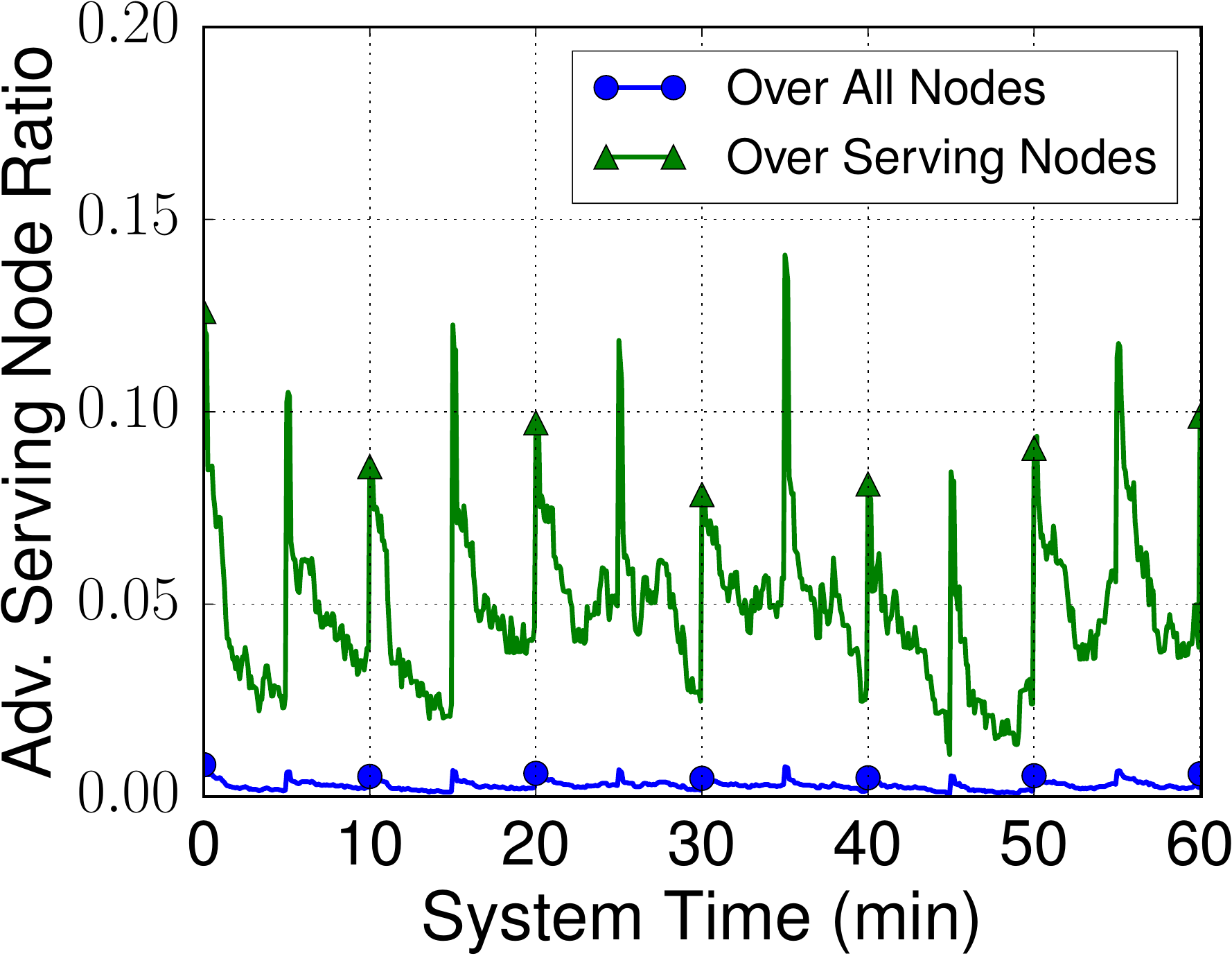}%
		\caption{}%
		\label{subfig_ratio_node_3_3km}%
	\end{subfigure}
	\caption{LuST: Ratio of malicious serving nodes with (\subref{subfig_ratio_node_2}) $N=2$, and $N=3$ when $L =$ (\subref{subfig_ratio_node_3_1km}) $1\ km$, (\subref{subfig_ratio_node_3_2km}) $2\ km$, and (\subref{subfig_ratio_node_3_3km}) $3\ km$.}
	\label{fig_ratio_node}
\end{figure}

\begin{figure}[tbp!]
	\centering
	\begin{subfigure}[b]{0.24\columnwidth}		
		\includegraphics[width=\columnwidth]{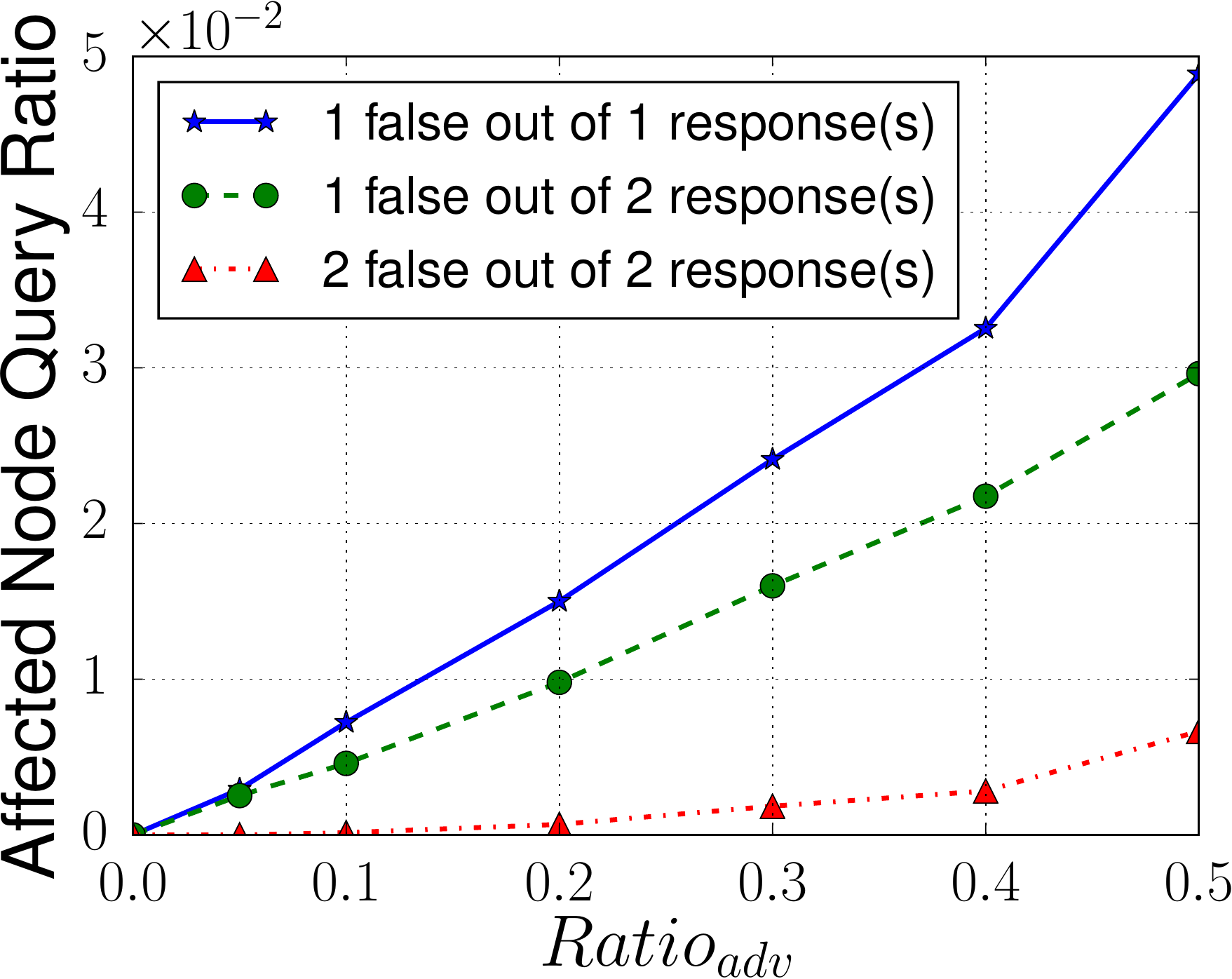}
		\caption{}%
		\label{subfig_ratio_resp_2}
	\end{subfigure}
	\begin{subfigure}[b]{0.24\columnwidth}
		\includegraphics[width=\columnwidth]{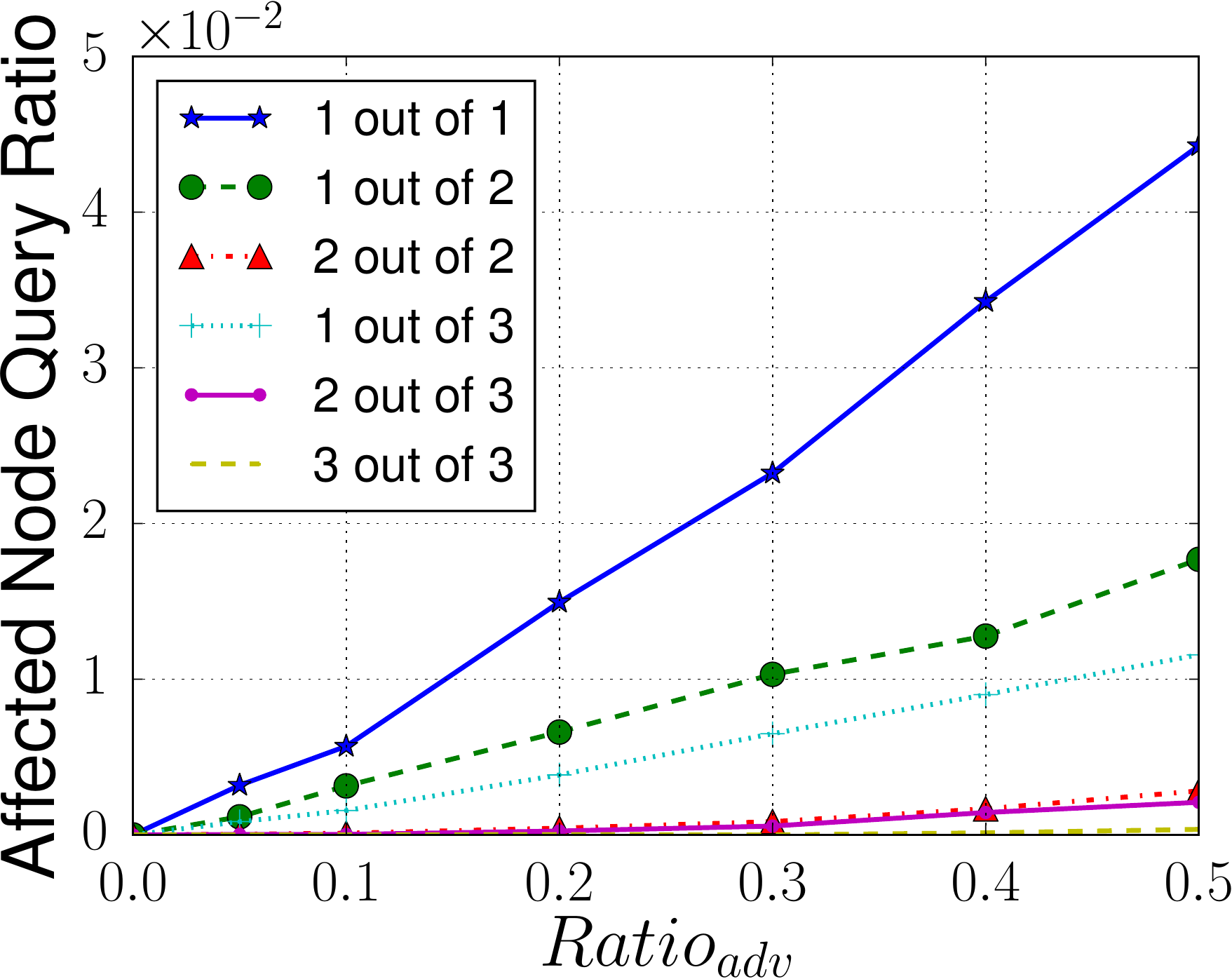}%
		\caption{}%
		\label{subfig_ratio_resp_3_1km}
	\end{subfigure}
	\begin{subfigure}[b]{0.24\columnwidth}
		\includegraphics[width=\columnwidth]{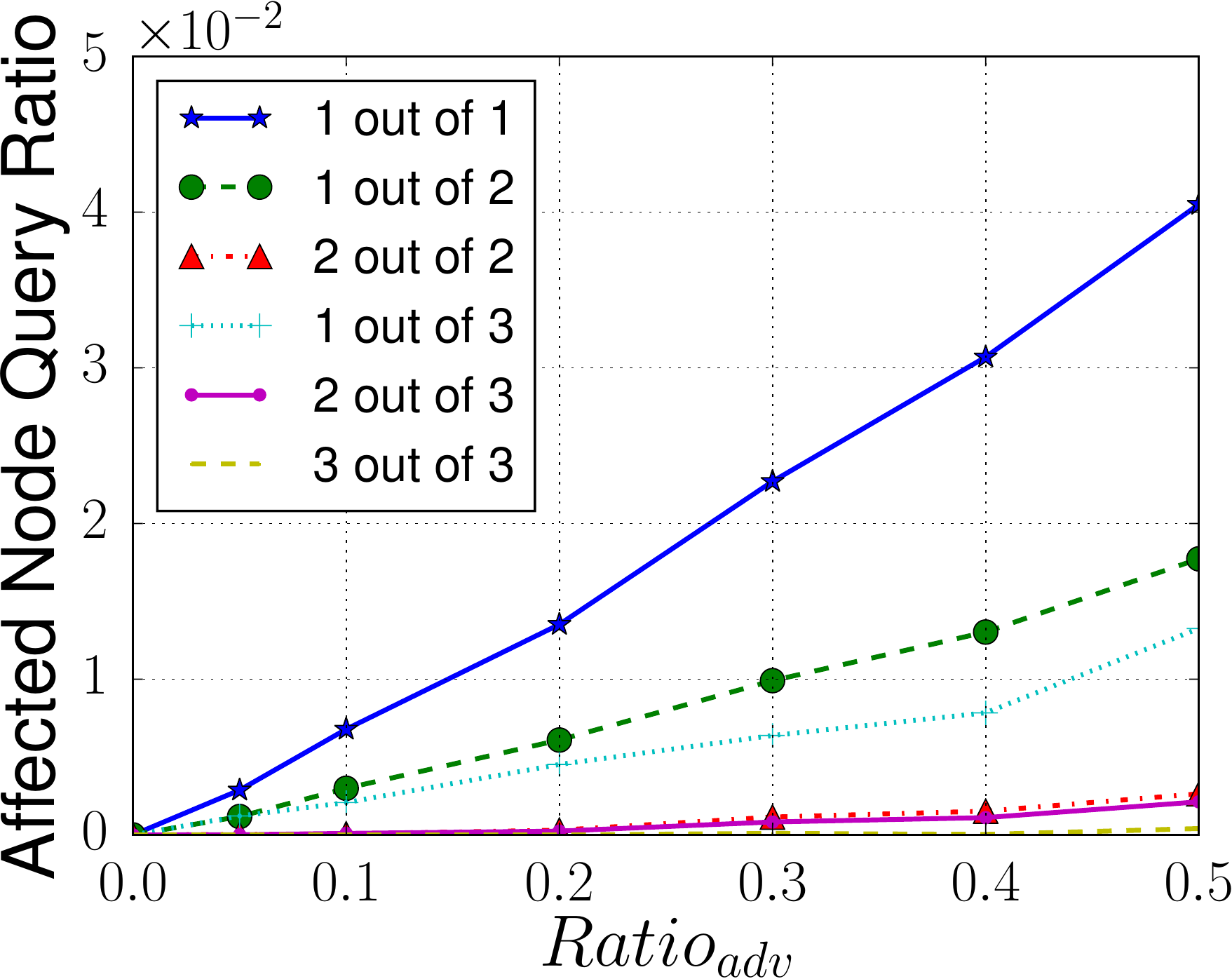}%
		\caption{}%
		\label{subfig_ratio_resp_3_2km}
	\end{subfigure}
	\begin{subfigure}[b]{0.24\columnwidth}
		\includegraphics[width=\columnwidth]{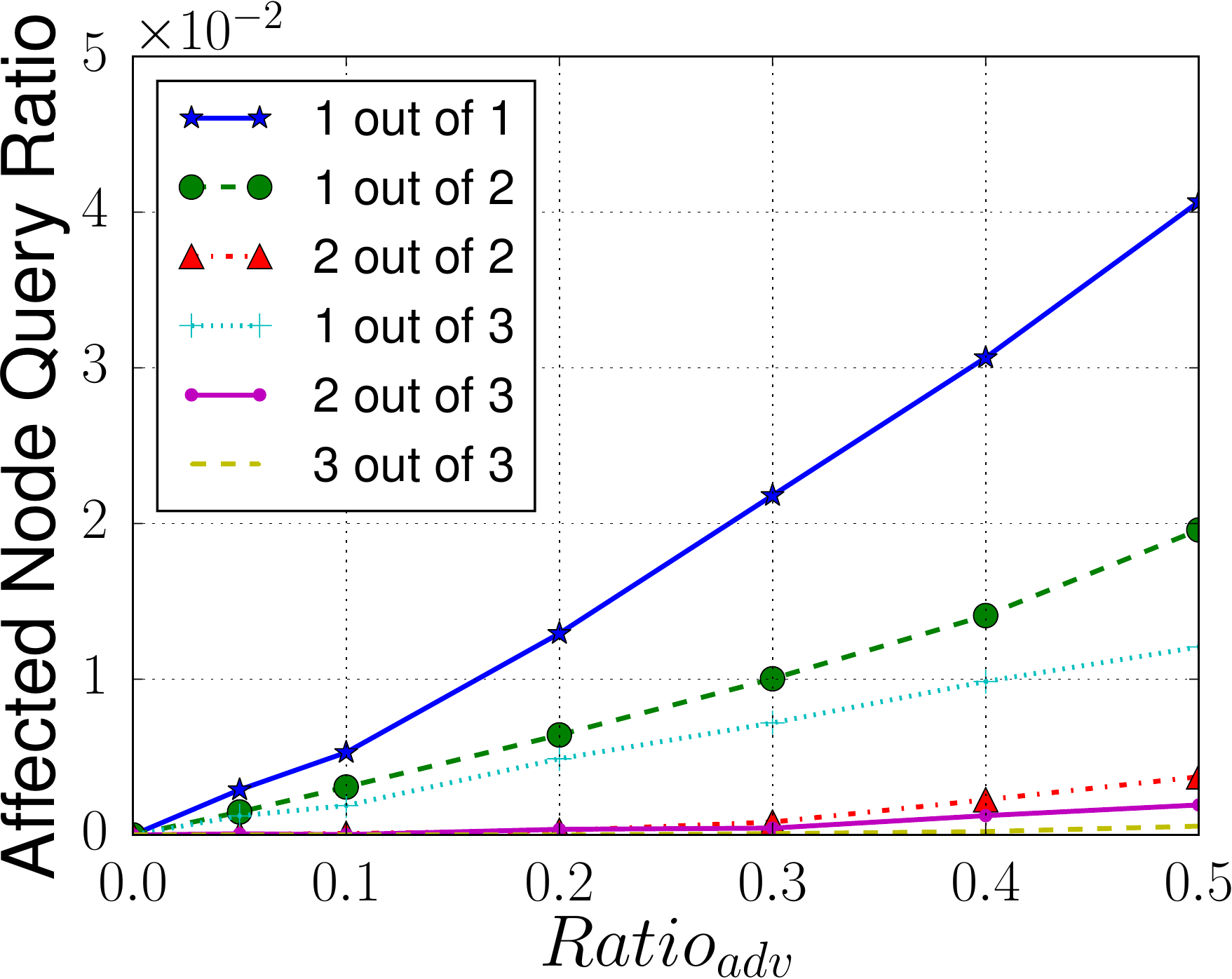}%
		\caption{}%
		\label{subfig_ratio_resp_3_3km}
	\end{subfigure}
	\caption{LuST: Affected node query ratio as a function of $Ratio_{adv}$ with (\subref{subfig_ratio_resp_2}) $N=2$, and $N=3$ when (\subref{subfig_ratio_resp_3_1km}) $L = 1$ $km$, (\subref{subfig_ratio_resp_3_2km}) $L = 2$ $km$, and (\subref{subfig_ratio_resp_3_3km}) $L = 3$ $km$.}
	\label{fig_ratio_resp}
\end{figure}

\begin{figure}[tbp!]
	\centering
	\begin{subfigure}[b]{0.24\columnwidth}
		\includegraphics[width=\columnwidth]{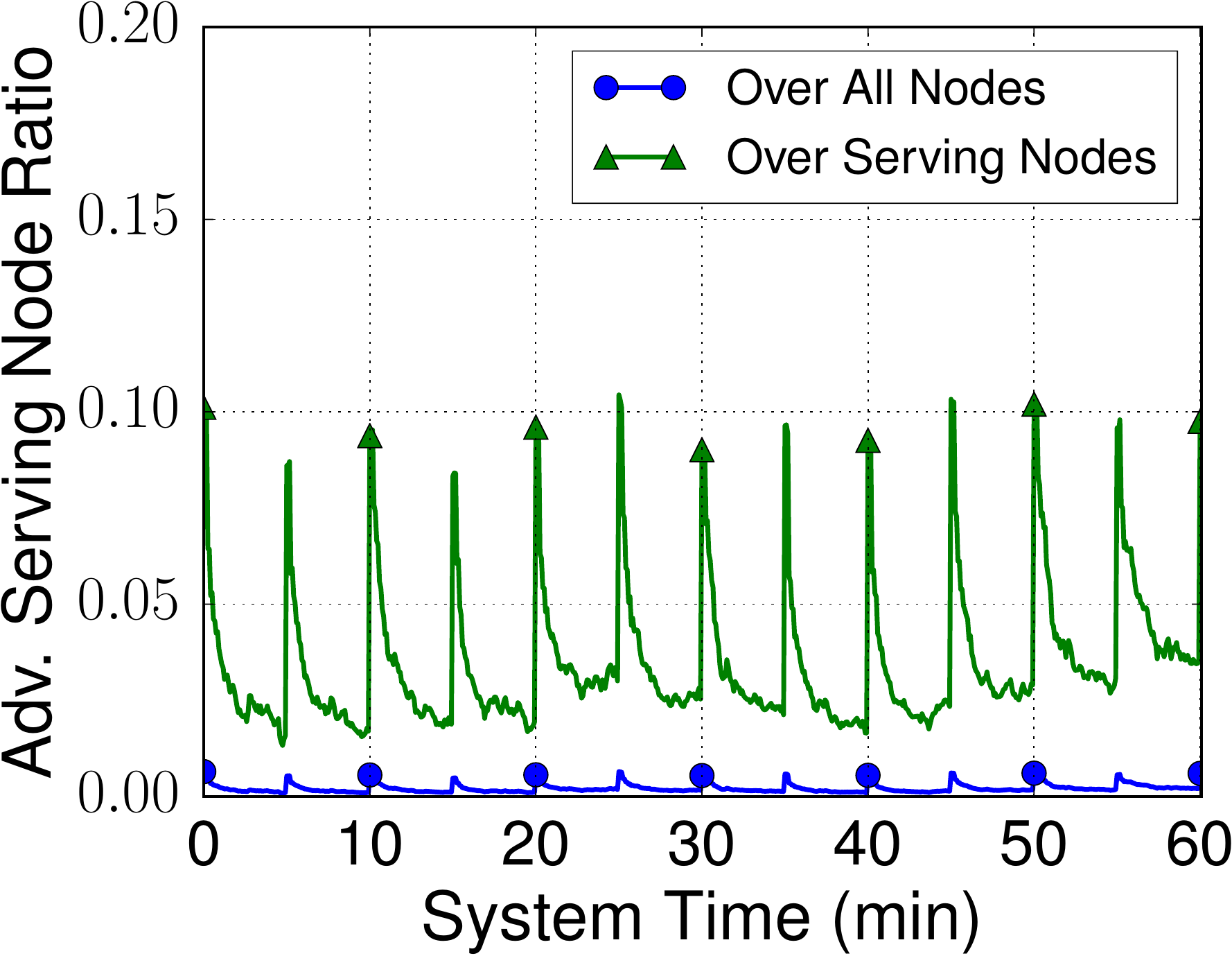}%
		\caption{}%
		\label{subfig_ratio_node_2_koln}
	\end{subfigure}
	\begin{subfigure}[b]{0.24\columnwidth}
		\includegraphics[width=\columnwidth]{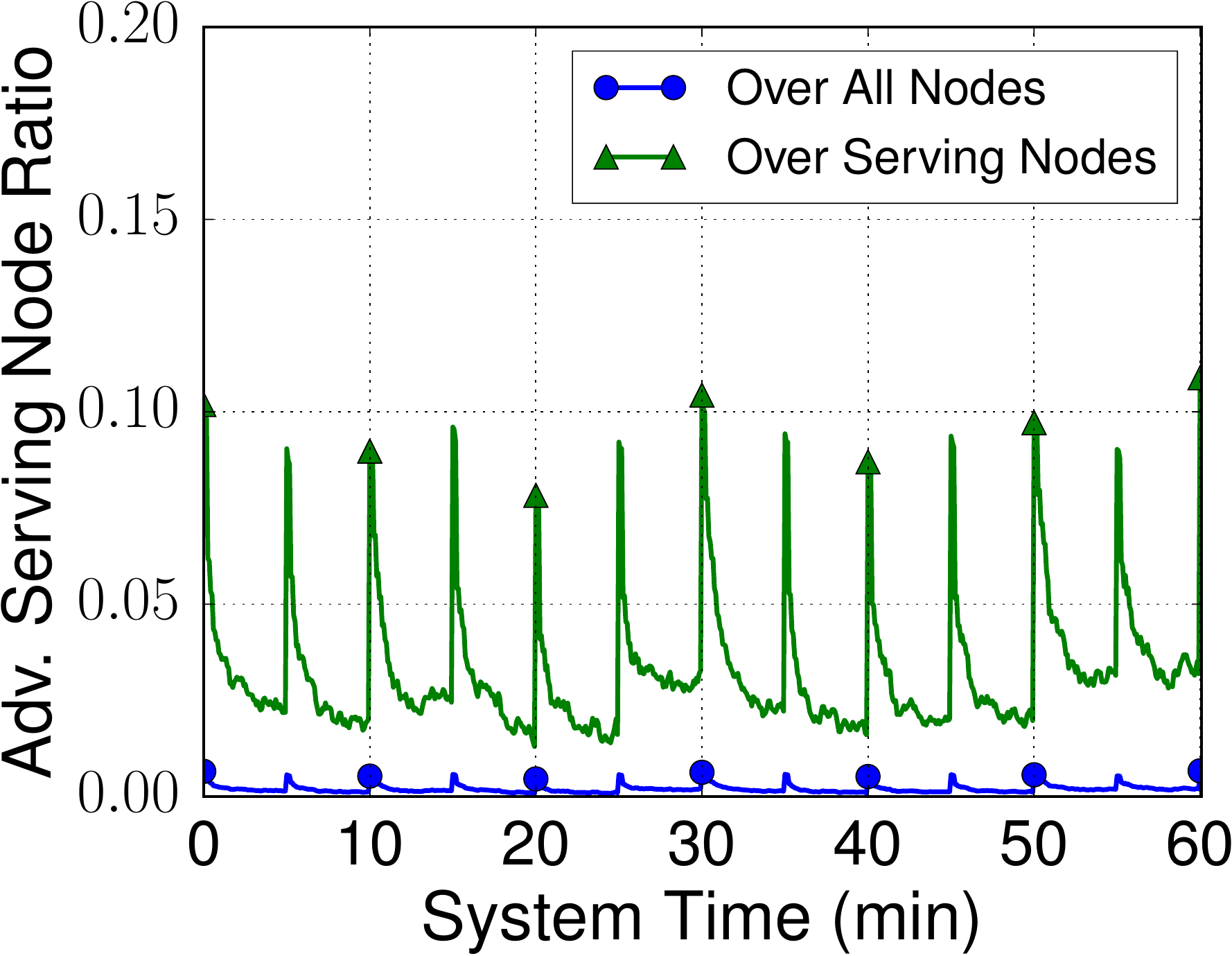}%
		\caption{}%
		\label{subfig_ratio_node_3_koln}%
	\end{subfigure}
	\begin{subfigure}[b]{0.24\columnwidth}		
		\includegraphics[width=\columnwidth]{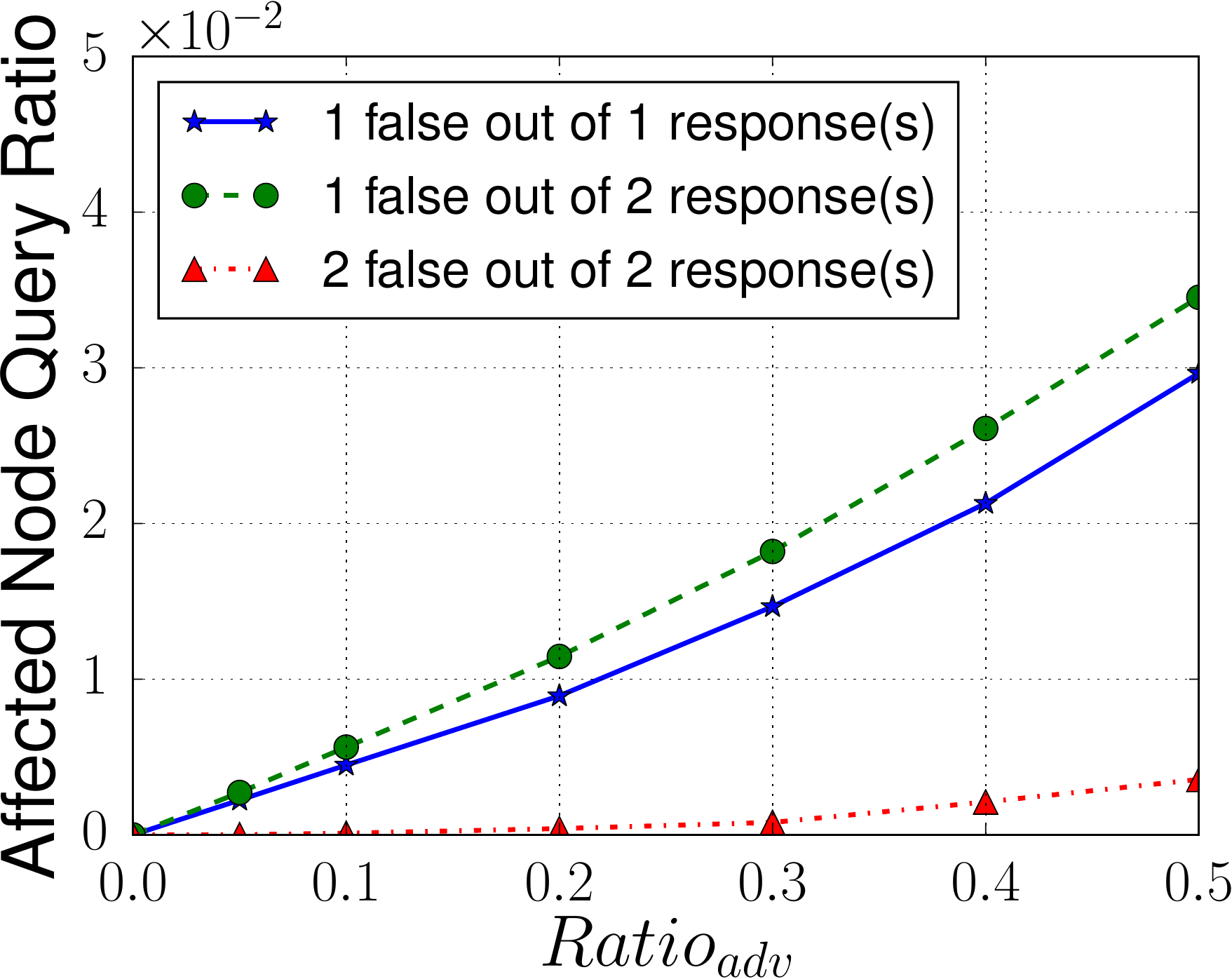}
		\caption{}%
		\label{subfig_resp_node_2_koln}
	\end{subfigure}
	\begin{subfigure}[b]{0.24\columnwidth}
		\includegraphics[width=\columnwidth]{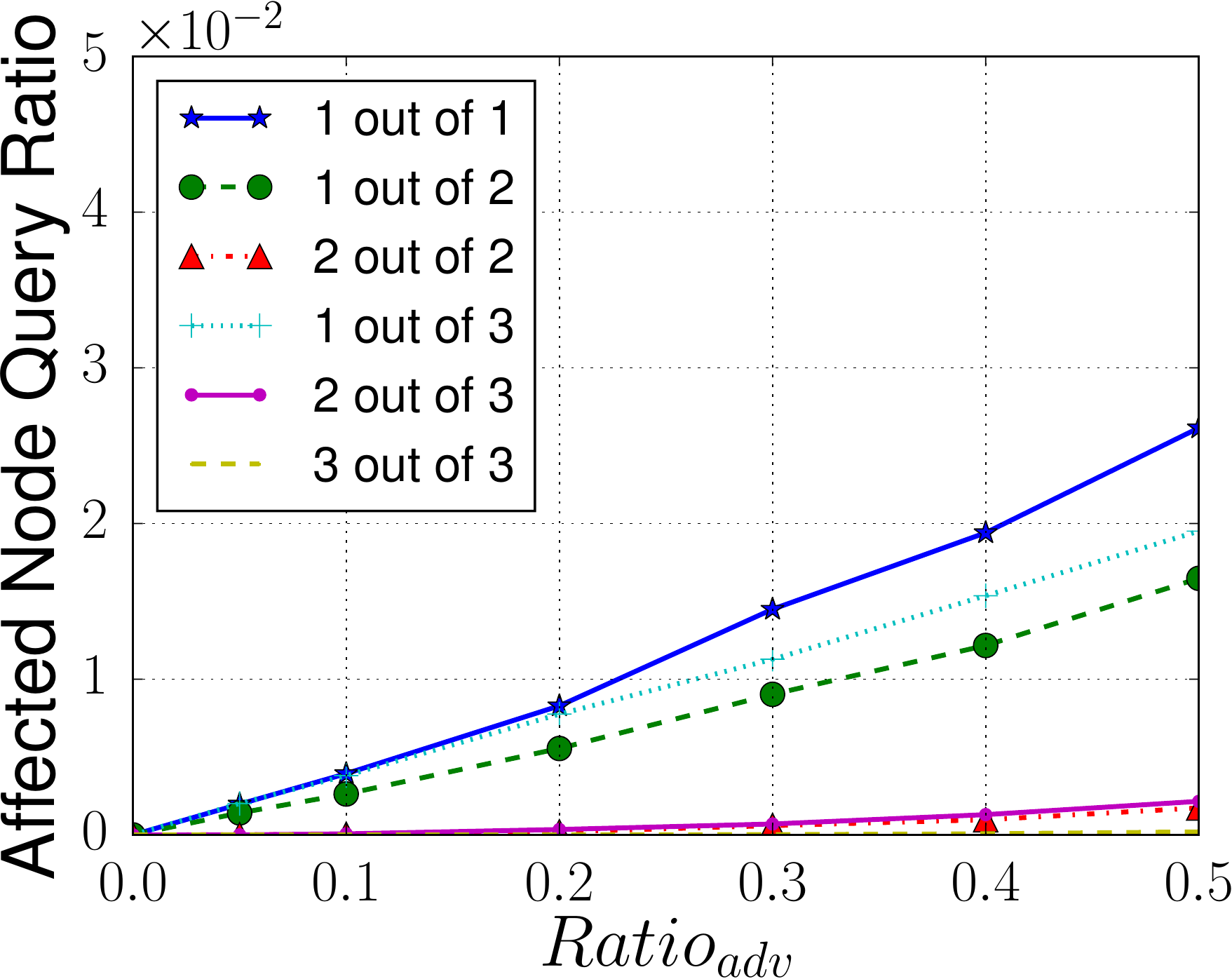}%
		\caption{}%
		\label{subfig_resp_node_3_koln}
	\end{subfigure}
	\caption{TAPASCologne: Ratio of malicious serving nodes with (\subref{subfig_ratio_node_2_koln}) $N=2$, and (\subref{subfig_ratio_node_3_koln}) $N=3$. Affected node query ratio as a function of $Ratio_{adv}$ with (\subref{subfig_resp_node_2_koln}) $N = 2$, and (\subref{subfig_resp_node_3_koln}) $N = 3$.}
	\label{fig_resp_node_koln}
\end{figure}

Fig.~\ref{fig_ratio_node} shows the ratio of the adversarial serving nodes over time, when $Pr_{check} = 0$. Even though $20\ \%$ of the nodes are compromised, the ratio of non-detected malicious serving nodes (over all serving nodes) is less than $10\ \%$, for the majority of time for both $N=2$ (\cref{subfig_ratio_node_2}) and $N=3$, with different region sizes (\cref{subfig_ratio_node_3_1km,subfig_ratio_node_3_2km,subfig_ratio_node_3_3km}). The periodical ratio peaks are due to the simultaneous reassignment of roles to the participating
nodes. Without misbehavior detection (i.e., cross-checking), the ratio would have roughly remained the same as $Ratio_{adv}$ (i.e., $20\ \%$). However, the controlled selection of serving nodes effectively limits active participation of malicious nodes, and the cross-checking mechanism further helps detecting and evicting them from the system. Moreover, the ratio of ``active'' malicious nodes (i.e., ones that could actually provide false responses) is always less than $1\ \%$ for all cases: a significant decrease from the original $Ratio_{adv}$. This shows that even if $20\ \%$ of the nodes are compromised (by a ``master'' adversary), the ratio of actual usable compromised nodes is considerably lower (i.e., less than $1\ \%$). Such low ratio of ``active'' malicious nodes is reflected on the ratio of the affected node queries in Fig.~\ref{fig_ratio_resp}: for example, when $Ratio_{adv}=0.2$, around only $1.5\ \%$ of the node queries get false peer responses that are not detected. This happens when all peer responses to a node query are given by malicious serving nodes, thus, there is no conflict. If conflicting responses are received, the querying node checks the correctness with the \ac{RA} and reports the misbehavior accordingly (Sec.~\ref{sec:scheme}). The adversarial serving node ratios and affected node query ratios are overall lower for the TAPASCologne scenario (\cref{fig_resp_node_koln}), because higher node (thus benign serving node) density facilitates the detection of false peer responses (thus the eviction of malicious serving nodes).

\begin{figure}[tbp!]
	\centering
	\begin{subfigure}[b]{0.24\columnwidth}
		\includegraphics[width=\columnwidth]{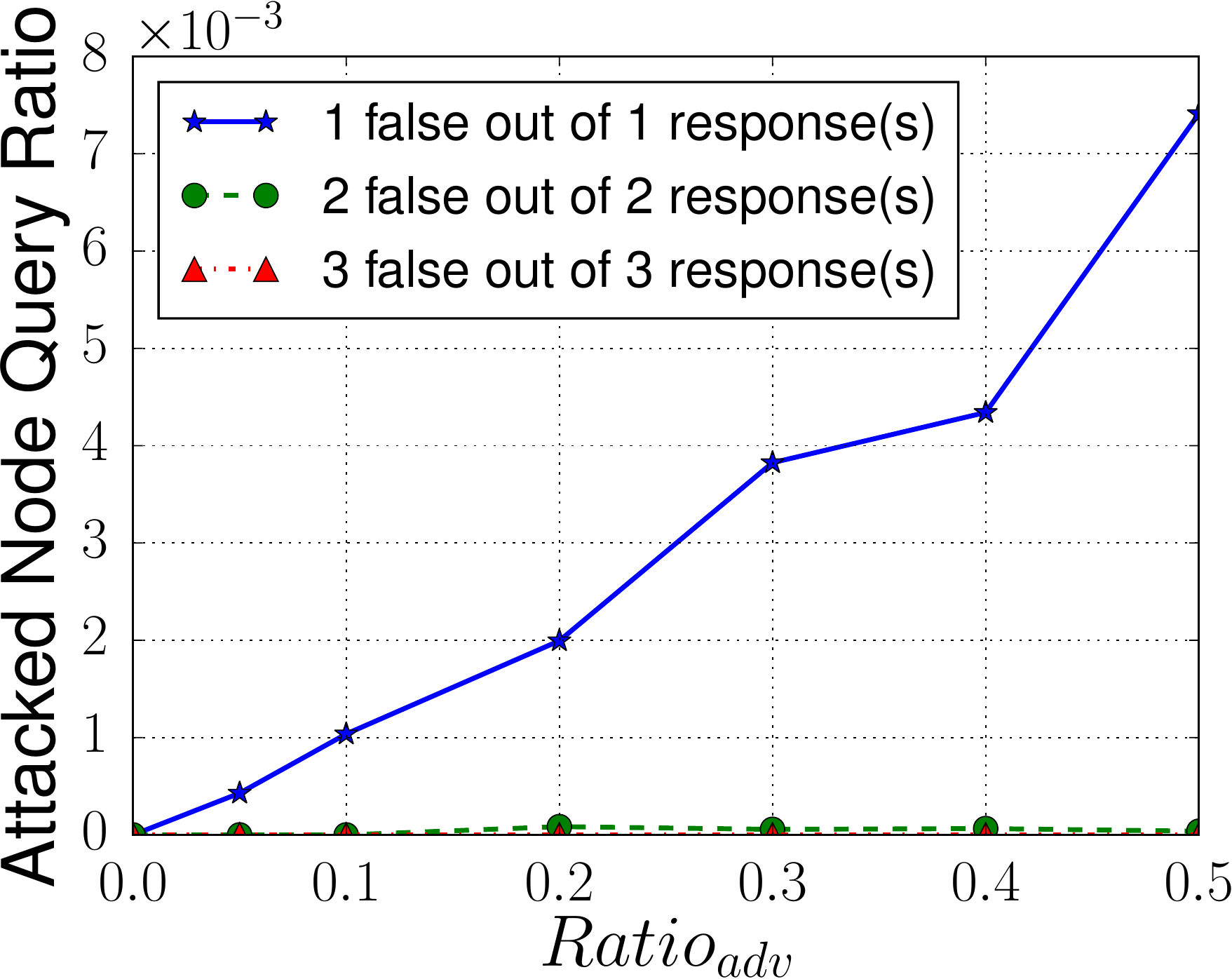}
		\caption{}%
		\label{subfig_ratio_acceted_resp_3}
	\end{subfigure}
	\begin{subfigure}[b]{0.24\columnwidth}
		\includegraphics[width=\columnwidth]{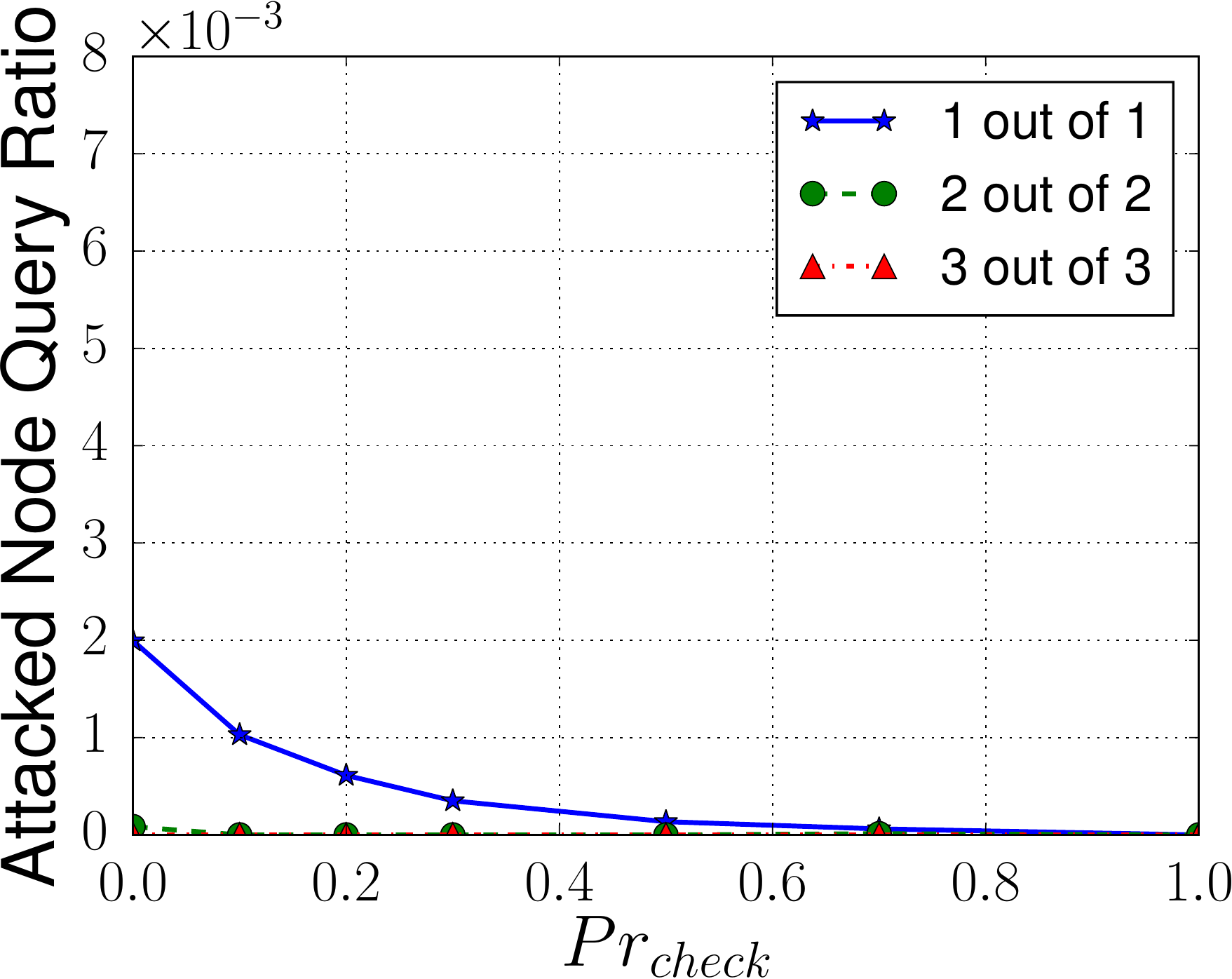}%
		\caption{}%
		\label{subfig_check_acceted_resp_3}
	\end{subfigure}
	\begin{subfigure}[b]{0.24\columnwidth}
		\includegraphics[width=\columnwidth]{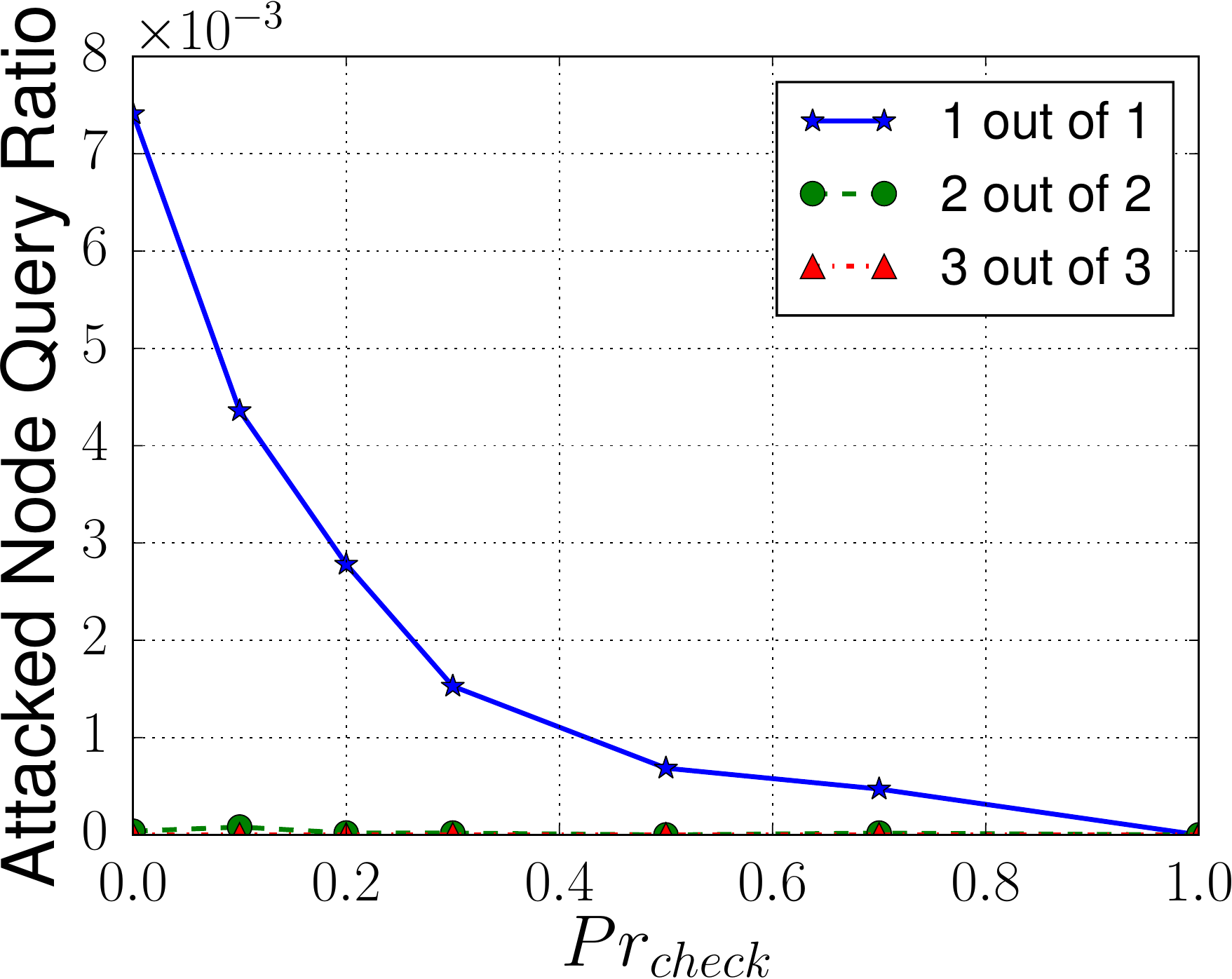}%
		\caption{}%
		\label{subfig_check05_acceted_resp_3}
	\end{subfigure}
	
	\begin{subfigure}[b]{0.24\columnwidth}
		\includegraphics[width=\columnwidth]{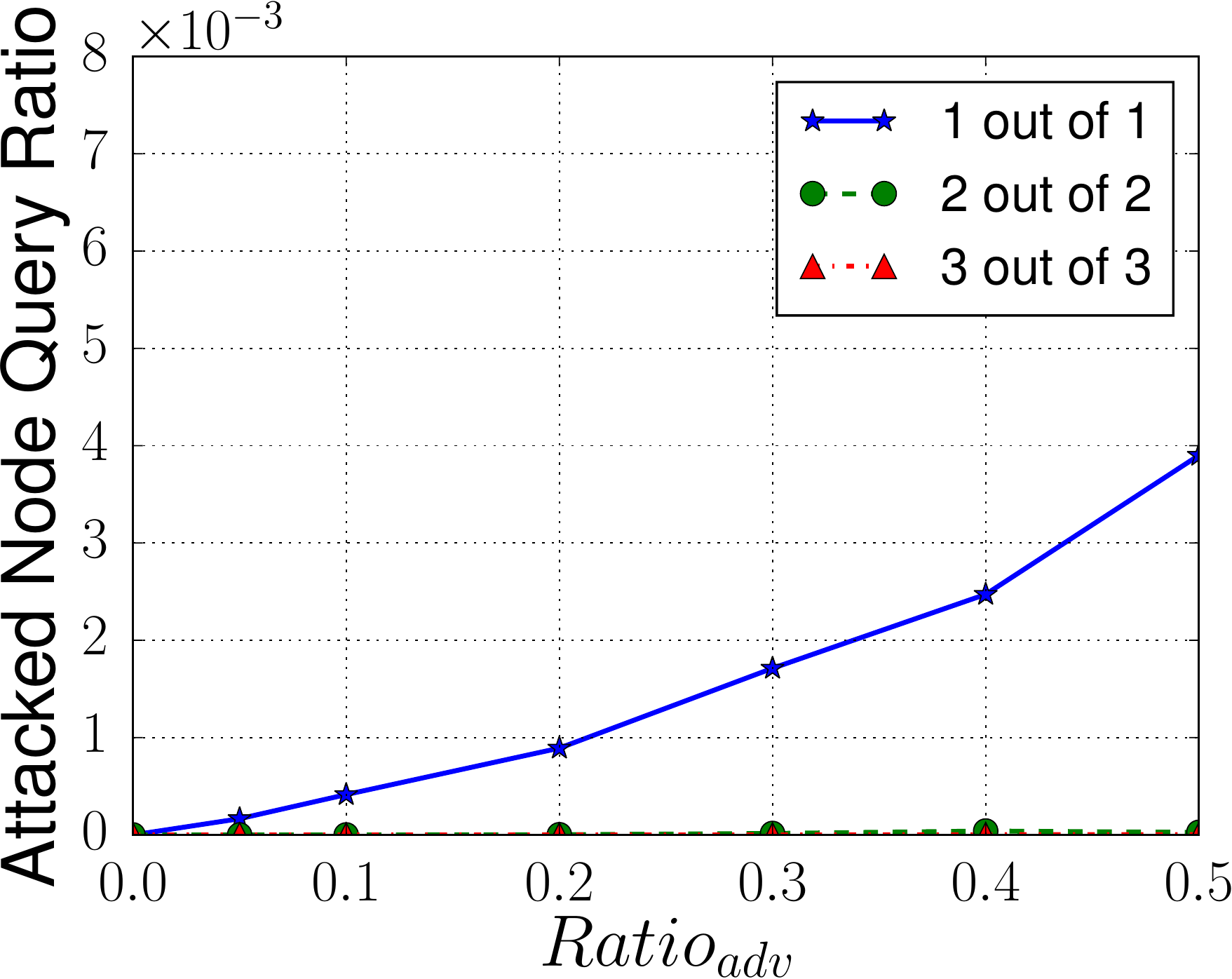}
		\caption{}%
		\label{subfig_ratio_acceted_resp_3_koln}
	\end{subfigure}
	\begin{subfigure}[b]{0.24\columnwidth}
		\includegraphics[width=\columnwidth]{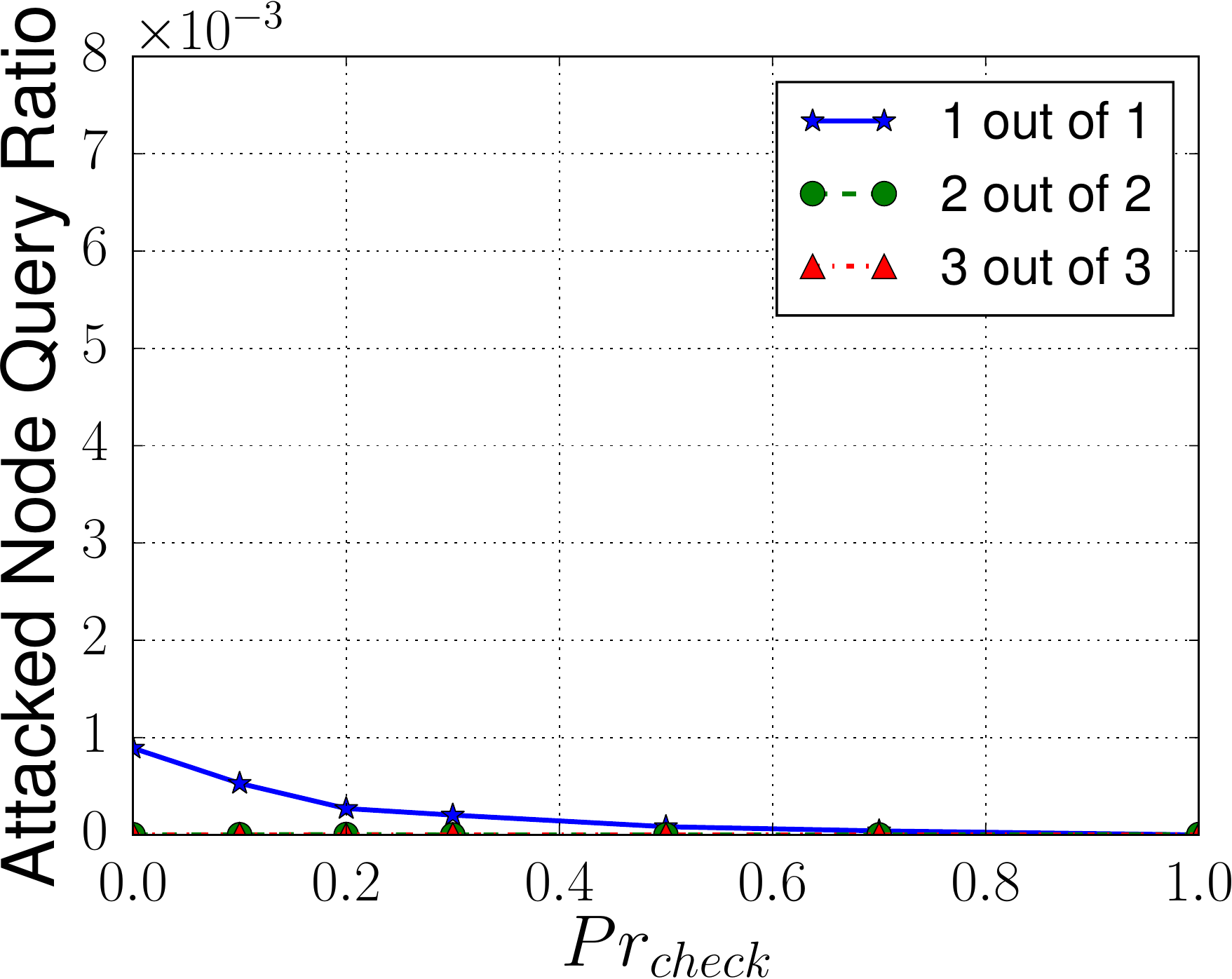}%
		\caption{}%
		\label{subfig_check_acceted_resp_3_koln}
	\end{subfigure}
	\begin{subfigure}[b]{0.24\columnwidth}
		\includegraphics[width=\columnwidth]{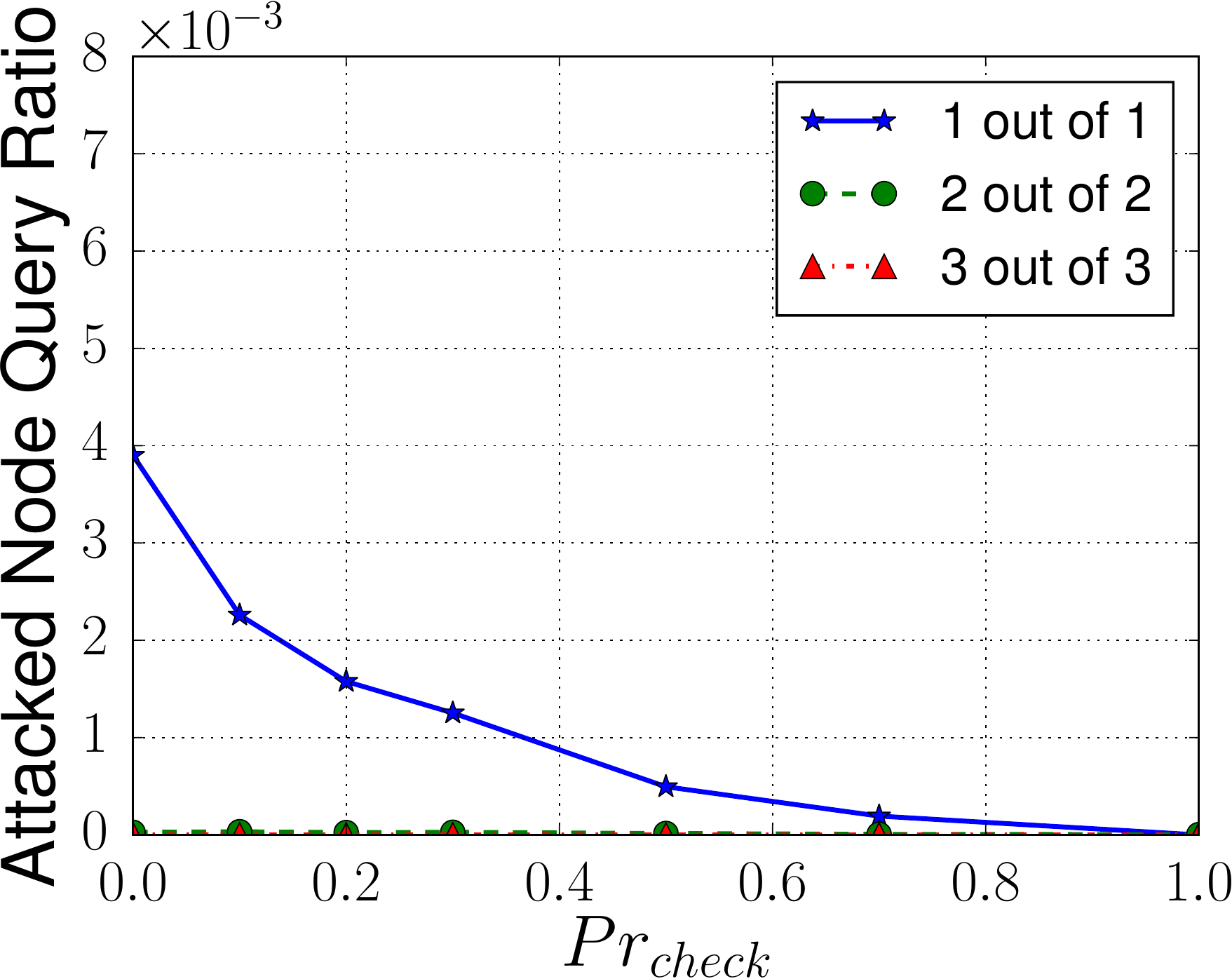}%
		\caption{}%
		\label{subfig_check05_acceted_resp_3_koln}
	\end{subfigure}
	
	\caption{LuST: (first row) and TAPASCologne (second row): Non-detected attacked node query ratio as a function of (first column) $Ratio_{adv}$, and $Pr_{check}$ when (second column) $Ratio_{adv} = 0.2$ and (third column) $Ratio_{adv} = 0.5$. (Default: $Ratio_{adv}=0.2$ and $Pr_{check}=0$.)}
	\label{fig_accepted_resp}
\end{figure}

For the evaluation above, we did not presume the availability of prior information on the security status of malicious peers. This can be available through a periodically publicized \ac{CRL} by the \ac{PKI}, and in turn, can further improve resilience to faulty peers. We consider next the effect of a \ac{CRL} on malicious node detection: for simplicity, assume that the \ac{CRL} is readily available to nodes, eliminating the effect of delay for obtaining the \ac{CRL} (which can be practically downloaded or distributed~\cite{khodaei2018efficient}). Consider, a query from $U_1$ was responded by two malicious serving nodes $U_{ms_1}$ and $U_{ms_2}$, while $U_{ms_2}$ was detected later as misbehaving node through a query from $U_2$ responded by $U_{ms_2}$ and a benign serving node $U_{bs_1}$. Once $U_1$ obtains the latest \ac{CRL}, it can do a post-checking for its queries responded by $U_{ms_2}$. This could further reveal $U_{ms_1}$ as a misbehavior, which, in turn, could disclose more node queries served by single false peer responses from $U_{ms_1}$. \cref{subfig_ratio_acceted_resp_3,subfig_ratio_acceted_resp_3_koln} show the ratio of non-detected attacked (refereed as \emph{attacked} in the rest of the paper) node queries when $N=3$ with the above mentioned post-checking. For example, when $Ratio_{adv} = 0.2$ for the LuST scenario, the ratio of \emph{attacked} queries is around $0.2\%$: a significant decrease from $1.5\%$, as shown in \cref{subfig_ratio_acceted_resp_3}, thanks to the access to the \ac{CRL}.

\cref{subfig_check_acceted_resp_3,subfig_check_acceted_resp_3_koln} show the ratio of attacked node queries as a function of $Pr_{check}$ with $Ratio_{adv} = 0.2$. We see the attacked node query ratio almost reaches $0$ when $Pr_{check}$ is higher than $0.5$. When $Ratio_{adv}=0.5$ (\cref{subfig_check05_acceted_resp_3,subfig_check05_acceted_resp_3_koln}), $Pr_{check}$ should be set to higher than $0.7$ (or even $1$), in order to achieve the same performance. With such a high ratio of serving nodes proven to be malicious, the system should instruct the benign nodes to proactively defend themselves (and the system) against the malicious nodes. Moreover, as described earlier, proactive \ac{LBS} checks do not expose benign querying nodes (i.e., do not harm their privacy), because the authenticators are the signatures generated by the serving nodes (Sec.~\ref{sec:privacy_lbs}).

\begin{figure}[tbp!]
	\centering
	\begin{subfigure}[b]{0.24\columnwidth}
		\includegraphics[width=\columnwidth]{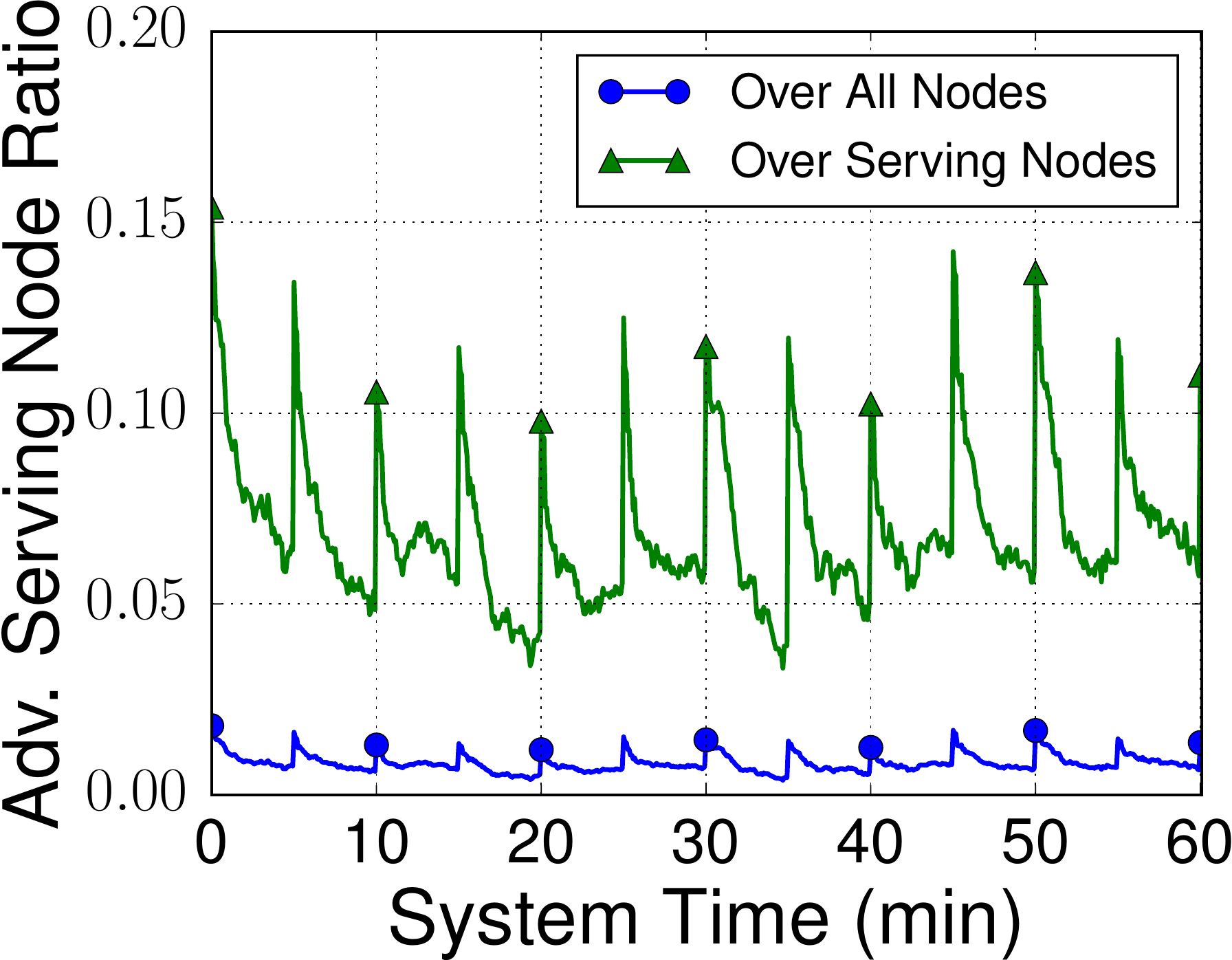}%
		\caption{}%
		\label{subfig_ratio_type_2_lust}
	\end{subfigure}
	\begin{subfigure}[b]{0.24\columnwidth}
		\includegraphics[width=\columnwidth]{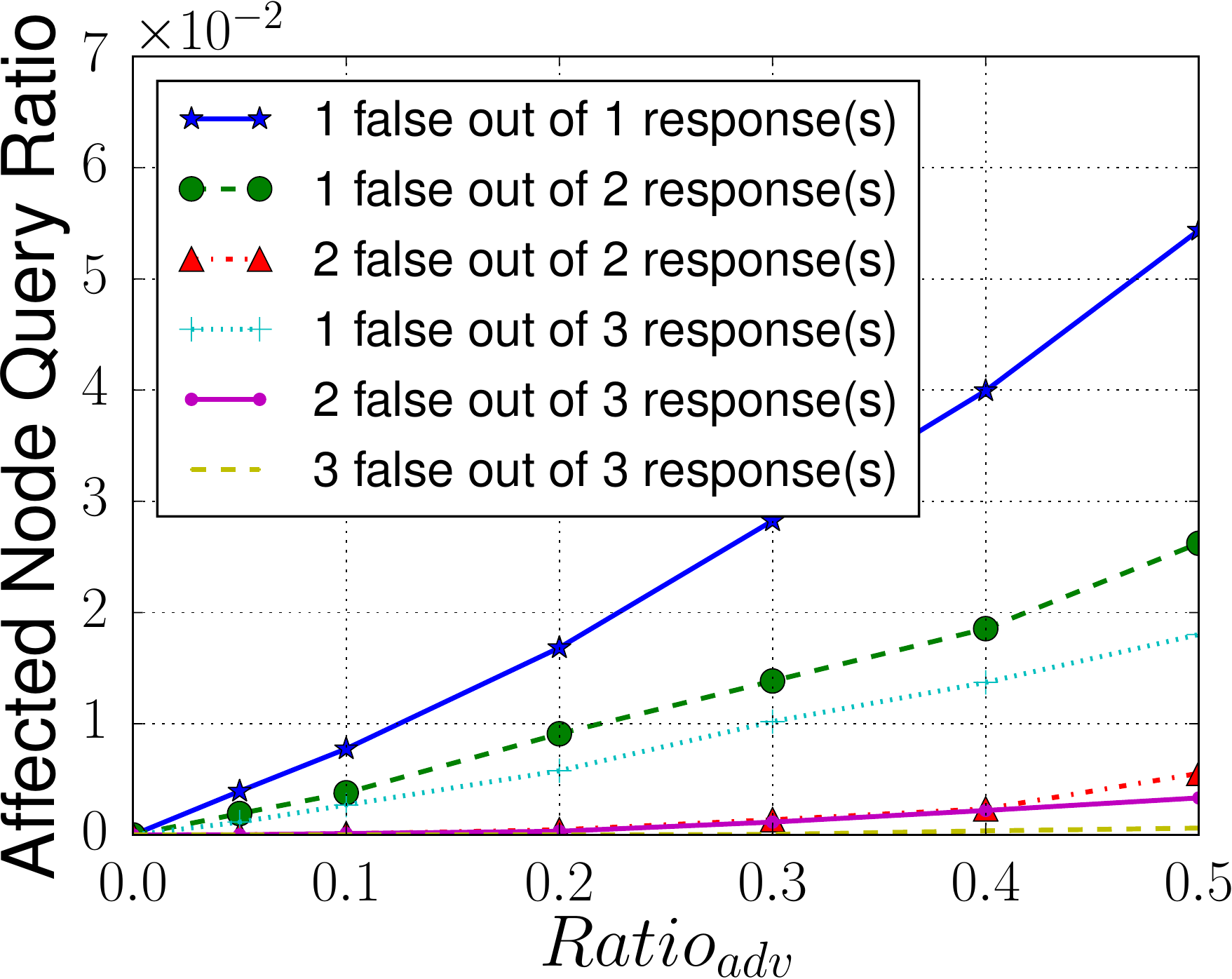}%
		\caption{}%
		\label{subfig_resp_type_2_lust}
	\end{subfigure}
	\begin{subfigure}[b]{0.24\columnwidth}		
		\includegraphics[width=\columnwidth]{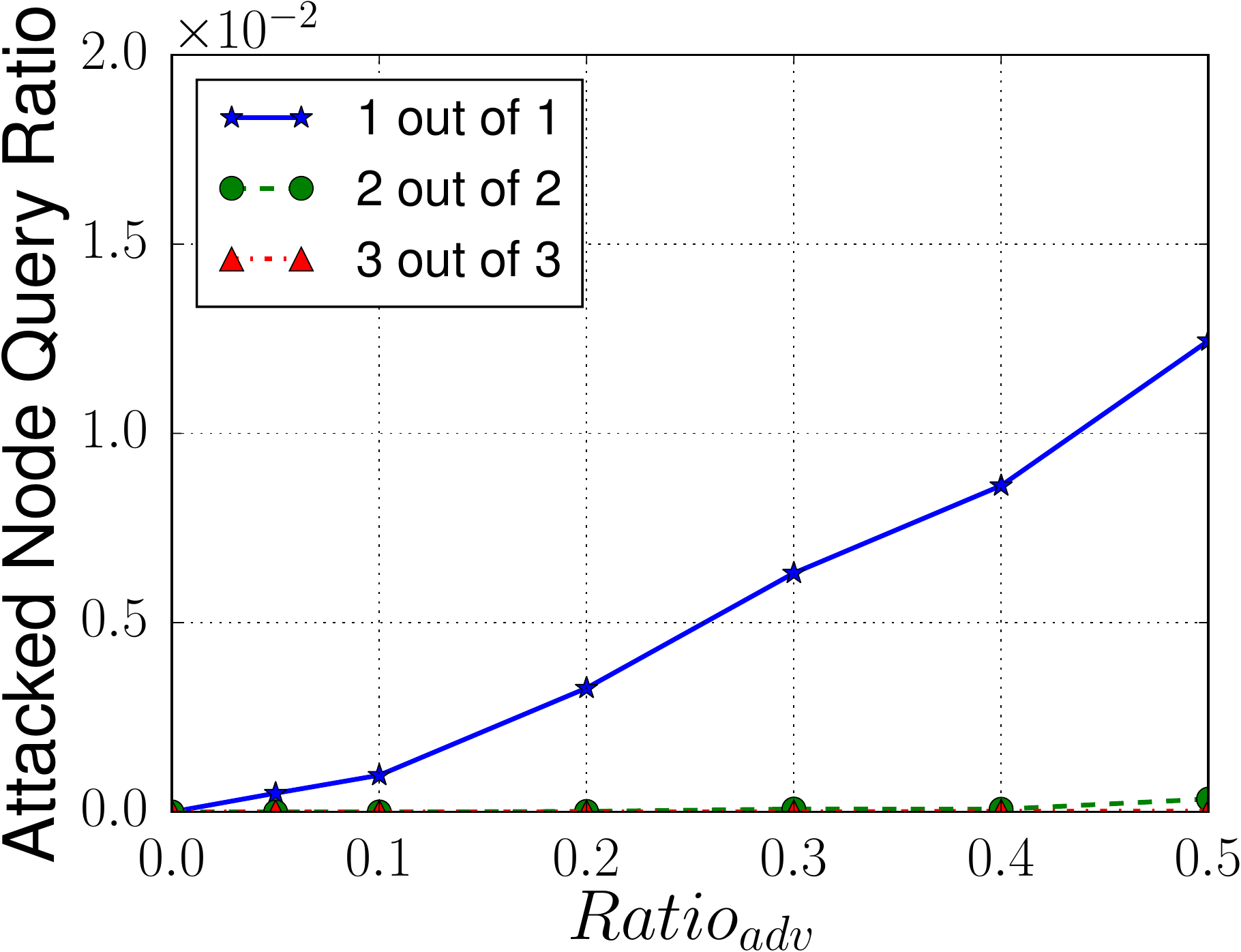}
		\caption{}%
		\label{subfig_accepted_type_2_lust}
	\end{subfigure}

	\begin{subfigure}[b]{0.24\columnwidth}
		\includegraphics[width=\columnwidth]{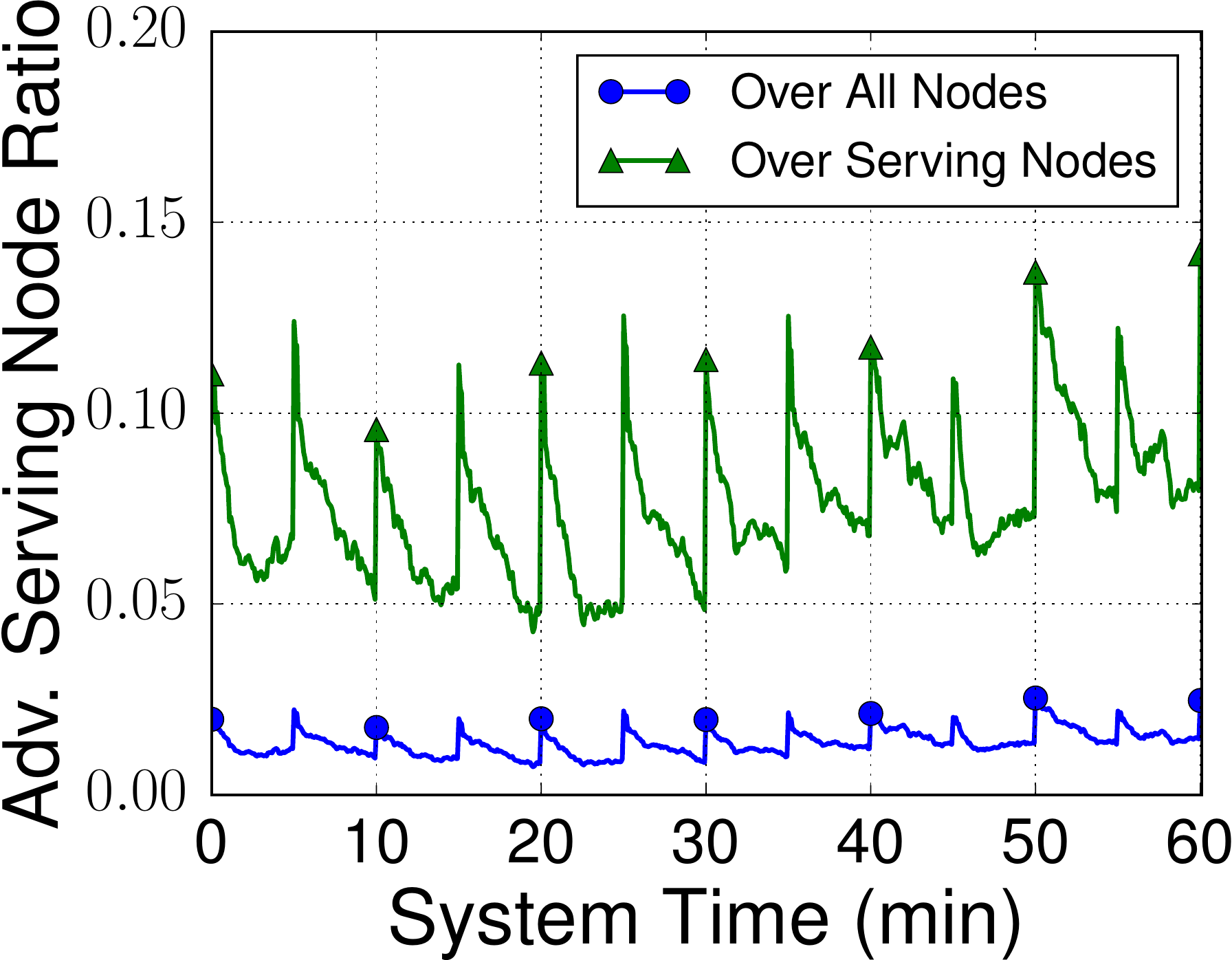}%
		\caption{}%
		\label{subfig_ratio_type_3_lust}%
	\end{subfigure}
	\begin{subfigure}[b]{0.24\columnwidth}		
		\includegraphics[width=\columnwidth]{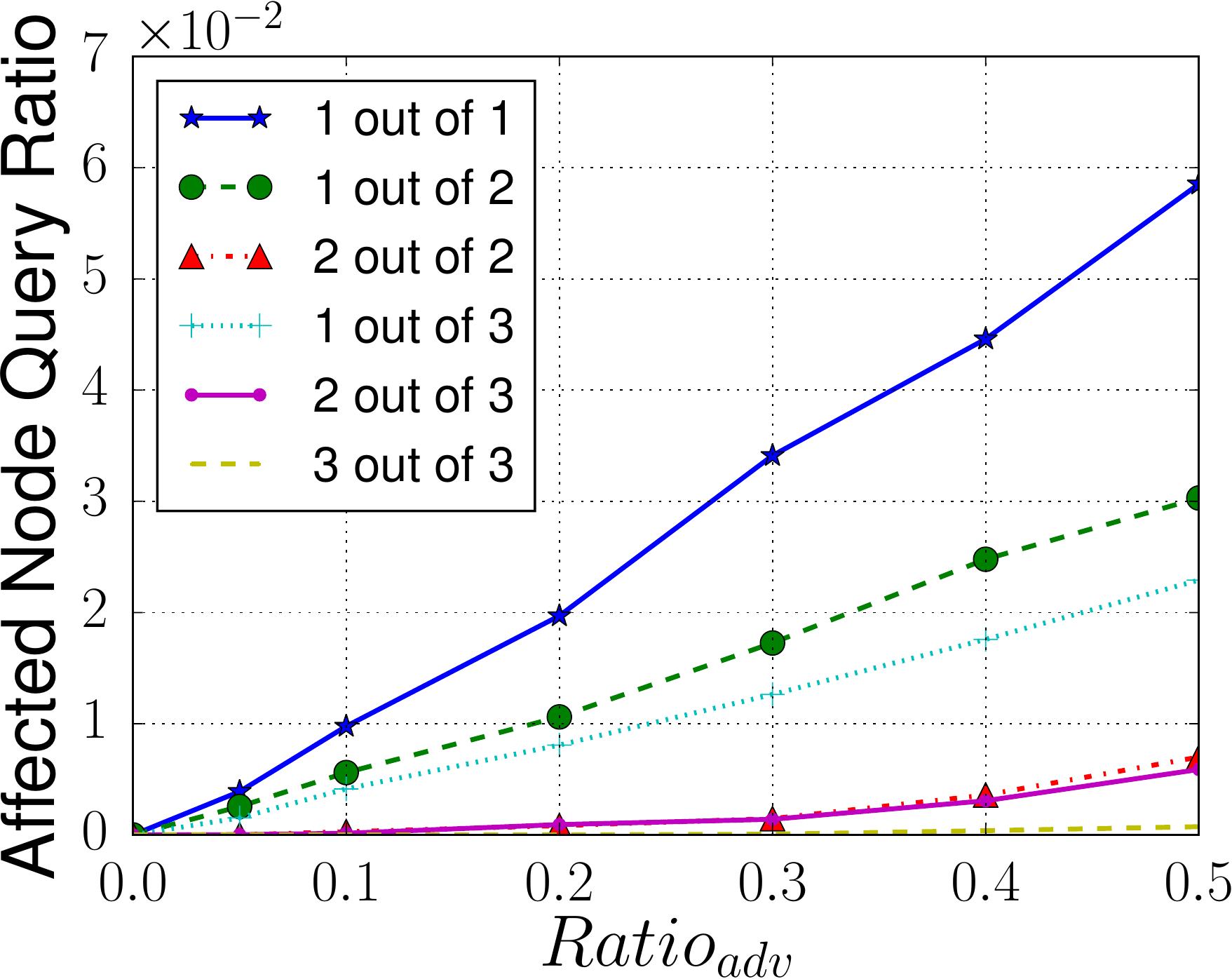}
		\caption{}%
		\label{subfig_resp_type_3_lust}
	\end{subfigure}
	\begin{subfigure}[b]{0.24\columnwidth}
		\includegraphics[width=\columnwidth]{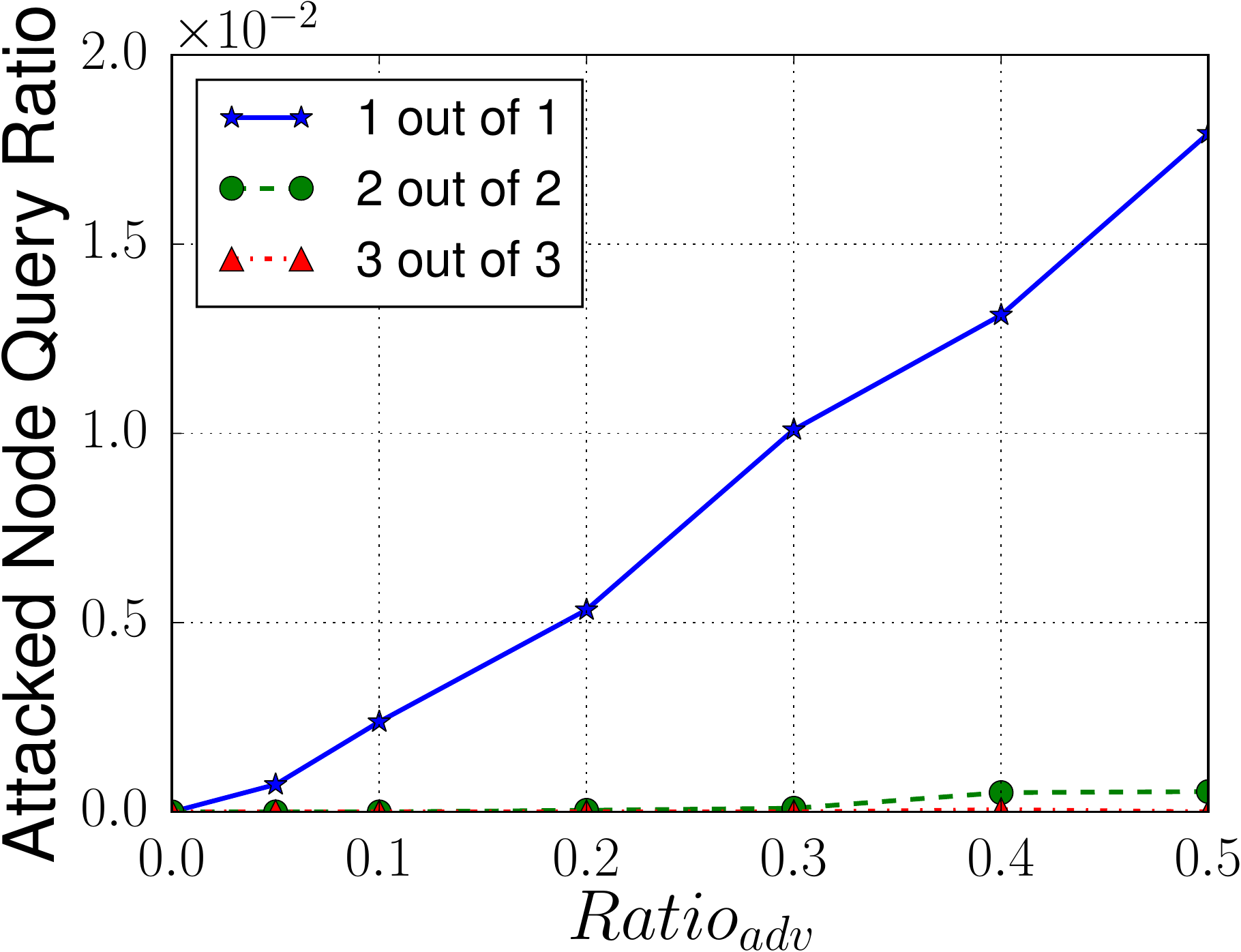}%
		\caption{}%
		\label{subfig_accepted_type_3_lust}
	\end{subfigure}
	\caption{LuST: (First column) Ratio of malicious serving nodes, (second column) affected node query ratio as a function of $Ratio_{adv}$, and (third column) non-detected attacked node query ratio as a function of $Ratio_{adv}$, with (first row) $G = 2$ and $Pr_{serve} = 0.12$, and (second row) $G = 3$ and $Pr_{serve} = 0.18$.}
	\label{fig_resp_type_lust}
\end{figure}

\cref{fig_resp_type_lust} shows the resultant resilience with $G>1$. From \cref{subfig_ratio_node_3_2km}, \cref{subfig_ratio_type_2_lust} and \cref{subfig_ratio_type_3_lust}, we see the trend that the overall adversarial serving node ratio is higher as $G$ grows, due to the higher $Pr_{serve}$ being set. As a result, affected and attacked node query ratios also increase with higher $G$. For example, the attacked node query ratio is around $0.2\%$, $0.3\%$ and $0.5\%$, with $Ratio_{adv}=0.2$, in \cref{subfig_ratio_acceted_resp_3}, \cref{subfig_accepted_type_2_lust} and \cref{subfig_accepted_type_3_lust} respectively.

\textbf{Communication Overhead:} We close this section with a comparative evaluation of the communication overhead based on simulation results for our scheme and MobiCrowd. In our scheme, only serving nodes need to request regional \ac{POI} data, and querying nodes only obtain responses with \ac{POI} entries that match their queries. However, in MobiCrowd, nodes always obtain regional \ac{POI} data (that is, not the subset of precise interest) either from their peers or from the \ac{LBS} server, in order to keep the \ac{POI} data verifiable with the \ac{LBS}-provided signatures.
	
\begin{figure*}[h]
	\begin{subfigure}[b]{.24\columnwidth}
		\includegraphics[width=\columnwidth]{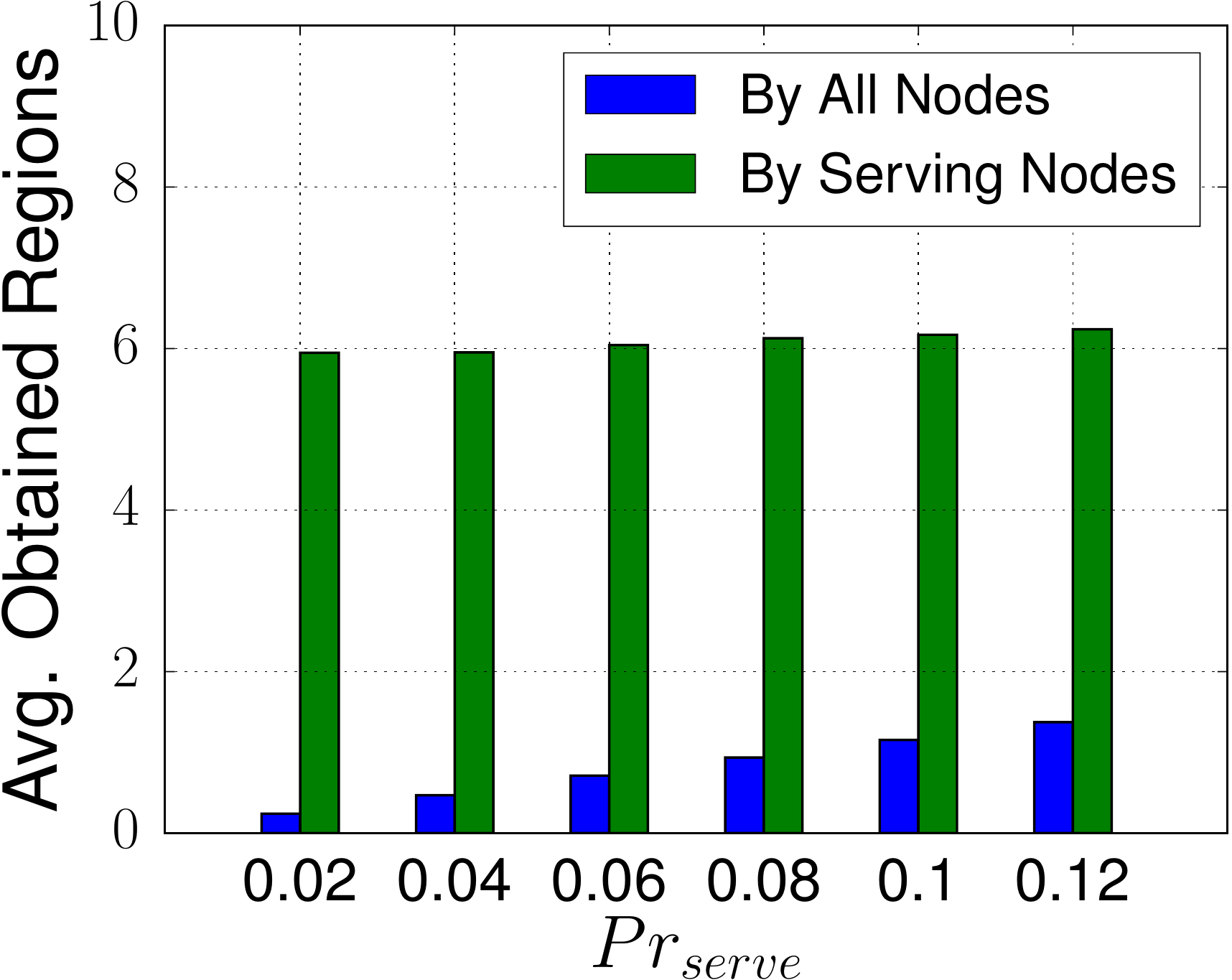}%
		\caption{}
		\label{subfig_oh_prserve_1km}
	\end{subfigure}
	\begin{subfigure}[b]{.24\columnwidth}
		\includegraphics[width=\columnwidth]{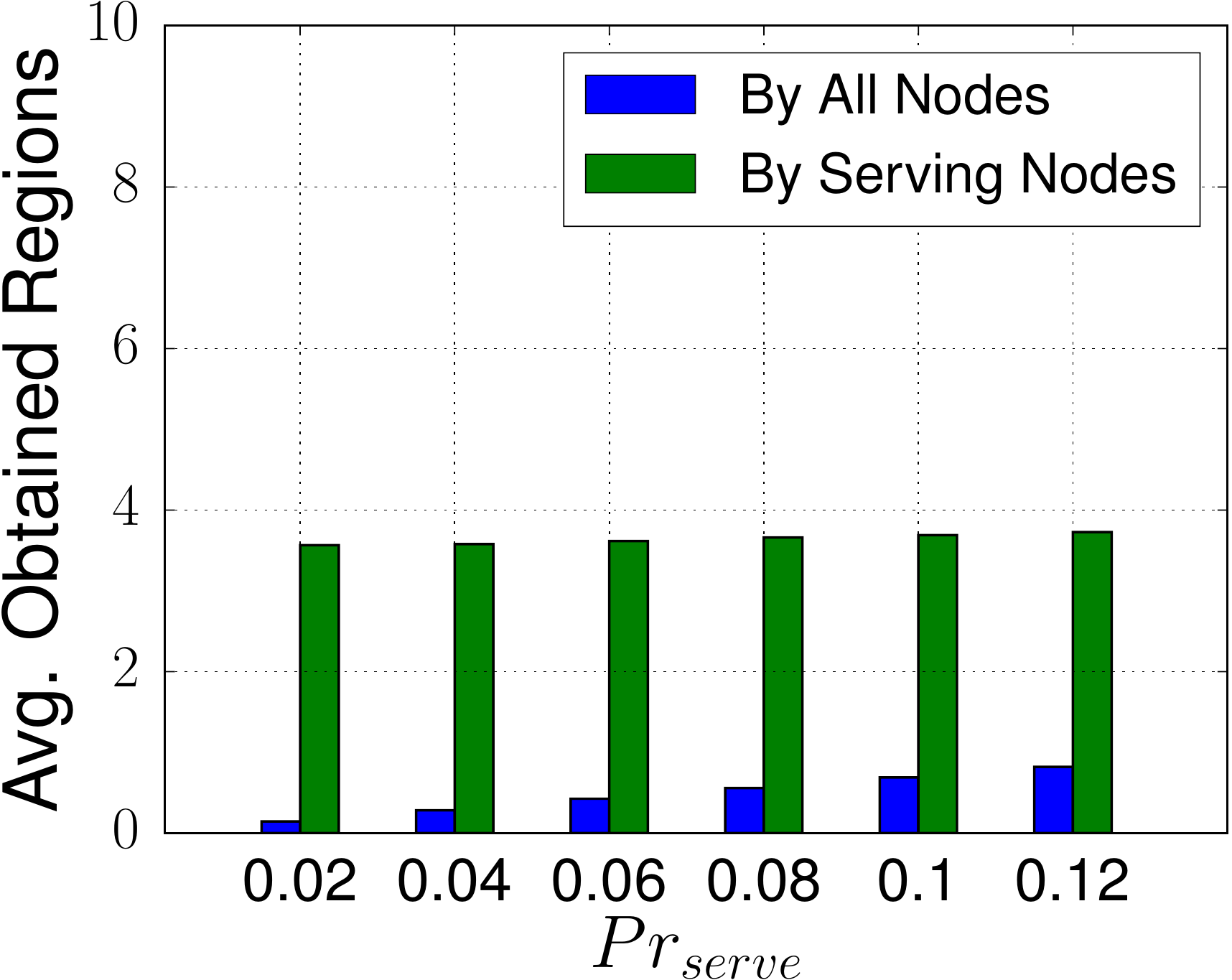}%
		\caption{}
		\label{subfig_oh_prserve_2km}
	\end{subfigure}
	\begin{subfigure}[b]{.24\columnwidth}
	\includegraphics[width=\columnwidth]{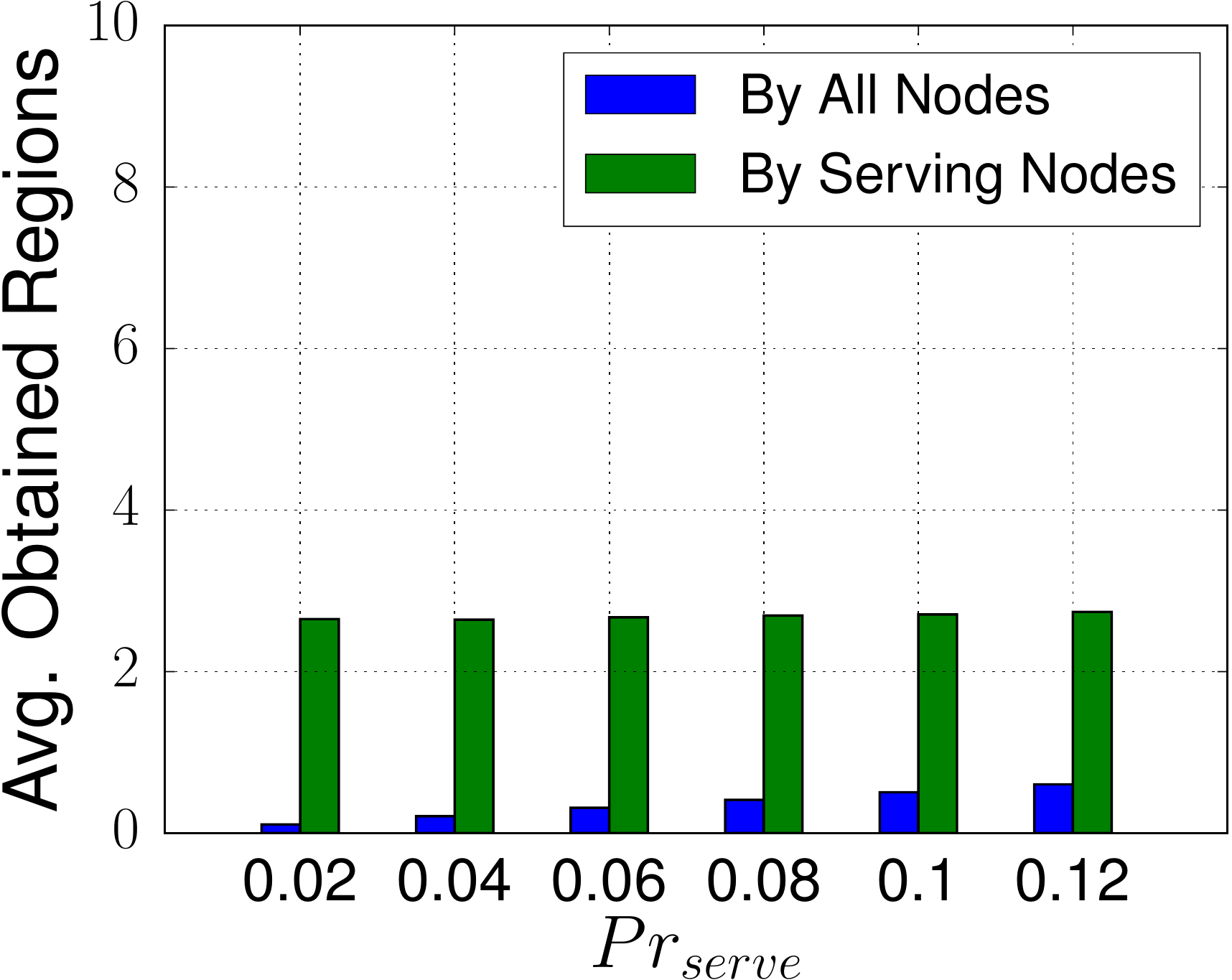}%
		\caption{}
		\label{subfig_oh_prserve_3km}
	\end{subfigure}
	
	\begin{subfigure}[b]{.24\columnwidth}
		\includegraphics[width=\columnwidth]{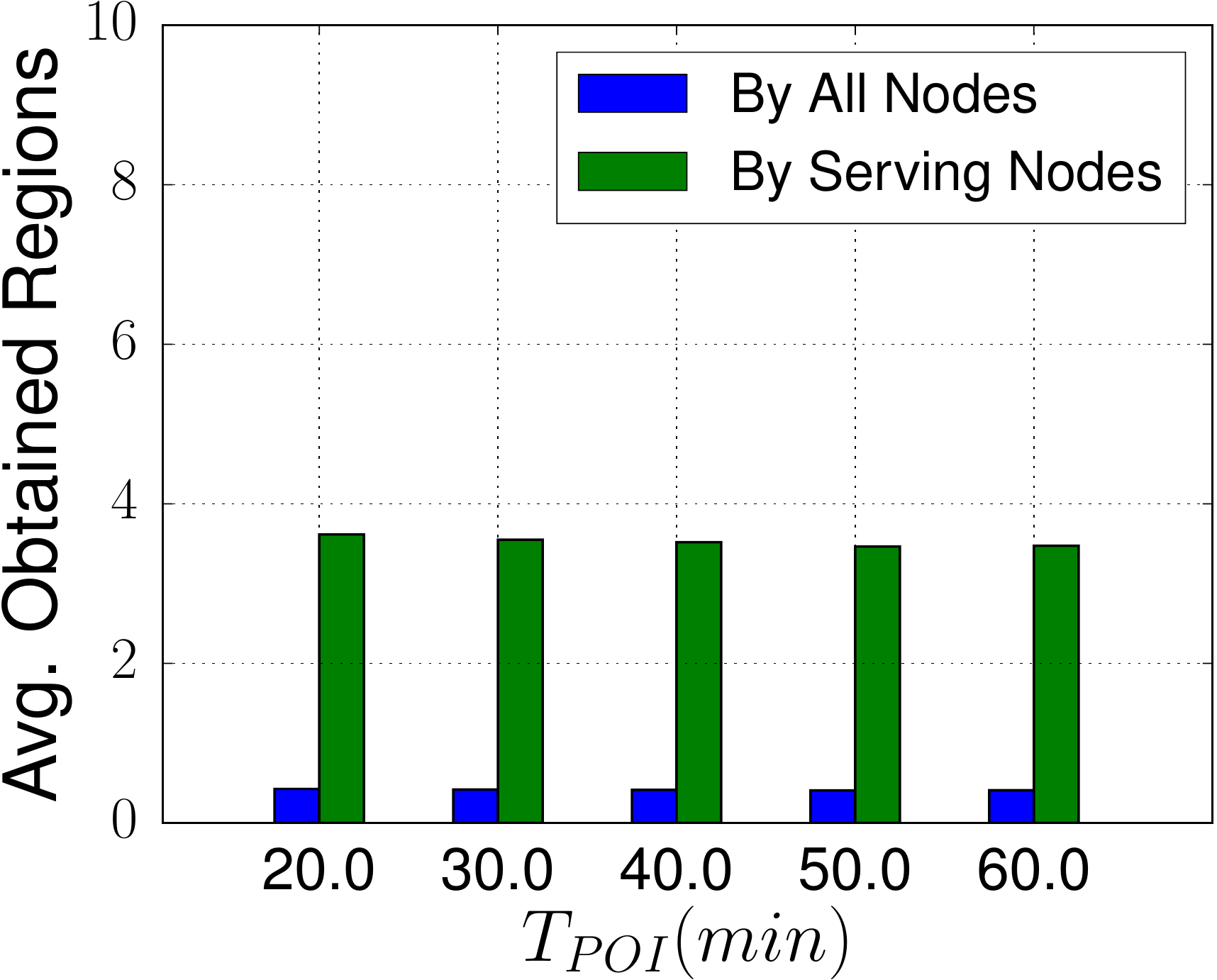}%
		\caption{}
		\label{subfig_oh_poi}
	\end{subfigure}
	\begin{subfigure}[b]{.24\columnwidth}
		\includegraphics[width=\columnwidth]{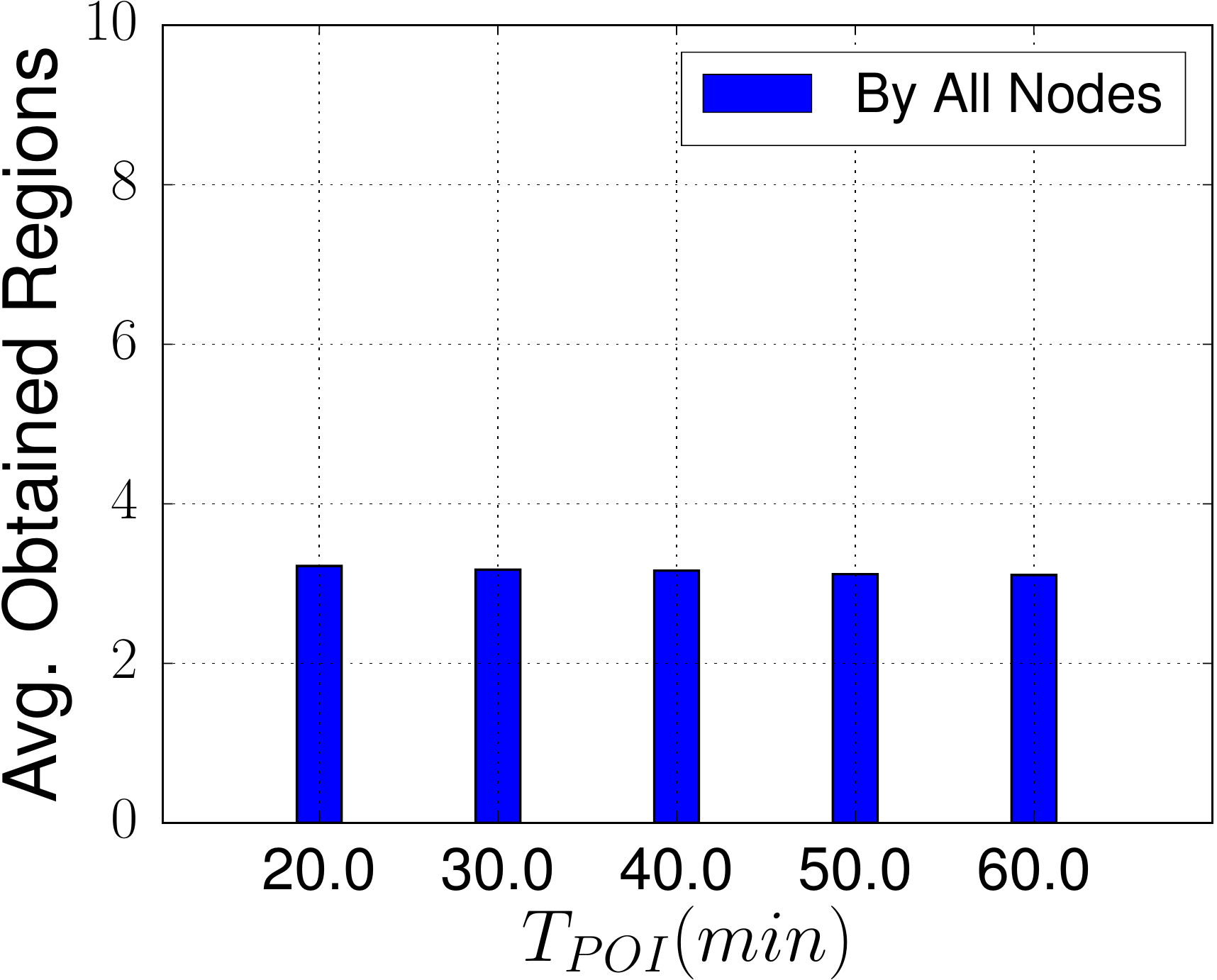}%
		\caption{}
		\label{subfig_oh_mc_poi}
	\end{subfigure}
	\begin{subfigure}[b]{.24\columnwidth}
		\includegraphics[width=\columnwidth]{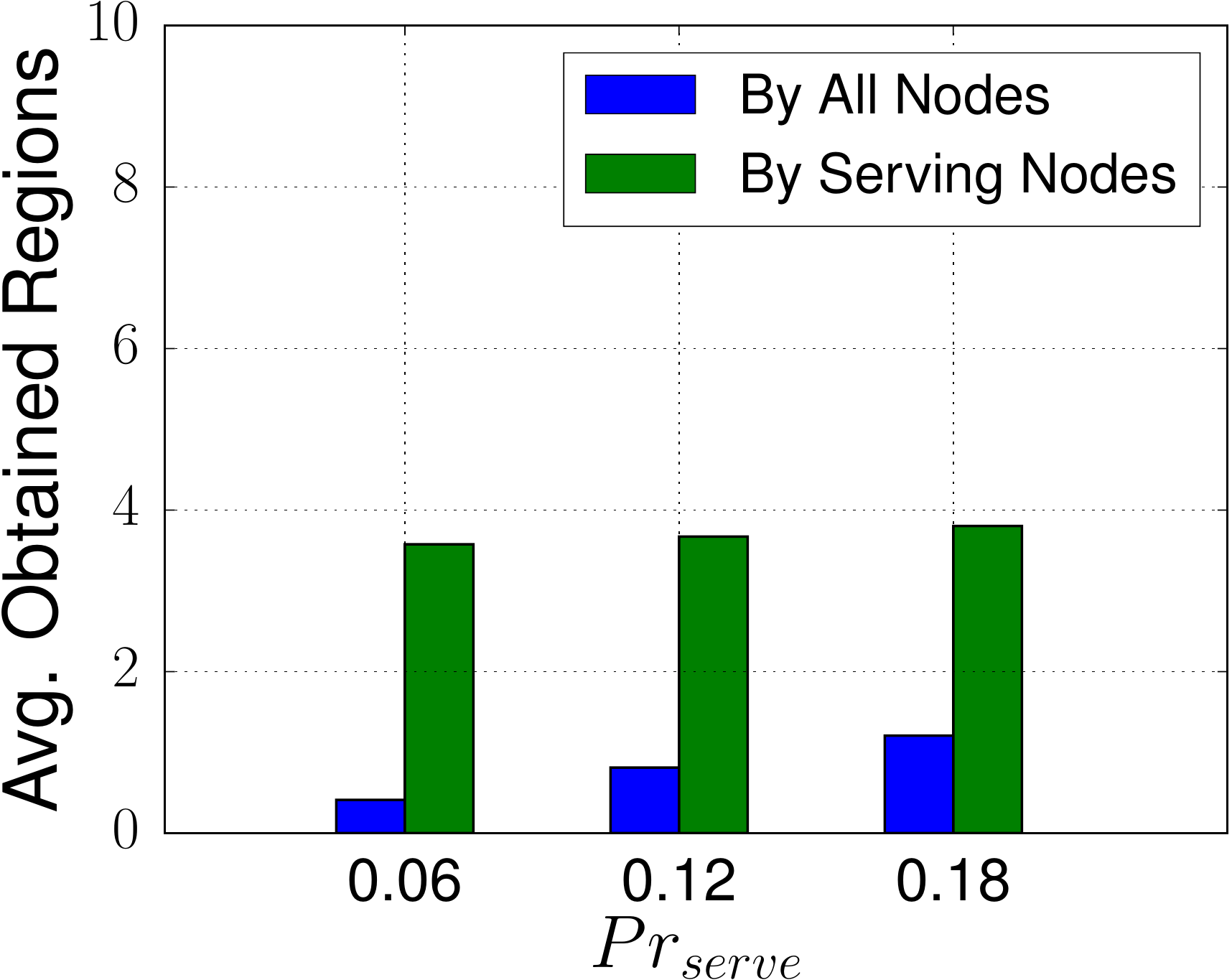}%
		\caption{}
		\label{subfig_oh_2}
	\end{subfigure}
	\begin{subfigure}[b]{.24\columnwidth}
		\includegraphics[width=\columnwidth]{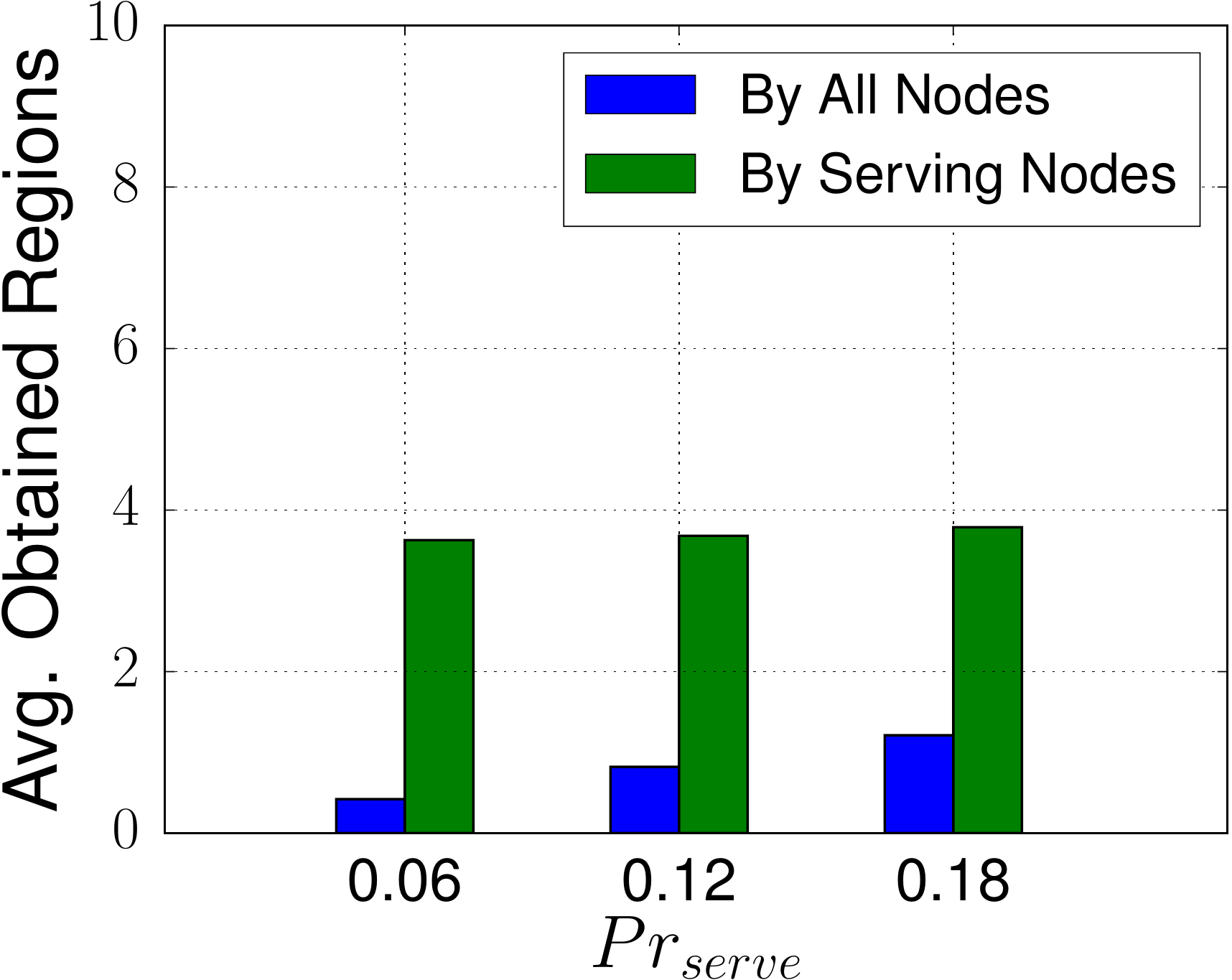}%
		\caption{}
		\label{subfig_oh_3}
	\end{subfigure}
	\caption{LuST: Average number of obtained regional \ac{POI} data as a function of (\subref{subfig_oh_prserve_1km}, \subref{subfig_oh_prserve_2km}, \subref{subfig_oh_prserve_3km}) $Pr_{serve}$  when $L = 1,2,3\ km$, as a function of $T_{POI}$ with (\subref{subfig_oh_poi}) our scheme and (\subref{subfig_oh_mc_poi}) MobiCrowd, and as a function of $Pr_{serve}$ with (\subref{subfig_oh_2}) $G=2$ and (\subref{subfig_oh_3}) $G=3$.}
	\label{fig_oh}
\end{figure*}

\cref{fig_oh} shows the communication overhead in terms of the average number of regional \ac{POI} data obtained from the \ac{LBS} server (and peers, for MobiCrowd only). \cref{subfig_oh_prserve_1km,subfig_oh_prserve_2km,subfig_oh_prserve_3km} show the number of regions with $L = 1,2,3\ km$ respectively. Around 6 regional \ac{POI} data are obtained by each serving node when $L = 1\ km$, while around 3.5 regional \ac{POI} data are obtained when $L = 2\ km$. This reduces to around around $0.6$ and $0.4$ for $L = 1\ km$ and $L = 2\ km$ if averaged over all nodes (when $Pr_{serve}=0.06$): a previously non-serving node could be chosen as the serving node in a future trip, thus the load is balanced among all nodes. A region size with $L = 2\ km$ ($L = 3\ km$) is 4 (9) times as large as region size with $L = 1\ km$. Therefore, larger $L$ results in higher communication overhead for obtaining regional \ac{POI} data, although the number of obtained regions is decreased.

The actual communication overhead depends on actual \ac{POI} entry size and \ac{POI} density in each region. Consider a typical value from the literature~\cite{olumofin2010achieving,liu2016silence}, e.g., 500 bytes for each \ac{POI} entry. If there are $10\ 000$ \ac{POI} entries in each regional \ac{POI} data set, then the communication overhead for obtaining regional \ac{POI} data is around $5$ $Mbytes$, while a region with $100\ 000$ \ac{POI} entries results in $50$ $Mbytes$. Consider regional \ac{POI} data of $5$ $Mbytes$. For the default settings, a serving node needs to proactively obtain around $17.5$ $Mbytes$ of regional \ac{POI} data (for around $3.5$ regions, as shown in \cref{subfig_oh_prserve_2km}) during their trips to effectively serve other nodes. This reduces to around $2$ $Mbytes$ (for around $0.4$ region) if averaged over all nodes. For peer queries and direct \ac{LBS} queries, only the required \ac{POI} entries need to be obtained. For example, if $10$ \ac{POI} entries are returned for each query, then around $5$ $Kbytes$ of \ac{POI} data needs to be obtained: significantly lower than the communication overhead for obtaining the entire regional \ac{POI} data. \cref{subfig_oh_poi,subfig_oh_mc_poi} show a comparison of communication overhead between our scheme and MobiCrowd. With MobiCrowd, each node need to obtain around 3 regional \ac{POI} data from the \ac{LBS} server and their peers: significantly higher than $0.4$ in our scheme. With higher $G$ (\cref{subfig_oh_2,subfig_oh_3}), the average number of obtained regional \ac{POI} data remains roughly the same as in \cref{subfig_oh_prserve_2km}, because the average node mobility remains the same with even if $G$ changes. However, the overhead for obtaining regional \ac{POI} data decreases by a factor of $1/G$ accordingly.

\section{Performance Evaluation}
\label{sec:experiment}

We further evaluate experimentally the performance of our scheme with vehicular \acp{OBU}. An \ac{OBU} has a 1.66 $GHz$ dual-core CPU and it is equipped with an IEEE 802.11p interface for \ac{P2P} communication. We implement our scheme with C++. \ac{EC} public/private key pairs are used for asymmetric cryptography and \ac{AES} with 256-bit session key is used for symmetric key encryption. We use Crypto++~\footnote{www.cryptopp.com} for the \ac{ECIES} and OpenSSL\footnote{www.openssl.org} for the rest of cryptographic algorithms (e.g., \ac{ECDSA} and \ac{AES}). Table~\ref{table:exp-param} shows the parameters we use in the experiment.

\begin{table}[htp!]
	\caption{Experimental Parameters}
	\centering
	\footnotesize
	\begin{tabular}{| l | c | c | c | c |}
		\hline
		& Serving & Querying & Query Generator & Beacon Generator \\\hline
		Node num. & 5 & 1 & 2 & 2 \\\hline
		$T_{beacon}$ & $Exp(\frac{1}{5s})$ & - & - & - \\\hline
		$T_{query}$ & - & $5\ s$ & $Exp(\frac{1}{0.5s})$, $Exp(\frac{1}{6s})$ & - \\\hline
		$\gamma$ & \multicolumn{2}{c|}{1, 2.5, 5 $Hz$} & - & 40, 100, 200 $Hz$ \\\hhline{-----}
		EC size & \multicolumn{3}{c|}{192, 224, 256 $bit$} & 256 $bit$ \\\hhline{-----}
		N & - & \multicolumn{2}{c|}{1, 2, 3} & - \\\hhline{-----}
		\hline
	\end{tabular}
	\renewcommand{\arraystretch}{1}
	\label{table:exp-param}
\end{table}

	We run experiments in two different settings. In the first setting, we use five \acp{OBU} as serving nodes, each broadcasting beacons at a rate of $\frac{1}{T_{beacon}}$. We use one \ac{OBU} as a querying node, with a node query rate of $\frac{1}{T_{query}}$. Two additional query generators are introduced to emulate a large number of querying nodes in the network, each with a much higher node query rate ($T_{query} = Exp(\frac{1}{0.5s})$) than the aforementioned querying node. The beacon interval and the query generator intervals are exponentially distributed, to emulate beacons and queries from mobile neighboring serving and querying nodes, from the perspective of the querying node we evaluate. We randomize the two aforementioned intervals so that network load changes over time and across experiments.

	We also run experiments in a \ac{VC} specific setting with each serving node and querying node running a transportation safety beacon application: broadcasting signed safety beacons (termed \acp{CAM}) and verifying received safety beacons. We consider three \ac{CAM} rate values, $1$, $2.5$ and $5$ $Hz$. Although a maximum $\gamma=10\ Hz$ is possible according to the standard~\cite{cam}, \ac{CAM} rate adaptation schemes~\cite{sommer2011traffic,nguyen2017mobility,schmidt2010exploration} show that an adaptive \ac{CAM} rates (generally lower than $10\ Hz$), considering channel load and vehicle mobility, leads to higher \ac{CAM} reception ratio. Therefore, we consider modest \ac{CAM} rates rather than the high-load approach ($10\ Hz$). We add in our experimental setup two \ac{CAM} generators, each broadcasting \acp{CAM} at a rate equivalent to an aggregate of 40 nodes (thus 80 nodes in total). $T_{query}$ is also adjusted to roughly match the aggregate query rate of 80 nodes. With $T_{query} = Exp(\frac{1}{6s})$ for each query generator, and $T_{query}=5\ s$ for the evaluated querying node, the average $T_{query}$ is roughly $2.5\ min$ for each node (among the 81 emulated nodes). In this setting, we only use EC-256 keys (which is the level in the corresponding standard~\cite{cam}).

	Moreover, we evaluate processing delays with an implementation of MobiCrowd. With MobiCrowd, once a querying node obtains and caches the regional \ac{POI} data, it no longer needs to query other nodes. With the limited number of \acp{OBU} in the testbed, all nodes would have the regional \ac{POI} data cached locally within a short period. However, to benchmark the processing delays with MobiCrowd in our experiment, the querying nodes continuously broadcast queries (without turning themselves into cached nodes, i,e, nodes that cached the local \ac{POI} data).
	
	We evaluate a setting with 3 querying nodes and 3 serving nodes (i.e., nodes that cached the regional \ac{POI} data), with the same settings for \ac{CAM} rates for the six evaluated nodes and the two \ac{CAM} generators. Essentially, we emulate a network of 86 nodes, among which three nodes cached the regional \ac{POI} data and three querying nodes are interested in the \ac{POI} data. We downloaded the OpenStreetMap \ac{POI} database for Luxembourg~\footnote{https://osm.kewl.lu/luxembourg.osm/}, and found that the highest number of \acp{POI} in a region with $L=2$ $km$ is around 5000. Consider a size of $500$ $bytes$ for each \ac{POI}, we assume the regional \ac{POI} data in our experiment is $2.5$ $Mbytes$.

\begin{table}[htp!]
	\newcommand{\tabincell}[2]{\begin{tabular}{@{}#1@{}}#2\end{tabular}}
	\caption{Computation operations for each message type}
	\centering
	\footnotesize
	\begin{tabular}{ | l | c | c | c | c |}
		\hline
		& \multicolumn{2}{c|}{\textbf{Our scheme}} & \multicolumn{2}{c|}{\textbf{MobiCrowd}} \\\hhline{-----}
		& \emph{Sender} & \emph{Receiver} & \emph{Sender} & \emph{Receiver} \\\hline
		\emph{Beacon} & ECDSA signing & ECDSA verification & - & - \\\hline
		\emph{Peer query} &  \tabincell{c}{ECIES encryption,\\ AES encryption,\\ ECDSA signing} & \tabincell{c}{ECIES decryption,\\ AES decryption,\\ ECDSA verification} & \tabincell{c}{ECDSA signing} & \tabincell{c}{ECDSA verification} \\\hline
		\emph{Peer response} & \tabincell{c}{AES encryption,\\ ECDSA signing} & \tabincell{c}{AES decryption,\\ ECDSA verification} & \tabincell{c}{None} & \tabincell{c}{ECDSA verification} \\\hline
	\end{tabular}
	\renewcommand{\arraystretch}{1}
	\label{table:operation}
\end{table}

\begin{table}[h]
	\newcommand{\tabincell}[2]{\begin{tabular}{@{}#1@{}}#2\end{tabular}}
	\centering
	\footnotesize
	\captionof{table}{EC cryptographic benchmarks on \ac{OBU} and \ac{P2P} query/response sizes}
	\begin{tabular}{ | c | c | c | c | c | c | c | }
		\hline
		\textbf{\emph{Security}} & \tabincell{c}{\textbf{\emph{Sign}}\\($ms$)} & \tabincell{c}{\textbf{\emph{Verify}}\\($ms$)} & \tabincell{c}{\textbf{\emph{Encryption}}\\($ms$)} & \tabincell{c}{\textbf{\emph{Decryption}}\\($ms$)} & \tabincell{c}{\textbf{\emph{Query}}\\($byte$)} & \tabincell{c}{\textbf{\emph{Response}}\\($byte$)} \\\hline
		Plaintext & - & - & - & - & 500 & 5000 \\\hline
		EC-192 & 1.75 & 1.95 & 6.52 & 4.73 & 1263 & 6877 \\\hline
		EC-224 & 2.65 & 2.94 & 8.13 & 5.54 & 1303 & 6897 \\\hline
		EC-256 & 3.03 & 3.42 & 9.43 & 6.22 & 1343 & 6921 \\\hline
	\end{tabular}
	\label{table:communication}
\end{table}

Table~\ref{table:operation} shows computational overhead for each message type generation and validation. Table~\ref{table:communication} shows the benchmark for \ac{EC} cryptographic operations on an \ac{OBU}, and query/response sizes in their plaintext, as well as signed and encrypted forms. We use UDP broadcasting for beacons from serving nodes and TCP for \ac{P2P} query-response process. In the experiment, the beacon size is 340 bytes (with a signature and a \ac{PC} attached) and we consider pure query and response sizes of 500 bytes and 5000 bytes (in accordance with the communication overhead in Sec.~\ref{sec:evaluation}) respectively. The extra communication overhead for secured formats is due to the attached (encrypted) symmetric key, the digital signature and the signer's \ac{PC}. We encode all binary data (including keys, signatures and cipher texts) into Base64 format, which further increases the size by $4/3$.

\begin{figure*}[t]
	\centering
	\begin{subfigure}[b]{0.24\columnwidth}
		\includegraphics[width=\columnwidth]{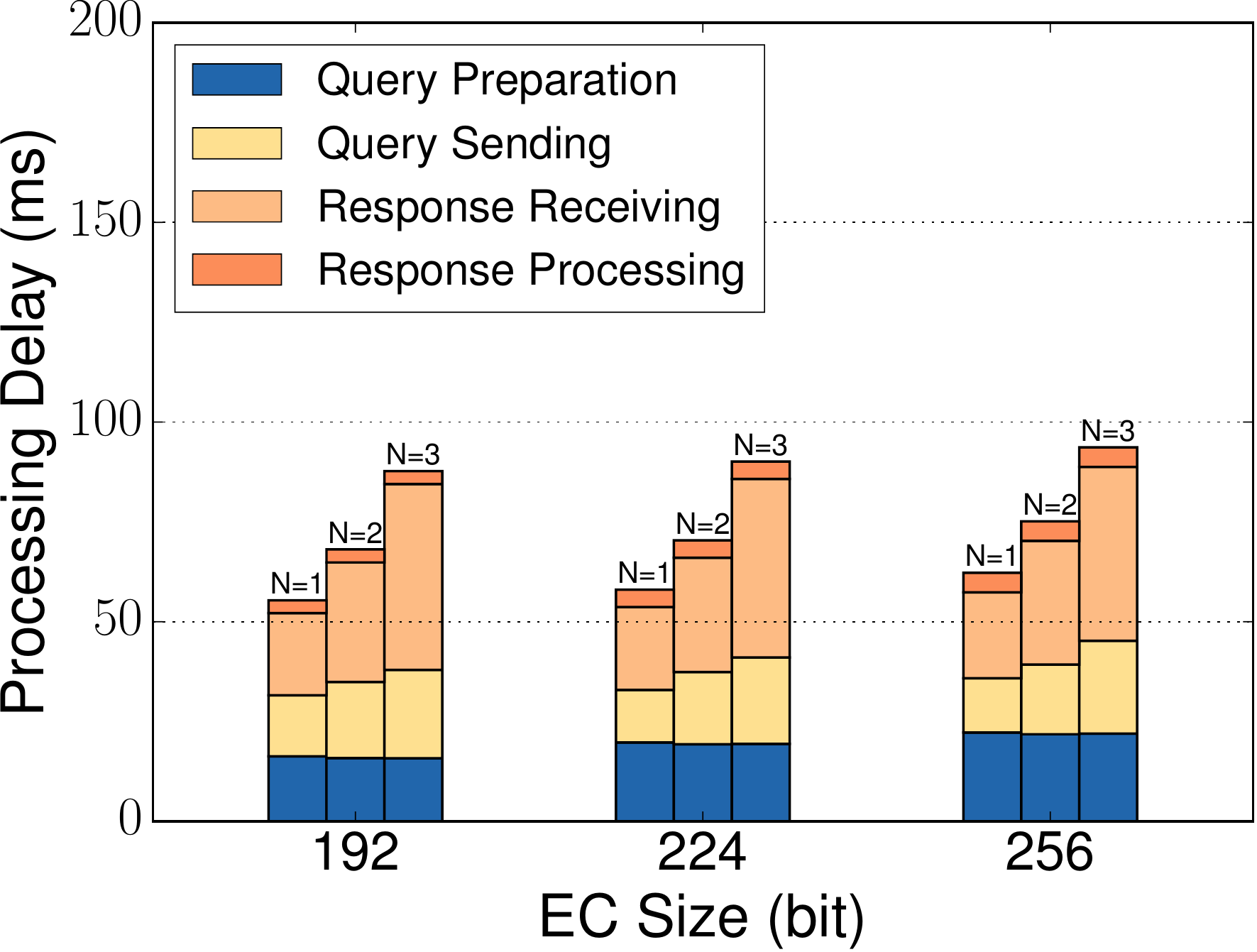}%
		\caption{}%
		\label{subfig_size_client}
	\end{subfigure}
	\begin{subfigure}[b]{0.24\columnwidth}
		\includegraphics[width=\columnwidth]{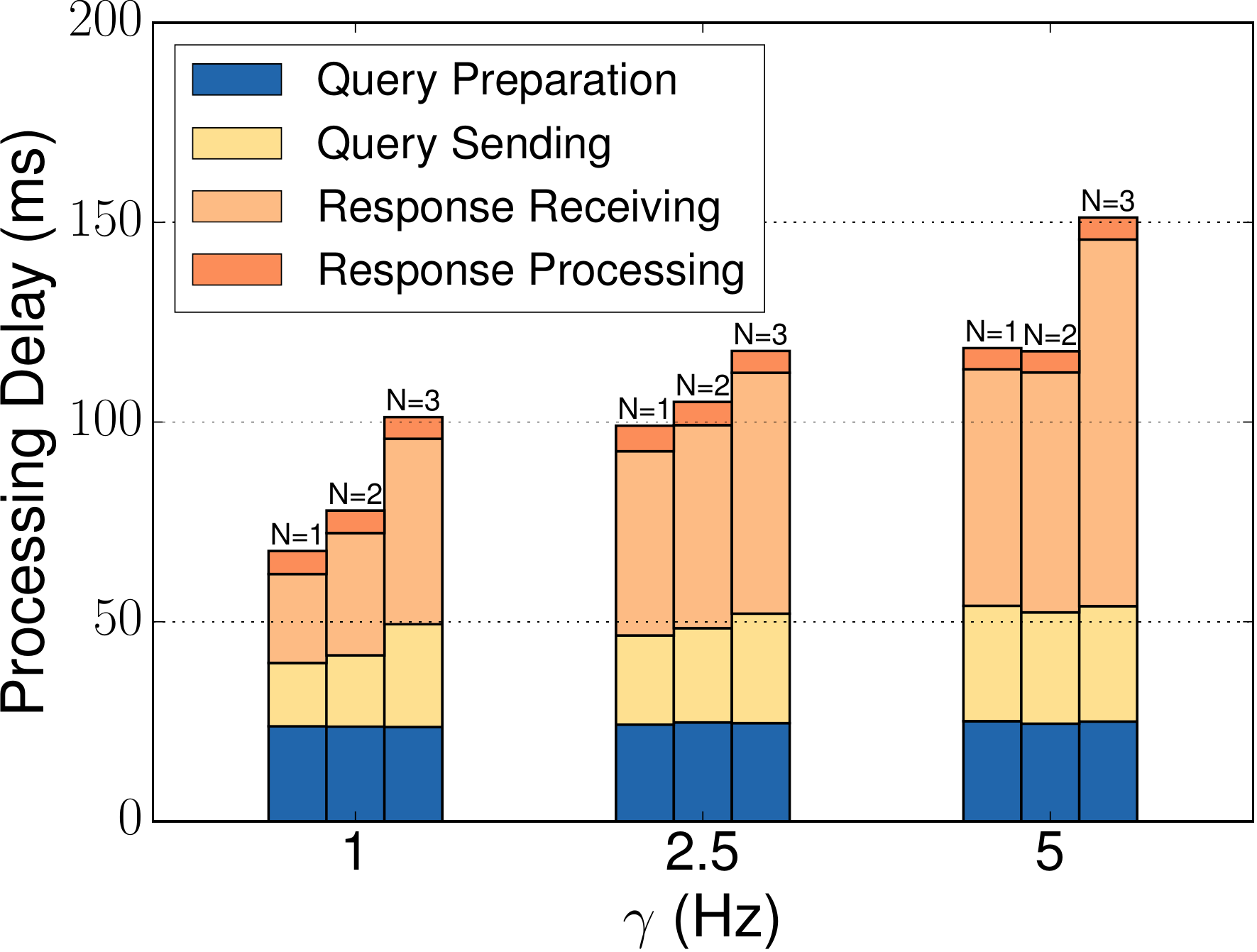}%
		\caption{}%
		\label{subfig_rate_client}
	\end{subfigure}
	\begin{subfigure}[b]{0.24\columnwidth}
		\includegraphics[width=\columnwidth]{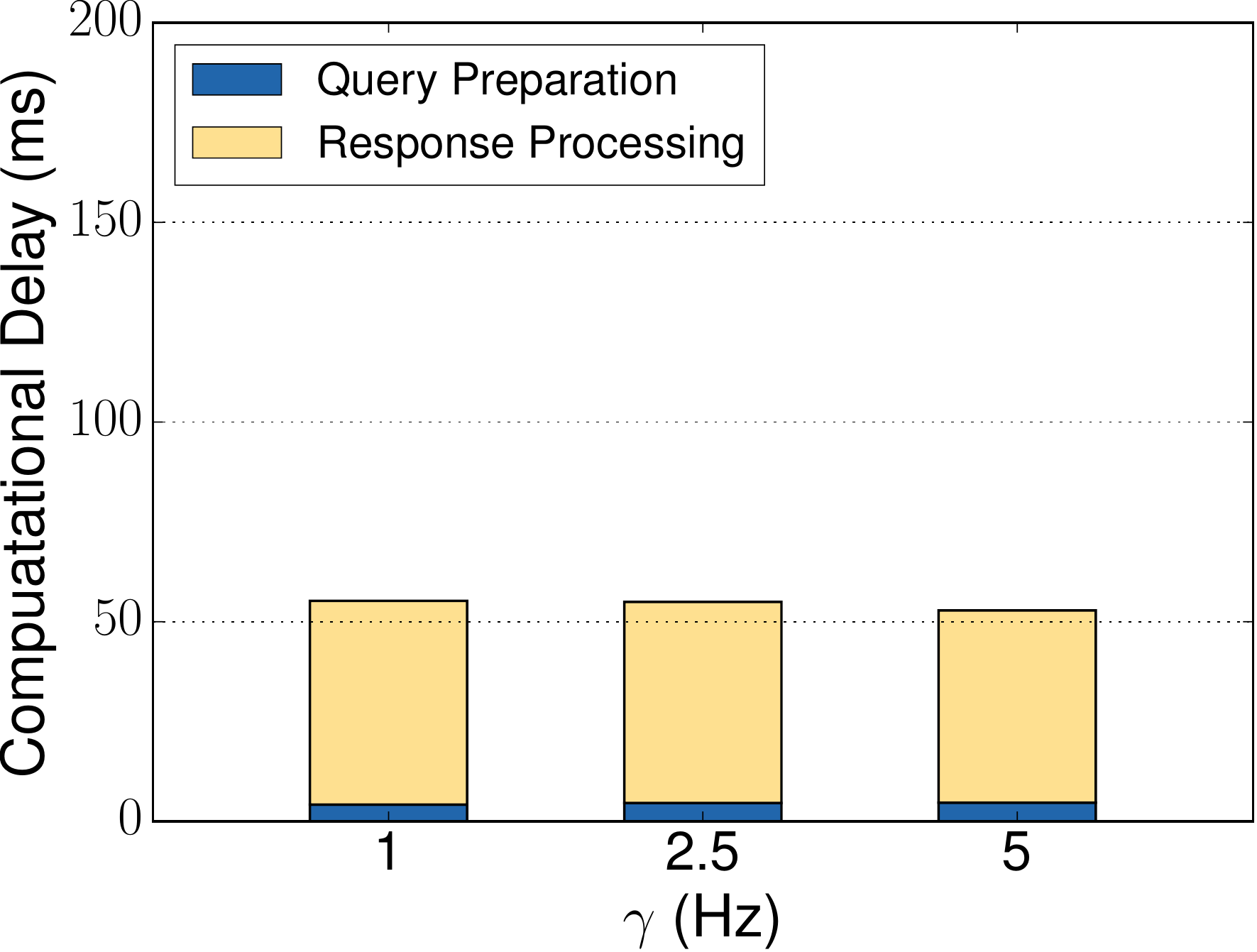}
		\caption{}%
		\label{subfig_mc_client}
	\end{subfigure}
	\begin{subfigure}[b]{0.24\columnwidth}
		\includegraphics[width=\columnwidth]{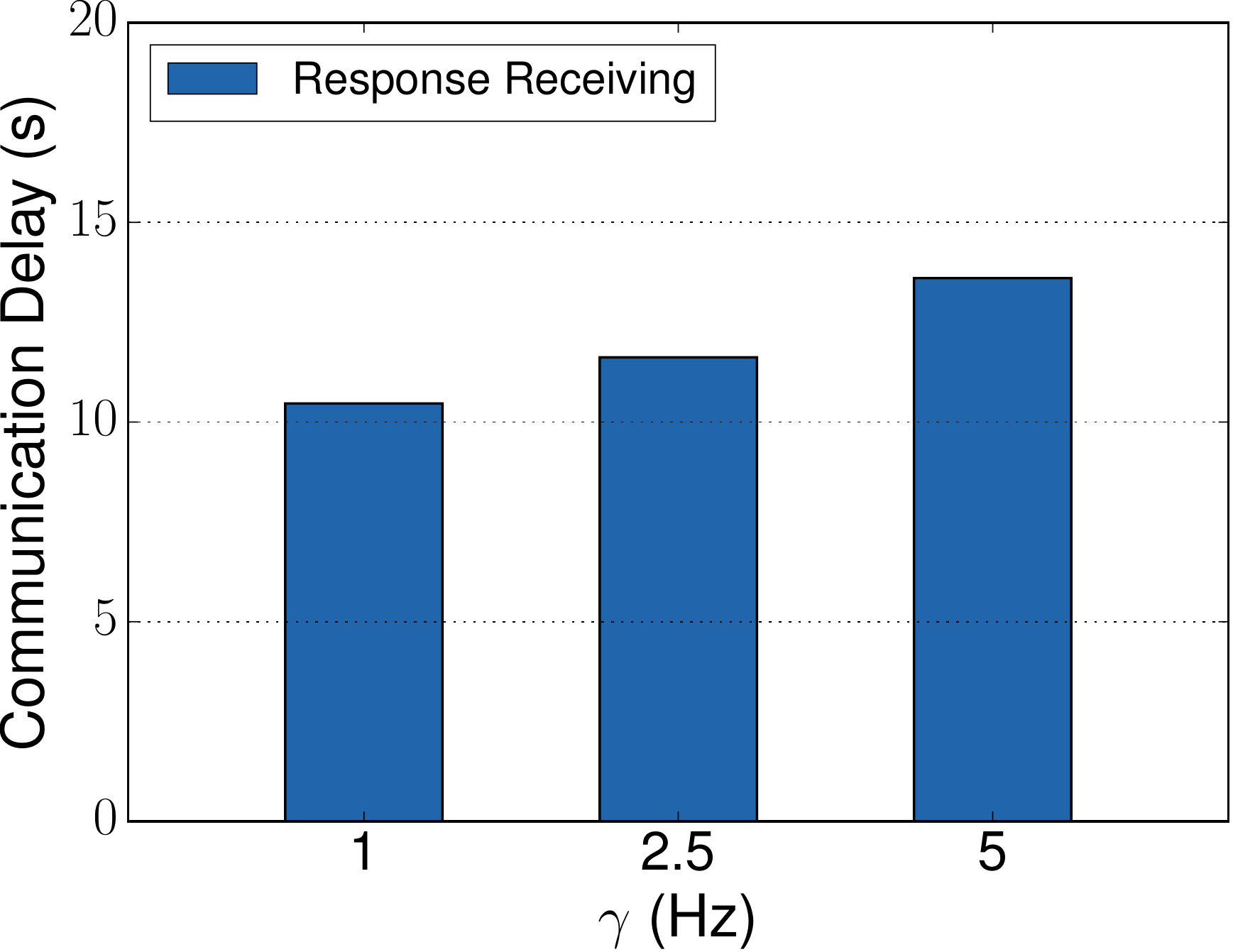}
		\caption{}%
		\label{subfig_mc_comm_client}
	\end{subfigure}
	
	\begin{subfigure}[b]{0.24\columnwidth}
		\includegraphics[width=\columnwidth]{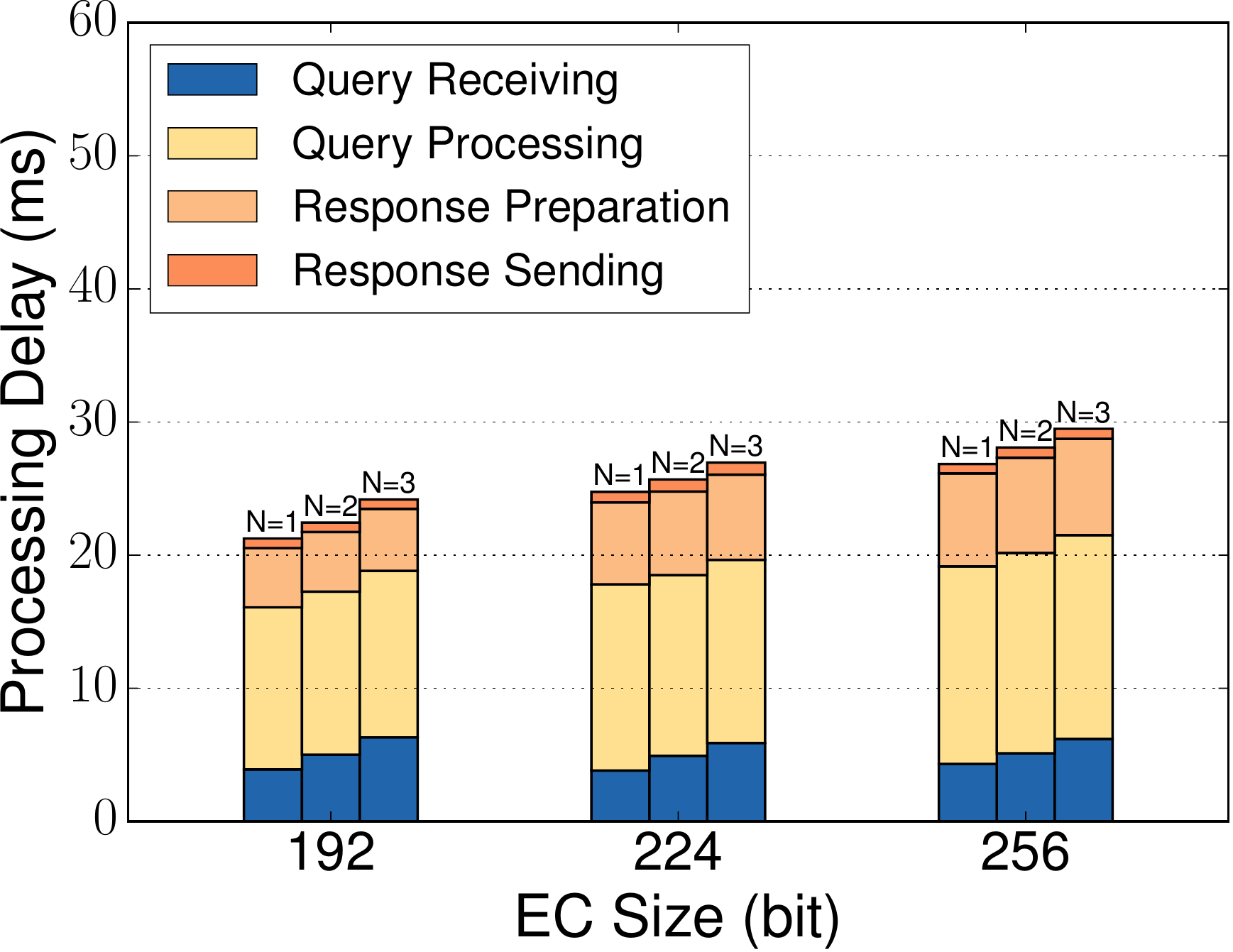}
		\caption{}%
		\label{subfig_size_serving}
	\end{subfigure}
	\begin{subfigure}[b]{0.24\columnwidth}
		\includegraphics[width=\columnwidth]{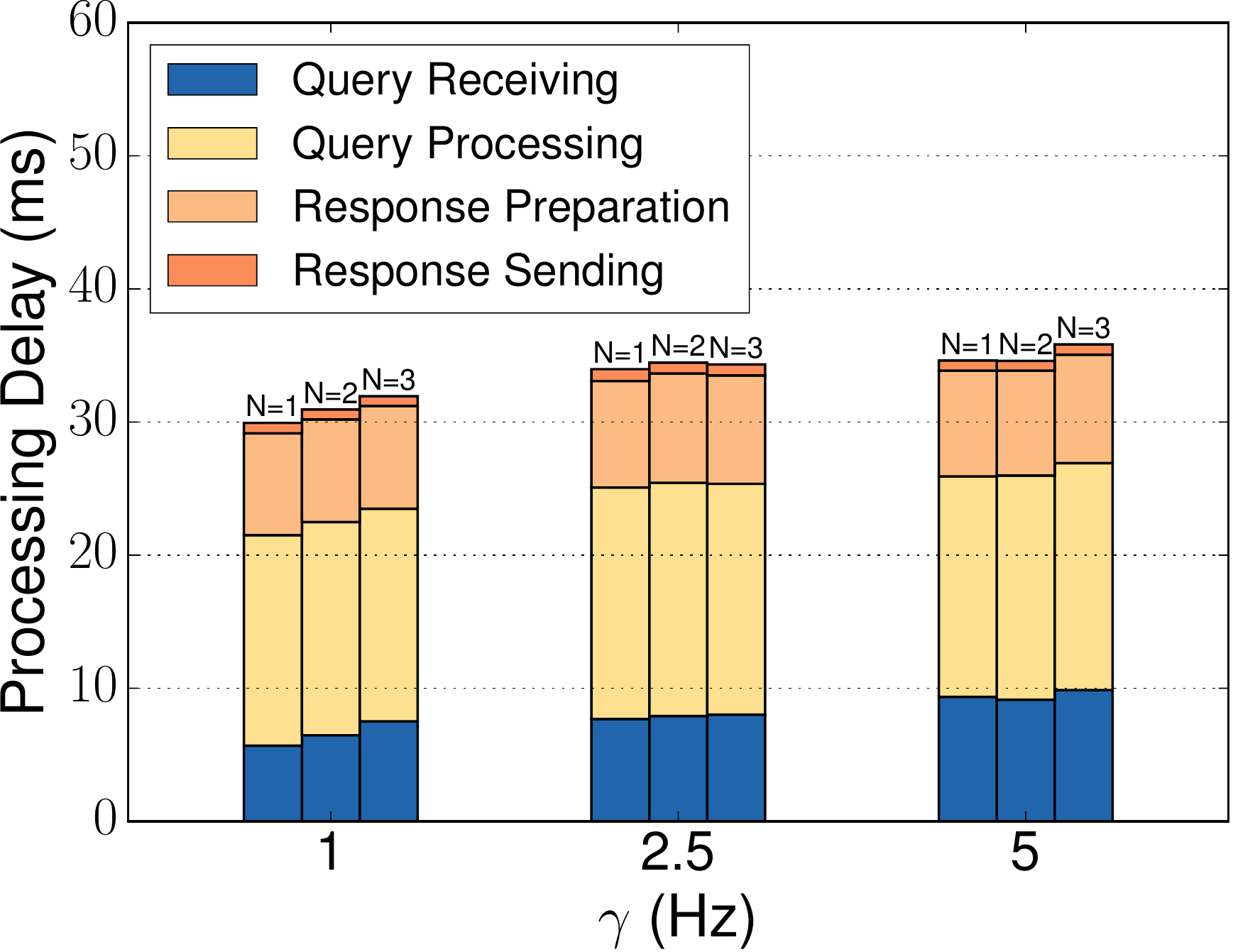}
		\caption{}%
		\label{subfig_rate_serving}
	\end{subfigure}
	\begin{subfigure}[b]{0.24\columnwidth}
		\includegraphics[width=\columnwidth]{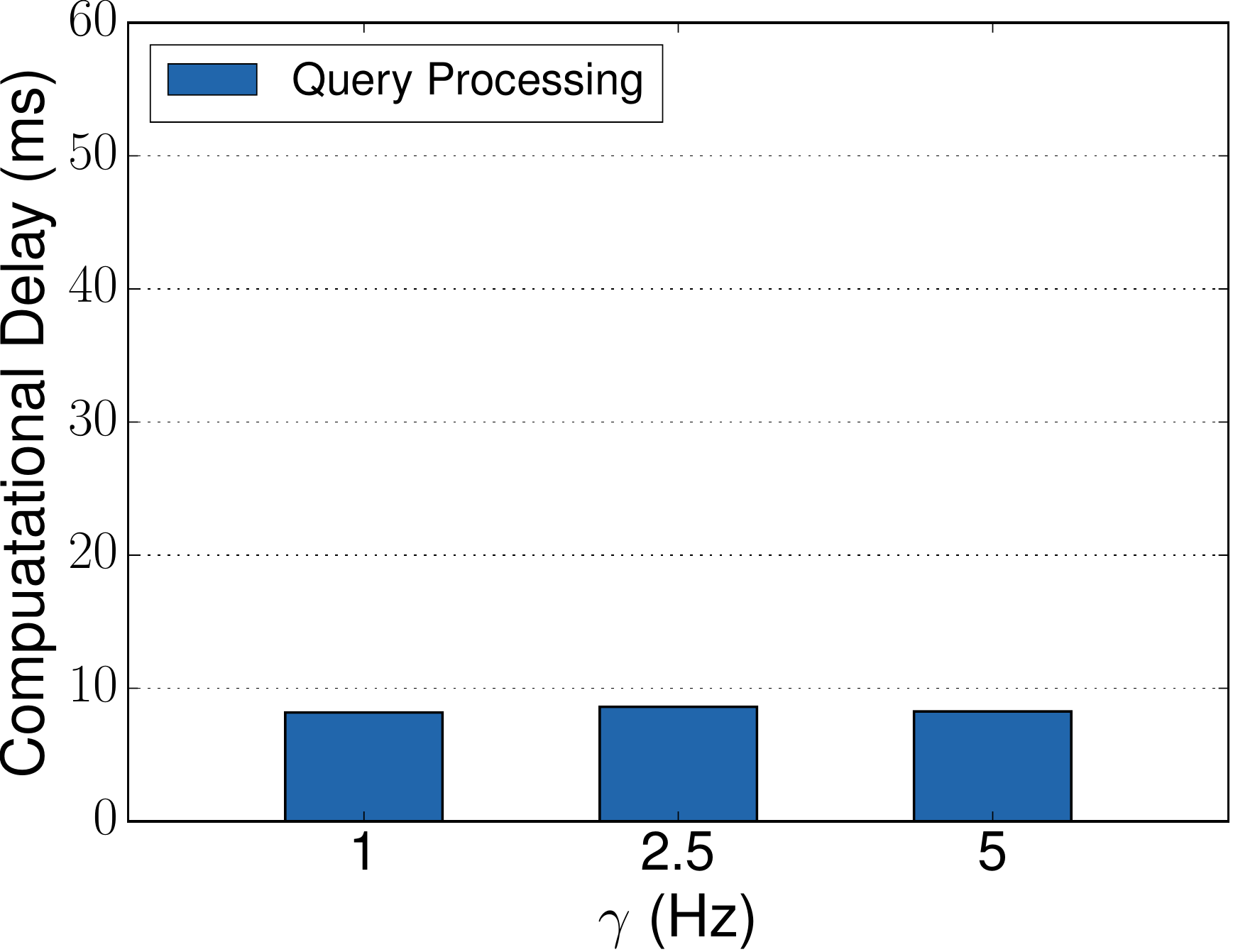}
		\caption{}%
		\label{subfig_mc_serving}
	\end{subfigure}
	\begin{subfigure}[b]{0.24\columnwidth}
		\includegraphics[width=\columnwidth]{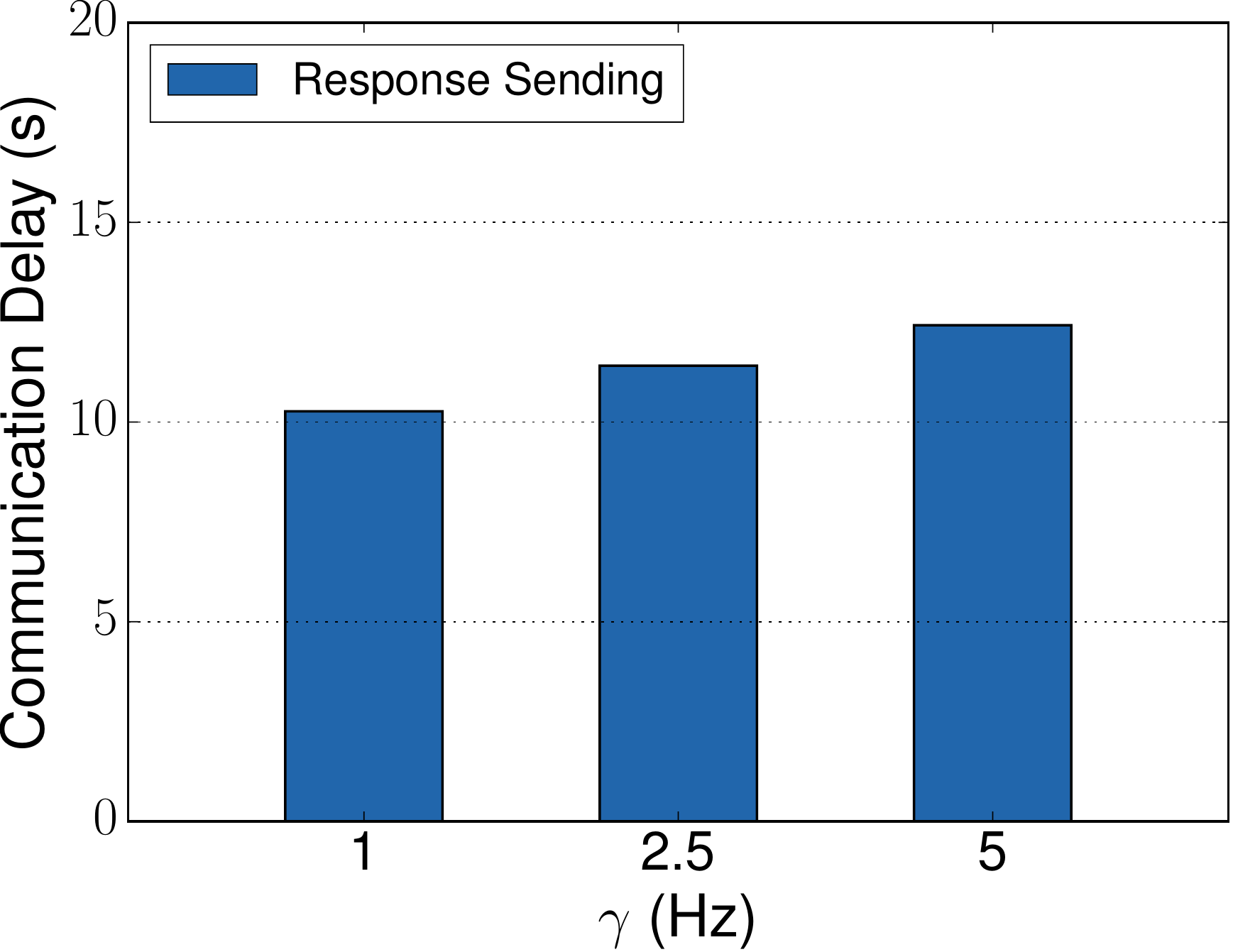}
		\caption{}%
		\label{subfig_mc_comm_serving}
	\end{subfigure}
	\caption{Average processing delay on (first row) querying node  and (second row) serving node. (First column) Without \ac{CAM} application. (Second column) EC-256 keys with \ac{CAM} application. (Third column) Computational and (fourth column) communication delays for MobiCrowd with EC-256 keys and \ac{CAM} application.}
	\label{fig_beacon_delay}
\end{figure*}

\cref{subfig_size_client,subfig_size_serving} show the average processing delay for the querying node and the serving node, for peer query and response with different combinations of $N$ and EC key size values. $T_{query} = Exp(\frac{1}{0.5s})$ is used for this experiment. The query sending delay includes the TCP handshake delay; the query reception delay on serving nodes does not include the TCP handshake delay, because it is not possible to record the starting time point of the TCP handshake (i.e., the starting time point of accepting a connection request by the \emph{listening socket} in C++) on the serving nodes.

For the querying node, we see the communication delays with different key sizes, given the same $N$, are roughly the same, because the difference in communication overhead (with different key sizes) is negligible compared to the total secured query/response size (see Table~\ref{table:communication}). For example, when $N=3$, the average communication delay (i.e., the sum of query sending and response receiving delays) of the querying node with EC-192, EC-224 and EC-256 keys are around 68 $ms$, 66 $ms$ and 67 $ms$ respectively. At the same time, the average total processing delay slightly increases with increased key size, due to higher cryptographic processing latencies.

However, given a key size, the average communication delay experienced by the querying node significantly increases as $N$ increases, because more peer responses are required for the same needed node queries. For example, with EC-192, the average processing delay at the querying node, with $N=1$, $N=2$, and $N=3$, is around 55 $ms$, 68 $ms$ and 88 $ms$ respectively.

For the serving nodes, the average processing delay does not change significantly for different $N$, because a serving node needs to handle high rates of incoming query even when $N=1$, thus may be saturated by high communication overload. From the experiments, we find that the TCP handshake delay is much higher than the networking delay for sending query/response messages. In our evaluation, the query delay includes the TCP handshake delay; while a response delay does not include a TCP handshake, because a TCP connection was already established upon receipt of the query. \cref{subfig_size_client,subfig_size_serving} show that the former is significantly higher than the latter one for any combination of $N$ and EC key size values.

The communication delays in \cref{subfig_rate_client,subfig_rate_serving} are higher than those in \cref{subfig_size_client,subfig_size_serving} even with lower query rate values for the query generators, because the network is loaded with \acp{CAM}. Moreover, the processing delay slightly increases because a safety beacon application is running in parallel with the \ac{LBS} application: \ac{CAM} generation and reception need frequent signature generations and verifications. Overall, we see only a modest increase in networking and processing delays even in such demanding network setting (note: we rely on software-only cryptographic operations, refraining from using a hardware accelerator that could validate several hundreds of signatures per second).

\cref{subfig_mc_client,subfig_mc_comm_client,subfig_mc_serving,subfig_mc_comm_serving} show processing delays for MobiCrowd. From \cref{subfig_mc_client,subfig_mc_serving}, we see the computational delays for MobiCrowd are lower than those in our scheme, because only signature verifications are required, without any encryption. However, we found the transmission of $2.5$ $Mbytes$ regional \ac{POI} data over a TCP connection requires more than $10$ $s$. In a realistic network with highly mobile nodes, transmission delays could be possibly higher, and essentially not workable with the current \acp{OBU} and protocol stack. Although such transmissions can be potentially done through a different type of connection, or over a \ac{RSU} as the proxy, our scheme or any scheme that requires exchange of much smaller data sizes is definitely advantageous, especially in a network with highly mobile nodes.

\textbf{Discussion:} Our scheme leverages \ac{P2P} interactions, requiring node-to-node communication within the network. In this experiment, our scheme is evaluated with vehicular \acp{OBU}, each equipped with an IEEE 802.11p interface, as user devices. IEEE 802.11p is designed mainly for safety applications in \ac{VC} domain. Apart from 802.11p, many network technologies support \ac{P2P} (or \ac{D2D}) communication. WiFi-Direct~\cite{camps2013device} and Bluetooth~\cite{gomez2012overview} can be used for \ac{P2P} communication, supported by most off-the-shelf smartphones nowadays. However, they do not support fully decentralized ad-hoc networking, i.e., a master node is needed and other nodes can join the network by connecting to the master. Therefore, they only support communication with a star topology, which is not fully decentralized. Moreover, the number of slaves a master can maintain is limited. LTE-Direct~\cite{lin2014overview} and upcoming 5G networks~\cite{tehrani2014device} support \ac{D2D} communication with the help of network infrastructure, while research on fully ad-hoc (direct) \ac{D2D} communication has been ongoing~\cite{lin2014overview}. While 802.11p is designed for \ac{VC}, \ac{D2D} communication in cellular networks could be supported by smartphones or any 4G/5G enabled device, which could promote the adoption of \ac{P2P}-based technologies, as is the proposed scheme here. An alternative approach, before the full deployment and support for direct \ac{D2D} communication, is using infrastructure mode WiFi networking, readily supported by off-the-shelf smartphones nowadays. Smartphone users can connect to an access point, and broadcast/multicast messages (e.g., beacons) and carry out \ac{P2P} interactions (e.g., query and response) via the access point. This requires users to connect to the same access point. In pedestrian settings, e.g., in a campus, this could be an option, but this is less so in a vehicular setting, considering the limited coverage of an access point and the mobility of vehicles: frequent switching from one access point to another would be necessary and users could be connected to different access points even if they are close to each other.

\section{Conclusion}
\label{sec:conclusion}

In this paper, we proposed a secure decentralized privacy protection scheme for \ac{LBS}. Our approach extends the recent \ac{P2P} \ac{LBS} privacy protection approach, addressing a number of practical open issues. More important, it ensures resilience to malicious nodes, and low exposure to honest-but-curious nodes and \ac{LBS} servers even if they collude with an honest-but-curious identity management facility. With a simulation-based quantitative analysis, we showed that the exposure to honest-but-curious nodes is low even if a high ratio (e.g., $20\ \%$) of nodes are honest-but-curious and collude; while the same ratio of malicious nodes could only affect less than $1.5\ \%$ of the peer-responded node queries. We showed the deployability of our scheme with an implementation on an automotive testbed and discussed the alternative approaches for deployment.

\begin{acks}
	This work has been supported by the Swedish Foundation for Strategic Research (SSF) SURPRISE project and the KAW Academy Fellow Trustworthy IoT project.
\end{acks}

%
\bibliographystyle{ACM-Reference-Format}
\bibliography{references.bib}

%

\end{document}